\newcommand*{\ATLASLATEXPATH}{latex/}
\documentclass[cernpreprint,UKenglish,texlive=2016,txfonts]{\ATLASLATEXPATH atlasdoc}

\usepackage{\ATLASLATEXPATH atlaspackage}
\usepackage{\ATLASLATEXPATH atlasbiblatex}

\usepackage{\ATLASLATEXPATH atlascontribute}

\usepackage{\ATLASLATEXPATH atlasphysics}

\graphicspath{{logos/}{figures/}{figaux/}}

\renewcommand{\sc}{\normalfont\scshape}
\renewcommand{\tt}{\normalfont\ttfamily}

\newcommand{\intlumi}{20.2}

\newcommand{\sxvt}{$\sqrt{s}=8$\,\TeV}
\newcommand{\sxwt}{$\sqrt{s}=7$\,\TeV}
\newcommand{\sxyt}{$\sqrt{s}=13$\,\TeV}

\newcommand{\xtt}{\mbox{$\sigma_{\ttbar}$}}
\newcommand{\mtop}{\mbox{$m_t$}}
\newcommand{\mtpole}{\mbox{$m_t^{\mathrm{pole}}$}}

\newcommand{\hdamp}{\mbox{$h_{\mathrm{damp}}$}}

\newcommand{\ptl}{\mbox{$p_{\mathrm T}^{\ell}$}}

\newcommand{\etal}{\mbox{$|\eta^{\ell}|$}}

\newcommand{\ptll}{\mbox{$p_{\mathrm T}^{e\mu}$}}
\newcommand{\mll}{\mbox{$m^{e\mu}$}}
\newcommand{\rapll}{\mbox{$|y^{e\mu}|$}}
\newcommand{\dphill}{\mbox{$\Delta\phi^{e\mu}$}}
\newcommand{\ptsum}{\mbox{$p_{\mathrm T}^{e}+p_{\mathrm T}^{\mu}$}}
\newcommand{\esum}{\mbox{$E^{e}+E^{\mu}$}}

\newcommand{\nxi}{\mbox{$N^i_1$}}
\newcommand{\nyi}{\mbox{$N^i_2$}}
\newcommand{\xtti}{\mbox{$\sigma^i_{\ttbar}$}}
\newcommand{\xttj}{\mbox{$\sigma^j_{\ttbar}$}}
\newcommand{\xntti}{\mbox{$\varsigma^i_{\ttbar}$}}
\newcommand{\xfid}{\mbox{$\sigma^{\ttbar}_{\mathrm{fid}}$}}
\newcommand{\gemi}{\mbox{$G^i_{e\mu}$}}
\newcommand{\epsbi}{\mbox{$\epsilon^i_{b}$}}
\newcommand{\epsbbi}{\mbox{$\epsilon^i_{bb}$}}
\newcommand{\cbi}{\mbox{$C^i_b$}}
\newcommand{\nibi}{\mbox{$N_1^{i,\mathrm{bkg}}$}}
\newcommand{\niibi}{\mbox{$N_2^{i,\mathrm{bkg}}$}}
\newcommand{\nidu}[2]{\mbox{$N_#1^{i,\mathrm{#2}}$}}
\newcommand{\fntaui}{\mbox{$f^i_{\bar{\tau}}$}}
\newcommand{\nijfake}{\nidu{j}{mis-id}}
\newcommand{\nijdss}{\nidu{j}{data,SS}}
\newcommand{\nijphss}{\nidu{j}{prompt,SS}}
\newcommand{\nijfakeos}{\nidu{j}{mis-id,OS}}
\newcommand{\nijfakess}{\nidu{j}{mis-id,SS}}
\newcommand{\rij}{\mbox{$R^i_j$}}

\newcommand{\xniexp}{\mbox{$\varsigma_i^{\mathrm{exp}}$}}
\newcommand{\xnith}{\mbox{$\varsigma_i^{\mathrm{th}}$}}
\newcommand{\xnjexp}{\mbox{$\varsigma_j^{\mathrm{exp}}$}}
\newcommand{\xnjth}{\mbox{$\varsigma_j^{\mathrm{th}}$}}
\newcommand{\vbexp}{\mbox{${\mathbf b}_{\mathrm{exp}}$}}
\newcommand{\vbth}{\mbox{${\mathbf b}_{\mathrm{th}}$}}
\newcommand{\vp}{\mbox{${\mathbf p}$}}
\newcommand{\gamexpij}{\mbox{$\gamma_{ij}^{\mathrm{exp}}$}}
\newcommand{\gamthik}{\mbox{$\gamma_{ik}^{\mathrm{th}}$}}
\newcommand{\betaexpj}{\mbox{$b_{j,{\mathrm{exp}}}$}}
\newcommand{\betathk}{\mbox{$b_{k,{\mathrm{th}}}$}}

\newcommand{\mmt}[1]{\mbox{$\mu^{(#1)}$}}
\newcommand{\mtheta}[1]{\mbox{$\Theta^{(#1)}$}}

\newcommand{\mtfoval}{173.2}
\newcommand{\mtfostat}{0.9}
\newcommand{\mtfoexp}{0.8}
\newcommand{\mtfoth}{1.2}
\newcommand{\mtfotot}{1.6}

\newcommand{\splitfigure}[4]{
\parbox{85mm}{
\includegraphics[width=78mm]{#1}
\vspace{-7mm}

\center{(#3)}
}
\parbox{85mm}{
\includegraphics[width=78mm]{#2}
\vspace{-7mm}

\center{(#4)}
}
\\
}

\AtlasTitle{Measurement of lepton differential distributions and the
top quark mass in \ttbar\
production in $pp$ collisions at $\sqrt{s}=8$\,\TeV\ with the ATLAS detector}

\AtlasRefCode{TOPQ-2015-02}
\AtlasNote{TOPQ-2015-02}
\AtlasDate{\today}
\AtlasJournal{Eur.\ Phys.\ J.\ C}
\AtlasJournalRef{Eur.\ Phys.\ J.\ C 77 (2017) 804}
\AtlasDOI{10.1140/epjc/s10052-017-5349-9}
\PreprintIdNumber{CERN-EP-2017-200}

\AtlasAbstract{%
This paper presents single lepton and dilepton kinematic distributions 
measured in dileptonic \ttbar\ events produced in \intlumi\,\ifb\ of
\sxvt\ $pp$ collisions
recorded by the ATLAS experiment at the LHC. Both absolute and normalised
differential cross-sections are measured, using events with an opposite-charge
$e\mu$ pair and one or two $b$-tagged jets. The cross-sections are 
measured in a fiducial region corresponding to the detector acceptance
for leptons, and are compared to the predictions from a variety of
Monte Carlo event generators, as well as fixed-order QCD calculations, exploring
the sensitivity of the cross-sections 
to the gluon parton distribution function. Some of the 
distributions are also sensitive to the top quark pole mass; a combined fit of
NLO fixed-order predictions to all the measured distributions 
yields a top quark mass value  of 
$\mtpole=\mtfoval\pm\mtfostat\pm\mtfoexp\pm\mtfoth$\,\GeV, where the three
uncertainties arise from data statistics, experimental systematics, and 
theoretical sources.
}

\hypersetup{pdftitle={ATLAS document},pdfauthor={The ATLAS Collaboration}}

\begin{document}

\maketitle



\section{Introduction}\label{s:intro}

The top quark is the heaviest known fundamental particle, with a mass (\mtop)
that is much larger than any of the other quarks, and close to the scale of
electroweak symmetry breaking. The study of its production and decay 
properties in proton--proton ($pp$) collisions forms an important part of the 
ATLAS physics program at the CERN Large Hadron Collider (LHC). Due to its large
mass and production cross-section, top quark production is also a significant
background to many searches for physics beyond the Standard Model, making
precise predictions of absolute rates and differential distributions for 
top quark production a vital tool in fully exploiting the discovery potential
of the LHC.

At the LHC, top quarks are primarily produced as quark-antiquark pairs
(\ttbar). The inclusive \ttbar\ production cross-section \xtt\ has been 
calculated at full next-to-next-to-leading-order (NNLO) accuracy in the strong
coupling constant \alphas, including the resummation of 
next-to-next-to-leading logarithmic (NNLL) soft gluon terms
\cite{topxtheo1,topxtheo2,topxtheo3,topxtheo4,topxtheo5}.
The resulting prediction at a centre-of-mass energy 
\sxvt\ is $\xtt=252.9\pm 11.7^{+6,4}_{-8.6}$\,pb for a top quark mass of 172.5\,\GeV,
calculated using the {\tt top++ 2.0} program \cite{toppp}.
The first uncertainty is due to parton distribution function (PDF)
and \alphas\ uncertainties,
calculated using the PDF4LHC prescription \cite{pdf4lhc} with the
MSTW2008 68\,\% \cite{mstwnnlo1,mstwnnlo2}, CT10 NNLO \cite{cttenpdf,cttennnlo}
and NNPDF~2.3 5f FFN \cite{nnpdfffn} PDF sets,
and the second to quantum chromodynamics (QCD) scale variations. 
This prediction, which has a relative precision of 5.5\,\%, 
agrees with measurements from ATLAS and CMS at \sxvt\ 
\cite{TOPQ-2013-04,ttxadd,CMS-TOP-13-004} which have reached a precision
of 3--4\,\%. Measurements in LHC $pp$ collisions at \sxwt\ 
\cite{TOPQ-2013-04,CMS-TOP-13-004} and more recently at 
\sxyt\ \cite{cms13dil,TOPQ-2015-09} 
are also in good agreement with the corresponding  NNLO\,+\,NNLL predictions.

Going beyond the inclusive production cross-section, measurements of 
\ttbar\ production as a function of the top quark and \ttbar\ system
kinematics properties allow the predictions of QCD calculations and Monte Carlo
event-generator programs to be probed in more detail. These comparisons are
typically more sensitive at the level of normalised differential cross-sections,
i.e. shape comparisons, where both experimental and theoretical uncertainties
are reduced. Measurements 
by ATLAS \cite{TOPQ-2012-08,TOPQ-2013-07,TOPQ-2015-06,TOPQ-2015-07} 
and CMS \cite{CMS-TOP-11-013,CMS-TOP-12-028,CMS-TOP-14-013upd}
have generally demonstrated good agreement with the 
predictions of leading-order (LO) multi-leg 
and next-to-leading-order (NLO) event
generators and calculations, though the top quark \pt\ spectrum is measured
to be softer than the predictions by both experiments; this distribution
appears to be sensitive to the additional corrections contributing at 
NNLO \cite{topptnnlo}. Measurements of jet activity in \ttbar\ events
\cite{TOPQ-2012-03,TOPQ-2015-04,CMS-TOP-12-018,CMS-TOP-12-041}
are also sensitive to gluon radiation and hence the \ttbar\ production
dynamics, without the need to fully reconstruct the kinematics of the \ttbar\
system.
However, all these measurements require  sophisticated unfolding procedures
to correct for the detector acceptance and resolution. This leads to
significant systematic uncertainties, especially due to modelling of the
showers and hadronisation of the quarks produced in the top quark decays, 
and the measurement of the resulting jets in the detector.

In the Standard Model (SM), the top quark decays almost exclusively
to a $W$ boson and a $b$ quark, and the final state topologies in \ttbar\
production are governed by the decay modes of the $W$ bosons. The channel
where one $W$ boson decays to an electron ($W\rightarrow e\nu$) and the
other to a muon ($W\rightarrow\mu\nu$), giving rise to the 
$e^+\mu^-\nu\bar{\nu}\bbbar$ final state\footnote{Charge-conjugate decay
modes are implied unless otherwise stated.}, is particularly clean and was
exploited to make the most precise ATLAS measurements of \xtt\
\cite{TOPQ-2013-04,TOPQ-2015-09}. The leptons carry information about the
underlying top quark kinematics, are free of the uncertainties related
to the hadronic part of the final state, and are precisely measured in the
detector. Measurements of the \ttbar\ differential cross-section as a function
of the lepton kinematics therefore have the potential to provide a complementary
view of \ttbar\ production and decay dynamics to that provided by 
the complete reconstruction of the \ttbar\ final state.

This paper reports such a measurement of the absolute and normalised 
differential cross-sections for $\ttbar\rightarrow e\mu\nu\bar{\nu}\bbbar$
produced in $pp$ collisions at \sxvt, as a function of the kinematics of the 
single leptons and of the dilepton system. Eight differential cross-section
distributions are measured: the transverse momentum \ptl\ and absolute
pseudorapidity \etal\ of
the single leptons (identical for electrons and muons), the \pt, 
invariant mass and absolute rapidity of the dilepton system 
(\ptll, \mll\ and \rapll),
the azimuthal angle in the transverse plane \dphill\ between the two leptons,
the scalar sum \ptsum\ of the \pt\ of the two leptons, and the sum \esum\
of the energies of the two leptons.\footnote{ATLAS uses a right-handed 
coordinate system with its origin at
the nominal interaction point in the centre of the detector, and the $z$ axis
along the beam line. Pseudorapidity is defined in terms of the polar angle
$\theta$ as $\eta=-\ln\tan{\theta/2}$, and transverse momentum and energy
are defined relative to the beamline as $\pt=p\sin\theta$ and
$\et=E\sin\theta$. The azimuthal angle around the beam line is denoted by 
$\phi$, and distances in $(\eta,\phi)$ space by 
$\Delta R=\sqrt{(\Delta\eta)^2+(\Delta\phi)^2}$.
The rapidity is defined as $y=\frac{1}{2}\ln\left(\frac{E+p_z}{E-p_z}\right)$,
where $p_z$ is the $z$-component of the momentum and  $E$ is the energy 
of the relevant object or system.}
The measurements are corrected to 
particle level and reported in a fiducial volume where both leptons
have $\pt>25$\,\GeV\ and $|\eta|<2.5$, avoiding extrapolations into regions of 
leptonic phase space which are not measured. The particle-level definition
includes the contribution of events where one or both $W$ bosons
decay to electrons or muons via leptonic decays of $\tau$-leptons 
($t\rightarrow W\rightarrow\tau\rightarrow e/\mu$), but an alternative set of
results is provided where the contributions of $\tau$-leptons are removed with 
a correction derived from simulation. The definition of the fiducial volume
does not make any requirement on the presence of jets from the 
hadronic decay products of the \ttbar\ system.
The measurements are made
using events with an opposite-charge $e\mu$ pair and one or two $b$-tagged 
jets, and extrapolated to the fiducial volume (without jet requirements),
using an extension of the double-tagging technique used in the inclusive
\ttbar\ cross-section measurement \cite{TOPQ-2013-04}. This approach 
minimises the systematic
uncertainties due to the use of jets and $b$-tagging in the experimental
event selection. Since the lepton kinematics are precisely measured in the ATLAS
detector, a simple bin-by-bin correction technique is adequate to correct
for efficiency and resolution effects, 
without the need for a full unfolding procedure.

The results are compared to the predictions of various NLO and LO multi-leg
\ttbar\ event generators, and to fixed-order perturbative QCD predictions from
the MCFM \cite{mcfm} program,
which is used to explore the sensitivity to PDFs and QCD scale uncertainties. 
These comparisons are complementary to previous ATLAS analyses exploring
how well \ttbar\ event generators can describe 
the jet activity \cite{TOPQ-2015-04} and production of extra heavy-flavour
jets \cite{TOPQ-2014-10} in the \sxvt\ \ttbar\ dilepton sample.

Some of the cross-section distributions 
are sensitive to the top quark mass, as suggested in Ref. \cite{topmassdiff},
and mass measurements are made by comparing the measured distributions to predictions from both NLO plus parton shower event generators and
fixed-order QCD calculations. The former are similar to traditional 
measurements where the top quark mass is reconstructed from its decay products
\cite{cdfmljets,d0mljets,TOPQ-2016-03,CMS-TOP-14-022},
but rely only on the leptonic decay products of the \ttbar\ system and are
less sensitive to experimental uncertainties related to the hadronic 
part of the final state. The measurements based on fixed-order QCD predictions
in a well-defined renormalisation scheme
correspond more directly to a measurement of the top quark
pole mass \mtpole, the mass definition corresponding to that of a free
particle, which may differ from that measured in direct reconstruction
of the decay products by $O(1\GeV)$ \cite{buckleypole,mochpole,hoangpole}.
Previous determinations of \mtpole\ from inclusive and differential
\ttbar\ cross-section measurements are compatible with the top quark
mass measured from direct reconstruction, with uncertainties of 2--3\,GeV
\cite{d0mtpoleincl,TOPQ-2013-04,CMS-TOP-13-004,TOPQ-2014-06}. 

The data and Monte Carlo simulation samples used in this analysis
are described in Section~\ref{s:datasim}, followed by the
event reconstruction and selection in Section~\ref{s:objevt}, definition and 
determination of the fiducial differential cross-sections in 
Section~\ref{s:xsecdet}
and systematic uncertainties in Section~\ref{s:syst}. Results and comparisons
with predictions are given in Section~\ref{s:res}. The ability of the
data to constrain the gluon PDF is investigated in Section~\ref{s:pdf}
and the determination
of the top quark mass is discussed in Section~\ref{s:mtop}. Finally,
conclusions are given in Section~\ref{s:conc}.


\section{Data and simulated samples}\label{s:datasim}

The ATLAS detector \cite{PERF-2007-01} at the LHC covers nearly the entire
solid angle around the collision point, and consists of an inner tracking
detector surrounded by a thin superconducting solenoid magnet producing
a 2\,T axial magnetic field, electromagnetic and hadronic calorimeters,
and an external muon spectrometer incorporating three large toroidal magnet
assemblies. The analysis was performed on a sample of proton--proton 
collision data at \sxvt\ recorded by the ATLAS detector in 2012,
corresponding to an integrated luminosity of \intlumi\,\ifb.
Events were required to pass a single-electron or single-muon trigger, with
thresholds set to be fully efficient for leptons with $\pt>25$\,\GeV\ passing
offline selections. Each triggered event also includes the signals from
on average 20 additional inelastic $pp$ collisions in the same
bunch crossing, referred to as pileup.

Monte Carlo simulated event samples were used to develop the analysis 
procedures, to compare with data, and to evaluate signal efficiencies and 
background contributions. 
An overview of the samples used for signal and background
modelling is shown in Table~\ref{t:simtune}, and further details are given
below. Samples were processed using either the full ATLAS 
detector simulation \cite{SOFT-2010-01} based on GEANT4 \cite{geant4}, or
a faster simulation making use of parameterised showers in
the calorimeters \cite{ATL-PHYS-PUB-2010-013}.
The effects of pileup were simulated by generating additional inelastic
$pp$ collisions with {\sc Pythia8} \cite{pythia8} using the A2 parameter
set (tune)
\cite{ATL-PHYS-PUB-2012-003} and overlaying them on the primary
simulated events. These combined events were then processed using the 
same reconstruction and analysis chain as the data. Small corrections were
applied to the lepton trigger and selection efficiencies better to model the 
performance measured in data.

\begin{table}[tp]
\centering

\begin{tabular}{l|ll|ll|l}\hline
Process & Matrix-element & PDF & Parton shower & Tune & Comments \\
\hline
\ttbar\ & {\sc Powheg} & CT10 & {\sc Pythia6} &  P2011C & $\hdamp=\mtop$ \\
 & {\sc Powheg} & CT10 & {\sc Herwig+Jimmy} & AUET2 & $\hdamp=\infty$ \\
 & {\sc MC@NLO} & CT10 & {\sc Herwig+Jimmy} & AUET2 & \\
 & {\sc Alpgen} & CTEQ6L1 & {\sc Herwig+Jimmy} & AUET2 & incl. \ttbar\bbbar, \ttbar\ccbar \\
& {\sc Powheg} & CT10 & {\sc Pythia6} &  P2012 radHi & $\hdamp=2\mtop$, $\frac{1}{2}\mu_{F,R}$ \\
& {\sc Powheg} & CT10 & {\sc Pythia6} &  P2012 radLo & $\hdamp=\mtop$, $2\mu_{F,R}$ \\
\hline
$Wt$ & {\sc Powheg} & CT10 & {\sc Pythia6} & P2011C & diagram removal \\
$Z,W$+jets & {\sc Alpgen} & CTEQ6L1 & {\sc Pythia6} & P2011C & incl. $Z\bbbar$ \\
$WW$, $WZ$, $ZZ$ & {\sc Alpgen} & CTEQ6L1 & {\sc Herwig} & AUET2  & \\
\ttbar+$W,Z$ & {\sc MadGraph} & CTEQ6L1 & {\sc Pythia6} & P2011C & \\
$W\gamma$+jets & {\sc Sherpa} & CT10 & {\sc Sherpa} & default & \\
$t$-channel top & {\sc AcerMC} & CTEQ6L1 & {\sc Pythia6} & AUET2B & \\
\hline
\end{tabular}
\caption{\label{t:simtune}Summary of simulated event samples used for 
\ttbar\ signal and background modelling, giving the matrix-element event 
generator,
PDF set, parton shower and associated tune parameter set. More details,
including generator version numbers and references, are given in the text.}
\end{table}

The baseline simulated \ttbar\ sample was produced using the NLO
matrix element event generator {\sc Powheg-Box} v1.0 (referred to hereafter as {\sc Powheg}) \cite{powheg,powheg2,powheg3,powheghvq}
using the CT10 PDFs \cite{cttenpdf},
interfaced to {\sc Pythia6} (version 6.426) \cite{pythia6} with the 
CTEQ6L1 PDF set \cite{ctsix} and the Perugia 2011C (P2011C) 
tune \cite{perugia} for parton shower, hadronisation and underlying
event modelling. This setup provides an NLO QCD prediction of the \ttbar\
production process, a leading-order prediction for the top quark decays, and
an approximate treatment of the 
spin correlations between the quark and antiquark.
The {\sc Powheg} parameter \hdamp, used in the
damping function that limits the resummation of higher-order effects 
incorporated into the Sudakov form factor, was set to \mtop. This value
was found to give a better modelling of the \ttbar\ system \pt\ at
\sxwt\ \cite{ATL-PHYS-PUB-2015-002} than the setting of $\hdamp=\infty$ 
 used for the baseline \ttbar\ sample in Ref. \cite{TOPQ-2013-04}, 
which corresponds to no damping.

Alternative \ttbar\ simulation samples used to evaluate systematic 
uncertainties were generated with {\sc Powheg} interfaced to 
{\sc Herwig} (version 6.520) \cite{herwig1,herwig2} with the ATLAS AUET2 tune
\cite{ATL-PHYS-PUB-2011-008} and {\sc Jimmy} (version 4.31) \cite{jimmy} 
for underlying event modelling, 
with {\sc MC@NLO} (version 4.01) \cite{mcatnlo1,mcatnlo2} interfaced to 
{\sc Herwig\,+\,Jimmy}, and with the leading-order `multi-leg' event generator
{\sc Alpgen} (version 2.13) \cite{alpgen}, also interfaced to 
{\sc Herwig}\,+ {\sc Jimmy}. The {\sc Alpgen} samples used leading-order matrix
elements for \ttbar\ production accompanied by up to three additional
light partons, and dedicated matrix elements for \ttbar\ plus \bbbar\ or
\ccbar\ production, together with the MLM parton-jet matching scheme 
\cite{mlmmatch}
to account for double-counting of configurations generated by both the
parton shower and matrix-element calculation. The effects of additional
radiation in \ttbar\ events were further studied using two additional
{\sc Powheg\,+\,Pythia6} samples, one using the Perugia 2012 radHi tune
\cite{perugia}, with \hdamp\ set to $2\mtop$ and
factorisation and renormalisation scales $\mu_F$ and $\mu_R$ reduced from their
event generator defaults by a factor of two, giving more parton shower 
radiation;
and one with the Perugia 2012 radLo tune \cite{perugia}, $\mu_F$ and $\mu_R$ 
increased by a factor of two and $\hdamp=\mtop$, giving less parton shower 
radiation. 
The parameters of these samples were chosen to span the uncertainties
in jet observables measured by ATLAS in \ttbar\ events at \sxwt\
\cite{ATL-PHYS-PUB-2015-002,TOPQ-2011-21,TOPQ-2012-03}.
The top quark mass was set to 172.5\,\GeV\ in all these samples, consistent
with recent measurements by ATLAS \cite{TOPQ-2016-03} and CMS
\cite{CMS-TOP-14-022}.
They were all normalised to the NNLO\,+\,NNLL cross-section prediction
discussed in Section~\ref{s:intro} when comparing simulation with data. Further
\ttbar\ simulation samples with different event generator setups were used for
comparisons with the measured differential cross-sections as discussed in
Section~\ref{ss:gencomp}, and in the extraction of the top quark mass
as discussed in Section~\ref{s:mtop}.

Backgrounds to the \ttbar\ event selection  are classified into two types:
those with two real prompt leptons from $W$ or $Z$ boson decays (including
those produced via leptonic $\tau$ decays), and those where one of the
reconstructed lepton candidates is misidentified, i.e. a non-prompt
lepton from the decay of a bottom or charm hadron, an electron from 
a photon conversion, hadronic jet activity misidentified as an electron,
or a muon produced from the decay in flight of a pion or kaon. The first
category is dominated by the associated production of a $W$ boson
and a single top quark, $Wt$, that is
simulated using {\sc Powheg\,+\,Pythia6} with the
CT10 PDFs and the P2011C tune. The `diagram removal' scheme 
was used to handle the interference between the \ttbar\ and $Wt$ final
states that occurs at NLO \cite{wtinter1,wtinter2}.
Smaller backgrounds result from $Z\rightarrow\tau\tau (\rightarrow e\mu)$+jets,
modelled using {\sc Alpgen\,+\,Pythia6} including leading-order matrix
elements for $Z\bbbar$ production, and diboson ($WW$, $WZ$ and $ZZ$) production
in association with jets, modelled with {\sc Alpgen\,+\,Herwig\,+\,Jimmy}.
The $Wt$ background was normalised to the approximate NNLO cross-section
of $22.4\pm 1.5$\,pb, determined as in Ref. \cite{wttheo}. The inclusive
$Z$ cross-section was set to the NNLO prediction from FEWZ \cite{fewz}, but 
the normalisation of the $Z\rightarrow\tau\tau$ background with $b$-tagged
jets was determined with the help of data control samples 
as discussed in Section~\ref{ss:backg}. The small
diboson background was normalised to the NLO QCD inclusive cross-section
predictions calculated with MCFM \cite{dibmcfm}, using the
{\sc Alpgen\,+\,Herwig} prediction for the fraction of diboson events
with extra jets. Production of \ttbar\ in 
association with a $W$ or $Z$ boson, which contributes to the control sample
with two same-charge leptons, was simulated with {\sc MadGraph} \cite{madgraph}
interfaced to {\sc Pythia6} with CTEQ6L1 PDFs, and normalised to NLO
cross-section predictions \cite{ttwznlo1,ttwznlo2}.

Backgrounds with one real and one misidentified lepton arise from \ttbar\ events
with one hadronically-decaying $W$; $W$+jets production, modelled as described
above for $Z$+jets; $W\gamma$+jets, modelled with {\sc Sherpa} 1.4.1 
\cite{sherpa}
with CT10 PDFs; and $t$-channel single top production, modelled with
{\sc AcerMC} \cite{acermc} with the AUET2B tune \cite{ATL-PHYS-PUB-2011-009} 
and CTEQ6L1 PDFs interfaced to {\sc Pythia6}.
The normalisations of these backgrounds in the opposite-charge $e\mu$ samples
were determined with the help of the corresponding same-charge $e\mu$ 
samples in data.
Other backgrounds, including processes with two misidentified leptons,
are negligible after the event selections used in this analysis.

\section{Event reconstruction and selection}\label{s:objevt}

The analysis makes use of reconstructed electrons, muons, and $b$-tagged jets,
selected exactly as described in Ref. \cite{TOPQ-2013-04}. In brief, electron
candidates \cite{PERF-2016-01} were required to satisfy $\et>25$\,\GeV\ and 
$|\eta|<2.47$, and to not lie within the transition region $1.37<|\eta|<1.52$ 
between the barrel and endcap electromagnetic calorimeters. 
Muon candidates \cite{PERF-2014-05} were required to satisfy $\pt>25$\,\GeV\ 
and $|\eta|<2.5$. In order to reduce background from non-prompt leptons,
electrons were required to be isolated
from nearby hadronic activity using both calorimeter and tracking information,
and muons were required to be isolated using tracking information alone.
Jets were reconstructed using the anti-$k_t$ algorithm \cite{antikt2,fastjet}
with radius parameter $R=0.4$ using calorimeter energy clusters calibrated
with the local cluster weighting method \cite{PERF-2011-03}. 
Jets were further calibrated using information from both simulation and data 
\cite{PERF-2012-01,PERF-2014-03}, 
and required to satisfy $\pt>25$\,\GeV\ and $|\eta|<2.5$. Jets satisfying
$\pt<50$\,\GeV\ and $|\eta|<2.4$ were additionally required to pass
pileup rejection criteria based on their associated tracks \cite{PERF-2014-03}.
To further suppress non-isolated leptons likely to originate from heavy-flavour
decays within jets, electron and muon candidates within $\Delta R<0.4$ of
selected jets were discarded.
Finally, jets likely to contain $b$-hadrons were $b$-tagged
using the MV1 algorithm \cite{PERF-2012-04}, a multivariate discriminant 
making use of track impact
parameters and reconstructed secondary vertices. A tagging working
point corresponding to a 70\,\% efficiency for tagging $b$-quark jets from
top decays in \ttbar\ events was used, giving a rejection factor of about 
140 against light-quark and gluon jets, and about five against jets 
originating from charm quarks.

As in Ref. \cite{TOPQ-2013-04}, events were required to have at least
one reconstructed primary vertex\footnote{The reconstructed vertex with the
largest sum of $\pt^2$ for the constituent tracks was selected as the 
primary vertex.} and to have no jets with $\pt>20$\,GeV\
failing jet quality requirements \cite{PERF-2012-01}.
Events having muons compatible with cosmic-ray interactions
or losing substantial energy following bremsstrahlung in the calorimeter
material were rejected. A preselection requiring exactly one electron and one
muon selected as described above was then applied, requiring at least one
selected lepton to be matched to a corresponding electron or muon trigger 
signature. Events
with an opposite-charge-sign $e\mu$ pair formed the main analysis sample, with
events having a same-sign pair being used to estimate the background from
misidentified leptons.

A total of 66453 data events passed the opposite-sign $e\mu$ preselection.
Events were then further sub-divided according to the number of $b$-tagged jets,
irrespective of the number of untagged jets, and events having one or
two $b$-tagged jets were retained for further analysis.
The numbers of one and two $b$-tagged jet events selected in data are shown in 
Table~\ref{t:evtcount}, compared with expected non-\ttbar\ contributions
from $Wt$ and dibosons evaluated from simulation, and 
$Z(\rightarrow\tau\tau\rightarrow e\mu)$+jets
and misidentified leptons evaluated from data and simulation, 
as discussed in detail in 
Section~\ref{ss:backg} and Section~\ref{s:syst} below.\footnote{The background
event counts and uncertainties shown in Table~\ref{t:evtcount}
differ from those in Ref. \cite{TOPQ-2013-04}
due to the use of different simulation samples and the estimation
of the background in bins of lepton kinematic variables.}
In simulation, the one 
$b$-tagged sample is about 88\,\% pure and the two $b$-tagged
sample 96\,\% pure in \ttbar\ events, with the largest backgrounds coming
from $Wt$ production in both cases. 
The distribution of the number of $b$-tagged jets in preselected
opposite-sign $e\mu$ events is shown in Figure~\ref{f:dmcjlept}(a), compared
to the predictions from simulation using {\sc Powheg\,+\,Pythia6} (PY6),
{\sc MC@NLO\,+\,Herwig} (HW) and {\sc Alpgen\,+\,Herwig} \ttbar\ samples, 
normalising the total simulation prediction 
in each case using the integrated luminosity
of the data sample. The distributions of the \pt\ of $b$-tagged jets, 
 and the reconstructed electron and muon \pt\ and $|\eta|$ 
in events with at least one $b$-tagged jet are shown in 
Figure~\ref{f:dmcjlept}(b--f), with the total simulation prediction
normalised to the same number of events as the data to facilitate shape
comparisons. The distributions of the reconstructed
dilepton variables \ptll, \mll, \rapll, \dphill, \ptsum\ and \esum\
are shown in Figure~\ref{f:dmcdilept}, with the simulation normalised as for
Figure~\ref{f:dmcjlept}(b--f).  In general the data are well described
by the predictions using the different \ttbar\ models, but a few differences
are visible. The lepton \pt\ spectra
are softer in data than in simulation, the lepton \etal\ and
dilepton \rapll\ distributions are more central than the 
{\sc Powheg\,+\,Pythia6} and {\sc MC@NLO\,+\,Herwig} predictions,
and the \dphill\ distribution is slightly flatter in data than in all the
predictions.

\begin{table}[tp]
\centering

\begin{tabular}{lcc}\hline
Event counts & $N_1$ & $N_2$ \\ \hline
Data & 21666 & 11739 \\
\hline
$Wt$ single top & $ 2080\pm   210$ & $ 350\pm   120$ \\
$Z(\rightarrow\tau\tau\rightarrow e\mu)$+jets & $  210\pm    40$ & $    7\pm     2$ \\
Diboson & $  120\pm    30$ & $    3\pm     1$ \\
Misidentified leptons & $  220\pm    80$ & $   78\pm    50$ \\
\hline
Total background & $ 2630\pm   230$ & $  440\pm   130$ \\
\hline
\end{tabular}
\caption{\label{t:evtcount}Observed numbers of opposite-sign $e\mu$ events
with one and two $b$-tagged jets ($N_1$ and $N_2$)
together with the estimates of backgrounds and associated total
uncertainties described in Section~\ref{s:syst}.}
\end{table}

\begin{figure}[htp]
\vspace{-7mm}
\splitfigure{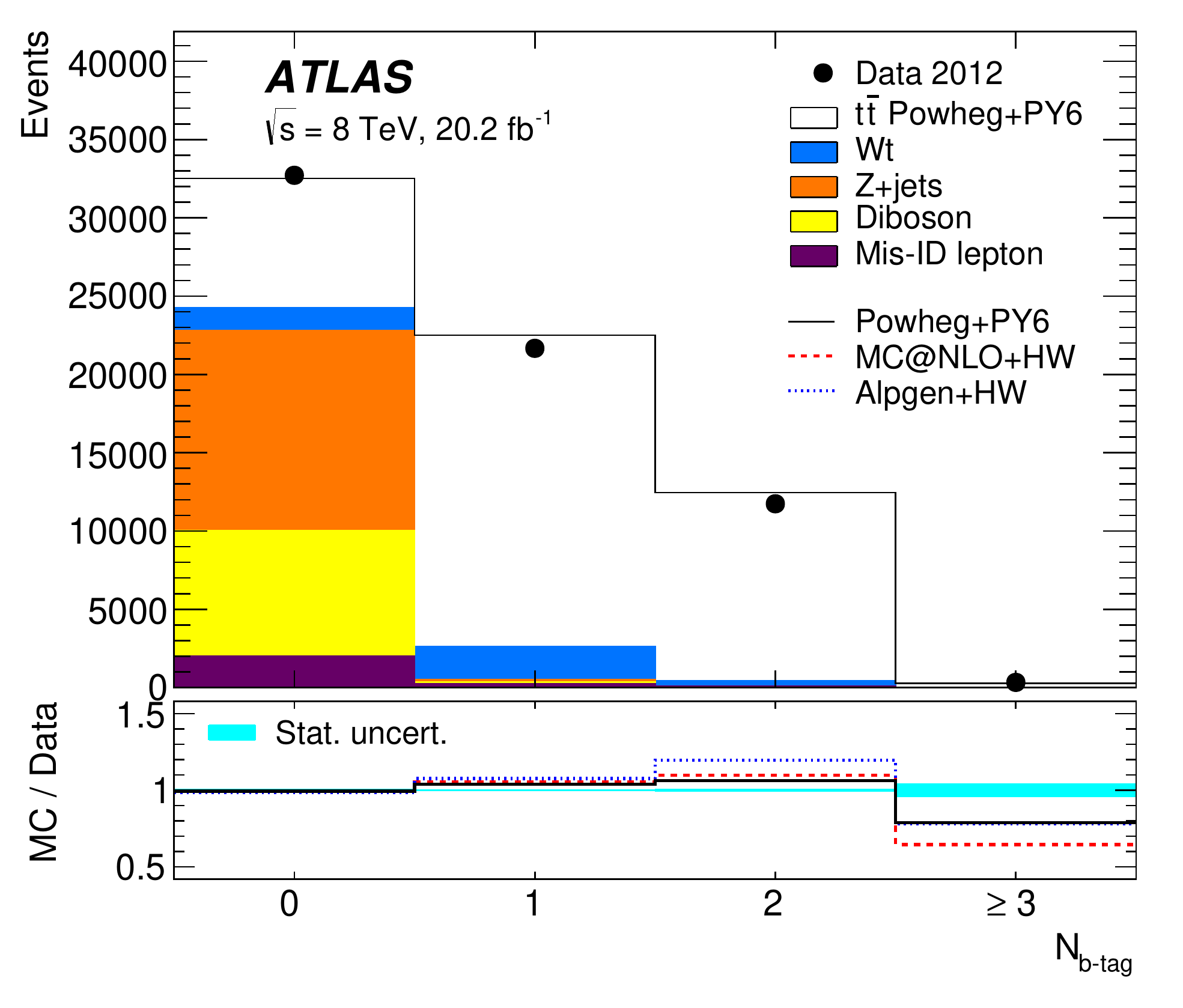}{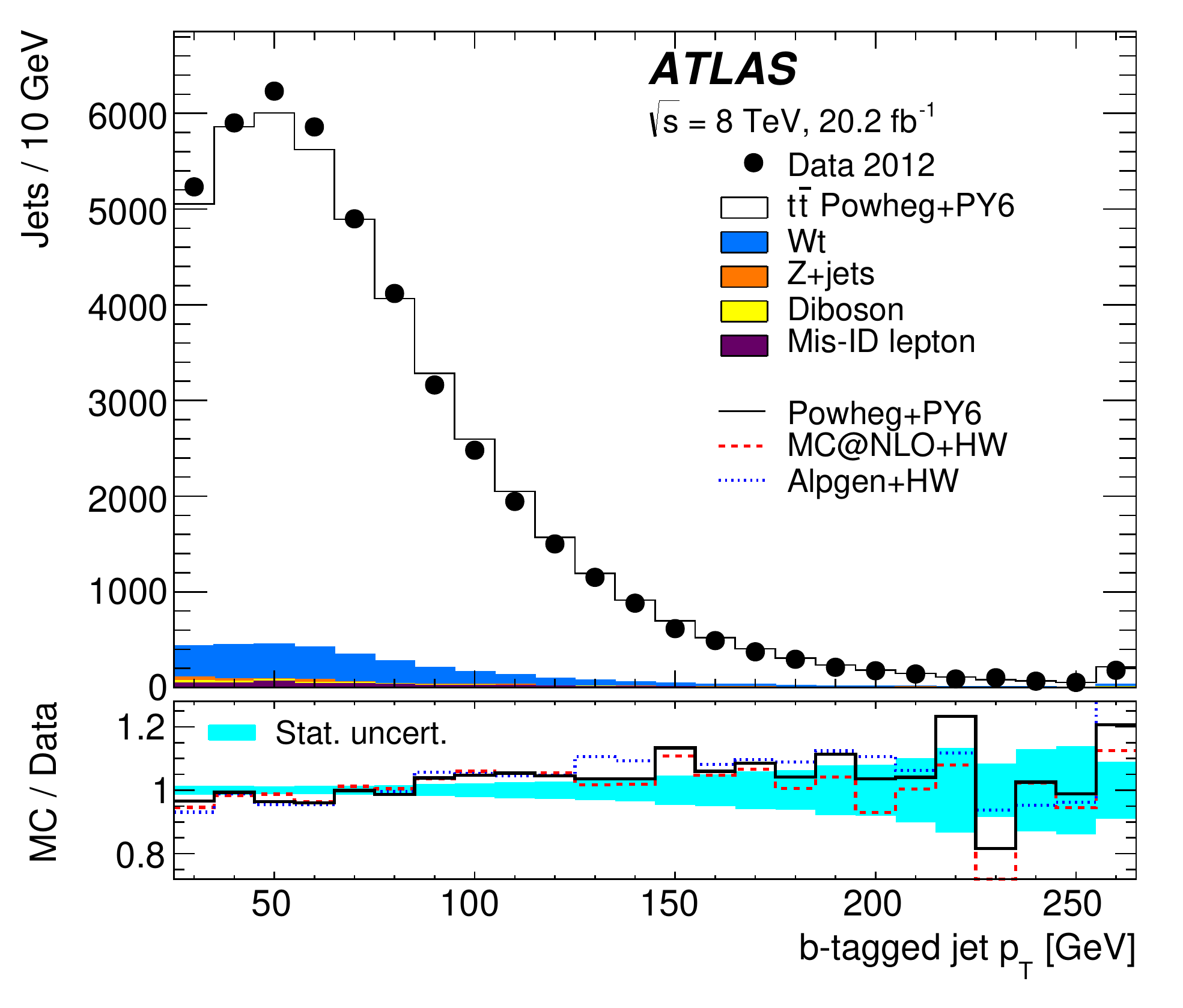}{a}{b}
\splitfigure{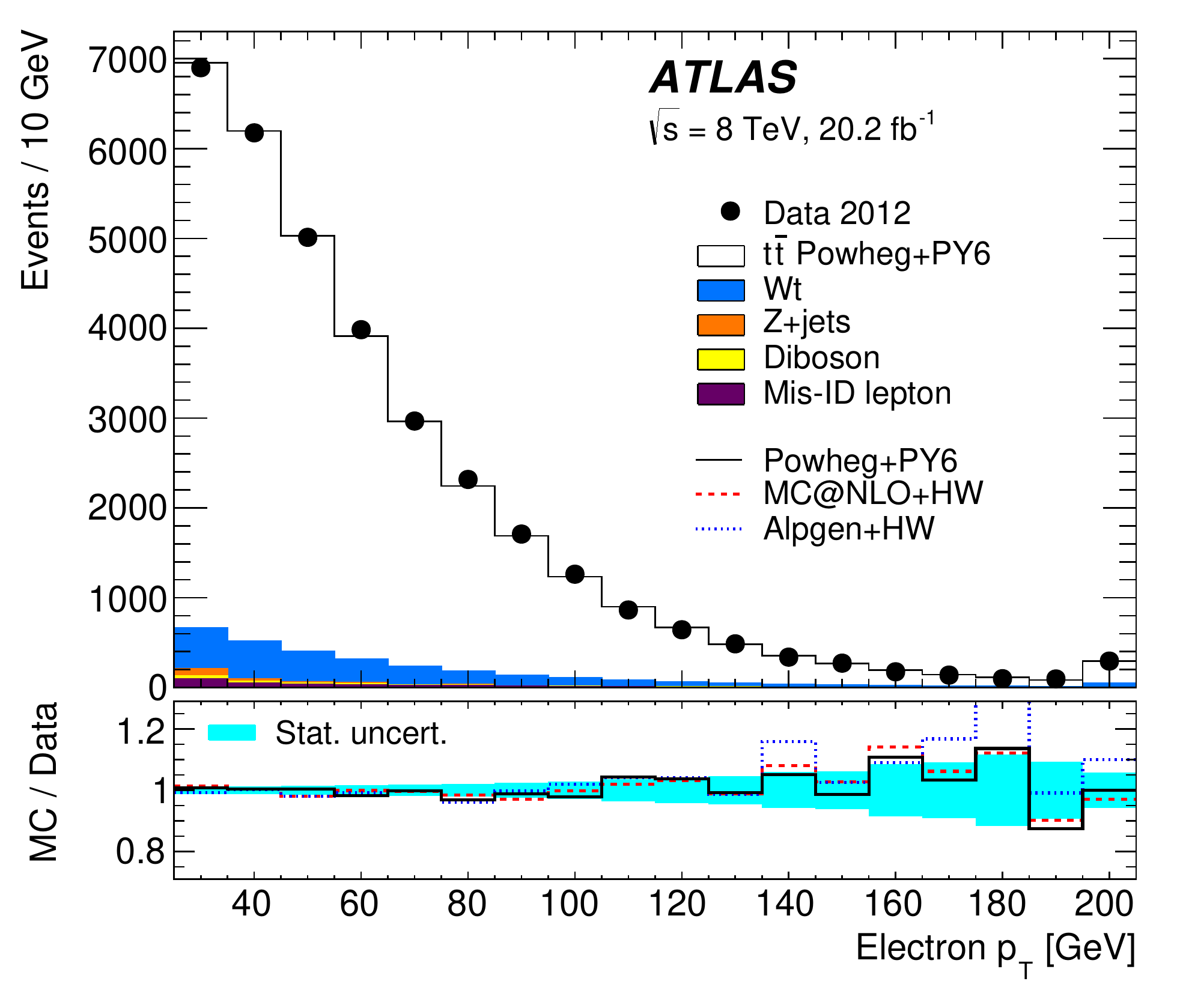}{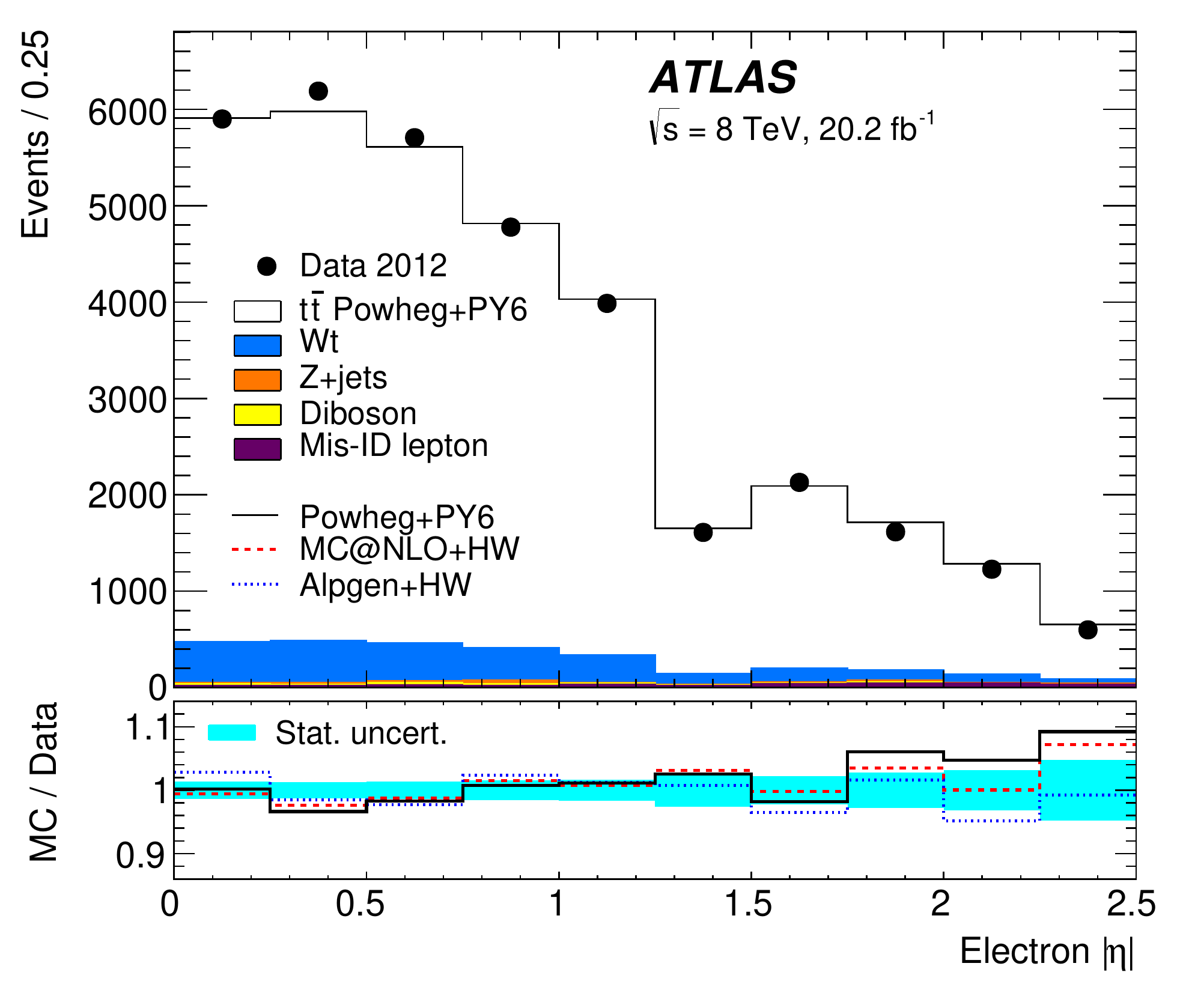}{c}{d}
\splitfigure{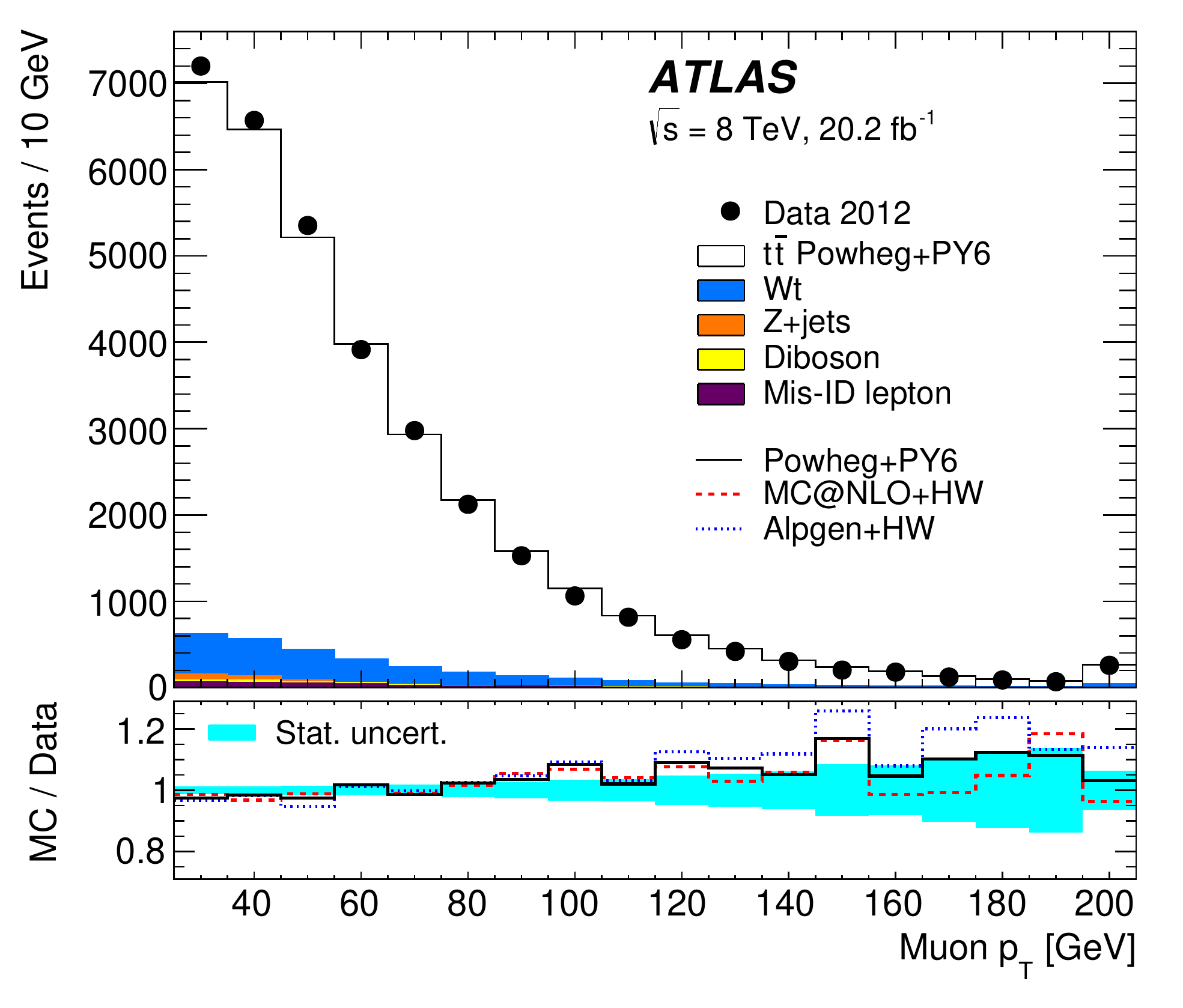}{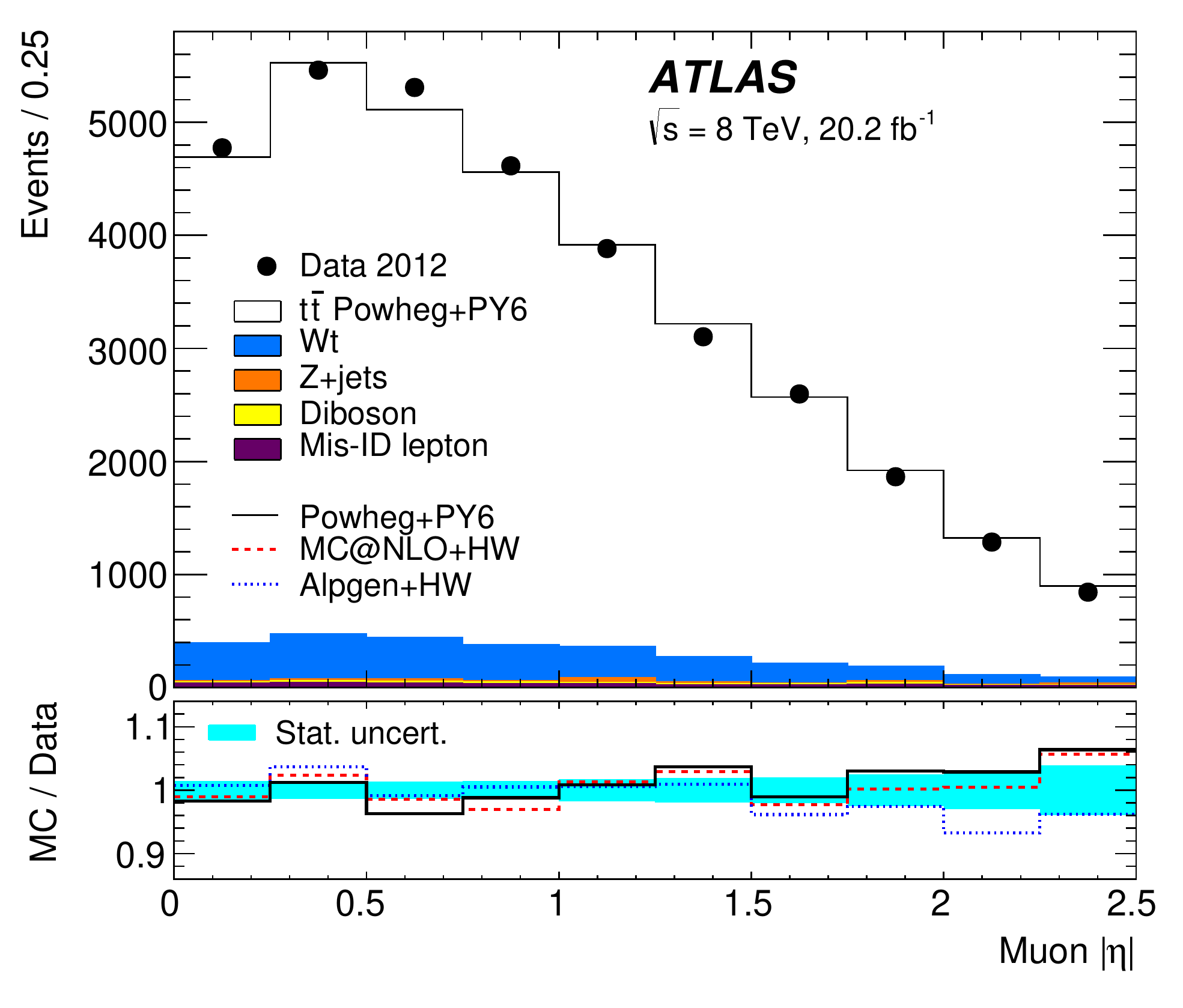}{e}{f}
\caption{\label{f:dmcjlept}Distributions of
(a) the number of $b$-tagged jets in preselected opposite-sign $e\mu$ events;
and (b) the \pt\ of $b$-tagged jets, 
(c) the \pt\ of the electron, (d) the $|\eta|$ of the electron,
(e) the \pt\ of the muon and 
(f) the $|\eta|$ of the muon, in events with an opposite-sign $e\mu$ pair and 
at least one $b$-tagged jet. The reconstruction-level
data are compared to the expectation from
simulation, broken down into contributions from 
\ttbar\ \,({\sc Powheg\,+\,Pythia6}), single top, $Z$+jets,
dibosons, and events with misidentified electrons or muons. The simulation
prediction is normalised to the same integrated luminosity as the data in
(a) and to the same number of entries as the data in (b--f).
The lower parts of the figure show the ratios of simulation to data, 
using various \ttbar\ signal samples and with the cyan band indicating the 
data statistical uncertainty. 
The last bin includes the overflow in panels (b), (c) and (e).}
\end{figure}

\begin{figure}
\vspace{-7mm}
\splitfigure{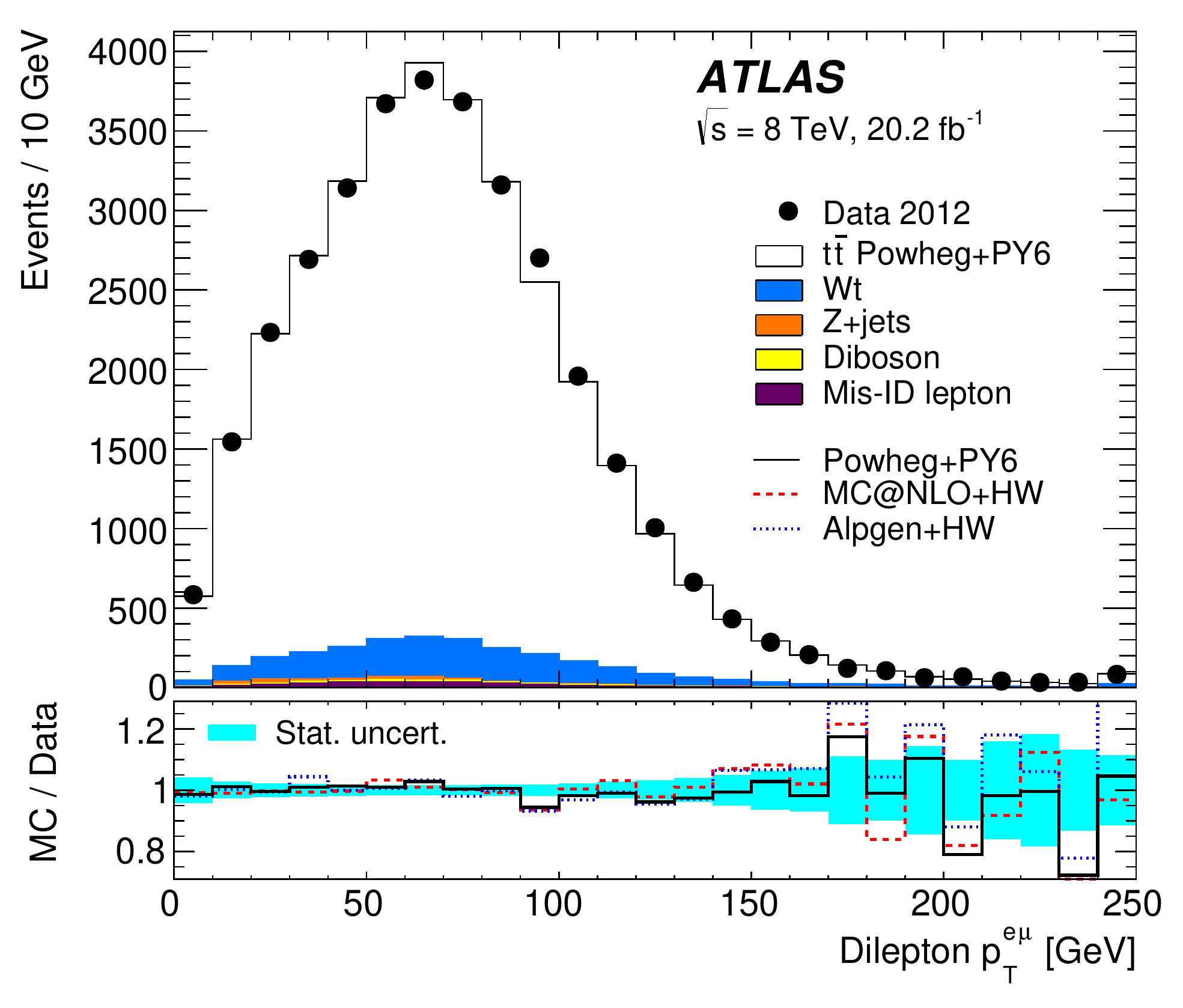}{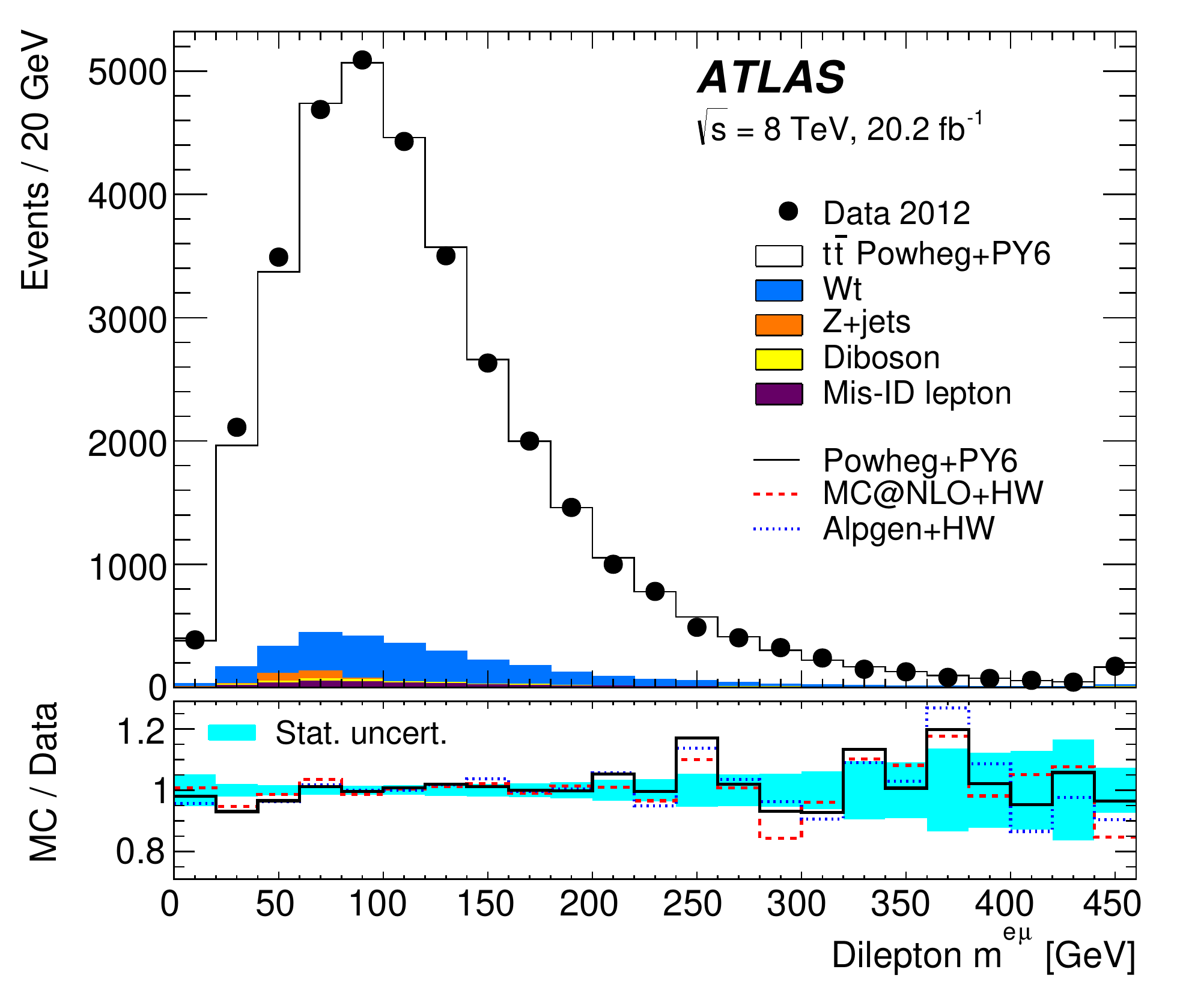}{a}{b}
\splitfigure{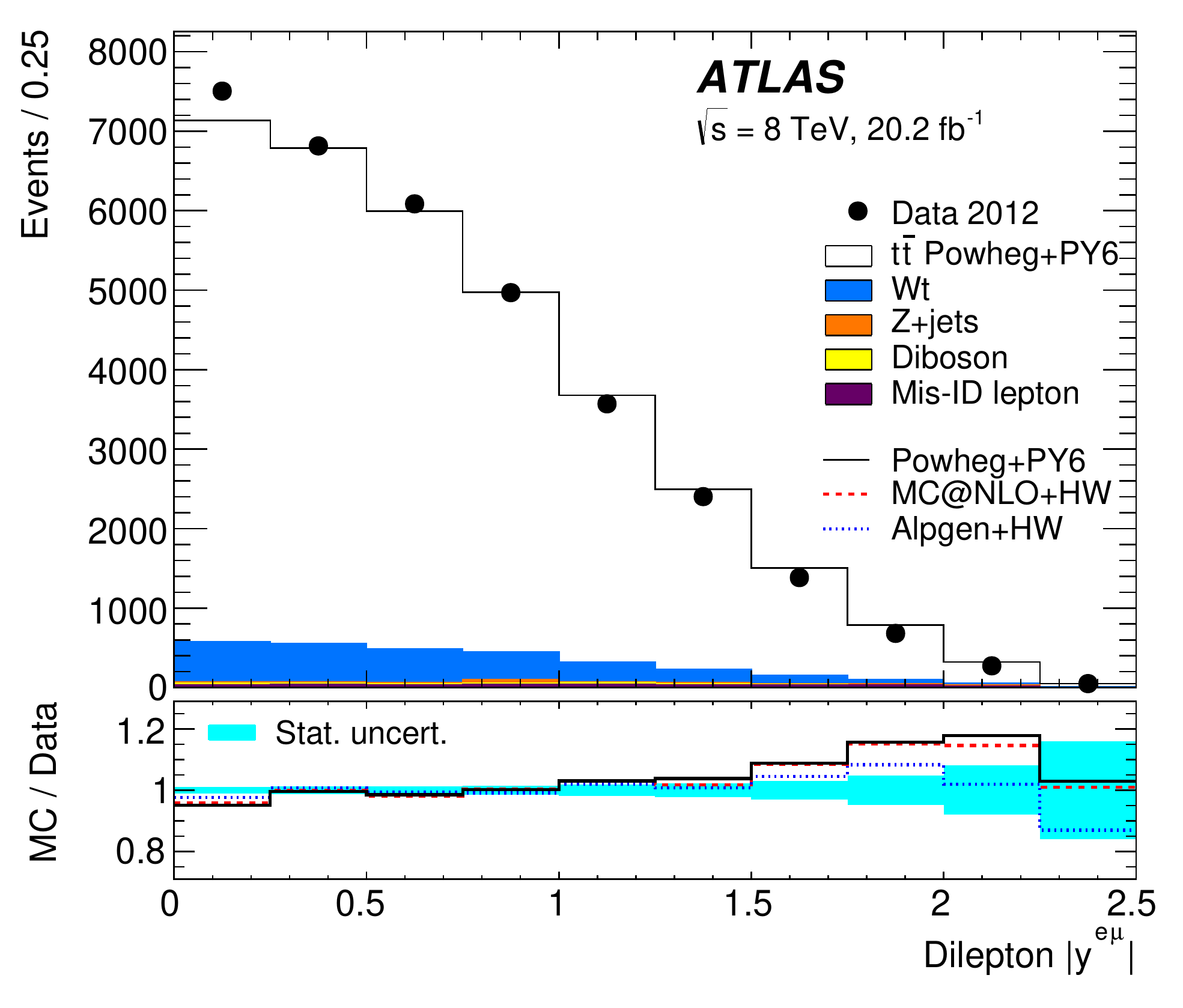}{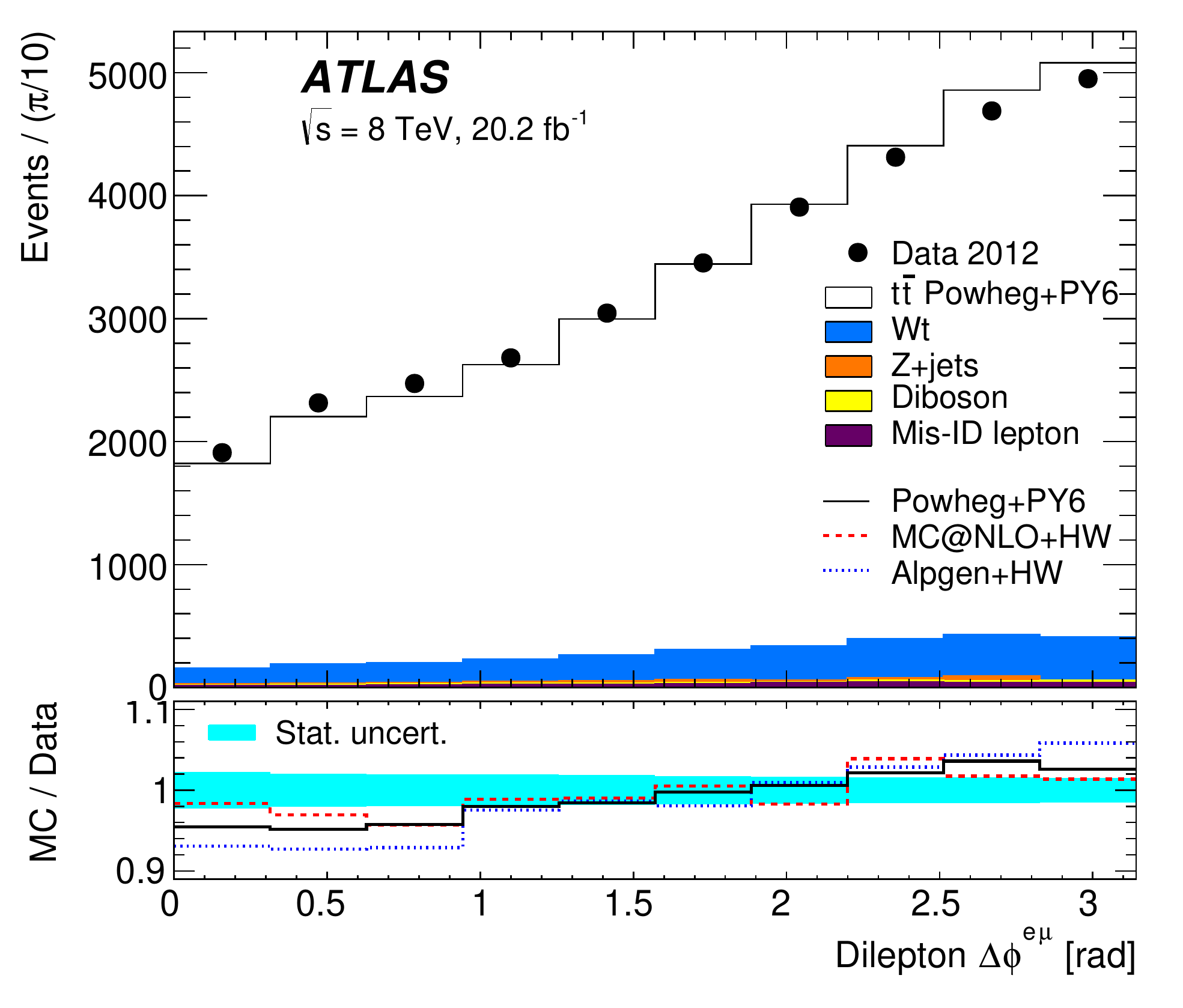}{c}{d}
\splitfigure{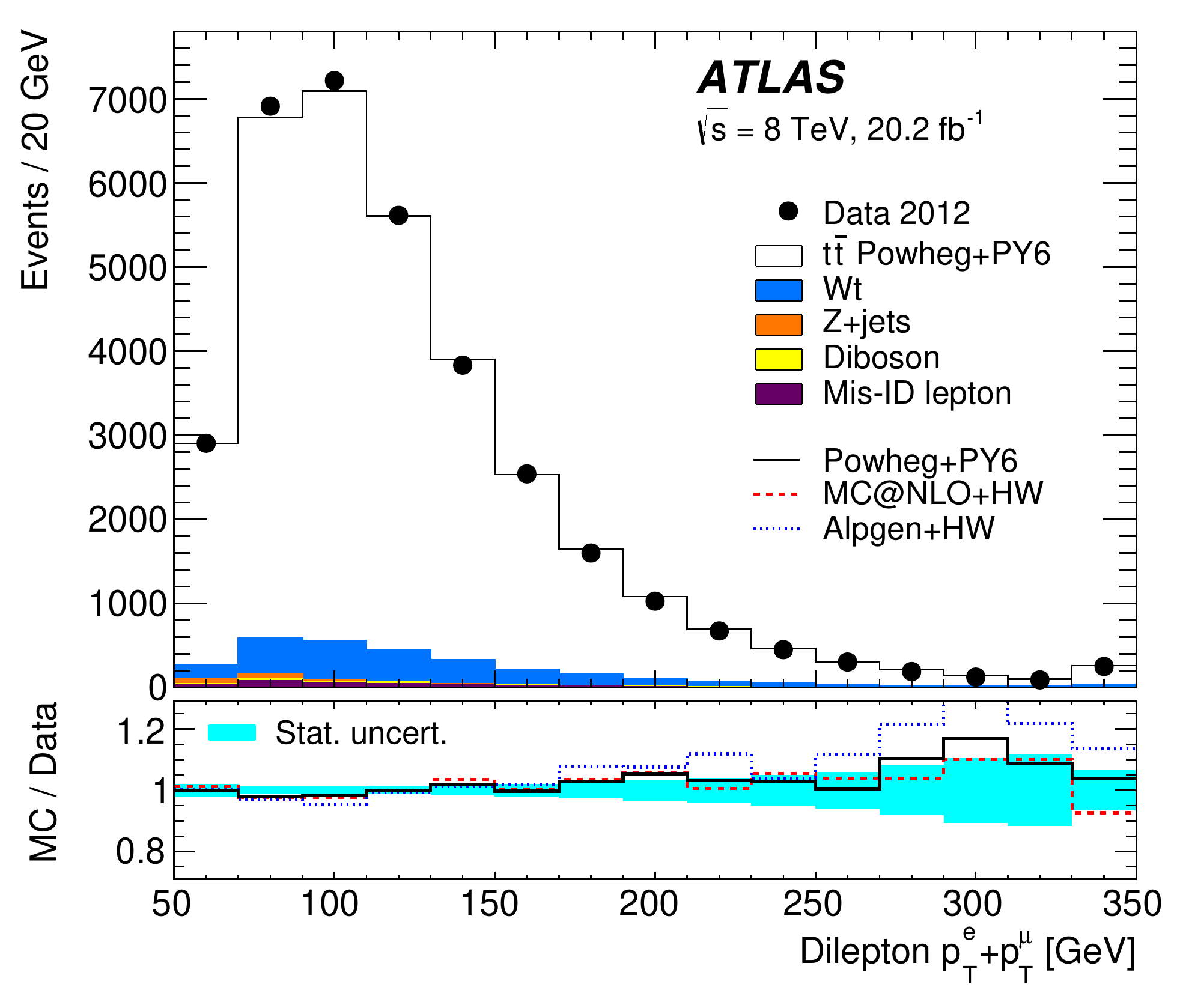}{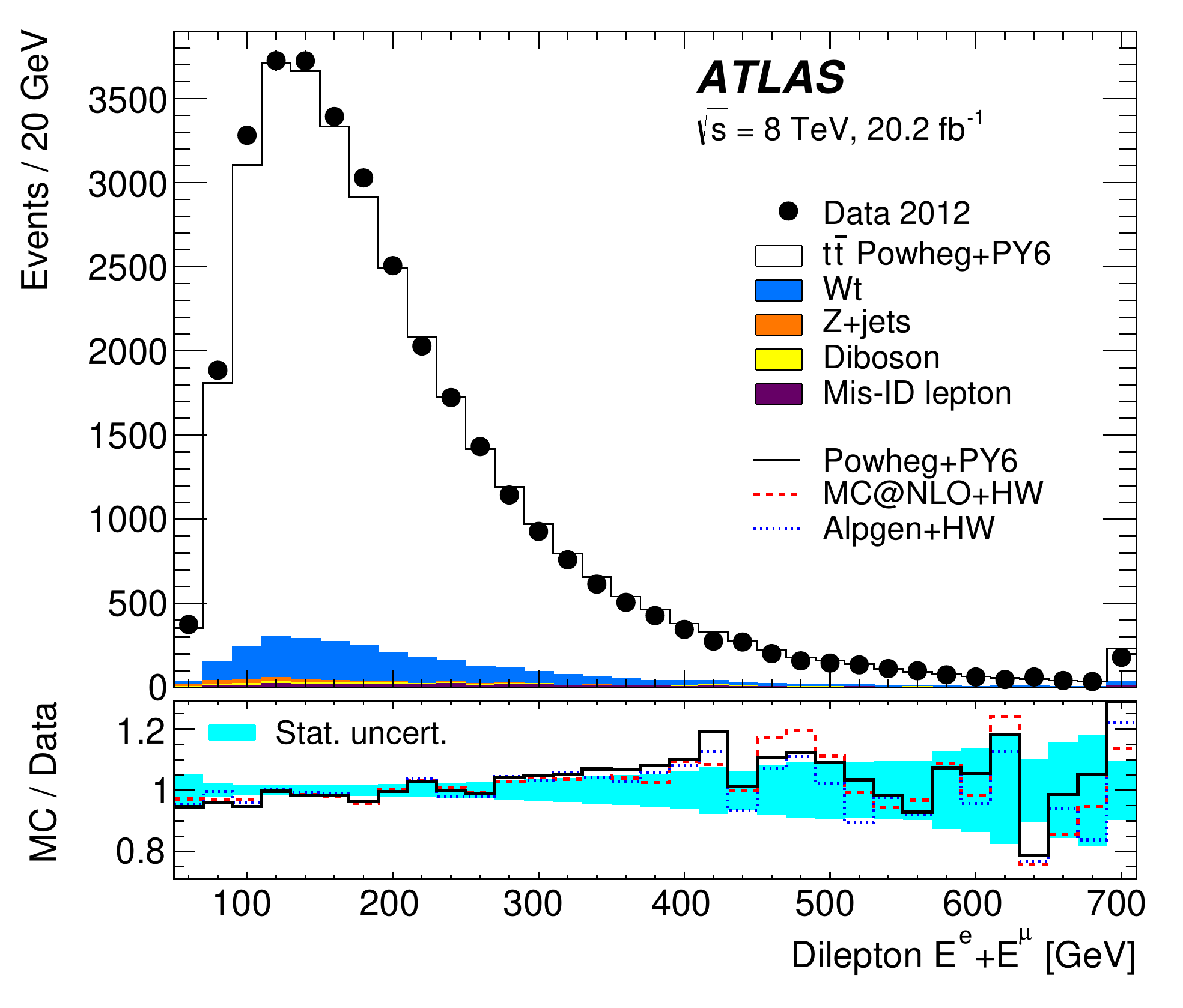}{e}{f}
\caption{\label{f:dmcdilept}Distributions of
(a) the dilepton \ptll, (b) invariant mass \mll, (c) rapidity \rapll,
(d) azimuthal angle difference \dphill, (e) lepton \pt\ sum \ptsum\ and
(f) lepton energy sum \esum, in events with an opposite-sign $e\mu$ pair and 
at least one $b$-tagged jet. The reconstruction-level data are 
compared to the expectation from
simulation, broken down into contributions from 
\ttbar\ \,({\sc Powheg\,+\,Pythia6}), single top, $Z$+jets,
dibosons, and events with misidentified electrons or muons, normalised to the
same number of entries as the data. 
The lower parts of the figure show the ratios of simulation to data, 
using various \ttbar\ signal samples and with the cyan band indicating the 
data statistical uncertainty. The last bin includes the overflow in panels 
(a), (b), (e) and (f).}
\end{figure}


\section{Fiducial cross-section determination}\label{s:xsecdet}

The cross-section measurements were made for a fiducial region,
where the particle-level electron and muon were required to have opposite
charge signs, to each come from $W$ decays either directly or via 
$W\rightarrow\tau\rightarrow e/\mu$ and to each satisfy 
$\pt>25$\,\GeV\ and $|\eta|<2.5$. The lepton four-momenta were taken
after final-state radiation, and `dressed' by including the four-momenta
of any photons within a cone of size $\Delta R=0.1$ around the lepton
direction, excluding photons produced from hadronic decays or interactions
with the detector material. The total cross-section within this fiducial
volume corresponds to the fiducial cross-section  
measured in Ref. \cite{TOPQ-2013-04}. According to the predictions
of the baseline {\sc Powheg\,+\,Pythia6} \ttbar\ simulation,
it is about 44\,\% of the 
total $\ttbar\rightarrow e\mu\nu\bar{\nu}\bbbar$ cross-section without
restrictions on the lepton acceptance and including
contributions via $W\rightarrow\tau\rightarrow e/\mu$.

\subsection{Cross-section extraction}

The differential cross-sections were measured using an extension of the 
technique used in Ref. \cite{TOPQ-2013-04}, counting the number of leptons
or events with one (\nxi) or two (\nyi) $b$-tagged jets
where the lepton(s) fall in bin $i$ of a differential distribution at
reconstruction level. For the single-lepton
distributions \ptl\ and \etal, there are two counts per event, in the 
two bins corresponding to the electron and muon. For the dilepton distributions,
each event contributes a single count corresponding to the bin in which
the appropriate dilepton variable falls.
For each measured distribution, these counts satisfy the tagging equations:
\begin{equation}
\begin{array}{lll}
\nxi & = &  L \xtti\ \gemi 2\epsbi (1-\cbi\epsbi) + \nibi , \\*[2mm]
\nyi & = &  L \xtti\ \gemi \cbi(\epsbi)^2 + \niibi  , 
\end{array}\label{e:fidtags}
\end{equation}
where \xtti\ is the absolute fiducial differential cross-section in bin $i$, 
and $L$ is the integrated luminosity of the sample. The reconstruction
efficiency \gemi\ represents the ratio of the number of reconstructed 
$e\mu$ events
(or leptons for \ptl\ and \etal) falling in bin $i$ at reconstruction level 
to the number of true $e\mu$ events (or leptons) falling in the same bin at
particle level, evaluated using \ttbar\ simulation without making any 
requirements on reconstructed or particle-level jets. It therefore corrects for
both the lepton reconstruction efficiency and bin migration, where events
corresponding to bin $j$ at particle level appear in a different 
bin $i\neq j$ at reconstruction level. The values of \gemi\ in simulation
are typically in the range 0.5--0.6, with some dependence on lepton kinematics
due to the varying reconstruction efficiencies with lepton $|\eta|$ and  \pt,
and the effect of isolation requirements when the leptons are close together
in the detector.

The efficiency \epsbi\ represents the combined probability for a jet from the
quark $q$ in the $t\rightarrow Wq$ decay to fall within the detector acceptance,
be reconstructed as a jet with $\pt>25$\,\GeV\ and be tagged as a $b$-jet. 
Although this quark is almost always a $b$-quark, \epsbi\ also accounts
for the 0.2\,\% of top quarks that decay to $Ws$ or $Wd$.  If the kinematics
of the two $b$ quarks produced in the top quark decays are uncorrelated,
the probability to tag both is given by $\epsbbi=(\epsbi)^2$. In practice,
small correlations are present, for example due to kinematic correlations 
between the $b$-jets from the top quark decays, or extra \bbbar\ or \ccbar\ 
pairs produced in association with the \ttbar\ system \cite{TOPQ-2013-04}.
Their effects are corrected via the tagging
correlation coefficient $\cbi=\epsbbi/(\epsbi)^2$, whose values are taken
from \ttbar\ simulation. They depend slightly
on the bin $i$ of the dilepton system but are always within 1--2\,\% of 
unity, even for the bins at the edges of the differential distributions.
The correlation \cbi\ also corrects for the small effects on \nxi, \nyi\ and 
\epsbi\ of the small fraction of \ttbar\ events
which have additional $b$ quarks produced in association with the \ttbar\
system, and the even smaller effects from mistagged light quark, charm or
gluon jets in \ttbar\ events. This formalism involving \epsbi\ and \cbi\
allows the fraction of top quarks where the jet was not reconstructed 
to be inferred from the counts \nxi\ and \nyi, 
minimising the exposure to systematic uncertainties from
jet measurements and $b$-tagging, and
allowing the fiducial cross-sections \xtti\ to be defined
with no requirements on the jets in the final state.

Backgrounds from sources other than $\ttbar\rightarrow e\mu\nu\bar{\nu}\bbbar$
events also contribute to the counts \nxi\ and \nyi, and are represented
by the terms \nibi\ and \niibi\ in Eqs.~(\ref{e:fidtags}). These contributions
were evaluated using a combination of simulation- and data-based methods
as discussed in Section~\ref{ss:backg} below.

The tagging equations were solved numerically in each bin $i$ of each 
differential distribution separately. The bin ranges for each distribution
were chosen according to the experimental resolution, minimising the bin-to-bin
migration by keeping the bin purities (the fractions of 
reconstructed events in bin $i$ that
originate from events which are also in bin $i$ at particle level) above
about 0.9. The resolution on the reconstructed kinematic quantities is dominated
by the electron energy and muon momentum measurements, and the purities for
the distributions which depend mainly on angular variables are higher,
around 0.96 for \rapll\ and 0.99 for \etal\ and \dphill. For these 
distributions, the bin ranges were chosen so as to give about ten
bins for each distribution. The bin range choices for all distributions
can be seen in Tables~\ref{t:insXSec1} to~\ref{t:insXSec4} 
in Section~\ref{s:res}, and the last bin
of the \ptl, \ptll, \mll, \ptsum\ and \esum\ distributions includes 
overflow events falling above the last bin boundary, indicated by the `+' sign
after the upper bin limit.

The normalised fiducial differential cross-section distributions \xntti\
were calculated from the absolute cross-sections \xtti\ determined from
Eqs.~(\ref{e:fidtags}) as follows:
\begin{eqnarray}\label{e:normx}
\xntti = \frac{\xtti}{\Sigma_j\ \xttj} = \frac{\xtti}{\xfid},
\end{eqnarray}
where \xfid\ is the total cross-section summed over all bins of the
fiducial region. The \xntti\ values are divided by the bin widths $W_i$, to 
produce the
cross-sections differential in the variable $x$ ($x=\ptl$, \etal, etc.):
\begin{eqnarray}\nonumber
\frac{1}{\sigma}\left( \frac{{\mathrm d}\sigma}{{\mathrm d}x}\right)_i =
\frac{\xntti}{W_i}\ . \label{e:diffxsec}
\end{eqnarray}
The normalisation condition in Eq.~(\ref{e:normx}) induces a statistical
correlation between the normalised measurements in each bin. The absolute
dilepton cross-section measurements are not statistically correlated between
bins, but kinematic correlations between the electron and muon in each event
induce small statistical correlations between bins of the absolute single lepton
\ptl\ and \etal\ distributions, as discussed in Section~\ref{ss:valid} below.

The measured cross-sections include contributions where one or both leptons
are produced via leptonic tau decays 
($t\rightarrow W\rightarrow\tau\rightarrow e/\mu$), 
but the fixed-order predictions discussed in Section~\ref{ss:fixedpred}
only include the direct decays $t\rightarrow W\rightarrow e/\mu$. To allow
comparison with such predictions, a second set of cross-section results
were derived with a bin-by-bin multiplicative correction \fntaui\ to remove
the $\tau$ contributions:
\begin{eqnarray}\label{e:notau}
\xtti\,(\mbox{no--$\tau$}) = \fntaui\xtti\ ,
\end{eqnarray}
and similarly for the normalised cross-sections $\xntti\,(\mbox{no-$\tau$})$. 
The corrections
\fntaui\ were evaluated from the baseline {\sc Powheg\,+\,Pythia6} \ttbar\
simulation and are typically close to 0.9, decreasing to 0.8--0.85 at 
low lepton \pt.

\subsection{Background estimates}\label{ss:backg}

The $Wt$ single top and diboson backgrounds were estimated from simulation
using the samples discussed in Section~\ref{s:datasim}, whilst the $Z$+jets
background (with $Z\rightarrow\tau\tau\rightarrow e\mu 4\nu$) and the
contribution from events with one real and one misidentified lepton were 
estimated using both simulation and data as discussed below. 
The backgrounds in both the one and two $b$-tagged
samples are dominated by $Wt$ (see Table~\ref{t:evtcount}). The total
background fraction (i.e. the predicted fraction of events in each bin
which do not come from \ttbar\ with two real prompt leptons)
varies significantly as a function of some of the 
differential variables, as shown in Figure~\ref{f:bgrate}. 
This variation is taken
into account by estimating the background contributions \nibi\ and \niibi\
separately in each bin of each differential distribution.

\begin{figure}
\vspace{-7mm}
\splitfigure{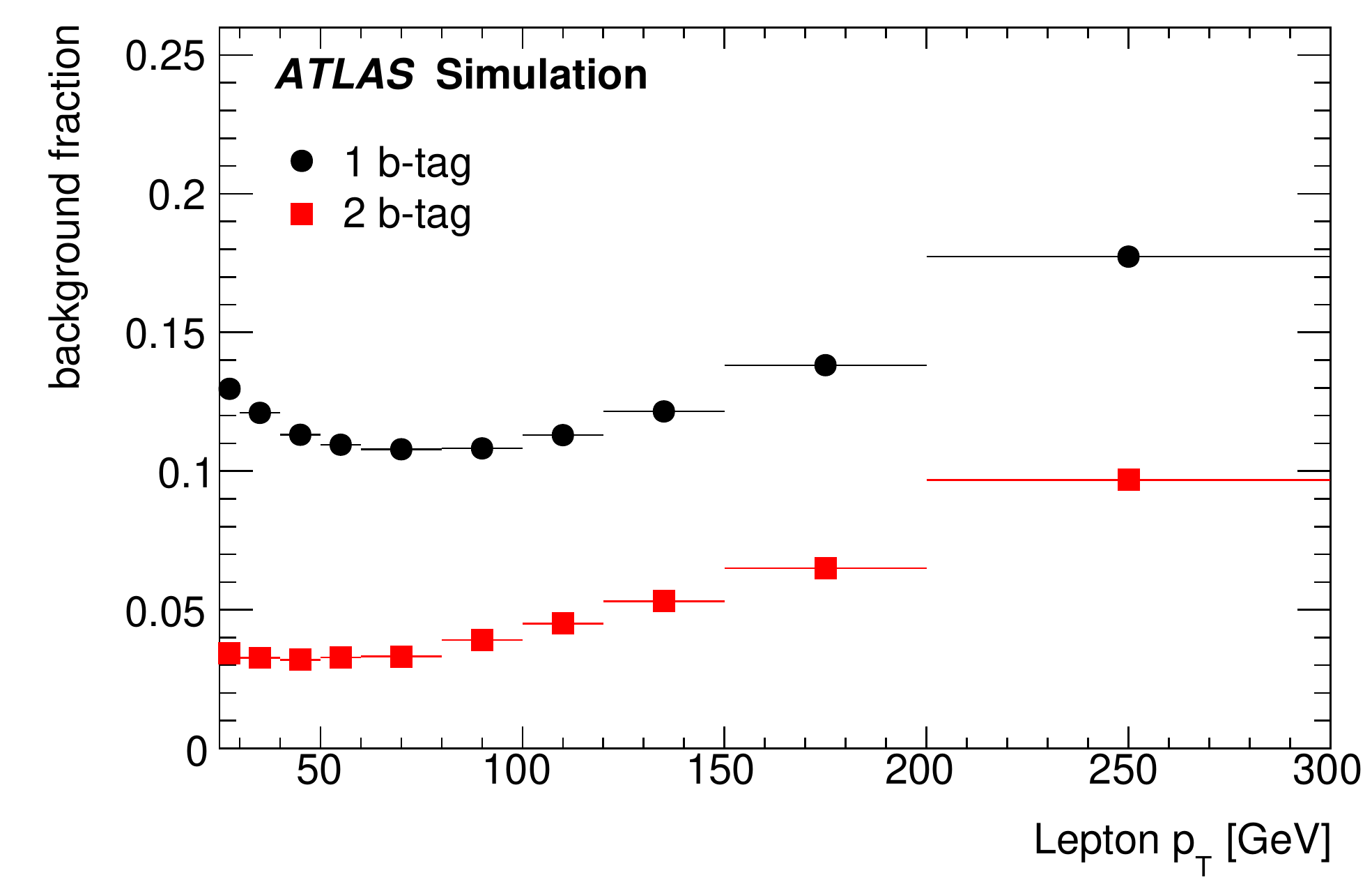}{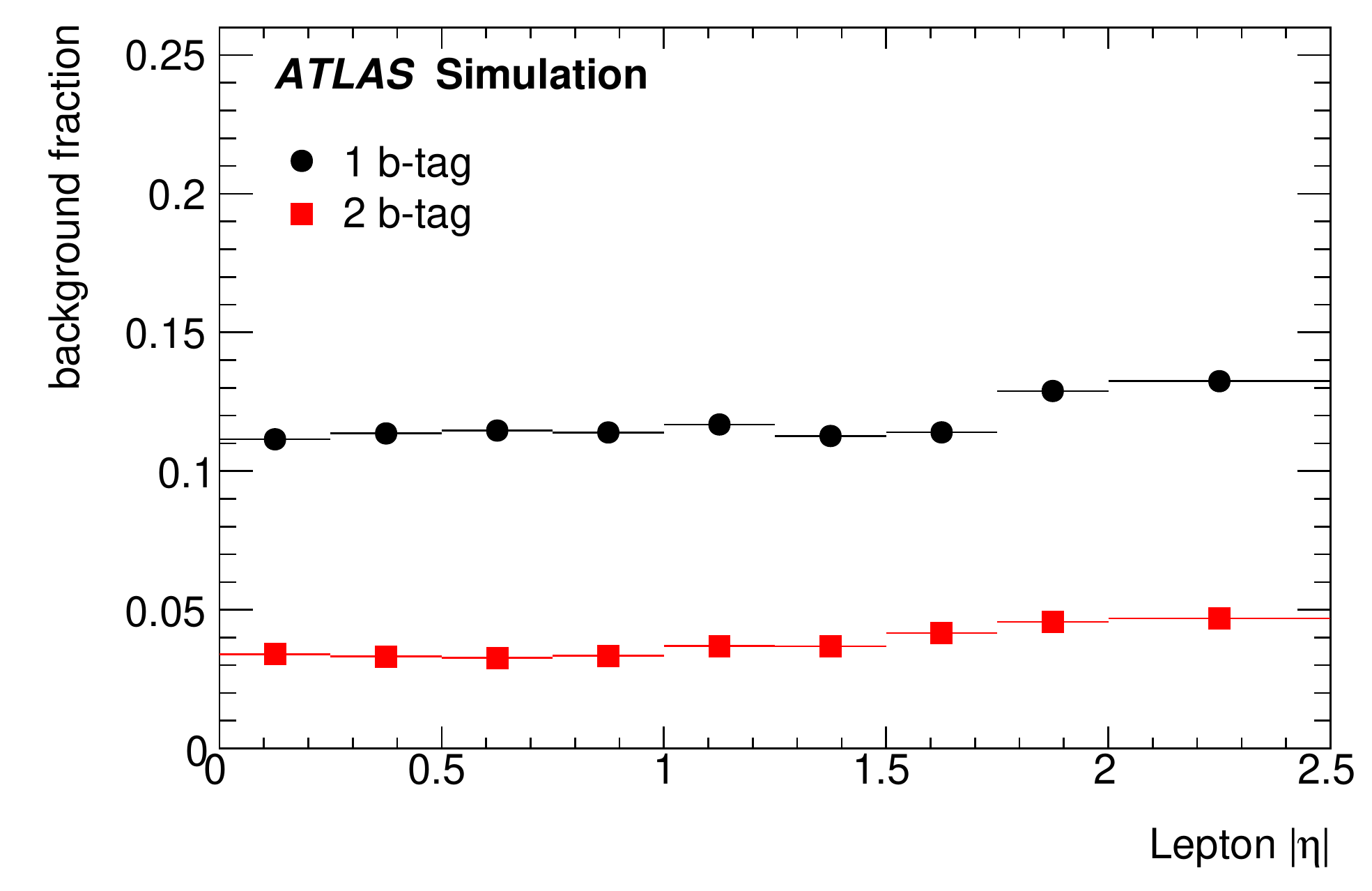}{a}{b}
\splitfigure{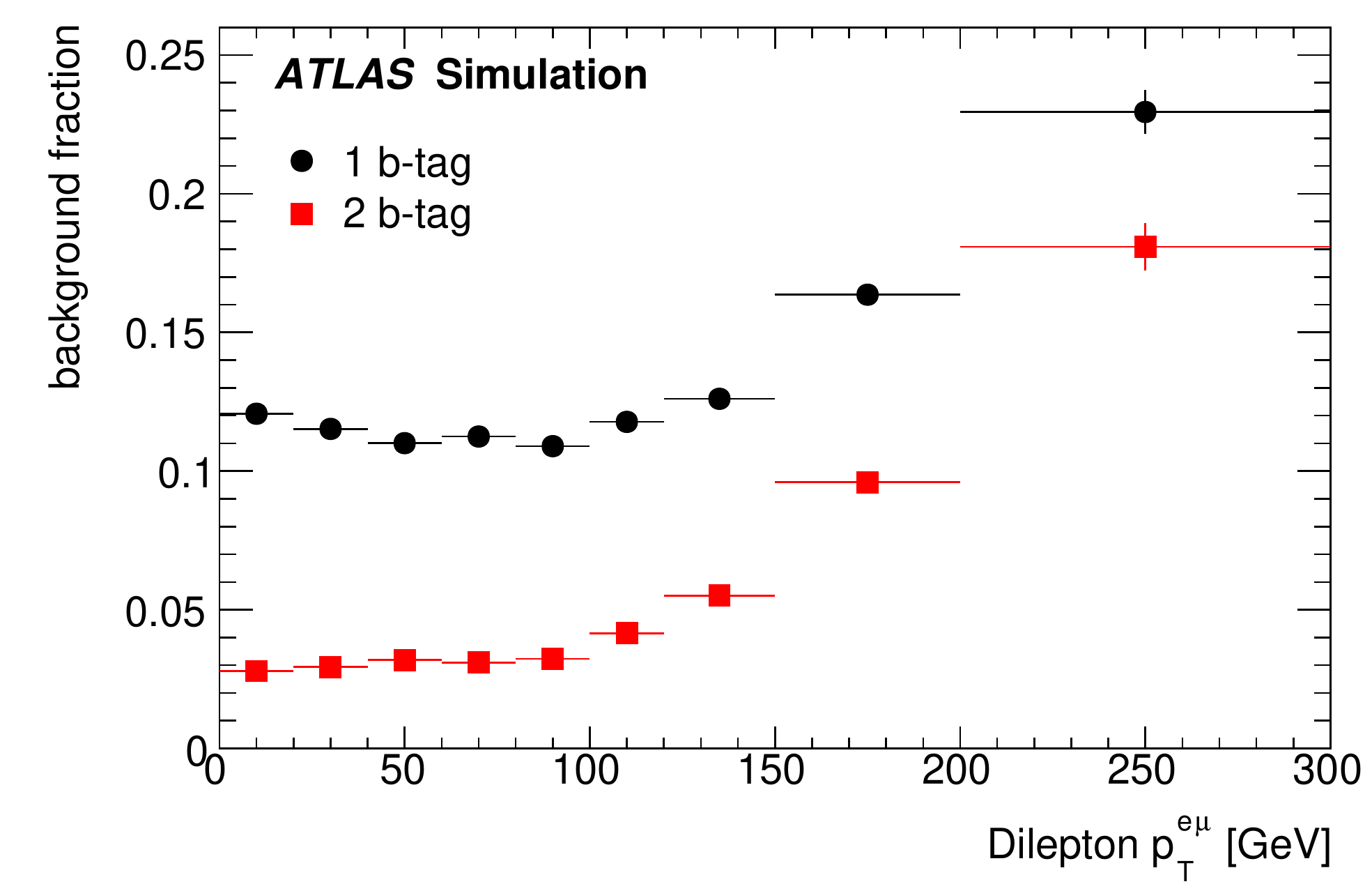}{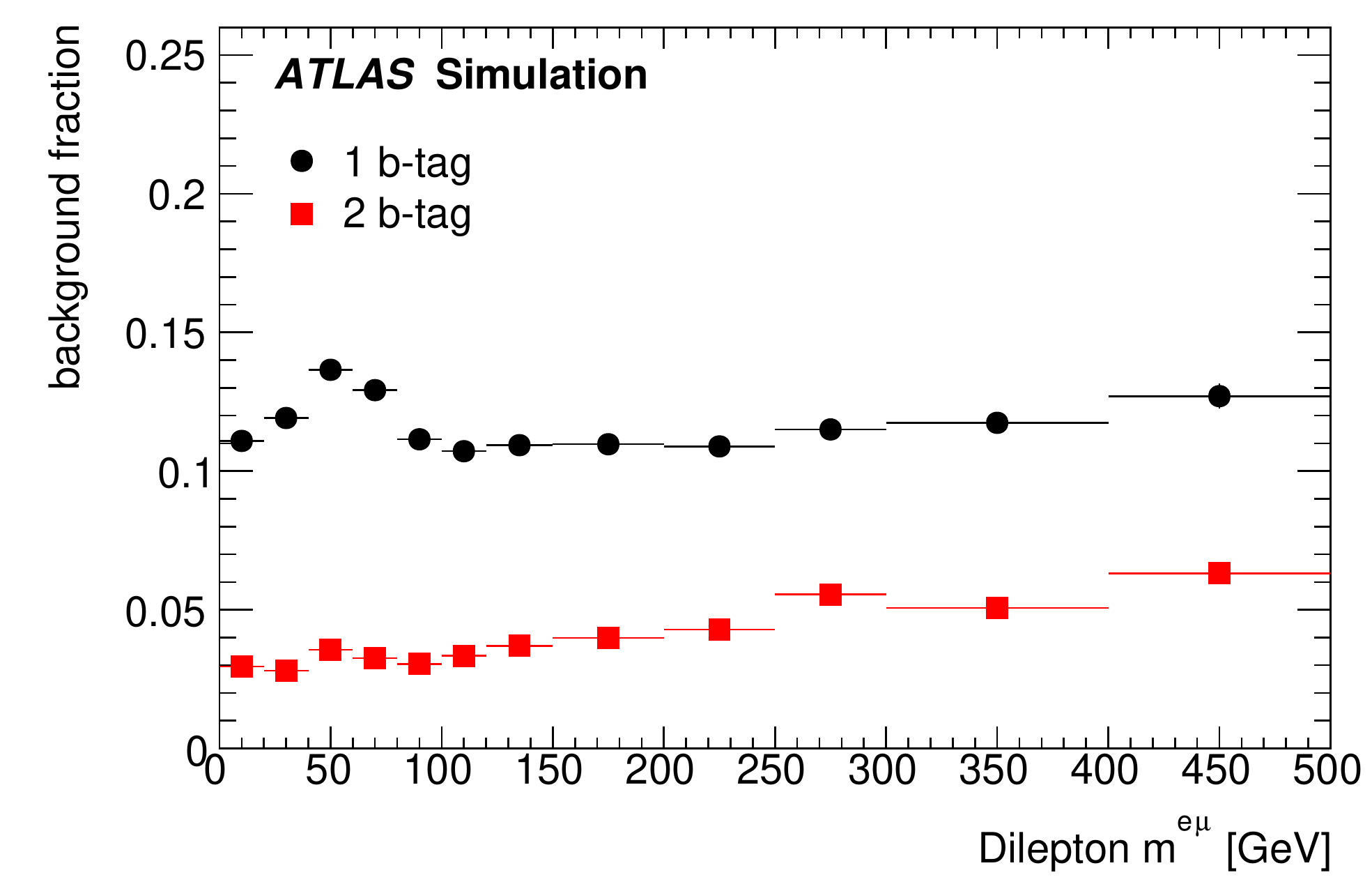}{c}{d}
\splitfigure{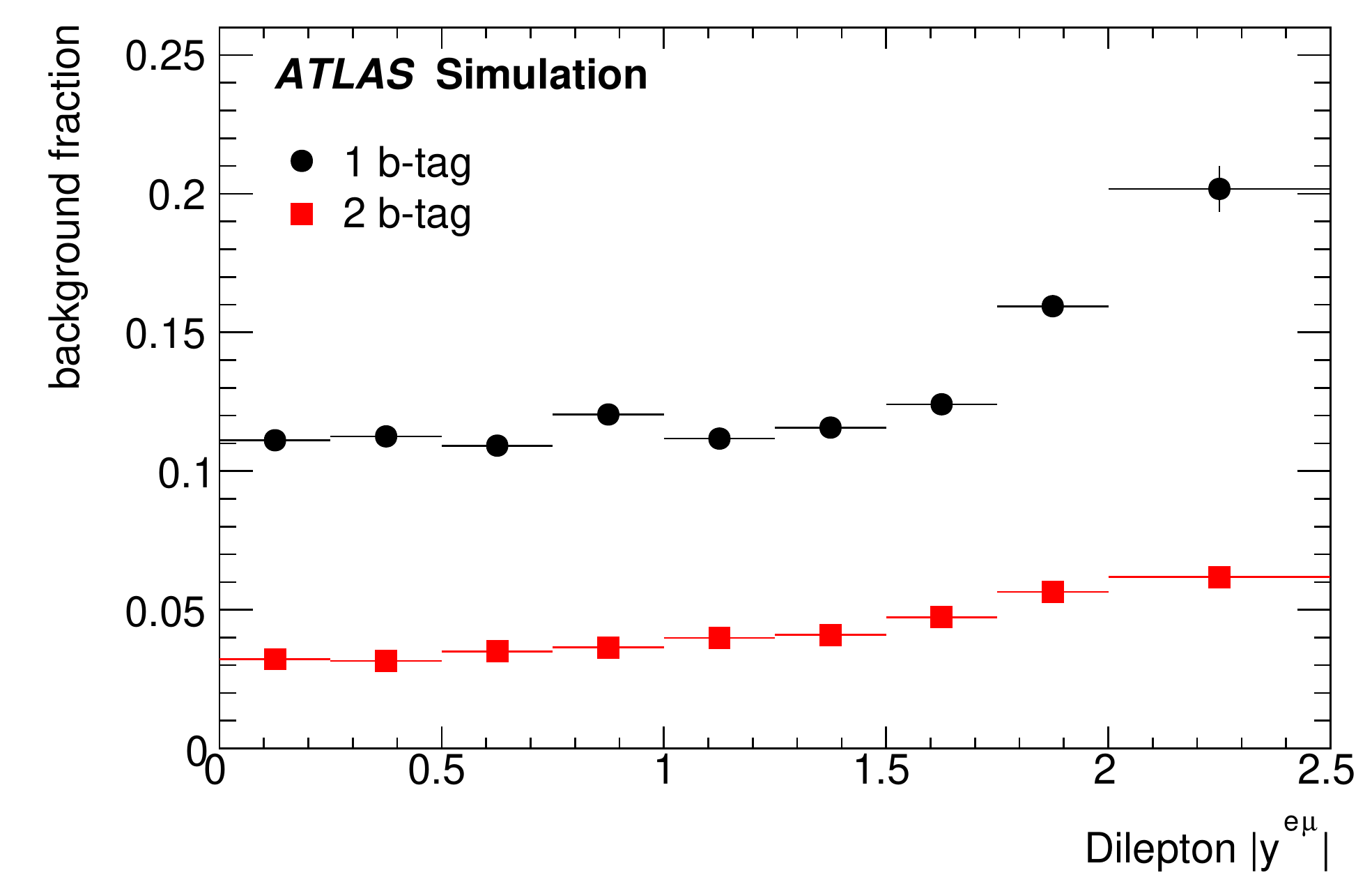}{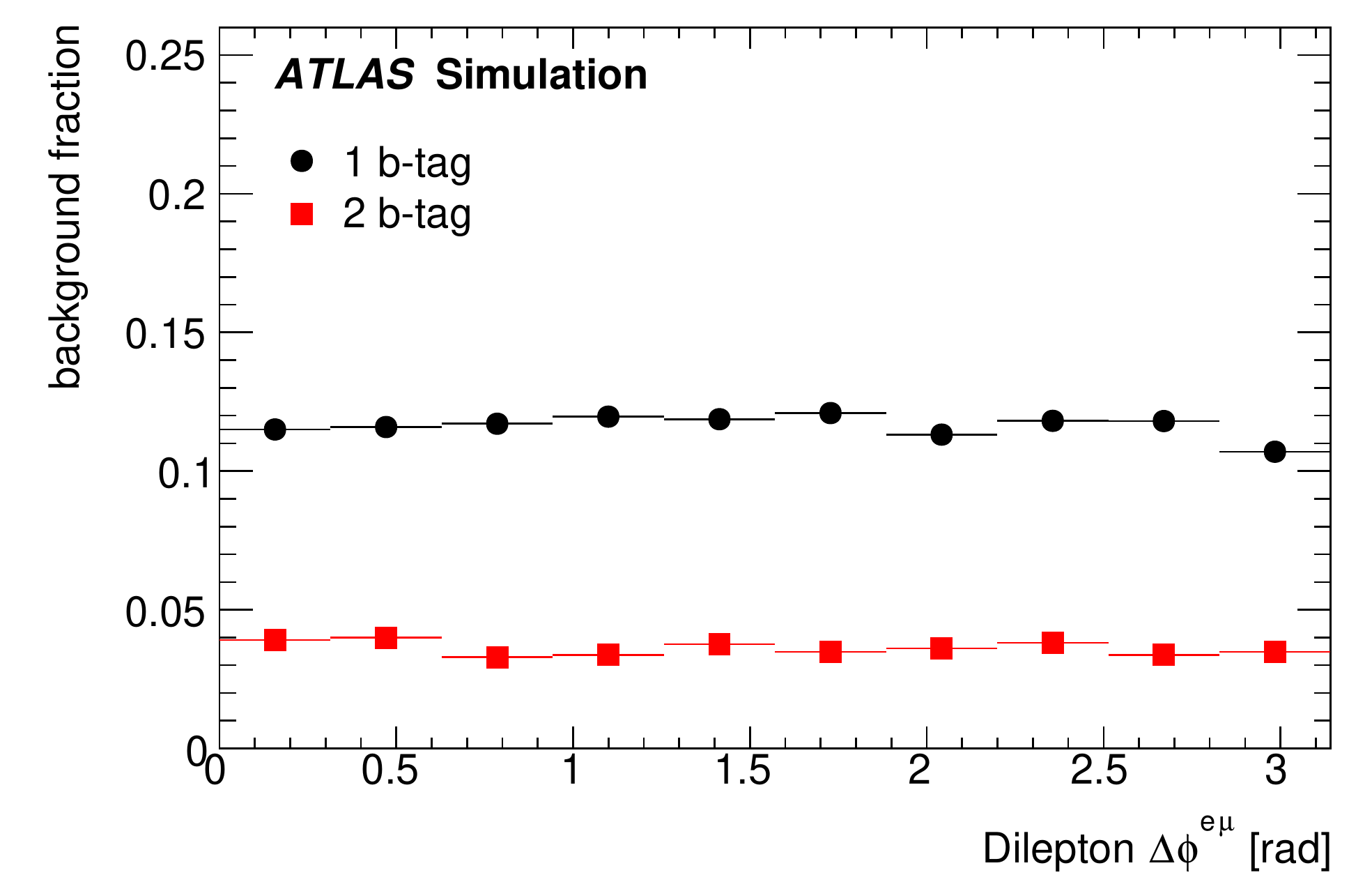}{e}{f}
\splitfigure{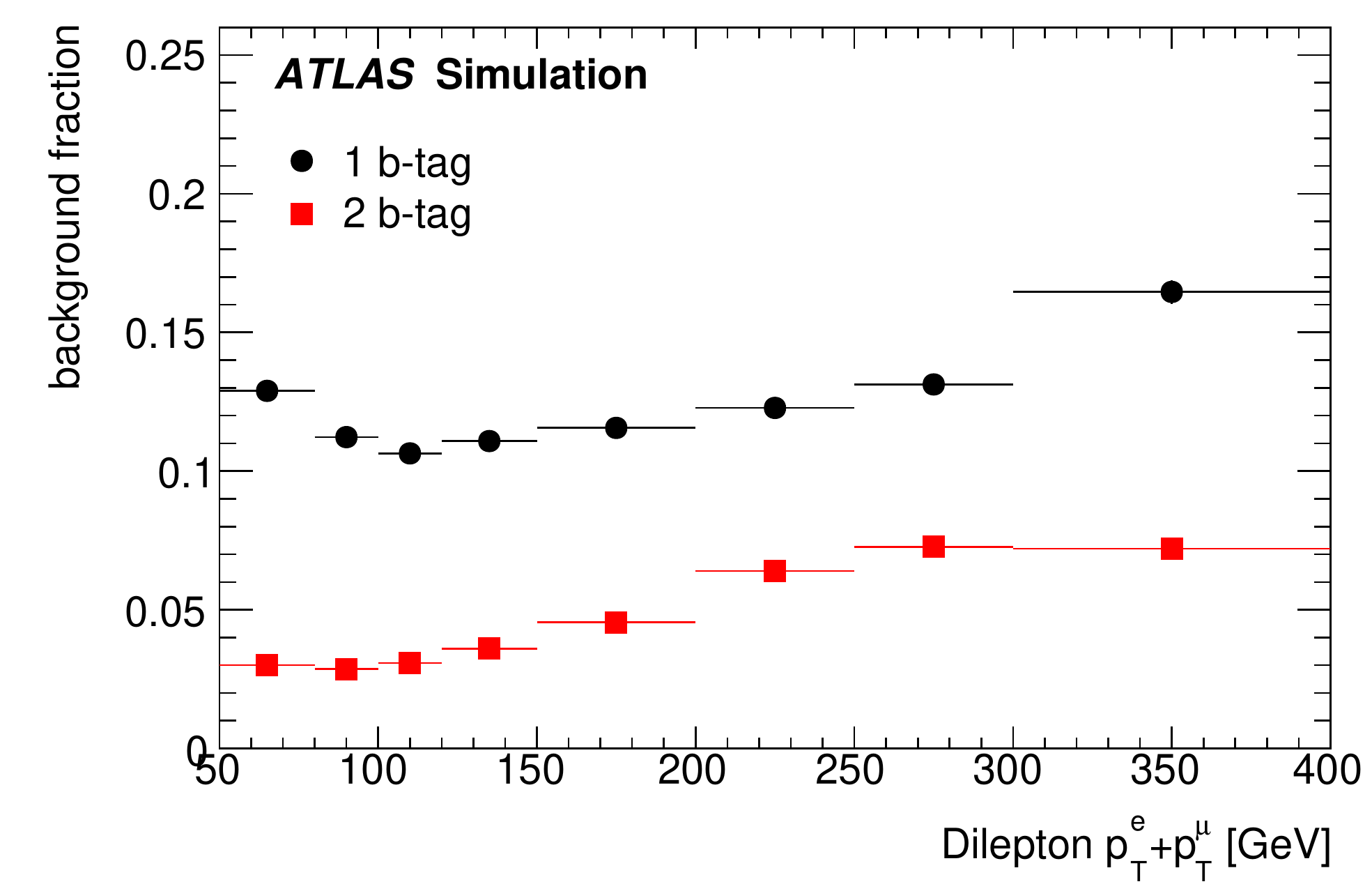}{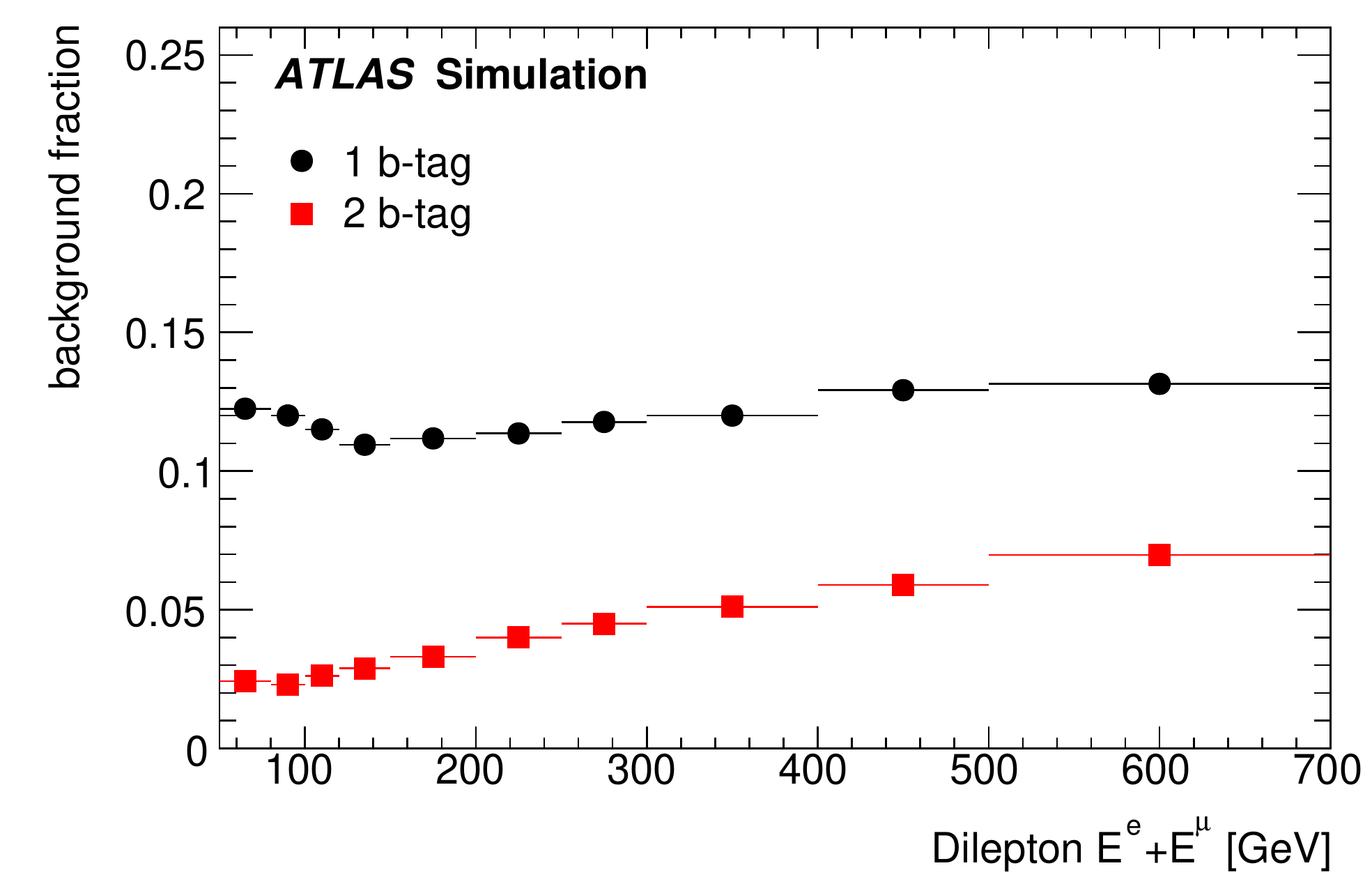}{g}{h}
\caption{\label{f:bgrate}Estimated background fractions in the one 
and two $b$-tagged samples as functions of each lepton and dilepton 
differential variable, estimated from simulation alone. The error bars
correspond to the statistical uncertainties of the simulation samples,
and are often smaller than the marker size.}
\end{figure}

The production cross-sections for $Z$ bosons accompanied by heavy-flavour
jets are subject to large theoretical uncertainties. The background
predictions from {\sc Alpgen\,+\,Pythia6} in each bin of each distribution
were therefore normalised from data, by multiplying them by constant scale 
factors of $1.4\pm 0.2$ for the one $b$-tagged jet sample and 
$1.1\pm 0.3$ for the two $b$-tagged jet sample. These
scale factors were derived from the comparison of data and simulated event 
yields for $Z\rightarrow ee$ and $Z\rightarrow\mu\mu$ plus one or two 
$b$-tagged jets, inclusively for all lepton pairs passing the
kinematic selections for electrons and muons \cite{TOPQ-2013-04}.
The uncertainties are dominated by the dependence of the scale factors
on lepton kinematics, investigated by studying their variation
with $Z$-boson \pt, reconstructed from the $ee$ or $\mu\mu$  system.  

The background from events with one real and one misidentified lepton was 
estimated using a combination of data and simulation in control regions with
an electron and muon of the same charge \cite{TOPQ-2013-04}. Simulation
studies showed that the samples with a same-sign $e\mu$ pair and one or two
$b$-tagged jets are dominated by events with a misidentified lepton, with
rates and kinematic distributions similar to those in the opposite-sign sample.
The distributions of the dilepton kinematic variables for same-sign events
with at least one $b$-tagged jet in data are shown in Figure~\ref{f:ssdilept},
and compared with the predictions from simulation. The expected 
contributions are shown separately for  events with two prompt leptons,  
events where the electron candidate
originates from a converted photon radiated from an electron produced in a
top quark decay, events with a converted photon from other sources, and events 
where the electron or muon originates from the decay of a bottom or
charm hadron. The analogous distributions for the electron and muon
\pt\ and $|\eta|$ are shown in Ref. \cite{TOPQ-2013-04}. 
In general, the simulation models the rates and kinematic distributions
of the same-sign events well. The modelling of misidentified leptons 
was further tested in control samples where either the electron or muon
isolation requirements were relaxed in order to enhance the contributions
from heavy-flavour decays, and similar levels of agreement were observed.

The contributions \nijfake\ of events with misidentified leptons to the 
opposite-sign samples with $j=1$, 2 $b$-tagged jets were 
estimated in each bin $i$ of each distribution using
\begin{equation}
\begin{array}{rll}
\nijfake & = & R^i_j (\nijdss-\nijphss) , \\*[1mm]
R^i_j & = & \frac{\nijfakeos}{\nijfakess},
\end{array}\label{e:fakeest}
\end{equation}
where \nijdss\ is the number of observed same-sign events in bin $i$ with
$j$ $b$-tagged jets, \nijphss\ is the estimated number of events in this 
bin with two prompt leptons, and \rij\ is the ratio of the number of opposite-
to same-sign events with misidentified leptons
in bin $i$ with $j$ $b$-tagged jets. This formalism uses
the observed data same-sign event rate in each bin to predict the 
corresponding opposite-sign contribution from misidentified leptons. It relies
on simulation to predict the ratios of opposite- to same-sign 
rates and the prompt same-sign contribution, but not the absolute normalisation 
of misidentified leptons.
The prompt-lepton contribution in Eq.~(\ref{e:fakeest}) comes mainly from 
semileptonic \ttbar\ events with an additional
$W$ or $Z$ boson, diboson events with two same-sign leptons, and 
$\ttbar\rightarrow e\mu\nu\bar{\nu}\bbbar$ events where the electron
charge was misreconstructed. These components were evaluated directly
from simulation in each bin $(i,j)$, and an uncertainty of $\pm 50$\,\% was 
assigned \cite{TOPQ-2013-04}. The values of \rij\ were taken from simulation,
separately for each differential distribution and 
$j=1$ and 2 $b$-tagged jets, and averaged over several consecutive bins $i$
in order to reduce statistical fluctuations. The values of $R^i_1$ range
from 0.8 to 1.5, and $R^i_2$ from 1.2 to 2.0, as the predicted background
composition changes across the kinematic distributions. As in Ref. 
\cite{TOPQ-2013-04}, uncertainties of $\pm 0.25$ and $\pm 0.5$ were assigned
to $R^i_1$ and $R^i_2$, based on the variation of $R^i_j$ for 
different components of the misidentified lepton background, and taken to be
correlated across all bins $(i,j)$.

\begin{figure}
\vspace{-7mm}
\splitfigure{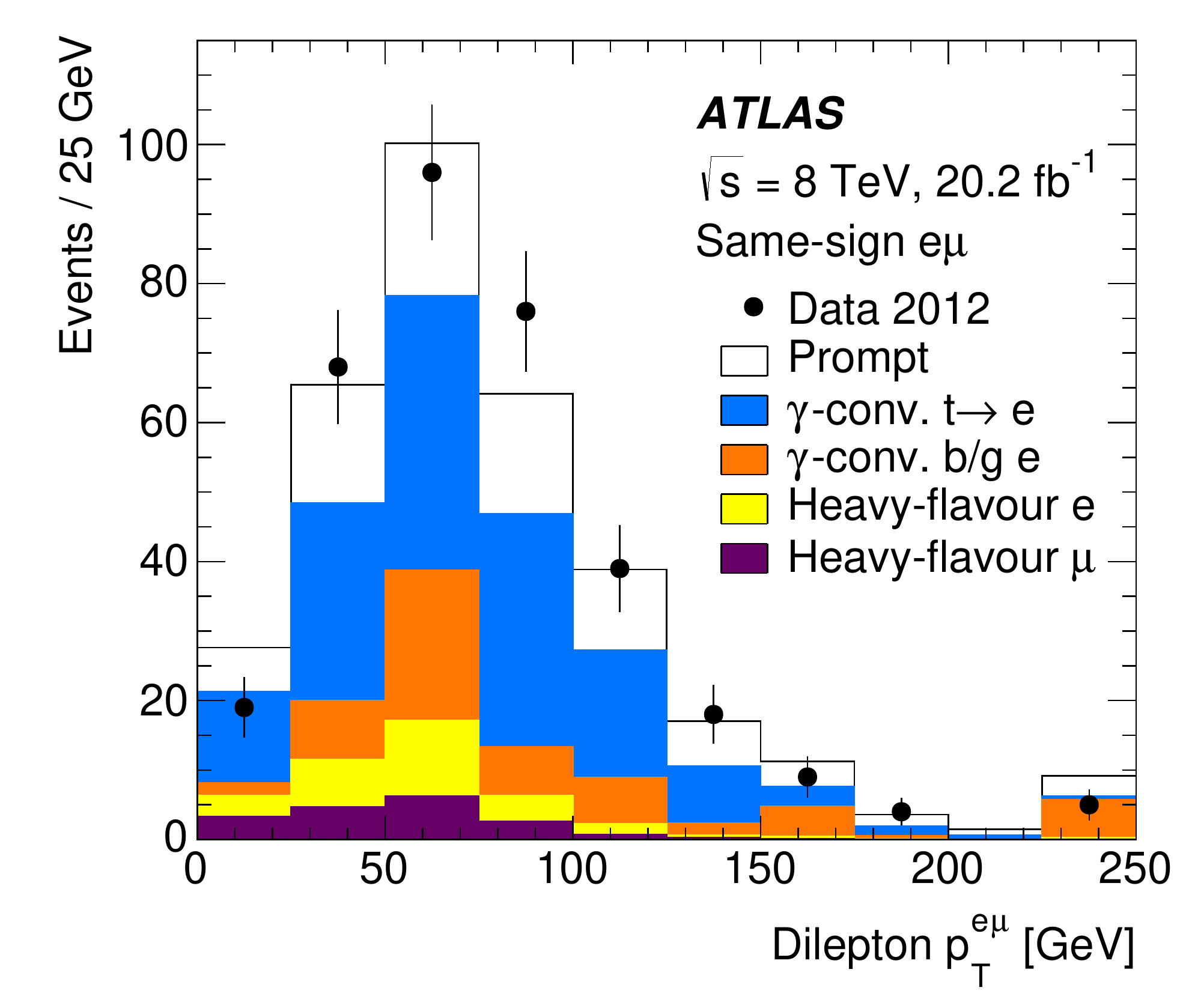}{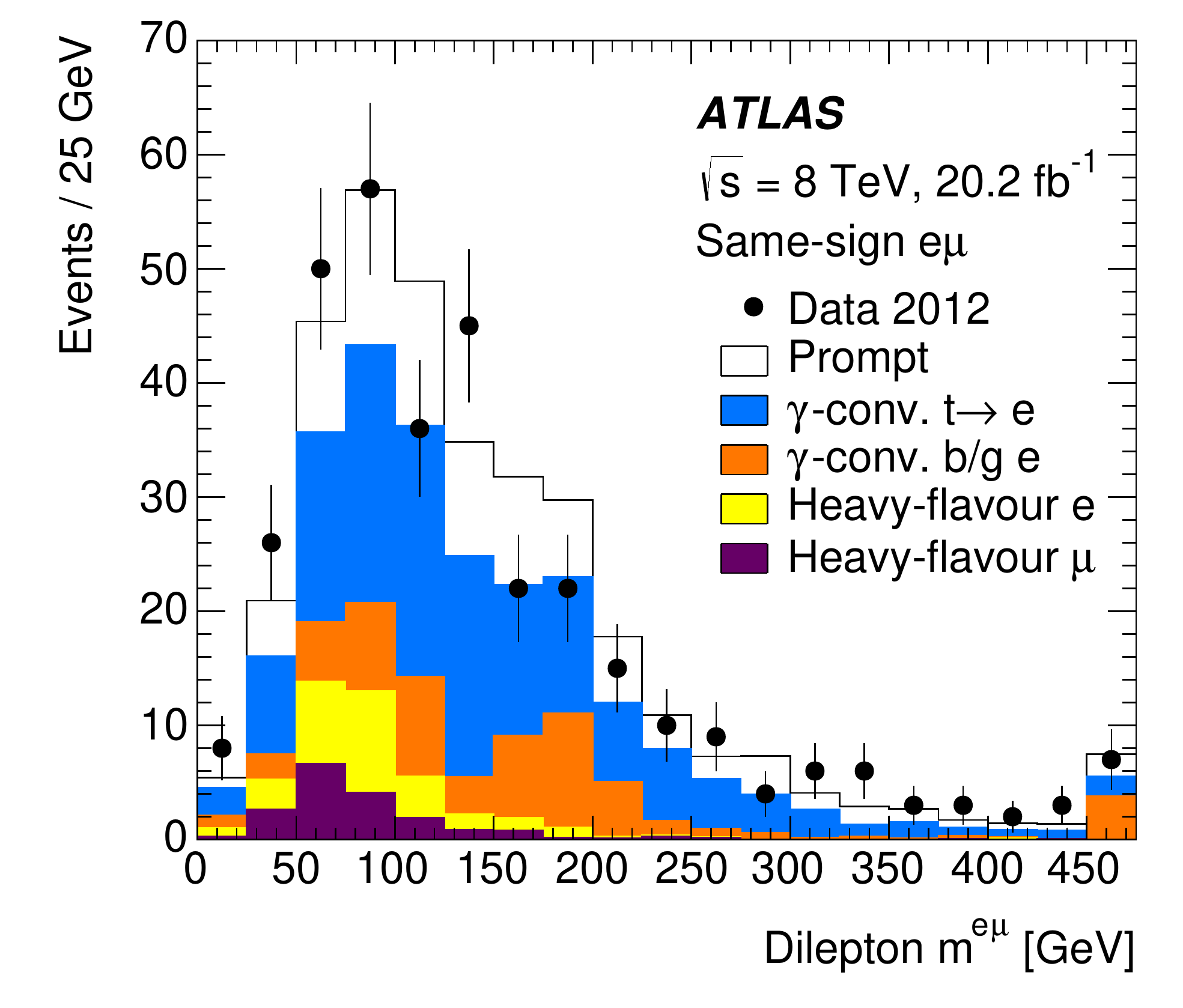}{a}{b}
\splitfigure{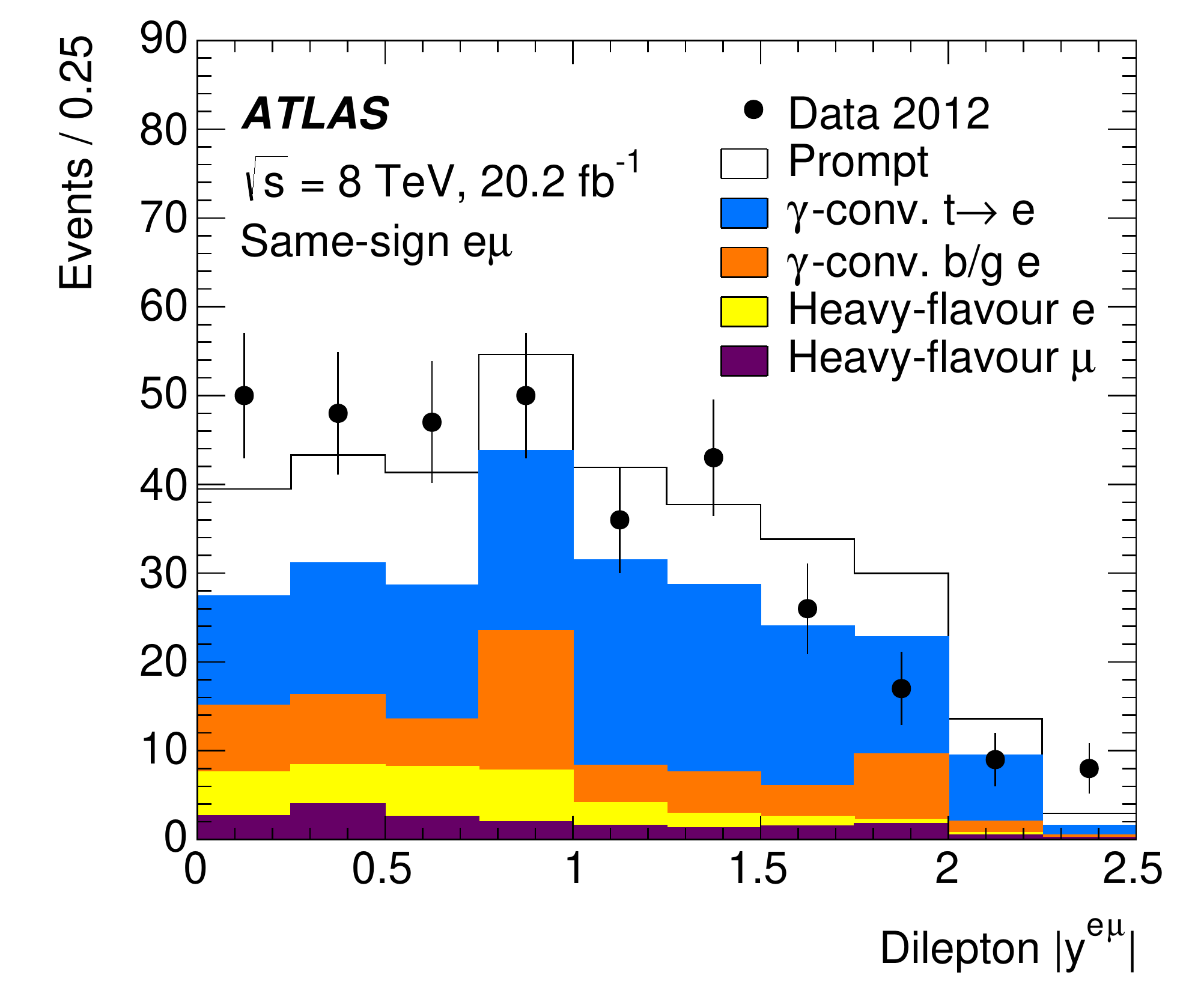}{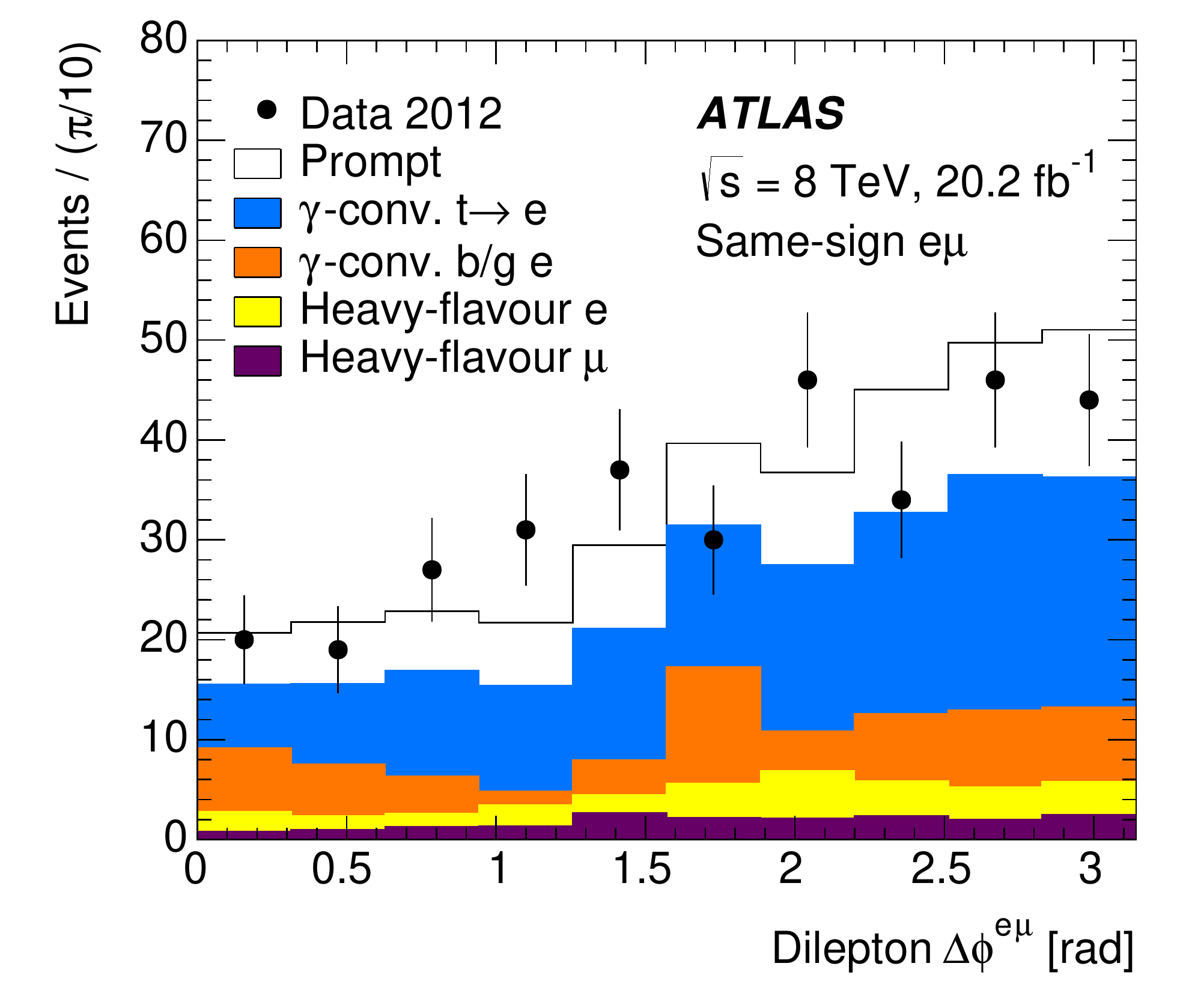}{c}{d}
\splitfigure{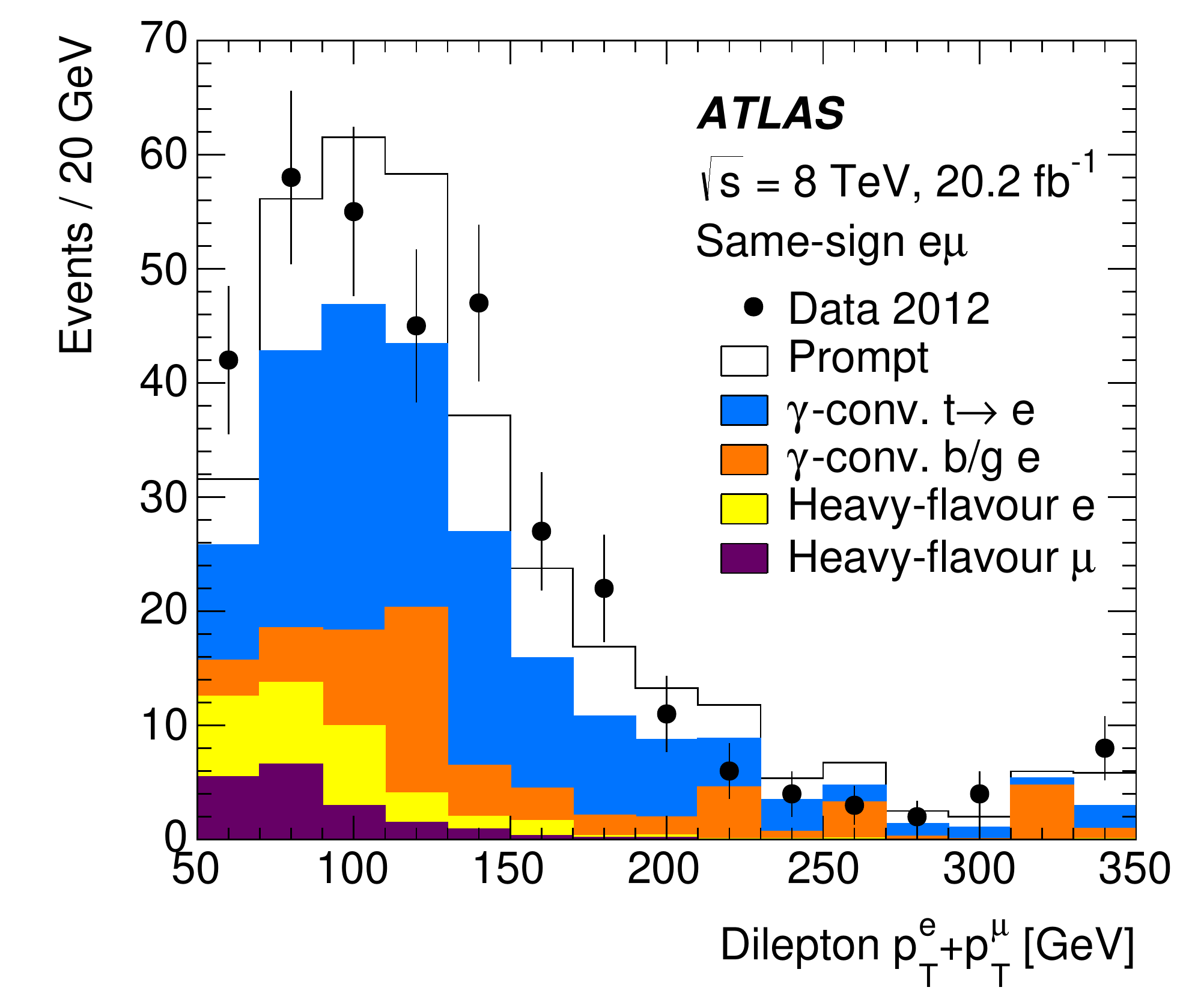}{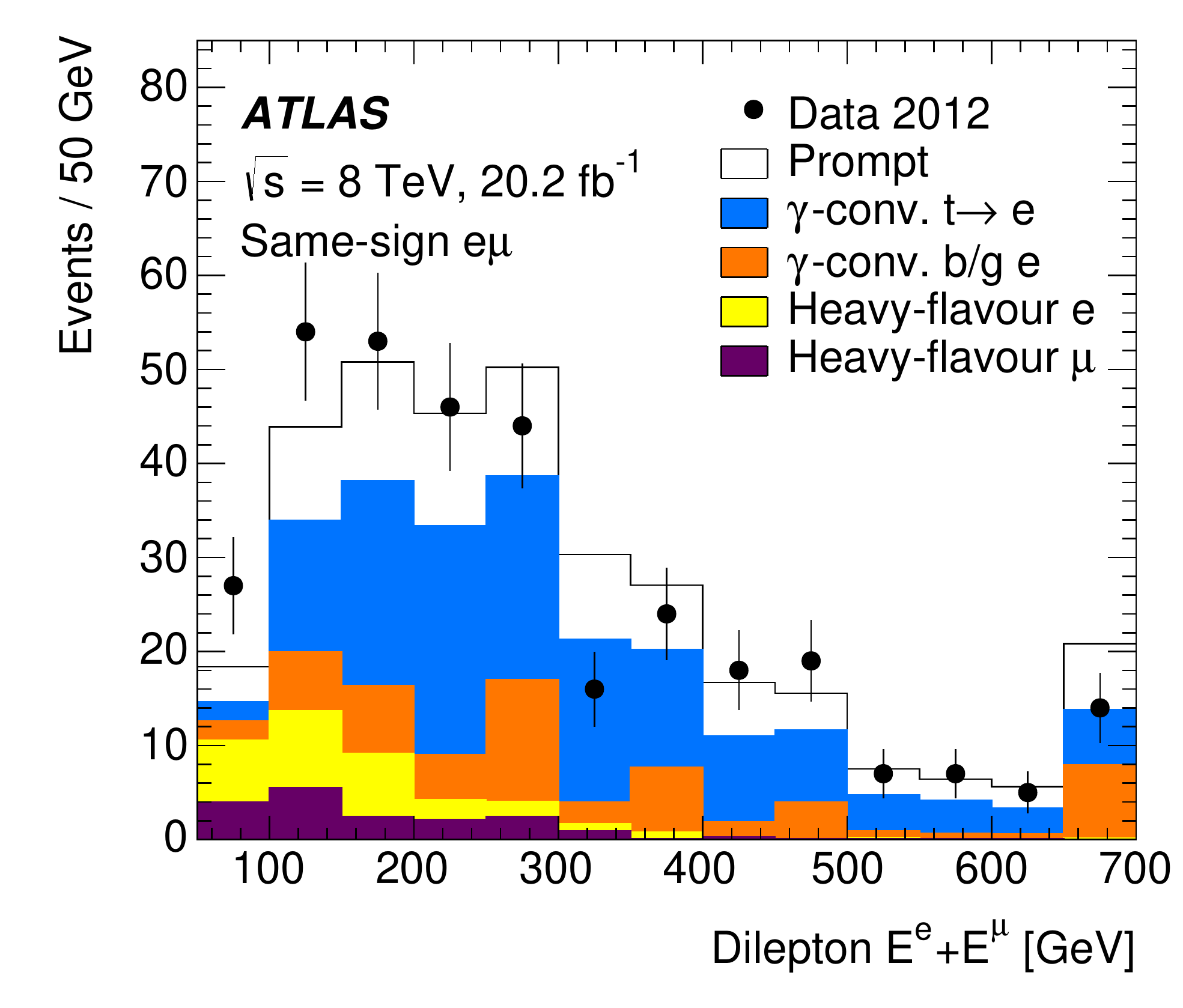}{e}{f}
\caption{\label{f:ssdilept}Distributions of
(a) the dilepton \ptll, (b) invariant mass \mll, (c) rapidity \rapll,
(d) azimuthal angle difference \dphill, (e) lepton \pt\ sum \ptsum\ and
(f) lepton energy sum \esum, in events with a same-sign $e\mu$ pair and 
at least one $b$-tagged jet. The simulation prediction is normalised to the
same integrated luminosity as the data, and broken down into contributions
where both leptons are prompt, or one is a misidentified lepton from a photon
conversion originating from a top quark decay or from background, or from
heavy-flavour decay. In the \ptll, \mll, \ptsum\ and \esum\ distributions, 
the last bin includes the overflows.}
\end{figure}

\subsection{Validation of the analysis procedure}\label{ss:valid}

The method for the differential cross-section determination was tested on 
simulated events in order to check for biases and determine the expected 
statistical uncertainties. Pseudo-data samples corresponding to the 
data integrated luminosity were produced by varying the 
event counts \nxi\ and \nyi\ in each bin $i$ independently, 
according to Poisson 
distributions with mean values predicted from 
a chosen \ttbar\ simulation sample plus
non-\ttbar\ backgrounds. The tagging equations Eqs.~(\ref{e:fidtags}) were then
solved for each pseudo-experiment using the values of \gemi, \cbi, \nibi\ 
and \niibi\ calculated with the baseline simulation samples. An initial set
of 1000 pseudo-experiments was performed using the baseline simulation sample
as a reference, and the mean and RMS width of the deviations of the result 
in each
bin from the reference values were used to validate the analysis procedure.
The black points in Figure~\ref{f:biastest} show the mean deviation of
the results (averaged over all pseudo-experiments) for four of
the measured normalised distributions, with error bars 
corresponding to the uncertainty in the mean due to the finite size of 
the simulation samples
(about 17 times the data integrated luminosity). The residual biases of the
mean deviations away from the reference are compatible with zero and in all 
cases much smaller than the expected statistical uncertainties in data, 
measured by the RMS widths and shown by the cyan bands. 
Similar results were obtained for the other normalised differential
cross-section distributions, and for the absolute distributions.
The pull distributions (i.e. the
distributions of deviations divided by the estimated statistical uncertainty
from each pseudo-experiment) were also found to have widths within a few 
percent of unity. The $\chi^2$ values for the compatibility of each measured 
distribution with the reference were also calculated for each 
pseudo-experiment and the distribution of the corresponding $p$-values 
across all pseudo-experiments was found to be uniform between zero and one.
These tests confirm
that the analysis procedure is unbiased and correctly estimates the 
statistical uncertainties in each bin of each distribution.

\begin{figure}[tp]
\splitfigure{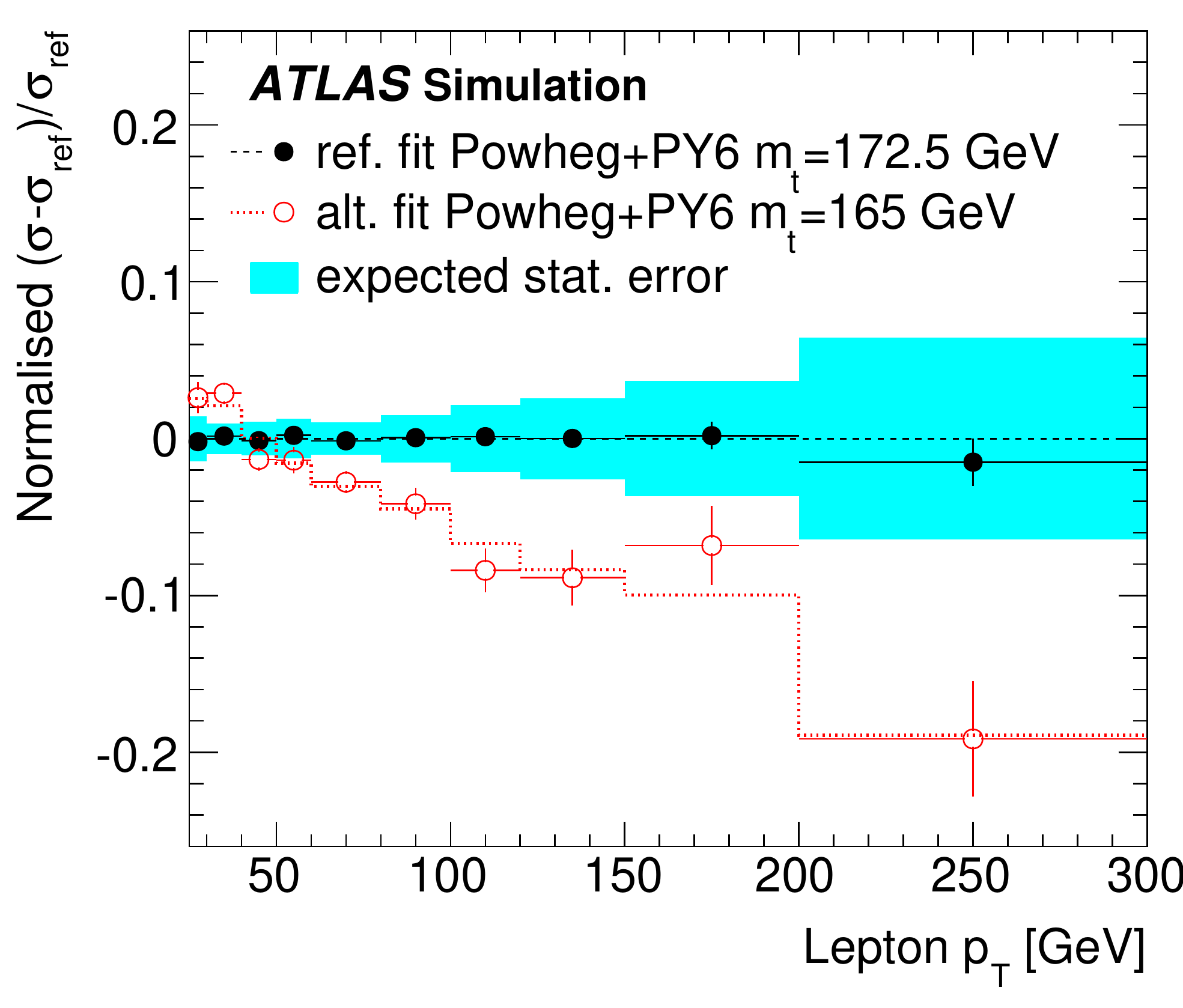}{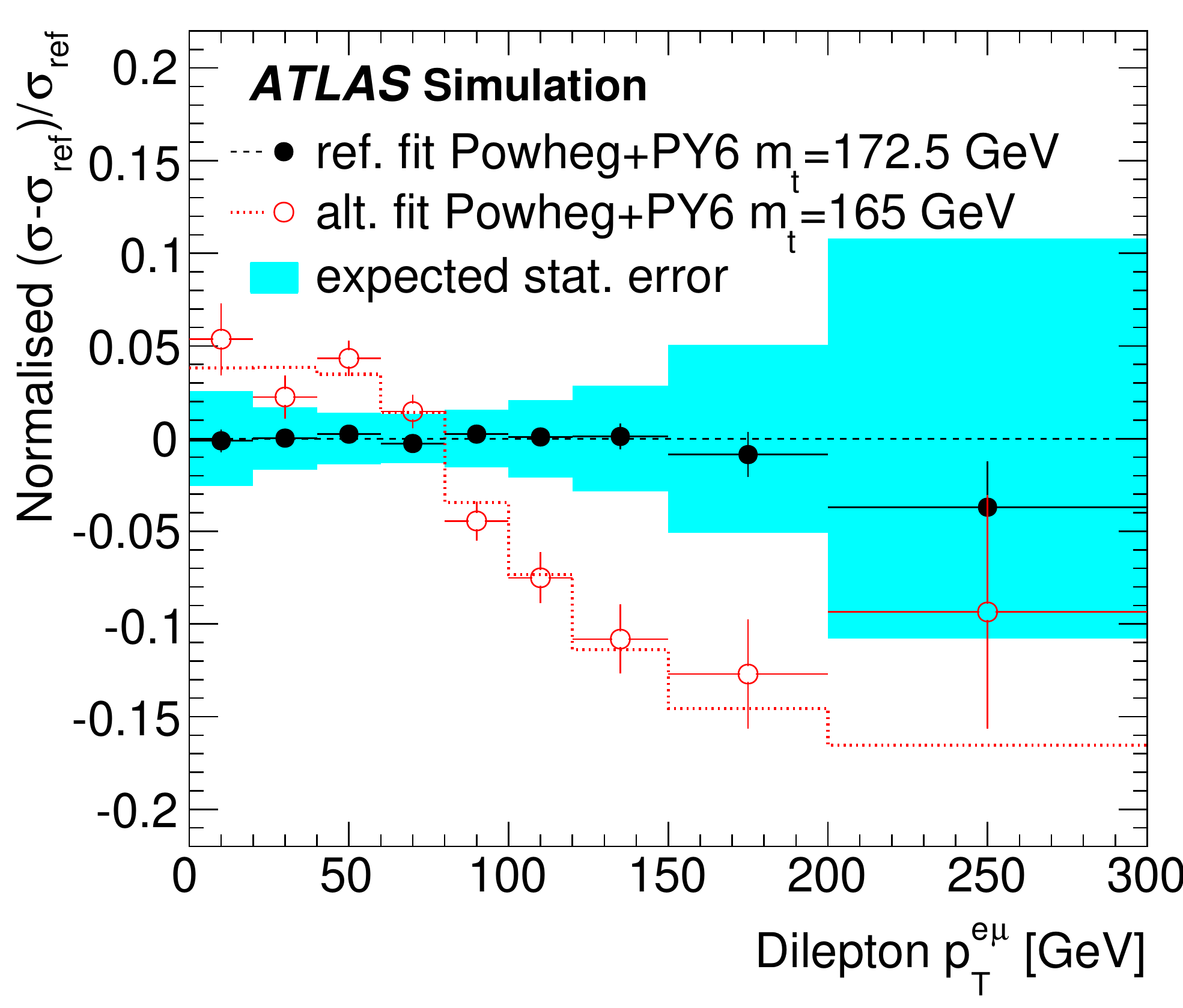}{a}{b}
\splitfigure{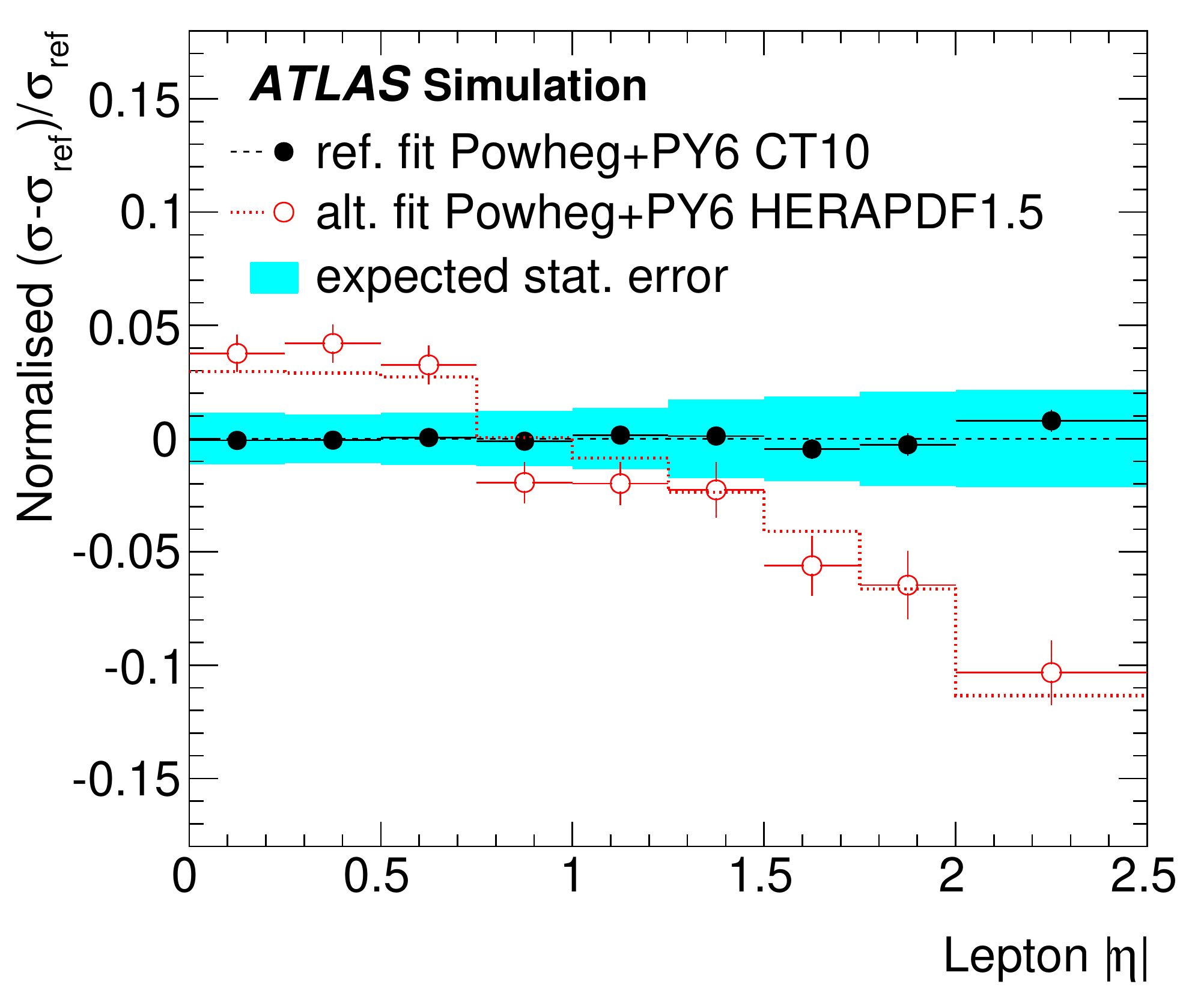}{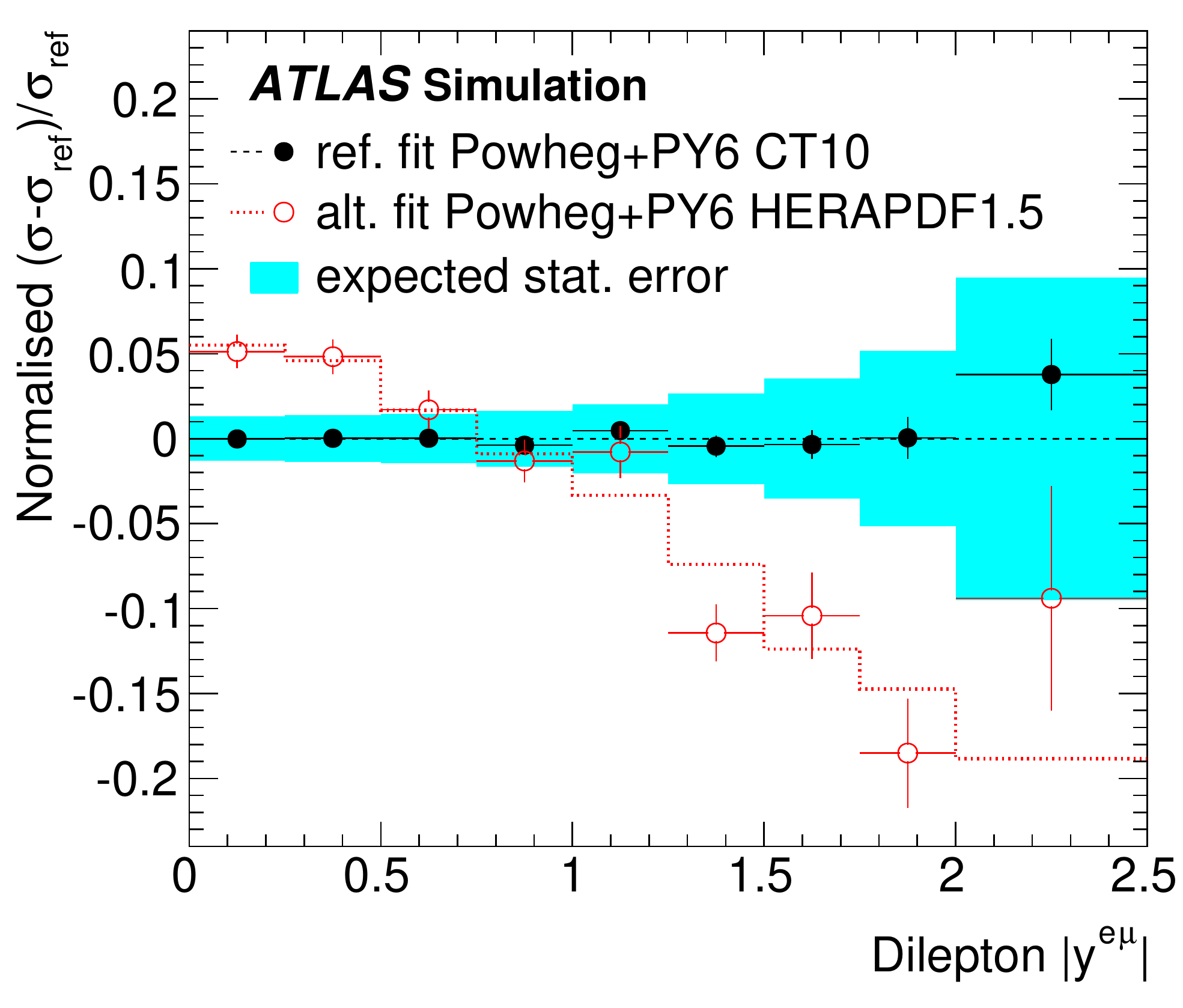}{c}{d}
\caption{\label{f:biastest}Results of pseudo-experiment studies on simulated
events for the extraction of the normalised differential cross-section
distributions for (a) \ptl, (b) \ptll, (c) \etal\ and (d) \rapll, shown as
relative deviations $(\sigma-\sigma_{\mathrm{ref}})/\sigma_{\mathrm{ref}} $
from the reference cross-section values in the baseline {\sc Powheg+Pythia6} 
CT10 sample with $\mtop=172.5$\,\GeV.
The black points show the mean deviations from the reference when fitting
pseudo-data samples generated with the baseline simulation sample, with
error bars indicating the uncertainties due to the limited number of simulated
events.
The cyan bands indicate the expected statistical uncertainties for a single
sample corresponding to the data integrated luminosity. The open red points 
show the mean deviations from the reference values when fitting 
pseudo-experiments
generated from alternative simulation samples with $\mtop=165$\,\GeV\ (a, b)
or with the HERAPDF 1.5 PDF (c, d), with error bars due to the limited size
of these alternative samples. The red dotted lines show the
true deviations from the reference in the alternative samples.}
\end{figure}

Additional pseudo-experiments were performed to test the ability of the
analysis procedure to reconstruct distributions different from the reference,
taking the values of \gemi, \cbi, \nibi\ and \niibi\ from the baseline
samples. Tests were conducted using simulated {\sc Powheg\,+\,Pythia6} and 
{\sc MC@NLO\,+\,Herwig} \ttbar\ samples with different top mass values, 
a {\sc Powheg\,+\,Pythia6} sample generated using the HERAPDF 1.5 
\cite{herapdf10,herapdf15}  PDF set
instead of CT10, and a {\sc Powheg\,+\,Pythia6}
sample reweighted to reproduce the top quark \pt\ distribution calculated
at NNLO from Ref. \cite{topptnnlo}. In all cases, the analysis procedure
recovered the true distributions from the alternative samples within the
statistical precision of the test, demonstrating the adequacy of the
bin-by-bin correction procedure without the need for iteration or a more
sophisticated matrix-based unfolding technique.  
Some examples are shown by the red points 
and dotted lines in Figure~\ref{f:biastest},
for an alternative sample with $\mtop=165$\,\GeV\ for \ptl\ and \ptll,
and for HERAPDF 1.5 for \etal\ and \rapll, both simulation samples having about
twice the statistics of the data. These figures also demonstrate the 
sensitivities of some of the measured distributions to \mtop\ 
and different PDFs.

For the single-lepton distributions \ptl\ and \etal, which have two entries per 
event, the formalism of Eqs.~(\ref{e:fidtags}) and the pseudo-experiments 
generated by fluctuating each bin independently do not take into account
correlations between the kinematics of the electron and muon in each event.
This effect was checked by generating pseudo-data samples 
corresponding to the data integrated luminosity from individual
simulated events, taken at random from a large \ttbar\ sample 
combining both full and fast simulation and corresponding to about 70 times the 
data integrated luminosity. The effect of neglecting the electron-muon 
correlations within an event was found to correspond to at most a 2\,\% 
fractional overestimate of the absolute and 2\,\% fractional underestimate 
of the normalised
cross-section uncertainties. Hence, no corresponding corrections to the 
statistical uncertainties were made.


\section{Systematic uncertainties}\label{s:syst}

Systematic uncertainties in the measured cross-sections arise from
uncertainties in the values of the input quantities \gemi, \cbi, \nibi, 
\niibi\ and $L$ used in
Eqs.~(\ref{e:fidtags}). Each source of systematic uncertainty was evaluated
by coherently changing the values of all relevant input quantities and
re-solving Eqs.~(\ref{e:fidtags}), thus taking into account correlations of
the uncertainties in e.g. \gemi\ and \cbi. The uncertainties are 
divided into five groups (\ttbar\ modelling, leptons, jets/$b$-tagging, 
background and luminosity/beam energy uncertainties) and are discussed
in Sections~\ref{ss:ttsyst} to~\ref{ss:elumisyst}. The resulting
relative uncertainties in each measured differential cross-section value 
are shown in the
results Tables~\ref{t:insXSec1} to~\ref{t:insXSec4}, and the grouped 
systematic
uncertainties for the normalised differential cross-sections are shown in 
Figure~\ref{f:fracsyst}, together with the statistical and total uncertainties.

\begin{figure}[tp]
\splitfigure{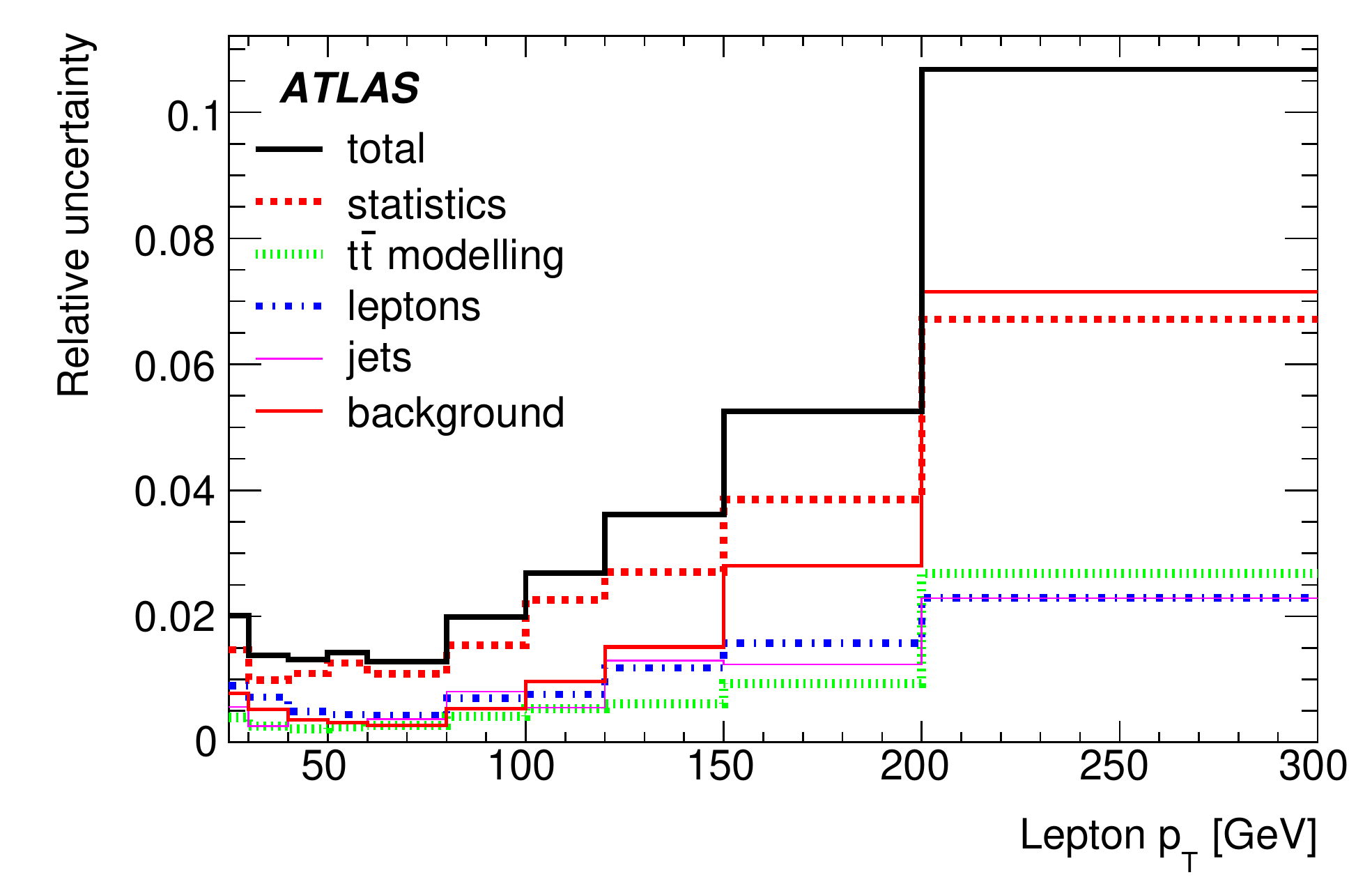}{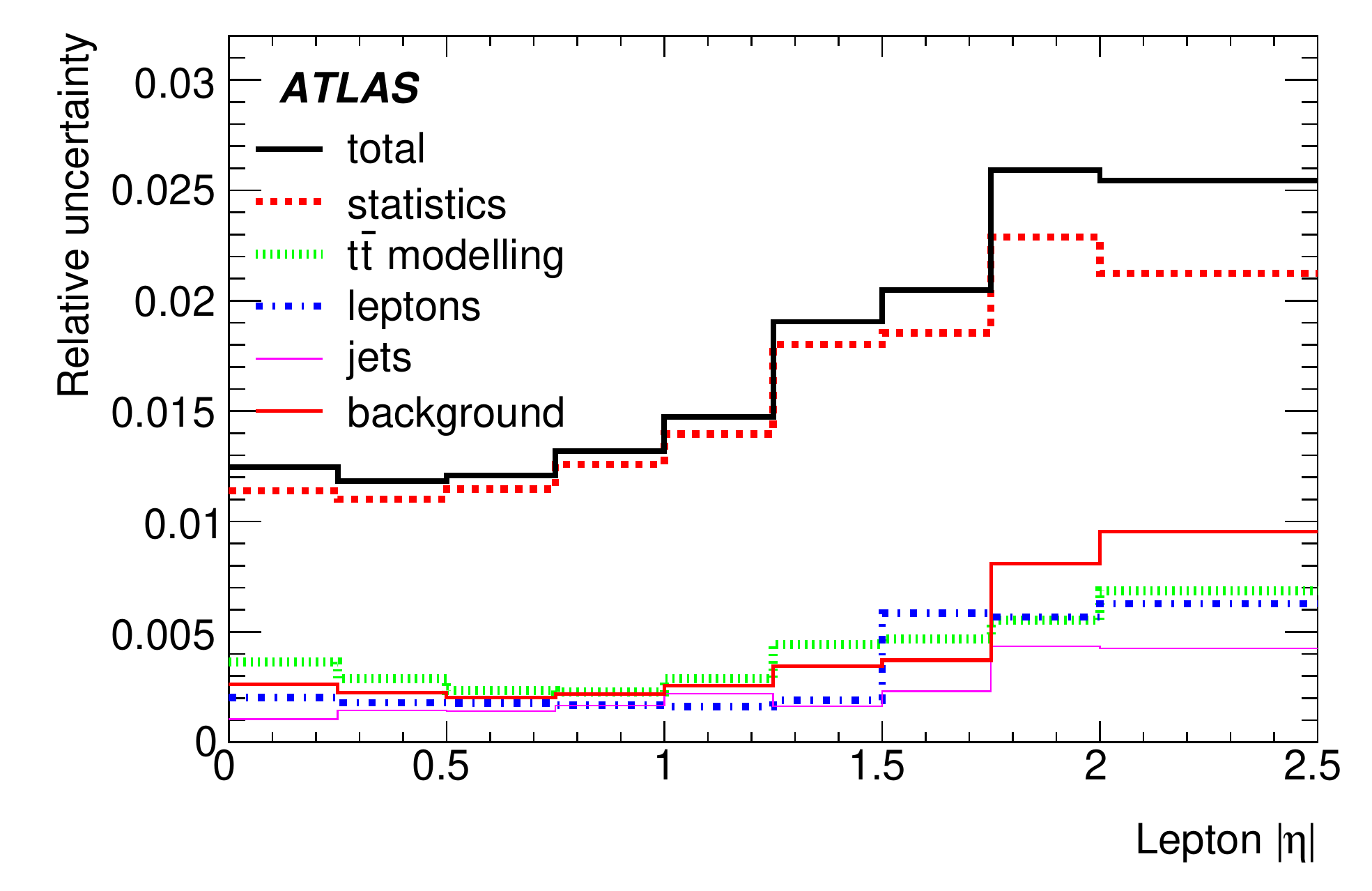}{a}{b}
\splitfigure{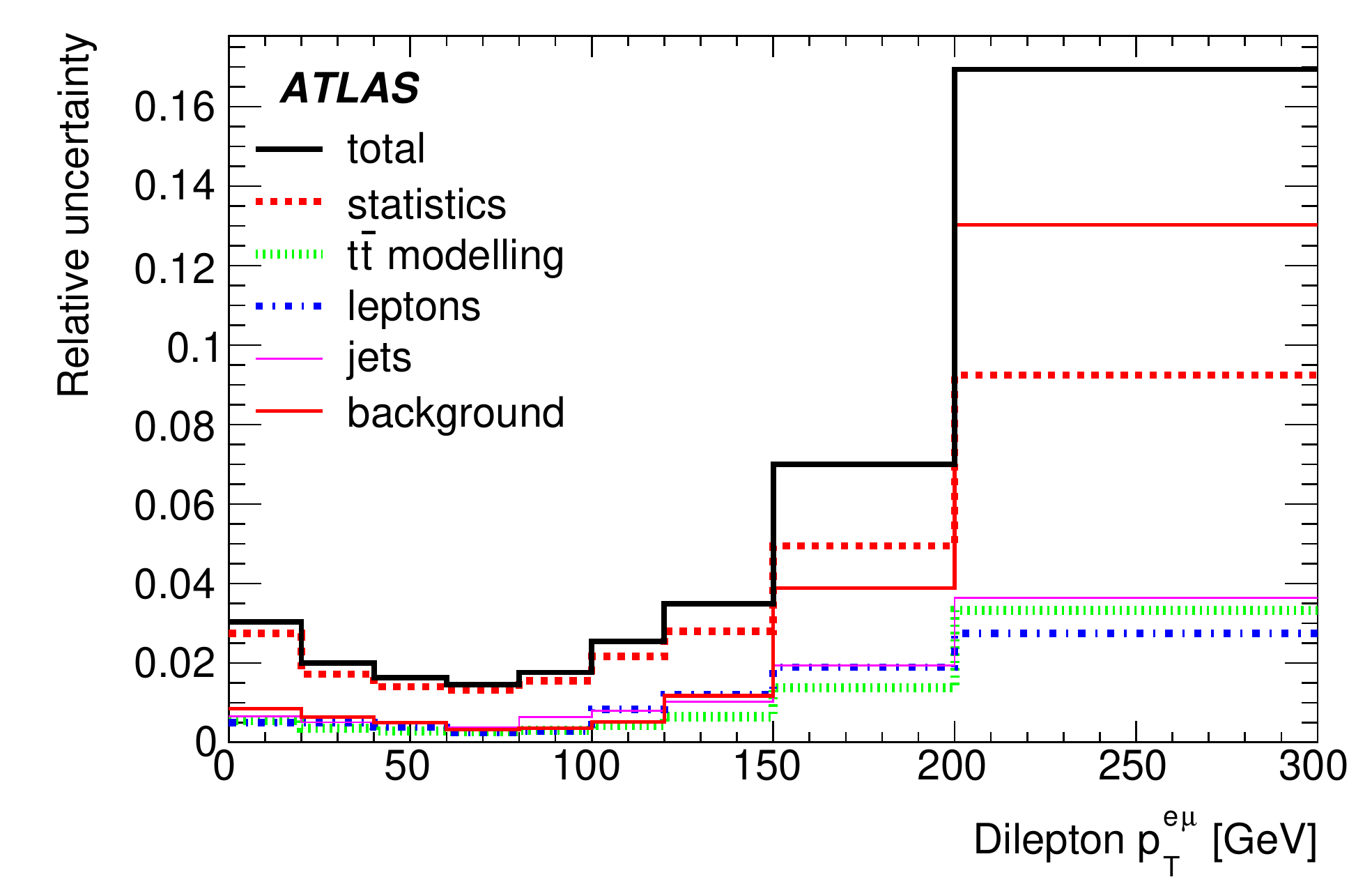}{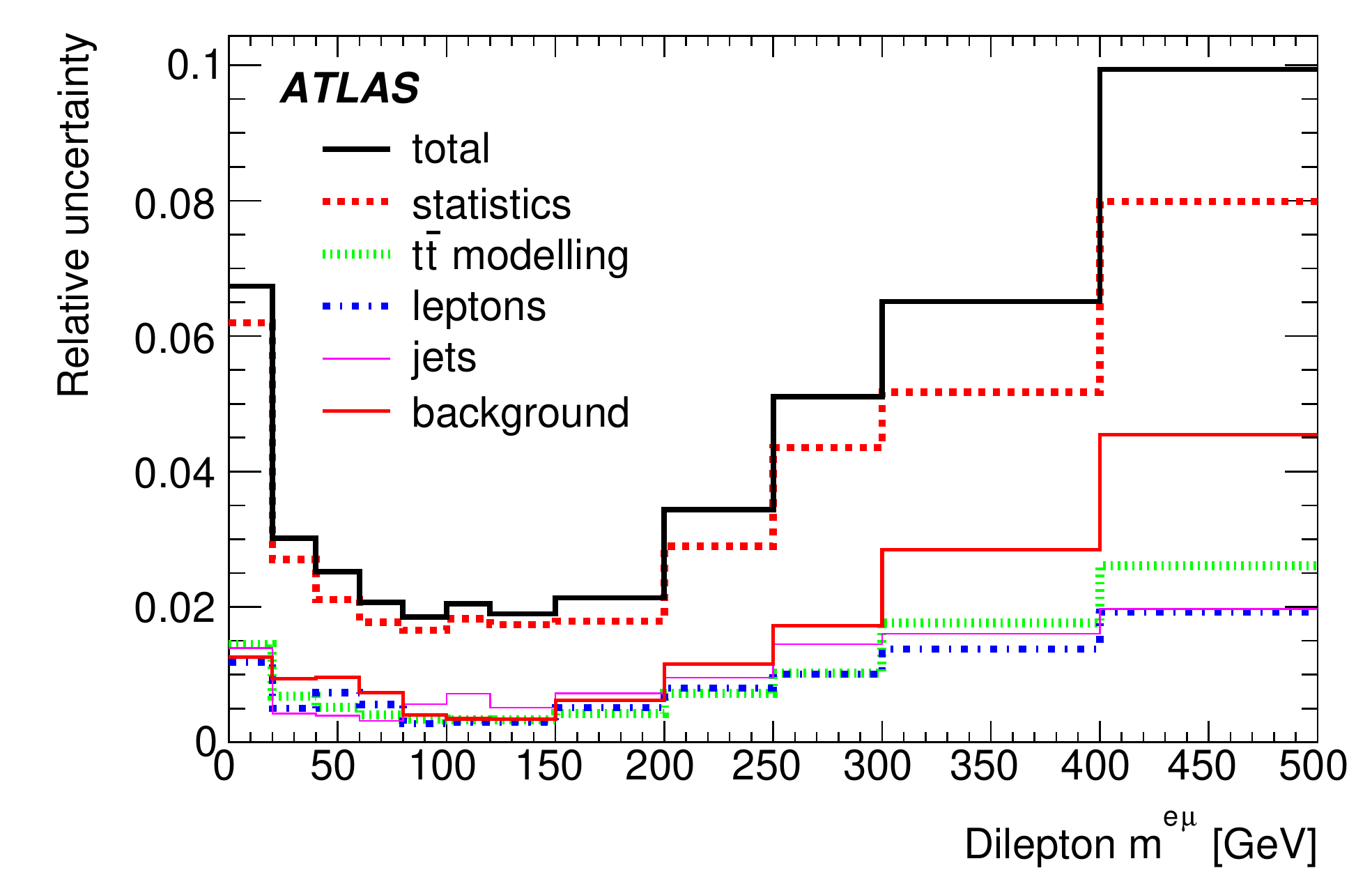}{c}{d}
\splitfigure{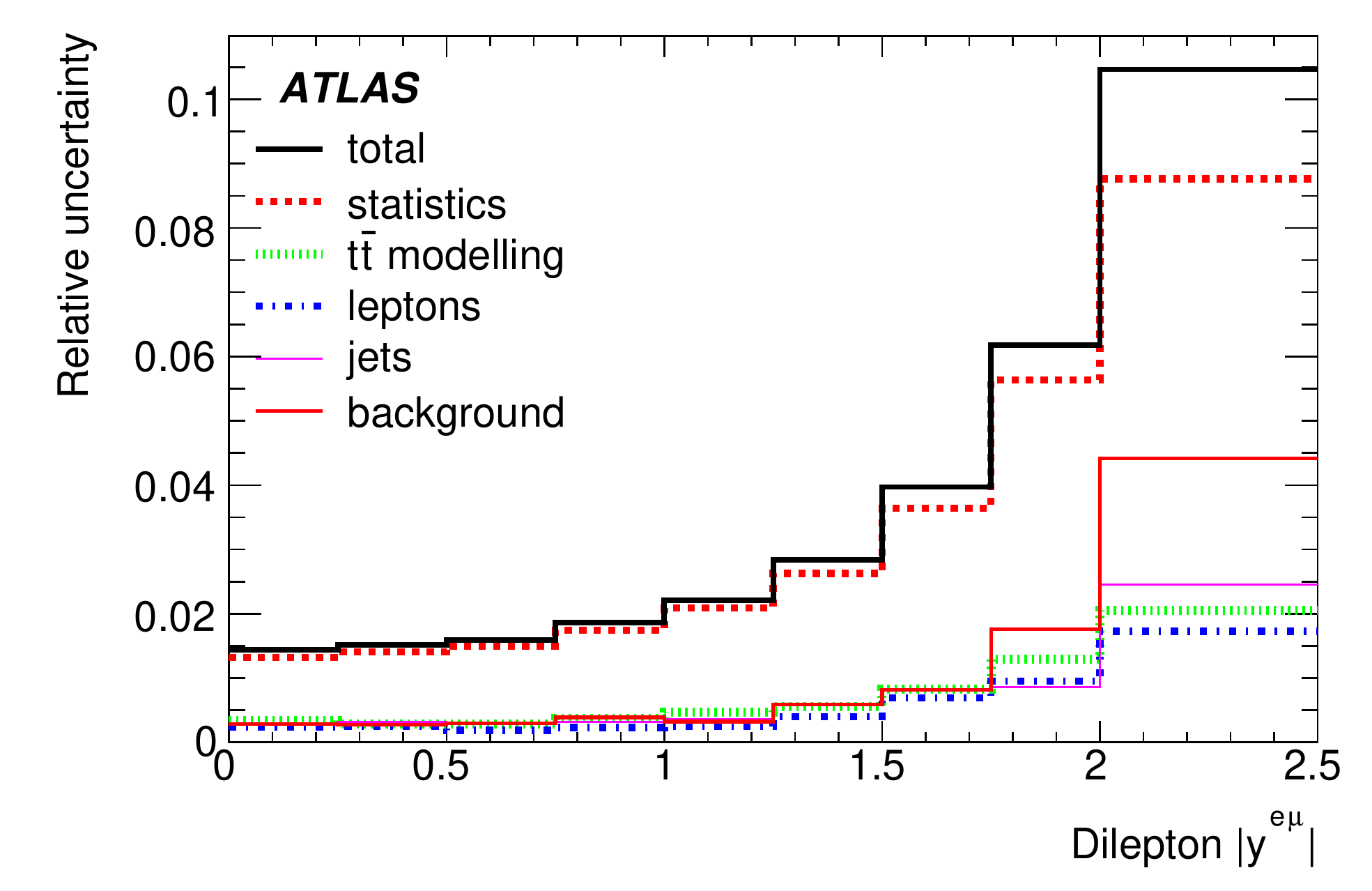}{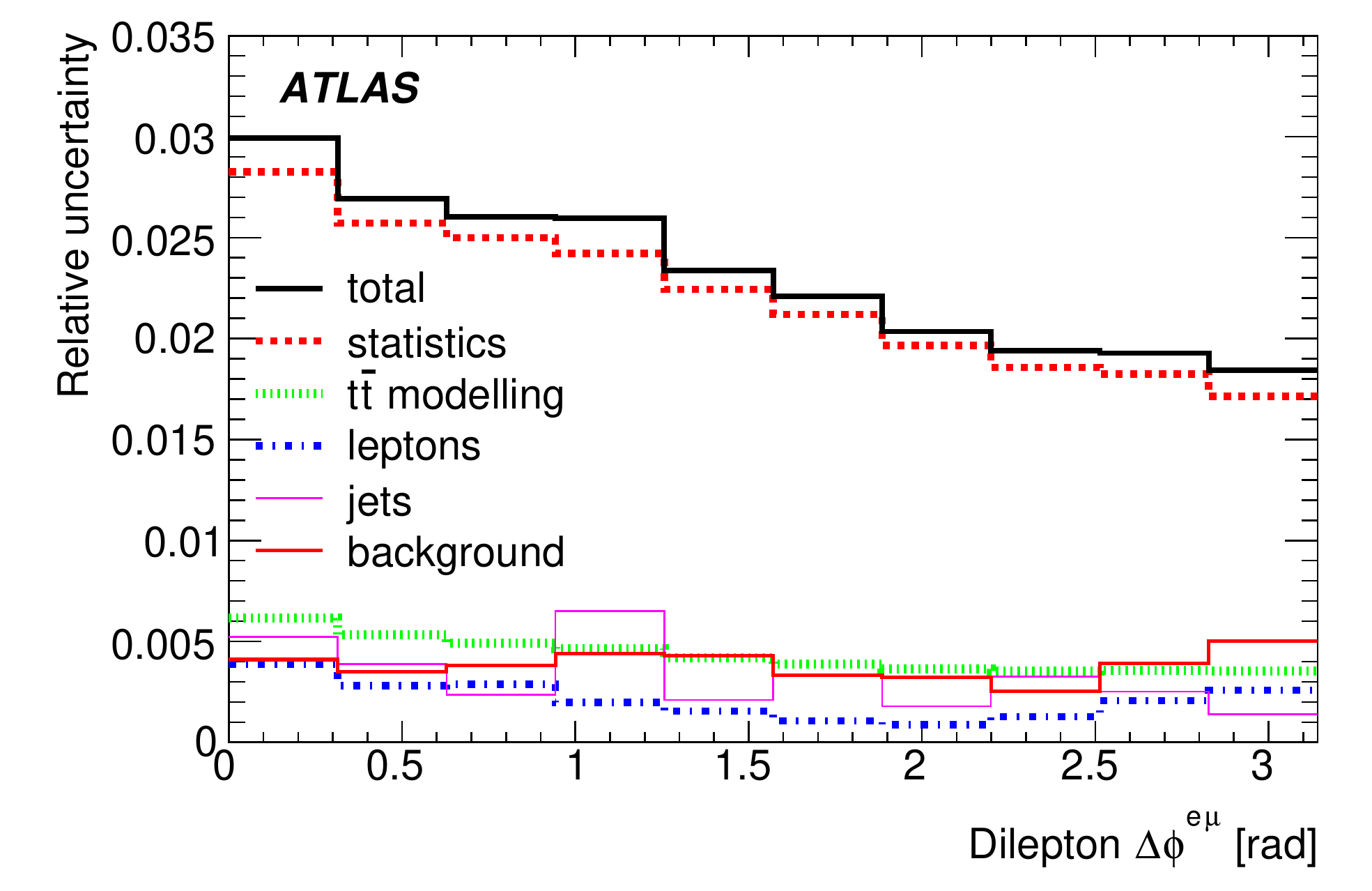}{e}{f}
\splitfigure{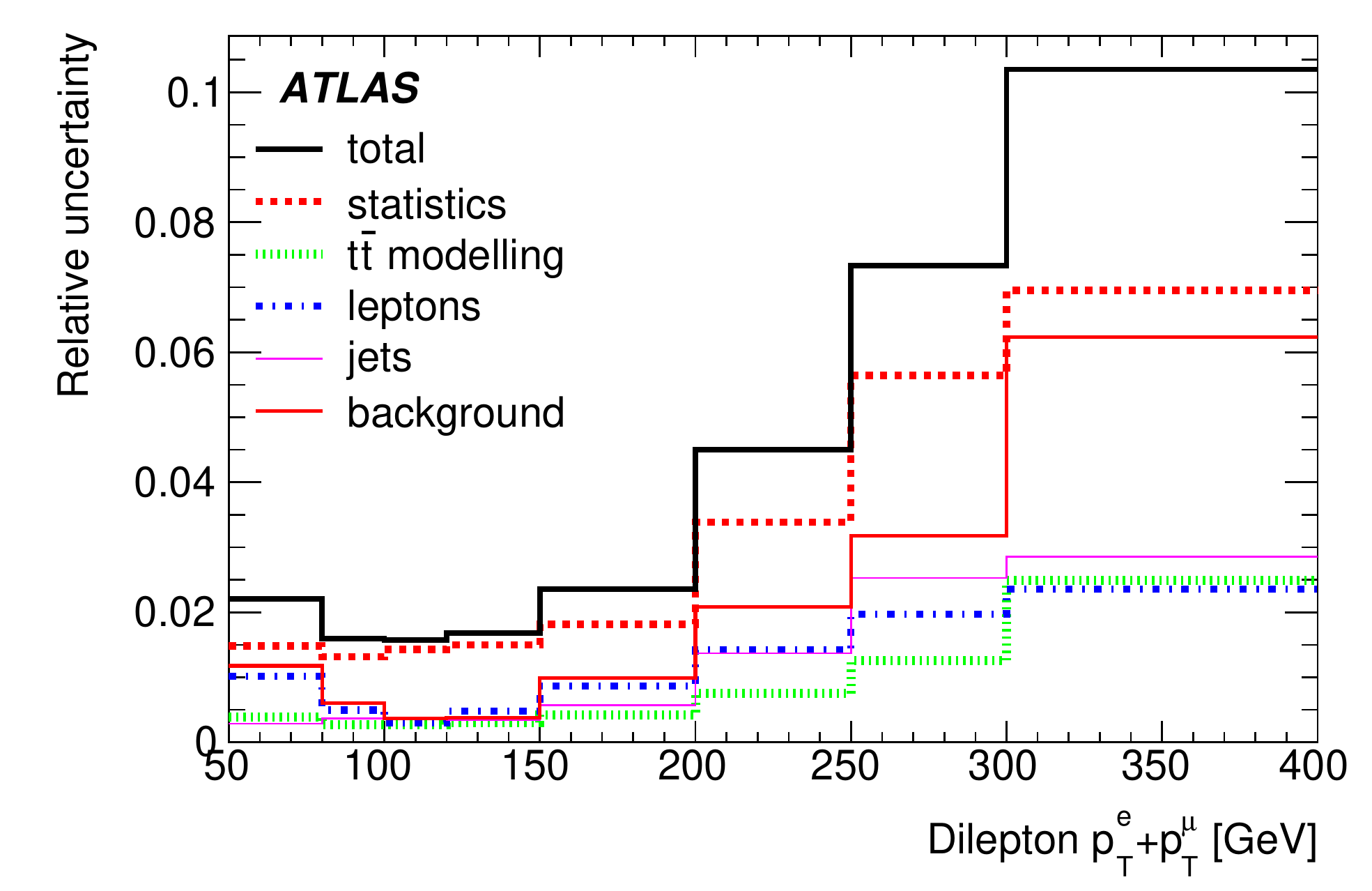}{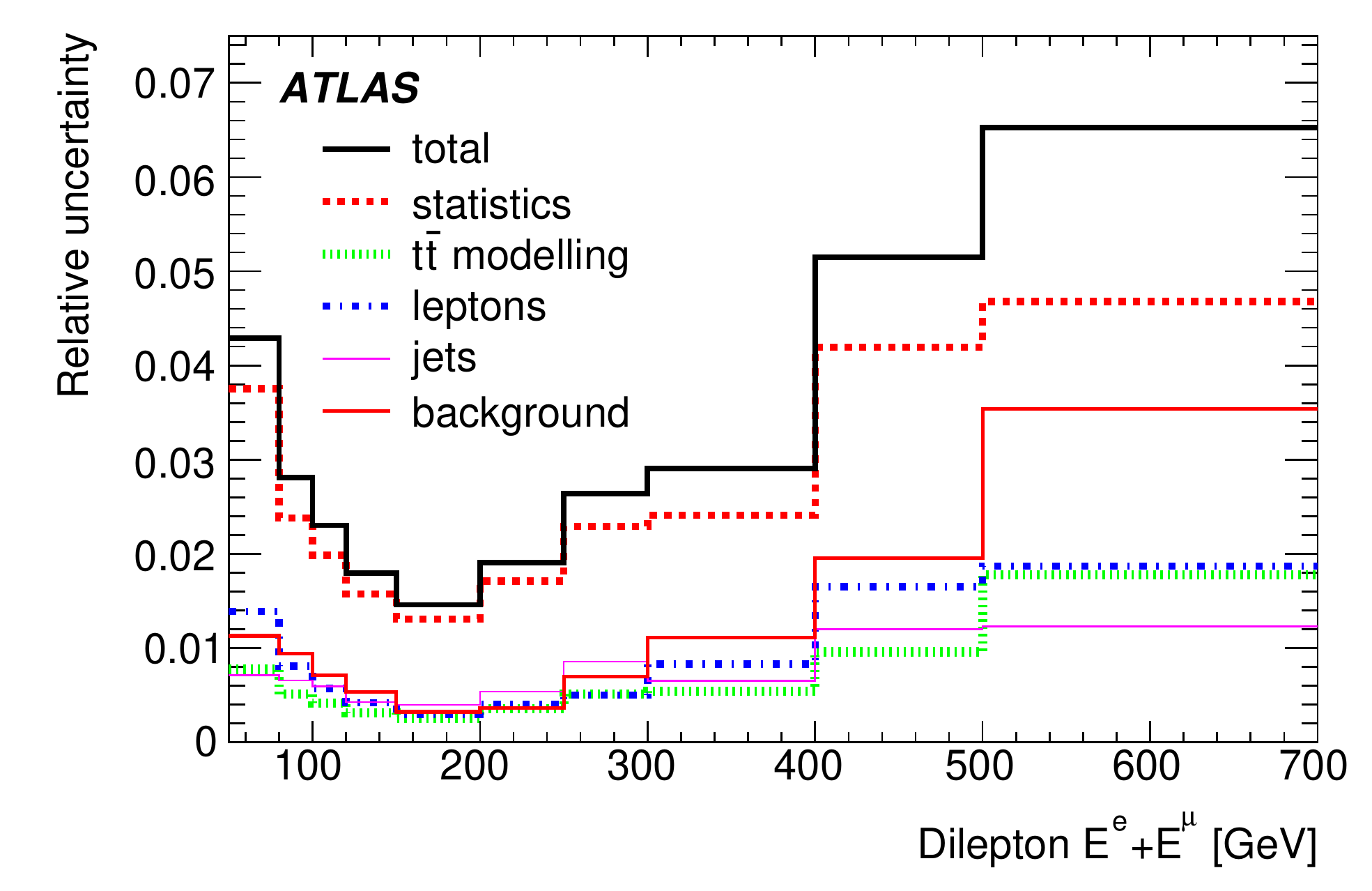}{g}{h}
\caption{\label{f:fracsyst}Relative uncertainties  on the measured normalised
differential cross-sections coming from data statistics, 
\ttbar\ modelling, leptons, jets and background,  as a function of each lepton 
or dilepton
differential variable. The total uncertainty is shown by the black lines,
and also includes small contributions from the integrated luminosity and
LHC beam energy uncertainties.}
\end{figure}

\subsection{\ttbar\ modelling}\label{ss:ttsyst}

The uncertainties in \gemi\ and \cbi\ (and \fntaui\ for the $\tau$-corrected 
cross-sections) were evaluated using the various alternative \ttbar\ simulation
samples detailed in Section~\ref{s:datasim}.

\begin{description}

\item[\ttbar\ generator:] Event generator uncertainties were evaluated by comparing
the baseline {\sc Powheg\,+\,Pythia6} \ttbar\ sample (with $\hdamp=\mtop$)
with alternative samples generated with {\sc MC@NLO} interfaced to {\sc Herwig}
(thus changing both the NLO 
hard-scattering event generator and the parton shower, hadronisation and underlying
event model), and with the LO multi-leg event generator {\sc Alpgen}, also 
interfaced to {\sc Herwig}.  The bin-by-bin shifts in \gemi\ and \cbi\ were 
fitted with polynomial functions in order to reduce statistical fluctuations 
caused by the limited size of the simulated samples, and 
the larger of the differences between the baseline
and the two alternative samples was taken in each bin to define the generator
uncertainty. As also found in the inclusive cross-section analysis 
\cite{TOPQ-2013-04},
a substantial part of the differences in \gemi\ in the various
samples arises from differences in the hadronic activity close to the leptons,
which affects the efficiency of the lepton isolation requirements. These
efficiencies were therefore measured in situ in \ttbar\ events selected in data
as discussed in Section~\ref{ss:leptsyst} below, and the simulation uncertainties
on \gemi\ evaluated by considering the lepton reconstruction, identification
and lepton-jet overlap requirements only. The resulting uncertainties
on \gemi\ are typically 0.5--1\,\% in most regions of the phase space, varying
only slightly as a function of the lepton and dilepton kinematics. The same
procedure was used to evaluate uncertainties in \cbi, and the predictions
of the three simulation samples were found to agree at the 0.5--1\,\% level,
giving similar predictions for the variations of \cbi\ across the bins
of the various measured distributions. 
Alternative \ttbar\ samples generated with {\sc Powheg\,+\,Pythia6}
and {\sc Powheg\,+\,Herwig} (both with $\hdamp=\infty$) were also considered,
but the resulting differences in \gemi\ and \cbi\ were found to be 
significantly less than those from the comparisons with {\sc MC@NLO\,+\,Herwig}
and thus no additional uncertainty was assigned. Variations in the predictions
of \fntaui\ from the three \ttbar\ samples were found to be at the 0.2\,\% 
level, and were also taken into account for the $\tau$-corrected cross-section
results.

\item[Initial/final-state radiation:] The effects on \gemi\, \cbi\ and \fntaui\
of uncertainties in the modelling of additional radiation in \ttbar\ events
were assessed as half the difference between {\sc Powheg\,+\,Pythia6} samples
tuned to span the uncertainties in jet activity measured in \sxwt\ 
ATLAS data \cite{ATL-PHYS-PUB-2015-002,TOPQ-2011-21,TOPQ-2012-03}, 
as discussed in Section~\ref{s:datasim}. The uncertainties
were taken as half the difference between the upward and downward variations,
and were substantially reduced by measuring the lepton isolation efficiencies
from data, in the same way as for the \ttbar\ generator uncertainties discussed
above.

\item[Parton distribution functions:] The uncertainties in \gemi\ due to 
limited knowledge of the proton PDFs were evaluated using the error sets of
the CT10 \cite{cttenpdf}, MSTW 2008 68\,\% CL \cite{mstwnnlo1} and
NNPDF 2.3 \cite{nnpdfffn} NLO PDF sets, by reweighting the 
{\sc MC@NLO\,+\,Herwig} \ttbar\ sample based on the $x$ and $Q^2$ values
of the partons participating in the hard scattering in each event. The 
final uncertainty in each bin was calculated as half the envelope encompassing
the predictions from all three PDF sets and their associated uncertainties,
following the PDF4LHC prescription \cite{pdf4lhc}. The resulting uncertainties
on \gemi\ are typically around 0.3\,\% except at the high ends of the 
distributions, and were taken to be fully correlated across all bins.

\item[Top quark mass:] The values of \gemi\ and the predicted levels of $Wt$ 
background depend weakly on the assumed value of \mtop. These effects were 
evaluated with \ttbar\ and $Wt$ samples simulated with \mtop\ values of
170 and 175\,\GeV, and scaled to a nominal $\pm 1$\,\GeV\ mass variation. The
resulting effects are at the level of 0.1-0.2\,\% on \gemi, and are 
partially cancelled
by the variations in the $Wt$ background, whose cross-section decreases with 
increasing \mtop. The residual uncertainties are typically around 0.1\,\% for
the absolute cross-sections except at the extreme ends of the distributions,
and smaller for the normalised cross-sections.

\end{description}

The total \ttbar\ modelling uncertainties in the normalised differential
cross-sections also include the small uncertainties
on \gemi\ and \cbi\ from the limited size of the simulated \ttbar\ samples,
and are shown by the green lines in
Figure~\ref{f:fracsyst}. They  are typically dominated by the \ttbar\
event generator comparisons.

\subsection{Lepton identification and measurement}\label{ss:leptsyst}

Uncertainties in the modelling of the detector response to electrons and 
muons affect both \gemi\ and the background estimates, with the largest 
uncertainties in the cross-section measurements coming via the former.

\begin{description}

\item[Lepton identification:] The modelling of the electron
and muon identification efficiencies, and the rate of electron charge 
misidentification, were studied using
$Z\rightarrow ee/\mu\mu$, $J/\psi\rightarrow ee/\mu\mu$ and 
$W\rightarrow e\nu$ events in data and simulation
\cite{PERF-2016-01,PERF-2014-05}, taking into account
the systematic correlations across different regions of the lepton
\pt\ and $\eta$ spectrum. The uncertainties in \gemi\ are typically
below 0.5\,\% for electron and below 0.3\,\% for muon efficiencies, with
significant cancellations in the normalised differential cross-sections.

\item[Lepton scales and resolution:] The electron and muon
energy/momentum scales and resolutions were determined using 
$Z\rightarrow ee/\mu\mu$, $Z\rightarrow (ee/\mu\mu)\gamma$,
$J/\psi\rightarrow ee/\mu\mu$ and $\Upsilon\rightarrow\mu\mu$ decays
\cite{PERF-2013-05,PERF-2014-05}. The largest uncertainty comes from the
limited knowledge of the electron energy scale, which gives uncertainties
varying from 0.2\,\% to over 2\% for the bins involving the highest energy
electrons. The muon momentum scale uncertainties are small in comparison.

\item[Lepton isolation:] Building on the studies described in Ref. 
\cite{TOPQ-2013-04}, the efficiencies of the lepton isolation requirements
were measured in data, using the fractions of selected 
opposite-sign $e\mu$ events with at least one $b$-tagged jet where either
the electron or the muon fails the isolation requirement. After correcting
for the contamination from events with a misidentified lepton, these fractions
give the inefficiency of the isolation requirements on signal \ttbar\ events. 
The misidentified lepton backgrounds were measured both by using the same-sign
$e\mu$ control samples discussed in Section~\ref{ss:backg} above,
and by using the distributions of lepton impact parameter significance
$|d_0|/\sigma_{d_0}$, where $d_0$ is the distance of closest approach
of the lepton track to the event primary vertex in the transverse plane,
and $\sigma_{d_0}$ its uncertainty. The isolation inefficiencies were measured
as functions of lepton \pt\ separately for the barrel ($|\eta|<1.5$) and
endcap regions of the detector. Consistent results were obtained using
both misidentified lepton estimation methods, and showed that the baseline
{\sc Powheg\,+\,Pythia6} \ttbar\ simulation sample overestimates the 
efficiencies of the lepton isolation requirements by up to 1\,\% for 
electrons with \pt\ in the range 40--80\,\GeV, and by up to 2\,\% for muons
at low \pt, decreasing rapidly to less than 0.5\,\% for 40\,\GeV. The values
of \gemi\ from the baseline simulation were corrected for these \pt-dependent
shifts using a reweighting technique. The corresponding uncertainties
are dominated by those on the misidentified lepton subtraction (including a
comparison of the same-sign and $|d_0|/\sigma_{d_0}$-based methods) and amount
to typically 0.5--1\,\% for electrons and 0.2--0.5\,\% for muons. 
The effect on the normalised cross-sections is about half
that on the absolute measurements, taking into account
systematic correlations across lepton \pt\ and $|\eta|$ bins.

\item[Lepton trigger:] The efficiencies of the single-lepton triggers
were measured in data using $Z\rightarrow ee/\mu\mu$ events 
\cite{TRIG-2012-03}. Since only
one lepton trigger was required to accept the $e\mu$ event, the trigger
efficiency with respect to the offline event selection is about 99\,\%, with 
a residual uncertainty of less than 0.2\,\%.

\end{description}

The lepton-related uncertainties are shown by the blue dot-dashed lines in 
Figure~\ref{f:fracsyst}, and the largest uncertainties typically come 
from the electron energy scale and electron isolation uncertainties.

\subsection{Jet measurement and $b$-tagging}\label{ss:jetbsyst}

Uncertainties in the selection and $b$-tagging of jets affect the background 
estimates \nibi\ and \niibi, and to a lesser extent, the correlation \cbi. 
The jet uncertainties also have a very small effect on \gemi, through the
requirement that leptons be separated from selected jets by $\Delta R>0.4$.

\begin{description}

\item[Jet-related uncertainties:] The jet energy scale was varied
according to the uncertainties derived from simulation and in situ
calibration measurements \cite{PERF-2012-01},
using a model with 22 orthogonal uncertainty components describing the
evolution with jet \pt\ and $|\eta|$. The effects of residual uncertainties 
in the modelling of the jet energy resolution \cite{PERF-2011-04}
were assessed by smearing jet energies in simulation. The jet reconstruction
efficiency was measured in data using track-based jets, and the effect of
residual uncertainties assessed in simulation by randomly discarding jets. The
modelling of the pileup rejection requirement applied to jets
was studied using $Z\rightarrow ee/\mu\mu$+jets events \cite{PERF-2014-03}.

\item[$b$-tagging uncertainties:] The efficiencies for $b$-tagging jets in 
\ttbar\ signal events were extracted from the data, but simulation was used
to predict the numbers of $b$-tagged jets in the $Wt$ single top and 
diboson backgrounds. The corresponding uncertainties were assessed using
studies of $b$-jets containing muons, charm jets containing $D^{*+}$ mesons
and inclusive jet events \cite{PERF-2012-04}.

\end{description}

The jet- and $b$-tagging-related uncertainties are shown by the purple lines
on Figure~\ref{f:fracsyst}, and are typically dominated by the effect of
the jet energy scale on the level of $Wt$ background.

\subsection{Background modelling}\label{ss:bgsyst}

As well as the detector-related uncertainties discussed above, the background
estimates depend on uncertainties in modelling the $Wt$ and diboson
processes taken from simulation, and uncertainties in the procedures used
for estimating the $Z$+jets and misidentified lepton backgrounds from data.

\begin{description}

\item[Single top modelling:] 
Uncertainties in the modelling of the $Wt$ background
were assessed by comparing the predictions from the
baseline {\sc Powheg\,+\,Pythia6} sample with those from
{\sc MC@NLO\,+\,Herwig}, and from two samples generated with 
{\sc AcerMC\,+\,Pythia6} utilising different tunes to vary the amount
of additional radiation, in all cases normalising the total production
cross-section to the approximate NNLO prediction based on Ref. \cite{wttheo}.
The uncertainty in this prediction was evaluated to be 6.8\,\%.
The $Wt$ background
with two $b$-tagged jets is sensitive to the production of $Wt$ with an 
additional $b$-jet, an NLO contribution which interferes with the 
\ttbar\ final state. The corresponding uncertainty was assessed by 
comparing the predictions of {\sc Powheg\,+\,Pythia6} with the 
diagram removal and diagram subtraction schemes for handling this
interference \cite{wtinter1,wtinter2}.
The latter predicts up to 25\,\% less $Wt$ background in the
one $b$-tagged and 60\,\% less in the two $b$-tagged channels at the
extreme high ends of the lepton \pt\ and dilepton \ptll, \mll, \ptsum\
and \esum\ distributions, but only 1--2\,\% and 20\,\% differences for 
one and two $b$-tagged $Wt$ events across the \etal, \rapll\ and \dphill\
distributions, similar to the differences seen for the inclusive 
analysis \cite{TOPQ-2013-04}. The uncertainties due to the limited
size of the $Wt$ simulation samples are negligible in comparison
to the modelling uncertainties.

\item[Diboson modelling:] The uncertainties in modelling the diboson background
events (mainly $WW$) with one and two additional $b$-tagged jets
were assessed by comparing the predictions from {\sc Alpgen}\,+ {\sc Herwig}
with those of {\sc Sherpa} 1.4.3 \cite{sherpa} including the effects of 
massive $b$ and $c$ quarks. The resulting uncertainties in the diboson
background are typically
in the range 20--30\,\%, substantially larger than the differences 
between recent predictions for the inclusive diboson cross-sections
at NNLO in QCD \cite{nnlovvxsec} and the NLO predictions from MCFM used
to normalise the simulated samples.
The background from SM Higgs production with  $H\rightarrow WW$ and
$H\rightarrow\tau\tau$ is smaller than the uncertainties assigned for 
diboson modelling, and was neglected.

\item[Z+jets extrapolation:] The backgrounds from
$Z\rightarrow\tau\tau\rightarrow e\mu$ accompanied by one or two $b$-tagged
jets were extrapolated from the analogous $Z\rightarrow ee/\mu\mu$ event rates,
with uncertainties of 20\,\% for one and 30\,\% for two additional
$b$-tagged jets, as discussed in Section~\ref{ss:backg}.

\item[Misidentified leptons:] Uncertainties in the numbers of events
with misidentified leptons arise from the statistical uncertainties in the
corresponding same-sign samples, together with systematic uncertainties in the 
opposite-to-same-sign ratios \rij\ and the estimated contributions of prompt
same-sign events. The total uncertainties in the measured cross-sections are
typically  0.2--0.5\,\%,
except at the extreme ends of distributions where the same-sign data
statistical uncertainties are larger.

\end{description}

The background uncertainties are shown by the solid red lines on 
Figure~\ref{f:fracsyst}, and are dominated by $Wt$ modelling uncertainties,
in particular from the $Wt$-\ttbar interference at the high ends of 
some distributions.

\subsection{Luminosity and beam energy}\label{ss:elumisyst}

Uncertainties in the integrated luminosity and LHC beam energy give rise
to additional uncertainties in the differential cross-section results.

\begin{description}

\item[Luminosity:] 
The uncertainty in the integrated luminosity is 1.9\,\%, derived from
beam-separation scans performed in November 2012 \cite{DAPR-2013-01}. The
corresponding uncertainty in the absolute cross-section measurements is
slightly larger, typically about 2.1\,\%, as the $Wt$ and diboson backgrounds 
were evaluated from simulation, thus becoming sensitive to the assumed 
integrated luminosity.
The sensitivity varies with the background fractions, leaving a residual
uncertainty of typically less than 0.1\,\% in the
normalised cross-section results.

\item[Beam energy:] 

The LHC beam energy during the 2012 $pp$ run was determined to be 
within 0.1\,\% of the nominal value of 4\,\TeV\ per beam, based on the 
LHC magnetic model together with measurements of the revolution 
frequency difference of proton and lead-ion beams \cite{ebeam2}.
Following the approach used in Ref.\ \cite{TOPQ-2013-04} with an 
earlier less precise determination of the LHC beam energy \cite{ebeam},
an additional uncertainty corresponding to the change in 
cross-sections for a 0.1\,\% change in $\sqrt{s}$ was applied to the final
results, allowing them to be interpreted as measurements at exactly
\sxvt. The changes in each differential cross-section bin were calculated
by scaling the differences seen in {\sc Powheg\,+\,Pythia6} samples
generated at \sxvt\ and \sxwt. The resulting values were cross-checked with 
an explicit NLO fixed-order
calculation using {\sc Sherpa 2.1} \cite{sherpa}, making use of the 
{\sc Applgrid} framework \cite{applgrid} to reweight an \sxvt\ prediction so as
to change the $\sqrt{s}$ value by $\pm 0.66$\,\% which was then rescaled
to correspond to a $\sqrt{s}$ change of 0.1\,\%.
The changes in the absolute cross-sections are in the range 
0.2--0.4\,\%, and largely cancel in the normalised cross-sections. 

\end{description}

These uncertainties are not shown separately in Figure~\ref{f:fracsyst},
but are included in the total uncertainties shown by the black lines,
and given in Tables~\ref{t:insXSec1} to~\ref{t:insXSec4}.


\section{Results}\label{s:res}

The absolute differential cross-sections were determined by solving 
Eqs.~(\ref{e:fidtags}) separately for each bin $i$  of each lepton and 
dilepton differential distribution, taking the effects of systematic
uncertainties into account as discussed in Section~\ref{s:syst}. The 
normalised differential cross-sections were determined from the absolute
results using Eq.~(\ref{e:normx}). The values of \epsbi, i.e.~the product 
of jet acceptance, reconstruction and $b$-tagging probabilities in each bin, 
were determined to be in the range 0.5--0.6, in 
agreement with the simulation prediction for each bin. The results were found
to be stable when changing the minimum jet \pt\ requirement from 25\,\GeV\ 
up to 55\,\GeV, and when using $b$-tagging working points corresponding
to $b$-jet efficiencies of 60--80\,\%. The electron and muon \pt\ and $|\eta|$
distributions were also measured separately, instead of combining them
into lepton distributions with two entries per event, and 
found to be compatible. The bin-by-bin
comparison of the electron and muon \pt\ ($|\eta|$) distributions 
has a $\chi^2$ per degree 
of freedom of 10.9/9 (12.5/8), in both cases taking into account statistical
and uncorrelated systematic uncertainties.

\subsection{Fiducial cross-section measurements}\label{ss:meas}

The measured absolute and normalised fiducial differential cross-sections
are shown in Table~\ref{t:insXSec1} (\ptl\ and \etal), 
Table~\ref{t:insXSec2} (\ptll\ and \mll), 
Table~\ref{t:insXSec3} (\rapll\ and \dphill) and
Table~\ref{t:insXSec4} (\ptsum\ and \esum).
Each table shows the measured cross-section values and uncertainties, together
with a breakdown of the total uncertainties into components due to data 
statistics (`Stat.'), \ttbar\ modelling uncertainties (`\ttbar\ mod.'), 
lepton-related uncertainties (`Lept'), jet and $b$-tagging uncertainties
(`Jet/$b$'), background uncertainties (`Bkg.') and luminosity/beam energy
uncertainties (`$L/E_\mathrm{b}$'), corresponding to the breakdown in 
Sections~\ref{ss:ttsyst} to~\ref{ss:elumisyst}. The rightmost columns show
the cross-sections corrected to remove the contributions where one or both
leptons result from $W\rightarrow\tau\rightarrow e/\mu$ decays using
Eq.~(\ref{e:notau}). 
As can also be seen from Figure~\ref{f:fracsyst}, the total uncertainties
on the normalised differential cross-sections
range from 1.2\,\% to around 10\,\%, typically smaller than those
for the measurements as a function of the 
\ttbar\ system kinematics in Ref. \cite{TOPQ-2015-07}.
The largest uncertainties are generally statistical (from 1.1\,\% to about
10\,\%), with the background uncertainties also becoming large at high
values of some kinematic variables. Other systematic uncertainties due
to \ttbar\ modelling, leptons and jets are significantly smaller
than the statistical uncertainties, benefiting from  cancellations
between bins. The cancellations are particularly important when leptons with
similar \pt\ contribute to all bins, as is the case for \dphill\ and the
bulk of the \etal\ and \rapll\ distributions.
The uncertainties in the absolute cross-sections are substantially larger,
with the systematic uncertainties due to \ttbar\ modelling and leptons
becoming comparable to the statistical uncertainties. The absolute 
cross-sections also have an uncertainty 
of 2.1--2.5\,\% from the integrated luminosity measurement, depending on the
background level in each bin.

The integrals of the differential cross-sections across 
all bins of a given distribution (\xfid\ in Eq.~(\ref{e:normx})) agree 
in all cases within 0.4\,\% of the
integrated fiducial cross-sections  of $3.455\pm 0.025$\,pb 
(or $3.043\pm 0.022$\,pb excluding $\tau$ contributions)
measured within the same fiducial region in Ref. \cite{TOPQ-2013-04,ttxadd}.
The quoted uncertainties are statistical.\footnote{The integrals of the \ptl\ 
and \etal\ distributions correspond to twice 
these values, as the definitions include two leptons per event.}

The normalised differential cross-sections are shown graphically
in Figures~\ref{f:distresa} and~\ref{f:distresb}; in these and later Figures,
the data points are plotted at the centre of each bin. The measured 
cross-sections are
compared to the particle-level predictions from the {\sc Powheg\,+\,Pythia6},
{\sc MC@NLO\,+\,Herwig} and {\sc Alpgen\,+\,Herwig} \ttbar\ sam\-ples within
the fiducial volume of the measurement, including the contributions
from $W\rightarrow\tau\rightarrow e/\mu$ decays. Similar trends in the 
description of the measured distributions by the predictions can be seen
as for the reconstructed distributions for events with at least one $b$-tagged
jet in Figures~\ref{f:dmcjlept} and~\ref{f:dmcdilept}.

\begin{table}

{\centering \small
\begin{tabular}{lrrrrrrrrr}
\hline
Absolute & $\mathrm{d}\sigma/\mathrm{d}\ptl$ & Stat. & \ttbar\ mod. & Lept. & Jet/$b$ & Bkg. & $L/E_\mathrm{b}$ & Total & $\mathrm{d}\sigma/\mathrm{d}\ptl$ (no $\tau$) \\
Bin [\GeV] & [fb/\GeV] & (\%) & (\%) & (\%) & (\%) & (\%) & (\%) & (\%) & [fb/\GeV] \\\hline
25--30 &$   154.8\pm     5.7$ & 1.6 &  1.3 &  1.8 &  0.8 &  1.2 &  2.1 &  3.7 & $   127.2\pm     4.8$\\
30--40 &$   146.1\pm     4.9$ & 1.1 &  1.2 &  1.5 &  0.8 &  1.0 &  2.1 &  3.3 & $   124.6\pm     4.2$\\
40--50 &$   118.8\pm     3.7$ & 1.2 &  1.1 &  1.0 &  1.0 &  0.9 &  2.1 &  3.1 & $   104.3\pm     3.3$\\
50--60 &$    93.5\pm     2.9$ & 1.4 &  1.0 &  1.0 &  0.8 &  0.9 &  2.1 &  3.1 & $    83.4\pm     2.6$\\
60--80 &$    60.0\pm     1.8$ & 1.2 &  0.9 &  0.9 &  0.6 &  0.9 &  2.1 &  3.0 & $    54.1\pm     1.6$\\
80--100 &$    32.4\pm     1.1$ & 1.6 &  0.8 &  1.1 &  1.4 &  1.1 &  2.1 &  3.5 & $    29.4\pm     1.0$\\
100--120 &$   16.23\pm    0.64$ & 2.3 &  0.9 &  1.1 &  1.1 &  1.5 &  2.2 &  3.9 & $   14.75\pm    0.58$\\
120--150 &$    7.61\pm    0.35$ & 2.7 &  1.1 &  1.4 &  1.5 &  1.9 &  2.2 &  4.6 & $    6.91\pm    0.32$\\
150--200 &$    2.41\pm    0.15$ & 3.9 &  1.6 &  1.7 &  1.6 &  3.2 &  2.2 &  6.2 & $    2.17\pm    0.13$\\
200--300+ &$    0.49\pm    0.06$ & 6.7 &  3.5 &  2.3 &  2.9 &  7.5 &  2.4 & 11.5 & $    0.44\pm    0.05$\\
\hline
Normalised & $\frac{1}{\sigma}\mathrm{d}\sigma/\mathrm{d}\ptl$ & Stat. & \ttbar\ mod. & Lept. & Jet/$b$ & Bkg. & $L/E_\mathrm{b}$ & Total & $\frac{1}{\sigma}\mathrm{d}\sigma/\mathrm{d}\ptl$ (no $\tau$) \\
Bin [\GeV] & [$10^{-2}/$\GeV] & (\%) & (\%) & (\%) & (\%) & (\%) & (\%) & (\%) & [$10^{-2}/$\GeV] \\\hline
25--30 &$   2.235\pm   0.045$ & 1.5 &  0.4 &  0.9 &  0.6 &  0.8 &  0.0 &  2.0 & $   2.090\pm   0.042$\\
30--40 &$   2.108\pm   0.029$ & 1.0 &  0.3 &  0.7 &  0.3 &  0.5 &  0.0 &  1.4 & $   2.048\pm   0.029$\\
40--50 &$   1.714\pm   0.023$ & 1.1 &  0.2 &  0.5 &  0.4 &  0.4 &  0.0 &  1.3 & $   1.714\pm   0.023$\\
50--60 &$   1.350\pm   0.019$ & 1.3 &  0.2 &  0.4 &  0.3 &  0.3 &  0.0 &  1.4 & $   1.370\pm   0.020$\\
60--80 &$   0.866\pm   0.011$ & 1.1 &  0.3 &  0.4 &  0.4 &  0.3 &  0.0 &  1.3 & $   0.890\pm   0.011$\\
80--100 &$  0.4673\pm  0.0093$ & 1.5 &  0.4 &  0.7 &  0.8 &  0.5 &  0.0 &  2.0 & $  0.4831\pm  0.0096$\\
100--120 &$  0.2343\pm  0.0063$ & 2.3 &  0.5 &  0.8 &  0.6 &  1.0 &  0.0 &  2.7 & $  0.2424\pm  0.0065$\\
120--150 &$  0.1098\pm  0.0040$ & 2.7 &  0.6 &  1.2 &  1.3 &  1.5 &  0.1 &  3.6 & $  0.1135\pm  0.0041$\\
150--200 &$  0.0348\pm  0.0018$ & 3.9 &  0.9 &  1.6 &  1.2 &  2.8 &  0.1 &  5.3 & $  0.0357\pm  0.0019$\\
200--300+ &$  0.0070\pm  0.0007$ & 6.7 &  2.7 &  2.3 &  2.3 &  7.2 &  0.3 & 10.7 & $  0.0072\pm  0.0008$\\
\hline
\end{tabular}
\vspace{3mm}

\begin{tabular}{lrrrrrrrrr}
\hline
Absolute & $\mathrm{d}\sigma/\mathrm{d}\etal$ & Stat. & \ttbar\ mod. & Lept. & Jet/$b$ & Bkg. & $L/E_\mathrm{b}$ & Total & $\mathrm{d}\sigma/\mathrm{d}\etal$ (no $\tau$) \\
Bin [unit $\eta$] & [fb/unit $\eta$] & (\%) & (\%) & (\%) & (\%) & (\%) & (\%) & (\%) & [fb/unit $\eta$] \\\hline
0.00--0.25 &$    4590\pm     140$ & 1.2 &  1.0 &  1.0 &  0.9 &  0.9 &  2.1 &  3.1 & $    4030\pm     130$\\
0.25--0.50 &$    4440\pm     140$ & 1.2 &  1.0 &  1.0 &  0.9 &  0.9 &  2.1 &  3.1 & $    3900\pm     120$\\
0.50--0.75 &$    4230\pm     130$ & 1.2 &  1.0 &  1.0 &  0.9 &  0.9 &  2.1 &  3.1 & $    3710\pm     120$\\
0.75--1.00 &$    3660\pm     110$ & 1.3 &  1.0 &  1.0 &  0.8 &  1.0 &  2.1 &  3.1 & $    3210\pm     100$\\
1.00--1.25 &$    3100\pm     100$ & 1.5 &  1.0 &  1.0 &  0.9 &  1.0 &  2.1 &  3.3 & $    2722\pm      89$\\
1.25--1.50 &$    2470\pm      87$ & 1.9 &  1.1 &  1.0 &  0.9 &  1.0 &  2.1 &  3.5 & $    2173\pm      77$\\
1.50--1.75 &$    2035\pm      73$ & 1.9 &  1.1 &  1.4 &  0.7 &  1.0 &  2.1 &  3.6 & $    1793\pm      65$\\
1.75--2.00 &$    1431\pm      57$ & 2.4 &  1.2 &  1.4 &  0.6 &  1.4 &  2.1 &  4.0 & $    1263\pm      50$\\
2.00--2.50 &$     844\pm      34$ & 2.2 &  1.3 &  1.4 &  0.7 &  1.4 &  2.1 &  4.0 & $     749\pm      30$\\
\hline
Normalised & $\frac{1}{\sigma}\mathrm{d}\sigma/\mathrm{d}\etal$ & Stat. & \ttbar\ mod. & Lept. & Jet/$b$ & Bkg. & $L/E_\mathrm{b}$ & Total & $\frac{1}{\sigma}\mathrm{d}\sigma/\mathrm{d}\etal$ (no $\tau$) \\
Bin [unit $\eta$] & [$10^{-1}/$unit $\eta$] & (\%) & (\%) & (\%) & (\%) & (\%) & (\%) & (\%) & [$10^{-1}/$unit $\eta$] \\\hline
0.00--0.25 &$   6.646\pm   0.083$ & 1.1 &  0.4 &  0.2 &  0.1 &  0.3 &  0.0 &  1.2 & $   6.632\pm   0.083$\\
0.25--0.50 &$   6.428\pm   0.076$ & 1.1 &  0.3 &  0.2 &  0.1 &  0.2 &  0.0 &  1.2 & $   6.416\pm   0.076$\\
0.50--0.75 &$   6.117\pm   0.074$ & 1.1 &  0.2 &  0.2 &  0.1 &  0.2 &  0.0 &  1.2 & $   6.103\pm   0.074$\\
0.75--1.00 &$   5.297\pm   0.070$ & 1.3 &  0.2 &  0.2 &  0.2 &  0.2 &  0.0 &  1.3 & $   5.286\pm   0.070$\\
1.00--1.25 &$   4.482\pm   0.066$ & 1.4 &  0.3 &  0.2 &  0.2 &  0.3 &  0.0 &  1.5 & $   4.484\pm   0.066$\\
1.25--1.50 &$   3.574\pm   0.068$ & 1.8 &  0.4 &  0.2 &  0.2 &  0.3 &  0.0 &  1.9 & $   3.579\pm   0.068$\\
1.50--1.75 &$   2.944\pm   0.060$ & 1.9 &  0.5 &  0.6 &  0.2 &  0.4 &  0.0 &  2.0 & $   2.954\pm   0.061$\\
1.75--2.00 &$   2.070\pm   0.054$ & 2.3 &  0.6 &  0.6 &  0.4 &  0.8 &  0.0 &  2.6 & $   2.080\pm   0.054$\\
2.00--2.50 &$   1.221\pm   0.031$ & 2.1 &  0.7 &  0.6 &  0.4 &  1.0 &  0.0 &  2.5 & $   1.233\pm   0.031$\\
\hline
\end{tabular}
\vspace{3mm}

}
\caption{\label{t:insXSec1}Absolute and normalised differential cross-sections as functions of \ptl\ (top) and \etal\ (bottom). The columns show the bin ranges, measured cross-section and total uncertainty, relative statistical uncertainty, relative systematic uncertainties in various categories (see text), total relative uncertainty, and differential cross-section corrected to remove contributions via $W\rightarrow\tau\rightarrow e/\mu$ decays. Relative uncertainties smaller than 0.05\,\% are indicated by `0.0'. The last bin includes overflows where indicated by the `+' sign.}
\end{table}

\begin{table}

{\centering \small
\begin{tabular}{lrrrrrrrrr}
\hline
Absolute & $\mathrm{d}\sigma/\mathrm{d}\ptll$ & Stat. & \ttbar\ mod. & Lept. & Jet/$b$ & Bkg. & $L/E_\mathrm{b}$ & Total & $\mathrm{d}\sigma/\mathrm{d}\ptll$ (no $\tau$) \\
Bin [\GeV] & [fb/\GeV] & (\%) & (\%) & (\%) & (\%) & (\%) & (\%) & (\%) & [fb/\GeV] \\\hline
0--20 &$   11.50\pm    0.49$ & 2.8 &  1.2 &  1.2 &  1.1 &  1.2 &  2.1 &  4.2 & $    9.62\pm    0.41$\\
20--40 &$   26.72\pm    0.94$ & 1.9 &  1.0 &  1.2 &  1.0 &  1.0 &  2.1 &  3.5 & $   22.62\pm    0.81$\\
40--60 &$    35.9\pm     1.2$ & 1.6 &  0.9 &  1.2 &  1.1 &  1.0 &  2.1 &  3.4 & $    30.6\pm     1.0$\\
60--80 &$    39.0\pm     1.3$ & 1.5 &  0.9 &  1.1 &  0.9 &  0.9 &  2.1 &  3.2 & $    34.4\pm     1.1$\\
80--100 &$   29.19\pm    0.96$ & 1.7 &  1.0 &  1.0 &  0.8 &  1.0 &  2.1 &  3.3 & $   26.48\pm    0.88$\\
100--120 &$   16.38\pm    0.65$ & 2.3 &  1.2 &  1.3 &  1.3 &  1.1 &  2.1 &  3.9 & $   15.11\pm    0.60$\\
120--150 &$    6.53\pm    0.30$ & 2.9 &  1.4 &  1.5 &  1.2 &  1.6 &  2.2 &  4.6 & $    6.06\pm    0.28$\\
150--200 &$    1.39\pm    0.11$ & 5.0 &  2.2 &  2.1 &  2.2 &  4.2 &  2.3 &  7.9 & $    1.27\pm    0.10$\\
200--300+ &$    0.23\pm    0.04$ & 9.3 &  4.2 &  2.9 &  4.1 & 13.2 &  2.5 & 17.6 & $    0.20\pm    0.04$\\
\hline
Normalised & $\frac{1}{\sigma}\mathrm{d}\sigma/\mathrm{d}\ptll$ & Stat. & \ttbar\ mod. & Lept. & Jet/$b$ & Bkg. & $L/E_\mathrm{b}$ & Total & $\frac{1}{\sigma}\mathrm{d}\sigma/\mathrm{d}\ptll$ (no $\tau$) \\
Bin [\GeV] & [$10^{-2}/$\GeV] & (\%) & (\%) & (\%) & (\%) & (\%) & (\%) & (\%) & [$10^{-2}/$\GeV] \\\hline
0--20 &$   0.332\pm   0.010$ & 2.7 &  0.5 &  0.5 &  0.7 &  0.8 &  0.0 &  3.0 & $   0.316\pm   0.010$\\
20--40 &$   0.772\pm   0.015$ & 1.7 &  0.4 &  0.5 &  0.5 &  0.6 &  0.0 &  2.0 & $   0.743\pm   0.015$\\
40--60 &$   1.036\pm   0.017$ & 1.4 &  0.3 &  0.4 &  0.5 &  0.5 &  0.0 &  1.6 & $   1.006\pm   0.017$\\
60--80 &$   1.127\pm   0.016$ & 1.3 &  0.3 &  0.3 &  0.4 &  0.3 &  0.0 &  1.5 & $   1.130\pm   0.016$\\
80--100 &$   0.843\pm   0.015$ & 1.5 &  0.3 &  0.3 &  0.6 &  0.4 &  0.0 &  1.8 & $   0.870\pm   0.015$\\
100--120 &$   0.473\pm   0.012$ & 2.2 &  0.4 &  0.8 &  0.8 &  0.5 &  0.0 &  2.5 & $   0.497\pm   0.013$\\
120--150 &$  0.1886\pm  0.0066$ & 2.8 &  0.6 &  1.2 &  1.0 &  1.2 &  0.0 &  3.5 & $  0.1993\pm  0.0069$\\
150--200 &$  0.0402\pm  0.0028$ & 4.9 &  1.4 &  1.9 &  1.9 &  3.9 &  0.2 &  7.0 & $  0.0419\pm  0.0029$\\
200--300+ &$  0.0066\pm  0.0011$ & 9.2 &  3.3 &  2.7 &  3.6 & 13.0 &  0.4 & 16.9 & $  0.0067\pm  0.0011$\\
\hline
\end{tabular}
\vspace{3mm}

\begin{tabular}{lrrrrrrrrr}
\hline
Absolute & $\mathrm{d}\sigma/\mathrm{d}\mll$ & Stat. & \ttbar\ mod. & Lept. & Jet/$b$ & Bkg. & $L/E_\mathrm{b}$ & Total & $\mathrm{d}\sigma/\mathrm{d}\mll$ (no $\tau$) \\
Bin [\GeV] & [fb/\GeV] & (\%) & (\%) & (\%) & (\%) & (\%) & (\%) & (\%) & [fb/\GeV] \\\hline
0--20 &$    3.37\pm    0.25$ & 6.3 &  2.0 &  1.9 &  1.3 &  1.3 &  2.1 &  7.4 & $    2.97\pm    0.22$\\
20--40 &$   10.94\pm    0.47$ & 2.8 &  1.5 &  1.4 &  0.9 &  1.2 &  2.1 &  4.3 & $    9.61\pm    0.41$\\
40--60 &$   17.66\pm    0.70$ & 2.2 &  1.4 &  1.5 &  0.7 &  1.3 &  2.1 &  4.0 & $   15.29\pm    0.61$\\
60--80 &$   23.98\pm    0.89$ & 1.9 &  1.3 &  1.4 &  0.8 &  1.2 &  2.1 &  3.7 & $   20.54\pm    0.76$\\
80--100 &$   26.00\pm    0.90$ & 1.8 &  1.2 &  1.1 &  0.8 &  0.9 &  2.1 &  3.4 & $   22.42\pm    0.78$\\
100--120 &$   23.03\pm    0.83$ & 2.0 &  1.1 &  1.0 &  1.2 &  0.9 &  2.1 &  3.6 & $   20.07\pm    0.73$\\
120--150 &$   16.71\pm    0.57$ & 1.9 &  1.0 &  0.9 &  1.0 &  1.0 &  2.1 &  3.4 & $   14.72\pm    0.51$\\
150--200 &$    9.38\pm    0.34$ & 1.9 &  0.8 &  0.9 &  1.3 &  1.2 &  2.1 &  3.6 & $    8.41\pm    0.30$\\
200--250 &$    4.09\pm    0.18$ & 3.0 &  0.8 &  1.1 &  1.1 &  1.6 &  2.1 &  4.4 & $    3.73\pm    0.16$\\
250--300 &$    1.95\pm    0.11$ & 4.4 &  1.0 &  1.2 &  1.4 &  2.1 &  2.2 &  5.7 & $    1.80\pm    0.10$\\
300--400 &$    0.66\pm    0.05$ & 5.2 &  1.3 &  1.5 &  2.0 &  3.1 &  2.2 &  7.1 & $    0.62\pm    0.04$\\
400--500+ &$    0.26\pm    0.03$ & 8.0 &  2.2 &  2.1 &  2.1 &  4.8 &  2.2 & 10.3 & $    0.25\pm    0.03$\\
\hline
Normalised & $\frac{1}{\sigma}\mathrm{d}\sigma/\mathrm{d}\mll$ & Stat. & \ttbar\ mod. & Lept. & Jet/$b$ & Bkg. & $L/E_\mathrm{b}$ & Total & $\frac{1}{\sigma}\mathrm{d}\sigma/\mathrm{d}\mll$ (no $\tau$) \\
Bin [\GeV] & [$10^{-3}/$\GeV] & (\%) & (\%) & (\%) & (\%) & (\%) & (\%) & (\%) & [$10^{-3}/$\GeV] \\\hline
0--20 &$   0.973\pm   0.066$ & 6.2 &  1.4 &  1.2 &  1.4 &  1.3 &  0.0 &  6.7 & $   0.977\pm   0.066$\\
20--40 &$   3.157\pm   0.095$ & 2.7 &  0.7 &  0.5 &  0.4 &  0.9 &  0.0 &  3.0 & $   3.156\pm   0.095$\\
40--60 &$    5.10\pm    0.13$ & 2.1 &  0.5 &  0.7 &  0.4 &  1.0 &  0.0 &  2.5 & $    5.02\pm    0.13$\\
60--80 &$    6.92\pm    0.14$ & 1.8 &  0.4 &  0.6 &  0.3 &  0.7 &  0.0 &  2.1 & $    6.75\pm    0.14$\\
80--100 &$    7.51\pm    0.14$ & 1.7 &  0.3 &  0.3 &  0.6 &  0.4 &  0.0 &  1.8 & $    7.37\pm    0.14$\\
100--120 &$    6.65\pm    0.14$ & 1.8 &  0.3 &  0.3 &  0.7 &  0.3 &  0.0 &  2.0 & $    6.60\pm    0.14$\\
120--150 &$   4.823\pm   0.092$ & 1.7 &  0.3 &  0.3 &  0.5 &  0.3 &  0.0 &  1.9 & $   4.839\pm   0.092$\\
150--200 &$   2.707\pm   0.058$ & 1.8 &  0.4 &  0.5 &  0.7 &  0.6 &  0.0 &  2.1 & $   2.763\pm   0.059$\\
200--250 &$   1.180\pm   0.041$ & 2.9 &  0.7 &  0.8 &  1.0 &  1.2 &  0.0 &  3.4 & $   1.224\pm   0.042$\\
250--300 &$   0.563\pm   0.029$ & 4.4 &  1.0 &  1.0 &  1.4 &  1.7 &  0.0 &  5.1 & $   0.590\pm   0.030$\\
300--400 &$   0.191\pm   0.012$ & 5.2 &  1.8 &  1.4 &  1.6 &  2.8 &  0.1 &  6.5 & $   0.203\pm   0.013$\\
400--500+ &$  0.0763\pm  0.0076$ & 8.0 &  2.6 &  1.9 &  2.0 &  4.5 &  0.1 &  9.9 & $  0.0820\pm  0.0081$\\
\hline
\end{tabular}
\vspace{3mm}

}
\caption{\label{t:insXSec2}Absolute and normalised differential cross-sections as functions of \ptll\ (top) and \mll\ (bottom). The columns show the bin ranges, measured cross-section and total uncertainty, relative statistical uncertainty, relative systematic uncertainties in various categories (see text), total relative uncertainty, and differential cross-section corrected to remove contributions via $W\rightarrow\tau\rightarrow e/\mu$ decays. Relative uncertainties smaller than 0.05\,\% are indicated by `0.0'. The last bin includes overflows where indicated by the `+' sign.}
\end{table}

\begin{table}

{\centering \small
\begin{tabular}{lrrrrrrrrr}
\hline
Absolute & $\mathrm{d}\sigma/\mathrm{d}\rapll$ & Stat. & \ttbar\ mod. & Lept. & Jet/$b$ & Bkg. & $L/E_\mathrm{b}$ & Total & $\mathrm{d}\sigma/\mathrm{d}\rapll$ (no $\tau$) \\
Bin [unit $y$] & [fb/unit $y$] & (\%) & (\%) & (\%) & (\%) & (\%) & (\%) & (\%) & [fb/unit $y$] \\\hline
0.00--0.25 &$    3007\pm      95$ & 1.5 &  0.9 &  1.0 &  1.0 &  0.8 &  2.1 &  3.2 & $    2639\pm      84$\\
0.25--0.50 &$    2681\pm      86$ & 1.5 &  0.9 &  1.0 &  0.8 &  0.9 &  2.1 &  3.2 & $    2353\pm      76$\\
0.50--0.75 &$    2419\pm      80$ & 1.6 &  1.0 &  1.0 &  1.0 &  1.0 &  2.1 &  3.3 & $    2123\pm      71$\\
0.75--1.00 &$    2026\pm      71$ & 1.9 &  1.1 &  1.1 &  0.8 &  1.1 &  2.1 &  3.5 & $    1780\pm      63$\\
1.00--1.25 &$    1536\pm      57$ & 2.2 &  1.2 &  1.1 &  0.8 &  1.0 &  2.1 &  3.7 & $    1351\pm      50$\\
1.25--1.50 &$    1038\pm      43$ & 2.7 &  1.3 &  1.3 &  0.7 &  1.2 &  2.1 &  4.2 & $     912\pm      38$\\
1.50--1.75 &$     637\pm      33$ & 3.7 &  1.6 &  1.6 &  1.3 &  1.3 &  2.2 &  5.2 & $     561\pm      29$\\
1.75--2.00 &$     321\pm      23$ & 5.7 &  2.0 &  1.8 &  1.2 &  2.1 &  2.2 &  7.1 & $     283\pm      20$\\
2.00--2.50 &$    69.1\pm     7.7$ & 8.8 &  2.7 &  2.3 &  2.3 &  4.7 &  2.2 & 11.1 & $    61.3\pm     6.8$\\
\hline
Normalised & $\frac{1}{\sigma}\mathrm{d}\sigma/\mathrm{d}\rapll$ & Stat. & \ttbar\ mod. & Lept. & Jet/$b$ & Bkg. & $L/E_\mathrm{b}$ & Total & $\frac{1}{\sigma}\mathrm{d}\sigma/\mathrm{d}\rapll$ (no $\tau$) \\
Bin [unit $y$] & [$10^{-1}/$unit $y$] & (\%) & (\%) & (\%) & (\%) & (\%) & (\%) & (\%) & [$10^{-1}/$unit $y$] \\\hline
0.00--0.25 &$    8.71\pm    0.13$ & 1.3 &  0.3 &  0.2 &  0.3 &  0.3 &  0.0 &  1.4 & $    8.71\pm    0.13$\\
0.25--0.50 &$    7.77\pm    0.12$ & 1.4 &  0.3 &  0.2 &  0.3 &  0.3 &  0.0 &  1.5 & $    7.76\pm    0.12$\\
0.50--0.75 &$    7.01\pm    0.11$ & 1.5 &  0.3 &  0.2 &  0.3 &  0.3 &  0.0 &  1.6 & $    7.00\pm    0.11$\\
0.75--1.00 &$    5.87\pm    0.11$ & 1.7 &  0.4 &  0.2 &  0.3 &  0.4 &  0.0 &  1.9 & $    5.87\pm    0.11$\\
1.00--1.25 &$   4.451\pm   0.099$ & 2.1 &  0.5 &  0.2 &  0.4 &  0.3 &  0.0 &  2.2 & $   4.458\pm   0.099$\\
1.25--1.50 &$   3.009\pm   0.085$ & 2.6 &  0.6 &  0.4 &  0.6 &  0.6 &  0.0 &  2.8 & $   3.009\pm   0.085$\\
1.50--1.75 &$   1.846\pm   0.073$ & 3.6 &  0.8 &  0.7 &  0.8 &  0.8 &  0.0 &  4.0 & $   1.850\pm   0.073$\\
1.75--2.00 &$   0.930\pm   0.057$ & 5.6 &  1.3 &  1.0 &  0.9 &  1.8 &  0.1 &  6.2 & $   0.935\pm   0.058$\\
2.00--2.50 &$   0.200\pm   0.021$ & 8.8 &  2.1 &  1.7 &  2.5 &  4.4 &  0.1 & 10.5 & $   0.202\pm   0.021$\\
\hline
\end{tabular}
\vspace{3mm}

\begin{tabular}{lrrrrrrrrr}
\hline
Absolute & $\mathrm{d}\sigma/\mathrm{d}\dphill$ & Stat. & \ttbar\ mod. & Lept. & Jet/$b$ & Bkg. & $L/E_\mathrm{b}$ & Total & $\mathrm{d}\sigma/\mathrm{d}\dphill$ (no $\tau$) \\
Bin [rad] & [fb/rad] & (\%) & (\%) & (\%) & (\%) & (\%) & (\%) & (\%) & [fb/rad] \\\hline
0.00--0.31 &$     696\pm      30$ & 2.9 &  1.4 &  1.3 &  0.7 &  0.9 &  2.1 &  4.2 & $     630\pm      27$\\
0.31--0.63 &$     735\pm      29$ & 2.7 &  1.3 &  1.3 &  0.6 &  0.9 &  2.1 &  4.0 & $     664\pm      26$\\
0.63--0.94 &$     780\pm      31$ & 2.6 &  1.3 &  1.2 &  0.7 &  0.9 &  2.1 &  3.9 & $     704\pm      28$\\
0.94--1.26 &$     850\pm      33$ & 2.5 &  1.2 &  1.2 &  0.7 &  0.9 &  2.1 &  3.9 & $     763\pm      30$\\
1.26--1.57 &$     947\pm      36$ & 2.3 &  1.1 &  1.2 &  0.8 &  1.0 &  2.1 &  3.8 & $     844\pm      32$\\
1.57--1.88 &$    1103\pm      41$ & 2.2 &  1.1 &  1.1 &  1.1 &  0.9 &  2.1 &  3.7 & $     977\pm      37$\\
1.88--2.20 &$    1235\pm      43$ & 2.1 &  1.0 &  1.0 &  0.7 &  0.9 &  2.1 &  3.5 & $    1084\pm      38$\\
2.20--2.51 &$    1410\pm      50$ & 2.0 &  1.0 &  1.0 &  1.1 &  1.0 &  2.1 &  3.5 & $    1226\pm      44$\\
2.51--2.83 &$    1575\pm      56$ & 2.0 &  0.9 &  0.9 &  1.1 &  1.1 &  2.1 &  3.6 & $    1353\pm      49$\\
2.83--3.14 &$    1696\pm      58$ & 1.9 &  0.9 &  0.9 &  0.9 &  1.2 &  2.1 &  3.4 & $    1449\pm      51$\\
\hline
Normalised & $\frac{1}{\sigma}\mathrm{d}\sigma/\mathrm{d}\dphill$ & Stat. & \ttbar\ mod. & Lept. & Jet/$b$ & Bkg. & $L/E_\mathrm{b}$ & Total & $\frac{1}{\sigma}\mathrm{d}\sigma/\mathrm{d}\dphill$ (no $\tau$) \\
Bin [rad] & [$10^{-1}/$rad] & (\%) & (\%) & (\%) & (\%) & (\%) & (\%) & (\%) & [$10^{-1}/$rad] \\\hline
0.00--0.31 &$   2.010\pm   0.060$ & 2.8 &  0.6 &  0.4 &  0.5 &  0.4 &  0.0 &  3.0 & $   2.068\pm   0.062$\\
0.31--0.63 &$   2.121\pm   0.057$ & 2.6 &  0.5 &  0.3 &  0.4 &  0.4 &  0.0 &  2.7 & $   2.179\pm   0.058$\\
0.63--0.94 &$   2.252\pm   0.059$ & 2.5 &  0.5 &  0.3 &  0.2 &  0.4 &  0.0 &  2.6 & $   2.311\pm   0.060$\\
0.94--1.26 &$   2.454\pm   0.064$ & 2.4 &  0.5 &  0.2 &  0.6 &  0.4 &  0.0 &  2.6 & $   2.506\pm   0.065$\\
1.26--1.57 &$   2.732\pm   0.064$ & 2.2 &  0.4 &  0.2 &  0.2 &  0.4 &  0.0 &  2.3 & $   2.773\pm   0.065$\\
1.57--1.88 &$   3.185\pm   0.070$ & 2.1 &  0.4 &  0.1 &  0.3 &  0.3 &  0.0 &  2.2 & $   3.207\pm   0.071$\\
1.88--2.20 &$   3.566\pm   0.073$ & 2.0 &  0.4 &  0.1 &  0.2 &  0.3 &  0.0 &  2.0 & $   3.559\pm   0.072$\\
2.20--2.51 &$   4.069\pm   0.079$ & 1.9 &  0.4 &  0.1 &  0.3 &  0.3 &  0.0 &  1.9 & $   4.028\pm   0.078$\\
2.51--2.83 &$   4.546\pm   0.088$ & 1.8 &  0.4 &  0.2 &  0.3 &  0.4 &  0.0 &  1.9 & $   4.443\pm   0.086$\\
2.83--3.14 &$   4.897\pm   0.090$ & 1.7 &  0.4 &  0.3 &  0.1 &  0.5 &  0.0 &  1.8 & $   4.757\pm   0.088$\\
\hline
\end{tabular}
\vspace{3mm}

}
\caption{\label{t:insXSec3}Absolute and normalised differential cross-sections as functions of \rapll\ (top) and \dphill\ (bottom). The columns show the bin ranges, measured cross-section and total uncertainty, relative statistical uncertainty, relative systematic uncertainties in various categories (see text), total relative uncertainty, and differential cross-section corrected to remove contributions via $W\rightarrow\tau\rightarrow e/\mu$ decays. Relative uncertainties smaller than 0.05\,\% are indicated by `0.0'. The bin boundaries for \dphill\ correspond to exact multiples of $\pi/10$ but are quoted to two decimal places.}
\end{table}

\begin{table}

{\centering \small
\begin{tabular}{lrrrrrrrrr}
\hline
Absolute & $\mathrm{d}\sigma/\mathrm{d}(\ptsum)$ & Stat. & \ttbar\ mod. & Lept. & Jet/$b$ & Bkg. & $L/E_\mathrm{b}$ & Total & $\mathrm{d}\sigma/\mathrm{d}(\ptsum)$ (no $\tau$) \\
Bin [\GeV] & [fb/\GeV] & (\%) & (\%) & (\%) & (\%) & (\%) & (\%) & (\%) & [fb/\GeV] \\\hline
50--80 &$   23.02\pm    0.89$ & 1.7 &  1.2 &  1.8 &  0.7 &  1.5 &  2.1 &  3.8 & $   18.90\pm    0.73$\\
80--100 &$    38.0\pm     1.2$ & 1.5 &  1.0 &  1.2 &  0.8 &  0.9 &  2.1 &  3.3 & $    33.0\pm     1.1$\\
100--120 &$    34.3\pm     1.1$ & 1.6 &  1.0 &  1.0 &  0.9 &  0.9 &  2.1 &  3.2 & $    30.5\pm     1.0$\\
120--150 &$   21.00\pm    0.69$ & 1.7 &  0.9 &  1.0 &  0.8 &  1.0 &  2.1 &  3.3 & $   18.95\pm    0.63$\\
150--200 &$    9.11\pm    0.34$ & 1.9 &  1.0 &  1.1 &  1.2 &  1.5 &  2.2 &  3.8 & $    8.27\pm    0.31$\\
200--250 &$    3.03\pm    0.16$ & 3.5 &  1.3 &  1.5 &  1.4 &  2.4 &  2.2 &  5.3 & $    2.78\pm    0.15$\\
250--300 &$    1.08\pm    0.09$ & 5.7 &  1.9 &  2.0 &  2.9 &  3.5 &  2.2 &  8.1 & $    0.99\pm    0.08$\\
300--400+ &$    0.38\pm    0.04$ & 7.0 &  3.3 &  2.3 &  3.3 &  6.5 &  2.3 & 11.1 & $    0.35\pm    0.04$\\
\hline
Normalised & $\frac{1}{\sigma}\mathrm{d}\sigma/\mathrm{d}(\ptsum)$ & Stat. & \ttbar\ mod. & Lept. & Jet/$b$ & Bkg. & $L/E_\mathrm{b}$ & Total & $\frac{1}{\sigma}\mathrm{d}\sigma/\mathrm{d}(\ptsum)$ (no $\tau$) \\
Bin [\GeV] & [$10^{-2}/$\GeV] & (\%) & (\%) & (\%) & (\%) & (\%) & (\%) & (\%) & [$10^{-2}/$\GeV] \\\hline
50--80 &$   0.664\pm   0.015$ & 1.5 &  0.4 &  1.0 &  0.3 &  1.2 &  0.0 &  2.2 & $   0.621\pm   0.014$\\
80--100 &$   1.097\pm   0.017$ & 1.3 &  0.3 &  0.5 &  0.4 &  0.6 &  0.0 &  1.6 & $   1.085\pm   0.018$\\
100--120 &$   0.990\pm   0.016$ & 1.4 &  0.3 &  0.3 &  0.4 &  0.4 &  0.0 &  1.6 & $   1.003\pm   0.016$\\
120--150 &$   0.606\pm   0.010$ & 1.5 &  0.3 &  0.5 &  0.3 &  0.4 &  0.0 &  1.7 & $   0.623\pm   0.010$\\
150--200 &$  0.2627\pm  0.0062$ & 1.8 &  0.4 &  0.9 &  0.6 &  1.0 &  0.0 &  2.4 & $  0.2716\pm  0.0063$\\
200--250 &$  0.0875\pm  0.0039$ & 3.4 &  0.7 &  1.4 &  1.4 &  2.1 &  0.1 &  4.5 & $  0.0912\pm  0.0041$\\
250--300 &$  0.0311\pm  0.0023$ & 5.6 &  1.3 &  2.0 &  2.5 &  3.2 &  0.1 &  7.3 & $  0.0326\pm  0.0024$\\
300--400+ &$  0.0110\pm  0.0011$ & 7.0 &  2.5 &  2.4 &  2.9 &  6.2 &  0.2 & 10.4 & $  0.0116\pm  0.0012$\\
\hline
\end{tabular}
\vspace{3mm}

\begin{tabular}{lrrrrrrrrr}
\hline
Absolute & $\mathrm{d}\sigma/\mathrm{d}(\esum)$ & Stat. & \ttbar\ mod. & Lept. & Jet/$b$ & Bkg. & $L/E_\mathrm{b}$ & Total & $\mathrm{d}\sigma/\mathrm{d}(\esum)$ (no $\tau$) \\
Bin [\GeV] & [fb/\GeV] & (\%) & (\%) & (\%) & (\%) & (\%) & (\%) & (\%) & [fb/\GeV] \\\hline
50--80 &$    4.05\pm    0.22$ & 3.8 &  1.3 &  2.2 &  1.1 &  1.2 &  2.1 &  5.3 & $    3.21\pm    0.17$\\
80--100 &$   13.68\pm    0.57$ & 2.5 &  1.2 &  1.6 &  1.2 &  1.1 &  2.1 &  4.2 & $   11.38\pm    0.48$\\
100--120 &$   18.36\pm    0.67$ & 2.1 &  1.1 &  1.3 &  0.8 &  0.9 &  2.1 &  3.6 & $   15.67\pm    0.57$\\
120--150 &$   19.10\pm    0.64$ & 1.7 &  1.1 &  1.1 &  0.8 &  0.9 &  2.1 &  3.3 & $   16.63\pm    0.56$\\
150--200 &$   15.79\pm    0.51$ & 1.5 &  1.0 &  0.9 &  1.0 &  0.9 &  2.1 &  3.2 & $   13.92\pm    0.45$\\
200--250 &$   10.04\pm    0.35$ & 1.8 &  1.0 &  0.9 &  1.1 &  1.1 &  2.1 &  3.5 & $    8.97\pm    0.31$\\
250--300 &$    6.24\pm    0.25$ & 2.4 &  1.0 &  1.0 &  1.2 &  1.4 &  2.1 &  4.0 & $    5.61\pm    0.22$\\
300--400 &$    3.04\pm    0.13$ & 2.5 &  1.1 &  1.3 &  0.9 &  1.7 &  2.2 &  4.2 & $    2.75\pm    0.12$\\
400--500 &$    1.20\pm    0.07$ & 4.3 &  1.5 &  2.0 &  1.3 &  2.4 &  2.2 &  6.1 & $    1.10\pm    0.07$\\
500--700+ &$    0.48\pm    0.04$ & 4.8 &  2.3 &  2.3 &  1.6 &  3.8 &  2.2 &  7.4 & $    0.44\pm    0.03$\\
\hline
Normalised & $\frac{1}{\sigma}\mathrm{d}\sigma/\mathrm{d}(\esum)$ & Stat. & \ttbar\ mod. & Lept. & Jet/$b$ & Bkg. & $L/E_\mathrm{b}$ & Total & $\frac{1}{\sigma}\mathrm{d}\sigma/\mathrm{d}(\esum)$ (no $\tau$) \\
Bin [\GeV] & [$10^{-3}/$\GeV] & (\%) & (\%) & (\%) & (\%) & (\%) & (\%) & (\%) & [$10^{-3}/$\GeV] \\\hline
50--80 &$   1.172\pm   0.050$ & 3.8 &  0.8 &  1.4 &  0.7 &  1.1 &  0.1 &  4.3 & $   1.058\pm   0.046$\\
80--100 &$    3.95\pm    0.11$ & 2.4 &  0.5 &  0.8 &  0.7 &  0.9 &  0.0 &  2.8 & $    3.75\pm    0.11$\\
100--120 &$    5.31\pm    0.12$ & 2.0 &  0.4 &  0.6 &  0.6 &  0.7 &  0.0 &  2.3 & $    5.16\pm    0.12$\\
120--150 &$   5.521\pm   0.099$ & 1.6 &  0.3 &  0.4 &  0.4 &  0.5 &  0.0 &  1.8 & $   5.478\pm   0.099$\\
150--200 &$   4.564\pm   0.067$ & 1.3 &  0.3 &  0.3 &  0.4 &  0.3 &  0.0 &  1.5 & $   4.585\pm   0.067$\\
200--250 &$   2.904\pm   0.055$ & 1.7 &  0.4 &  0.4 &  0.5 &  0.4 &  0.0 &  1.9 & $   2.955\pm   0.056$\\
250--300 &$   1.803\pm   0.048$ & 2.3 &  0.5 &  0.5 &  0.9 &  0.7 &  0.0 &  2.6 & $   1.849\pm   0.049$\\
300--400 &$   0.878\pm   0.026$ & 2.4 &  0.5 &  0.8 &  0.7 &  1.1 &  0.0 &  2.9 & $   0.907\pm   0.026$\\
400--500 &$   0.348\pm   0.018$ & 4.2 &  1.0 &  1.7 &  1.2 &  2.0 &  0.1 &  5.2 & $   0.362\pm   0.019$\\
500--700+ &$  0.1393\pm  0.0091$ & 4.7 &  1.8 &  1.9 &  1.2 &  3.5 &  0.1 &  6.5 & $  0.1463\pm  0.0095$\\
\hline
\end{tabular}
\vspace{3mm}

}
\caption{\label{t:insXSec4}Absolute and normalised differential cross-sections as functions of \ptsum\ (top) and \esum\ (bottom). The columns show the bin ranges, measured cross-section and total uncertainty, relative statistical uncertainty, relative systematic uncertainties in various categories (see text), total relative uncertainty, and differential cross-section corrected to remove contributions via $W\rightarrow\tau\rightarrow e/\mu$ decays. Relative uncertainties smaller than 0.05\,\% are indicated by `0.0'. The last bin includes overflows where indicated by the `+' sign.}
\end{table}

\begin{figure}[tp]
\splitfigure{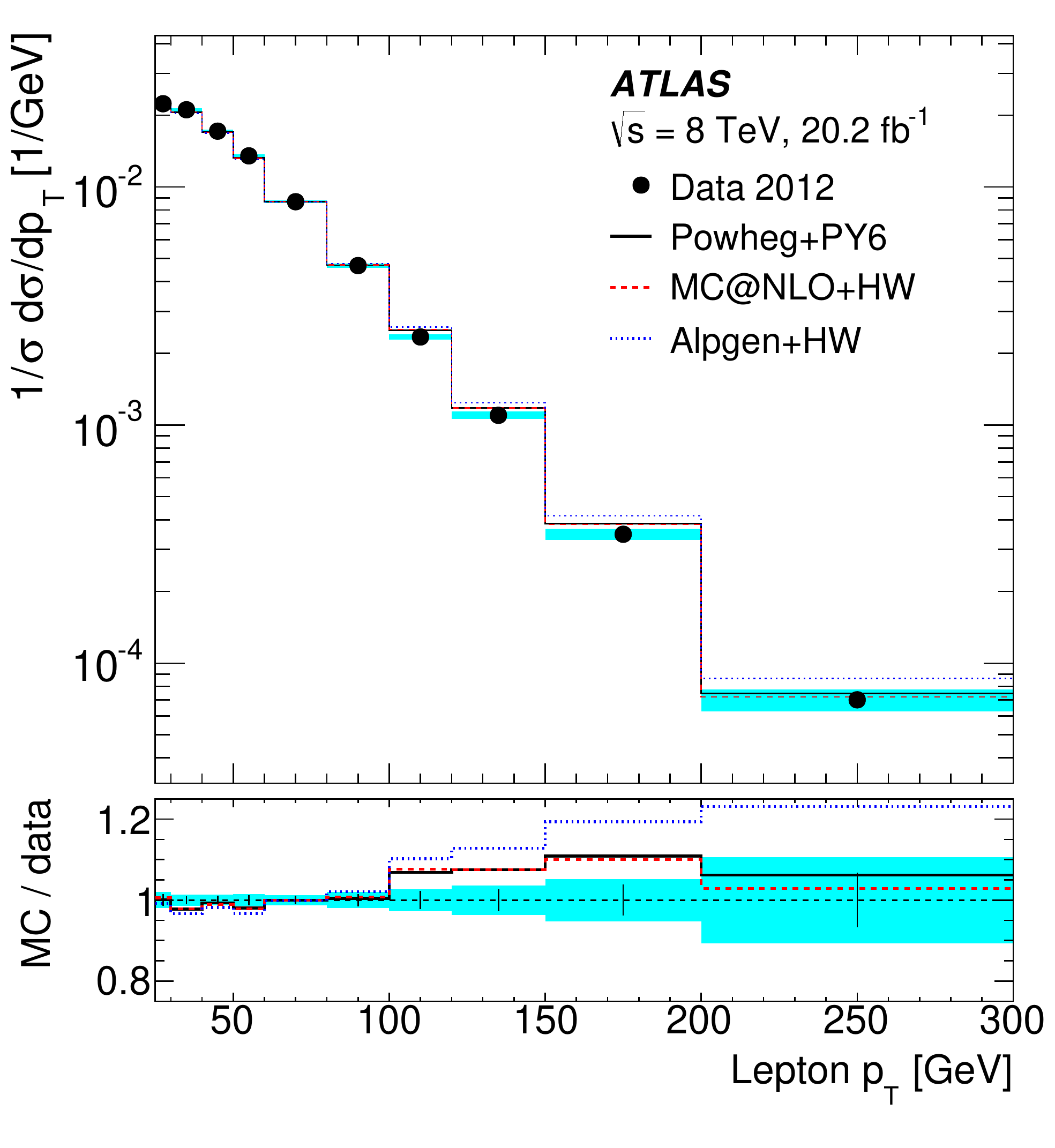}{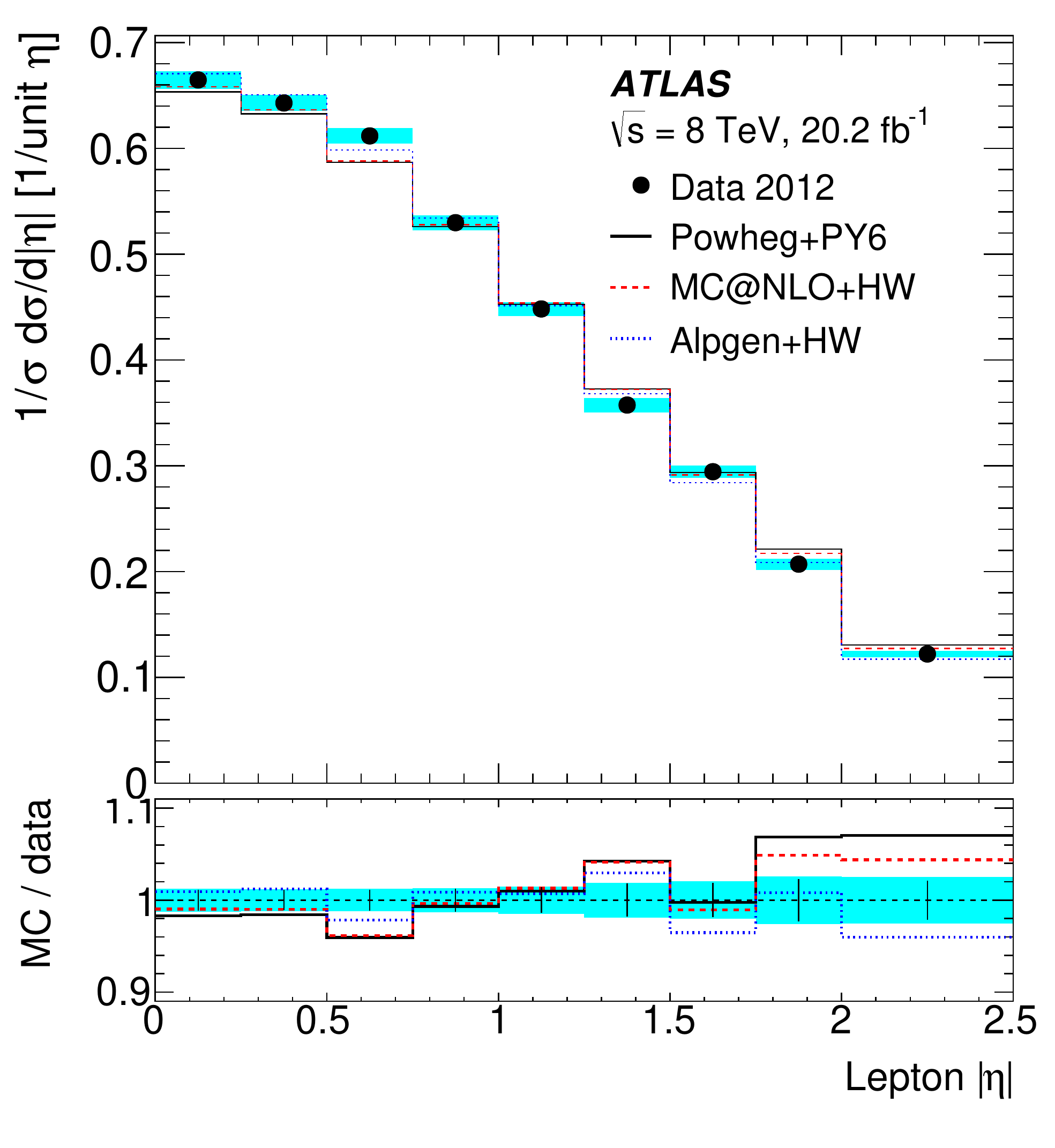}{a}{b}
\splitfigure{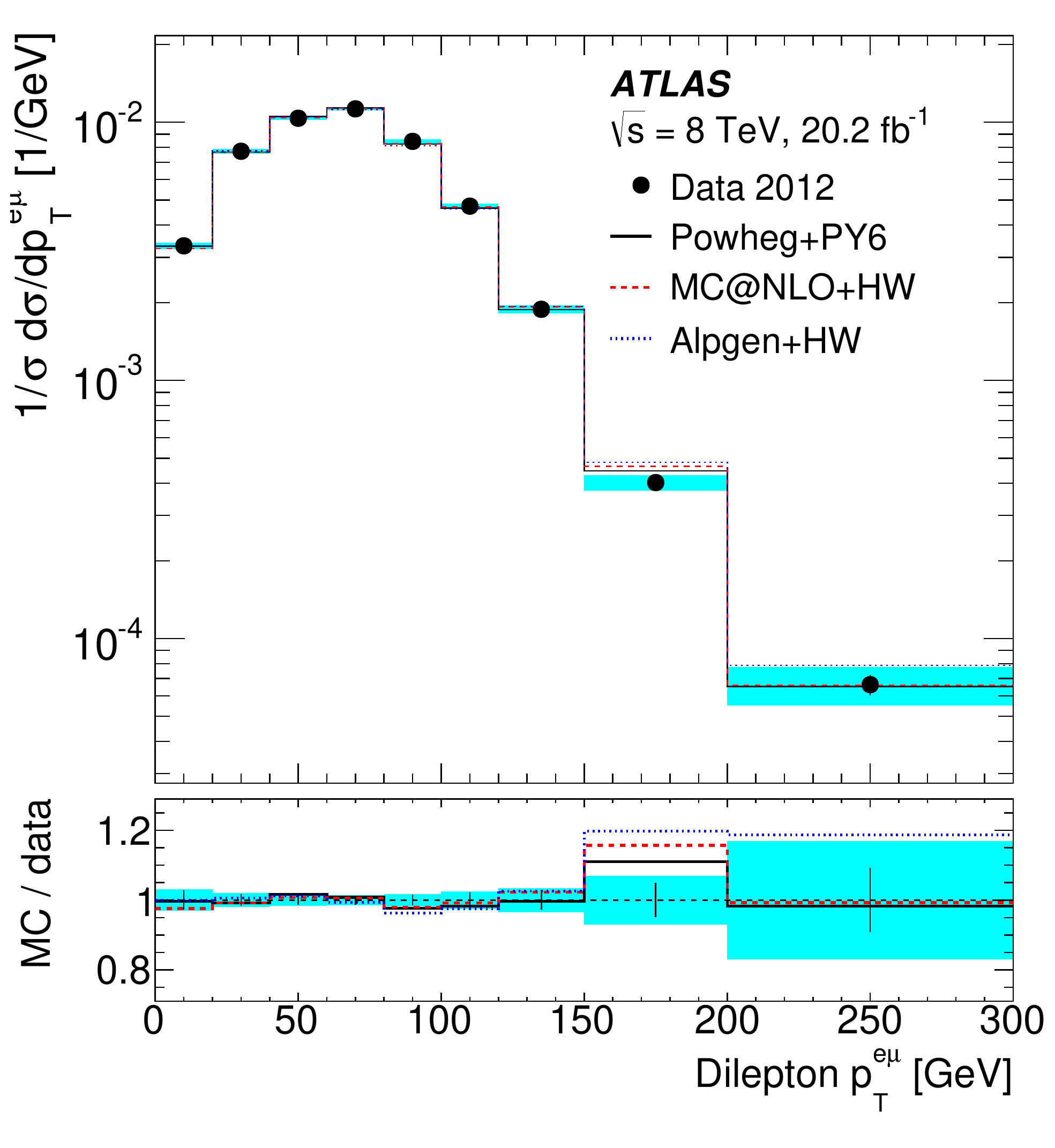}{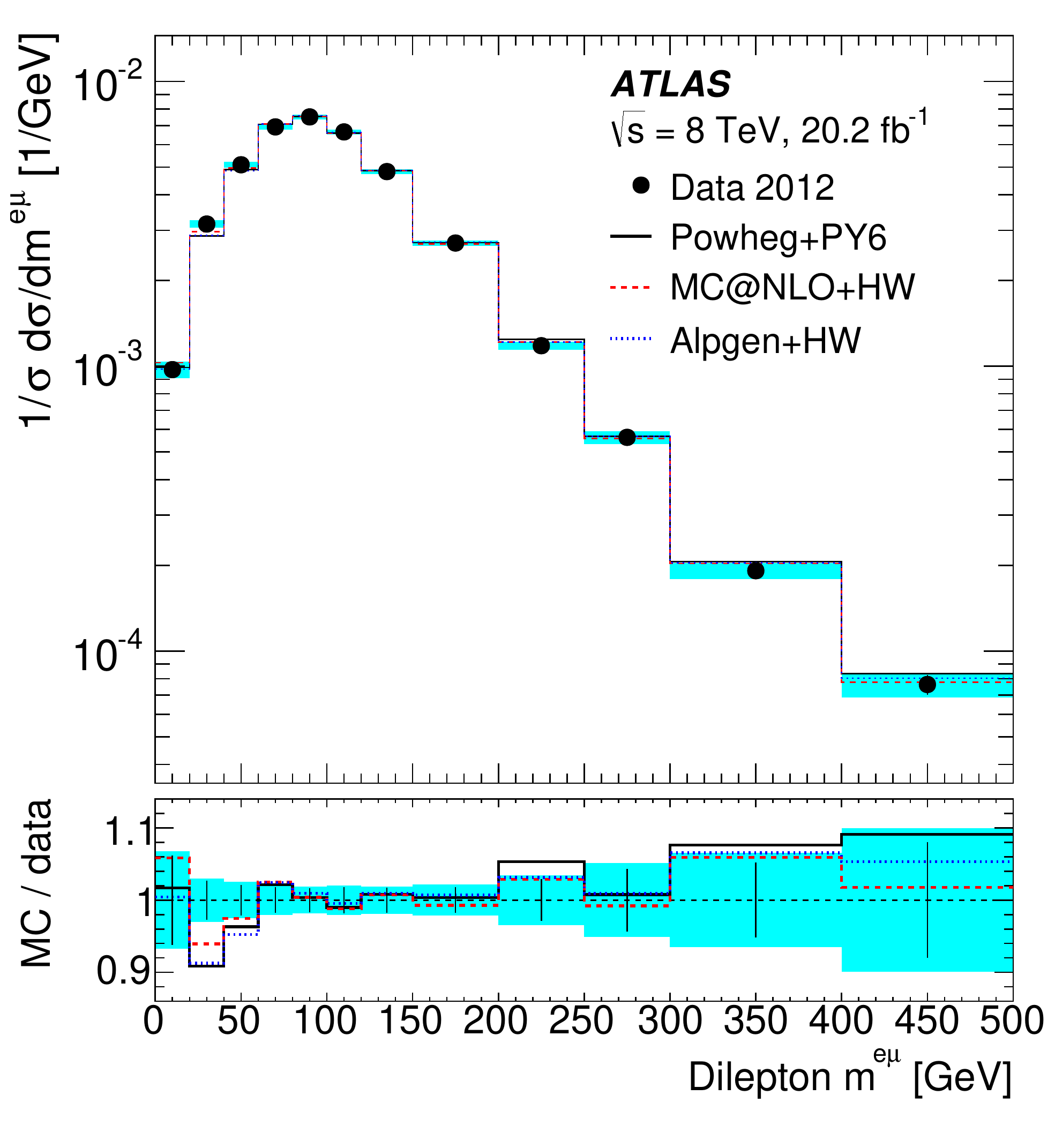}{c}{d}
\caption{\label{f:distresa}Normalised differential cross-sections as a function
of (a) \ptl, (b) \etal, (c) \ptll\ and (d) \mll. The measured values are
shown by the black points with error bars corresponding to the data statistical
uncertainties and cyan bands corresponding to the total uncertainties in each 
bin, and include the contributions via $W\rightarrow\tau\rightarrow e/\mu$ 
decays. The results are compared to the predictions from the
{\sc Powheg\,+\,Pythia6}, {\sc MC@NLO\,+\,Herwig} and {\sc Alpgen\,+\,Herwig}
\ttbar\ simulation samples. The lower plots show the ratios of predictions
to data, with the error bars indicating the data statistical uncertainties 
and the 
cyan bands indicating the total uncertainties in the measurements.}
\end{figure}


\begin{figure}[tp]
\splitfigure{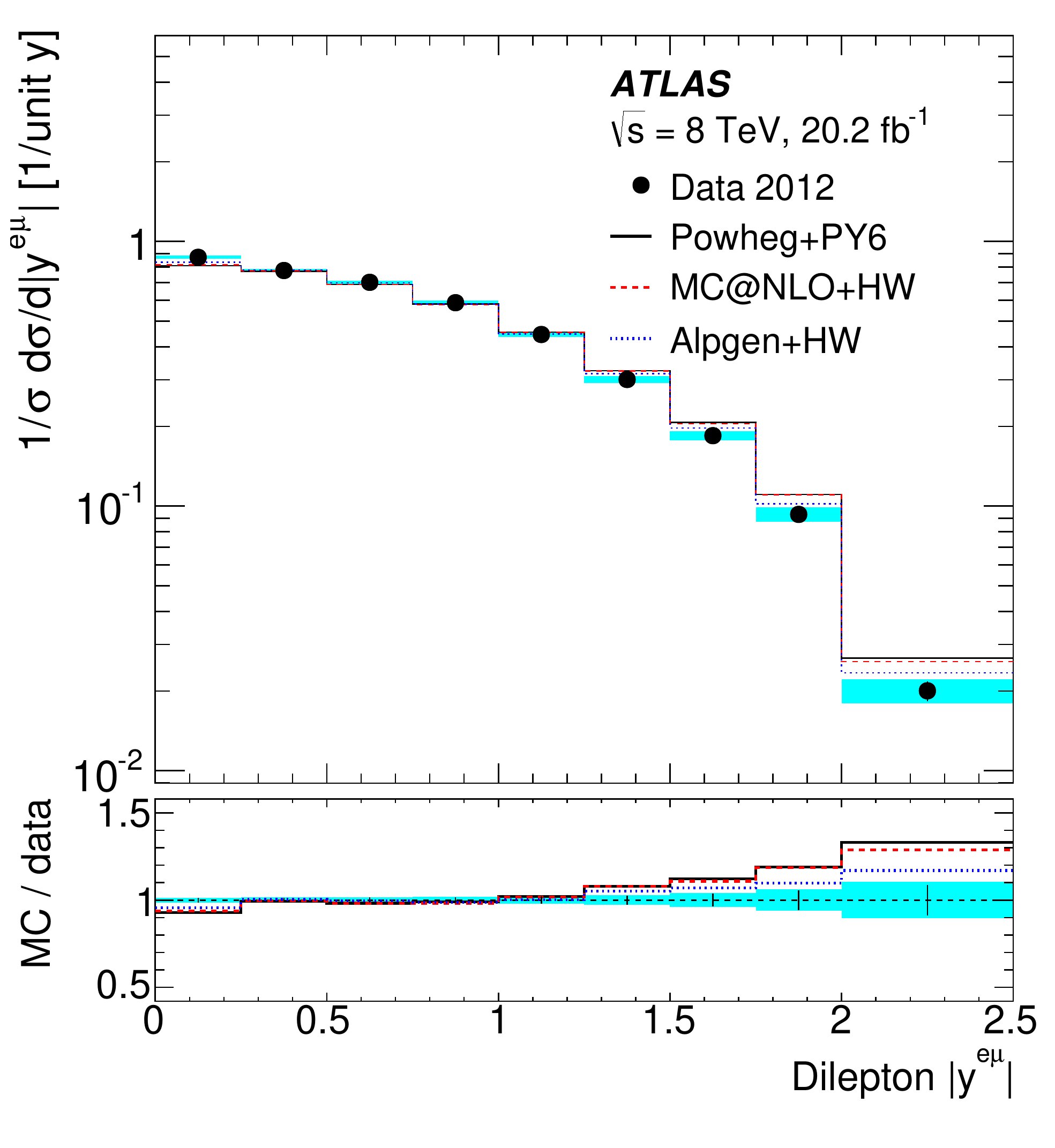}{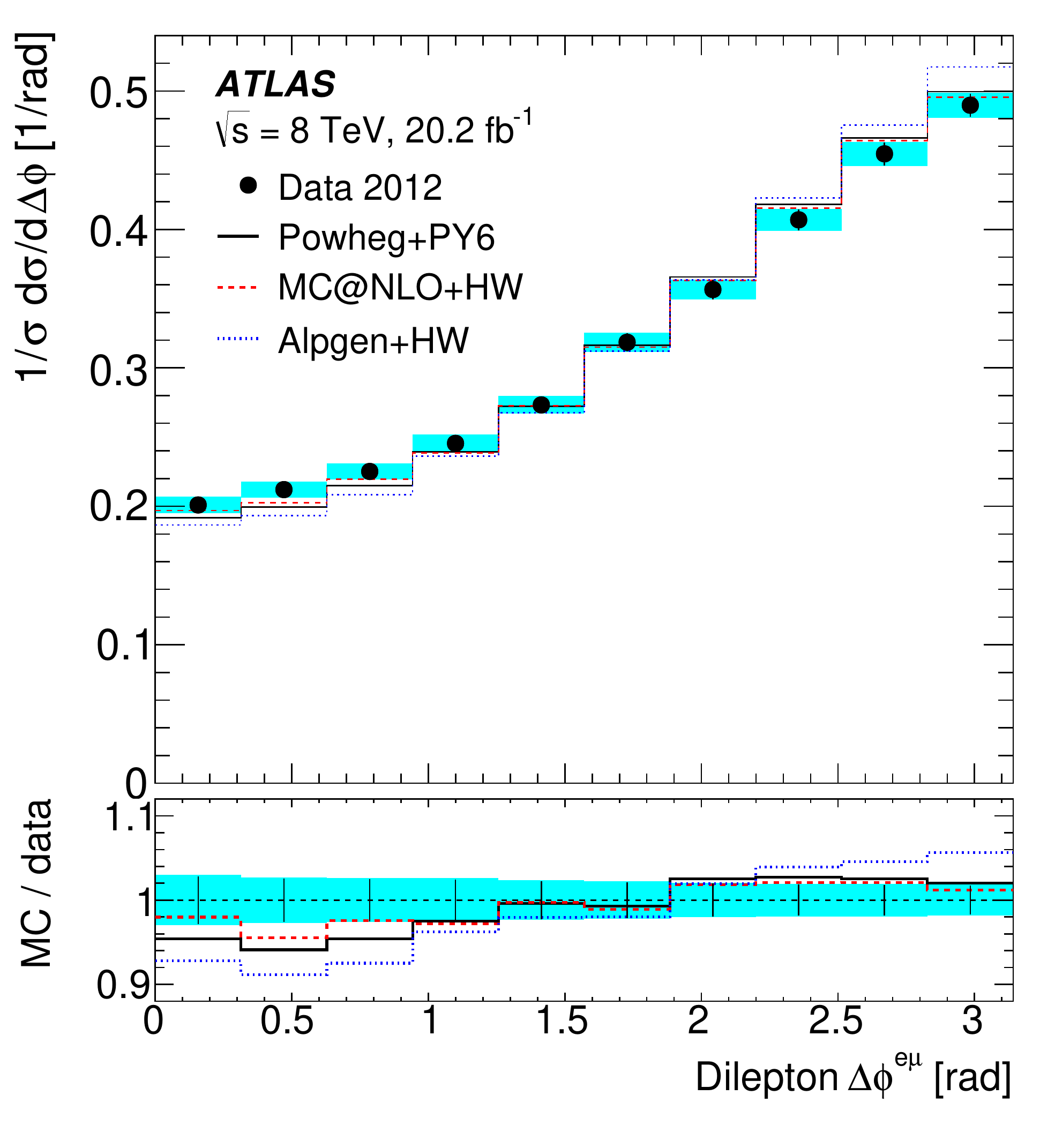}{a}{b}
\splitfigure{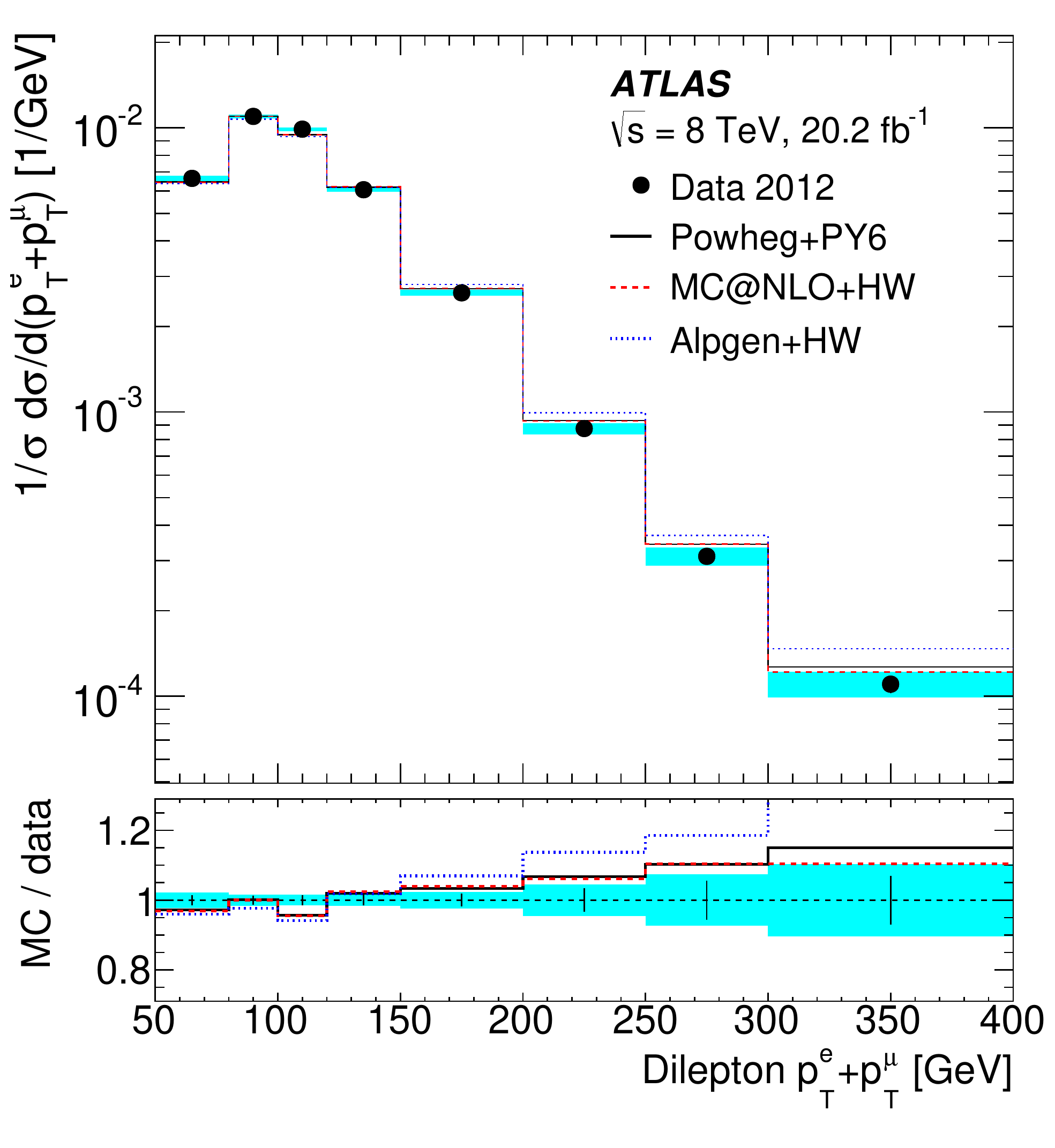}{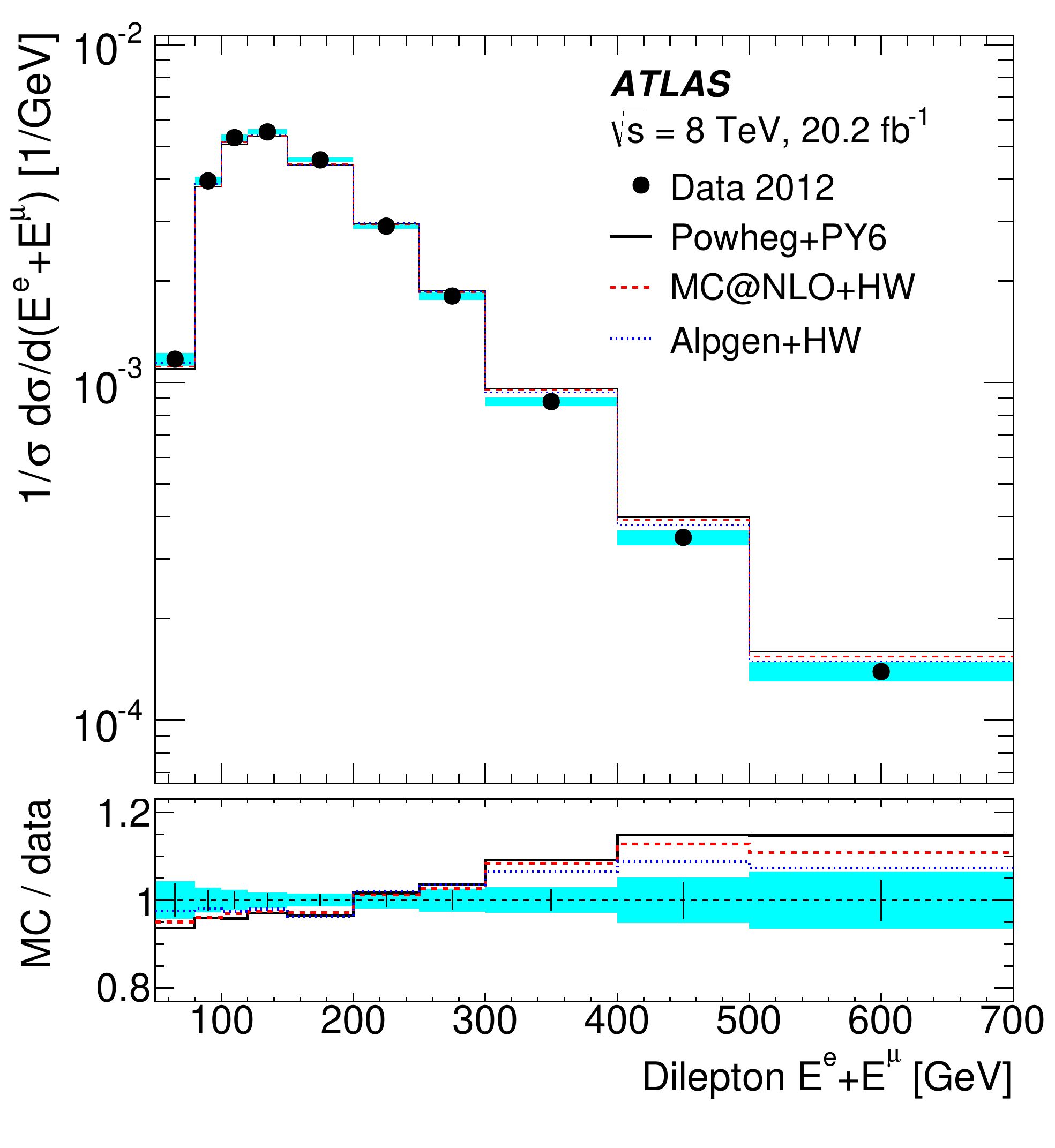}{c}{d}
\caption{\label{f:distresb}Normalised differential cross-sections as a function
of (a) \rapll, (b) \dphill, (c) \ptsum\ and (d) \esum. The measured values are
shown by the black points with error bars corresponding to the data statistical
uncertainties and cyan bands corresponding to the total uncertainties in each
bin, and include the contributions via $W\rightarrow\tau\rightarrow e/\mu$ 
decays. The results are compared to the predictions from the
{\sc Powheg\,+\,Pythia6}, {\sc MC@NLO\,+\,Herwig} and {\sc Alpgen\,+\,Herwig}
\ttbar\ simulation samples. The lower plots show the ratios of predictions
to data, with the error bars indicating the data statistical uncertainties 
and the cyan bands indicating the total uncertainties in the measurements.}
\end{figure}

\subsection{Comparison with event generator predictions}\label{ss:gencomp}

The measured normalised differential cross-sections are compared to a larger
set of predictions from different \ttbar\ Monte Carlo event generator 
configurations in Figures~\ref{f:rdistresa} to~\ref{f:rdistresd}. The 
figures show the ratios of
each prediction to the data as a function of the differential variables, 
organised into four groups of samples as summarised in Table~\ref{t:gencomp}.
These event generator setups and tunes were used in ATLAS top physics analyses at 
\sxwt\ and \sxvt, or have been studied in preparation for analyses at
\sxyt\ \cite{ATL-PHYS-PUB-2015-002,ATL-PHYS-PUB-2015-011,ATL-PHYS-PUB-2016-004}.

\begin{table}[tp]
\centering

\begin{tabular}{l|ll|ll|l}\hline
& Matrix-element & PDF & Parton shower & Tune & Comments \\
\hline
1 & {\sc Powheg} & CT10 & {\sc Pythia6} &  P2011C & $\hdamp=\mtop$ \\
& {\sc Powheg} & CT10 & {\sc Pythia6} &  P2012 radHi & $\hdamp=2\mtop$, $\frac{1}{2}\mu_{F,R}$ \\
& {\sc Powheg} & CT10 & {\sc Pythia6} &  P2012 radLo & $\hdamp=\mtop$, $2\mu_{F,R}$ \\
\hline
2 & {\sc Powheg} & CT10 & {\sc Pythia6} & P2011C & $\hdamp=\infty$ \\
& {\sc Powheg} & CT10 & {\sc Pythia6} &  P2011C & $\hdamp=\mtop$, NNLO top \pt
\\
& {\sc Powheg} & CT10 & {\sc Pythia8} & A14 & $\hdamp=\mtop$ \\
\hline
3 & {\sc Powheg} & HERAPDF 1.5 & {\sc Pythia6} & P2011C & $\hdamp=\mtop$ \\
& {\sc Powheg} & CT10 & {\sc Pythia6} & P2011C & $\hdamp=\infty$, no spin corl.\\
\hline
4 & {\sc Alpgen} & CTEQ6L1 & {\sc Herwig+Jimmy} & AUET2 & incl. \ttbar\bbbar, \ttbar\ccbar \\
& {\sc MC@NLO} & CT10 & {\sc Herwig+Jimmy} & AUET2 & \\
& {\sc MG5\_aMC@NLO} & CT10 & {\sc Herwig++} & UE-EE-5 & \\
\hline
\end{tabular}
\caption{\label{t:gencomp}Summary of particle-level simulation samples used in 
the comparison to the corrected data distributions in Section~\ref{ss:gencomp},
giving the matrix-element event generator,
PDF set, parton shower and associated tune parameter set. The four groups
shown correspond to the four panels for each measured distribution shown 
in Figures~\ref{f:rdistresa} to~\ref{f:rdistresd}.}
\end{table}

\begin{description}
\item[The first group] shows the baseline {\sc Powheg\,+\,Pythia6} \ttbar\ 
sample with \hdamp=\mtop\ (which is also shown in 
Figures~\ref{f:distresa}--\ref{f:distresb}),
together with the two tunes giving more or less 
parton shower radiation---the Perugia 2012 radHi and radLo tunes \cite{perugia}
coupled to scale and \hdamp\ parameter variations as discussed in 
Section~\ref{s:datasim}. 

\item[The second group] shows
a {\sc Powheg\,+\,Pythia6} sample with $\hdamp=\infty$ (i.e. no damping
of the first emission), the baseline {\sc Powheg\,+\,Pythia6} sample
with the top quark \pt\ spectrum reweighted to the NNLO prediction
of Ref. \cite{topptnnlo}, and a sample generated with {\sc Powheg} 
and $\hdamp=\mtop$ interfaced to {\sc Pythia8} (version 8.186) 
\cite{pythia8} with the A14 tune 
\cite{ATL-PHYS-PUB-2014-021} and the CTEQ6L1 PDF set for the parton shower, 
hadronisation and underlying event modelling as described in 
Ref. \cite{ATL-PHYS-PUB-2015-011}. 

\item[The third group] shows a 
{\sc Powheg\,+\,Pythia6} sample with $\hdamp=\mtop$ generated with
the HERAPDF 1.5 PDF set \cite{herapdf10,herapdf15} instead of 
CT10\footnote{Although HERAPDF 1.5 has been superseded by HERAPDF 2.0 
\cite{herapdf20}, which uses the final combined DIS data from the
H1 and ZEUS experiments, HERAPDF 1.5 is used here due to availability
of the corresponding simulation sample.}, and a {\sc Powheg\,+\,Pythia6} sample
with $\hdamp=\infty$ and no simulation of spin correlations between the
top and antitop quarks. 

\item[The fourth group] shows alternative matrix-element event
generators---the {\sc Alpgen\,+\,Herwig} and {\sc MC@NLO\,+\linebreak[0] Herwig} samples 
described in Section~\ref{s:datasim} and shown in 
Figures~\ref{f:distresa}--\ref{f:distresb}, 
together with a sample generated using
{\sc MadGraph5\_aMC@NLO} 2.2.1 \cite{amcnlo} (referred to as {\sc aMC@NLO} 
below) and CT10 PDFs, 
interfaced to {\sc Herwig++} \cite{herwigpp} with the UE-EE-5 
{\sc Herwig++} author tune.
\end{description}

\begin{figure}[tp]
\splitfigure{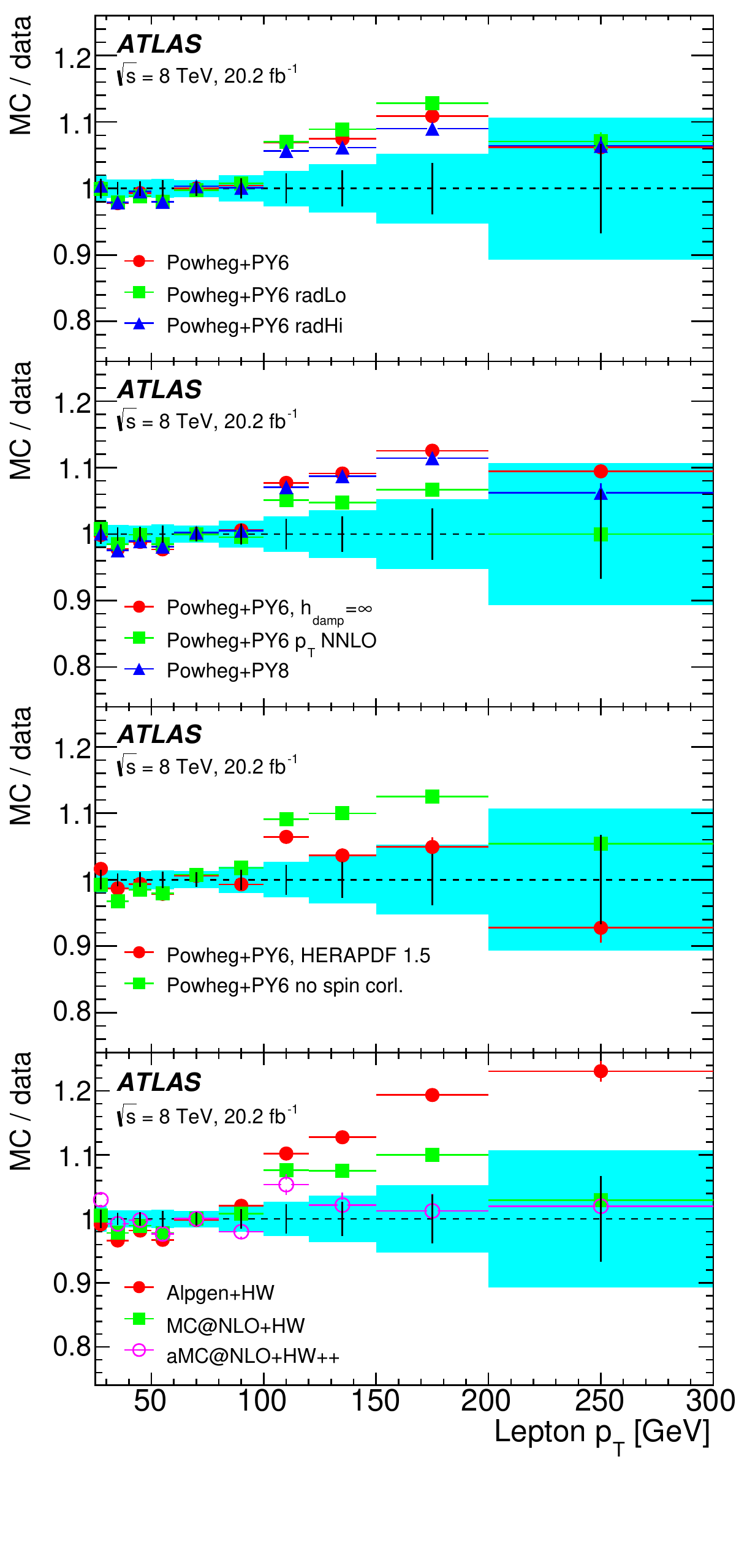}{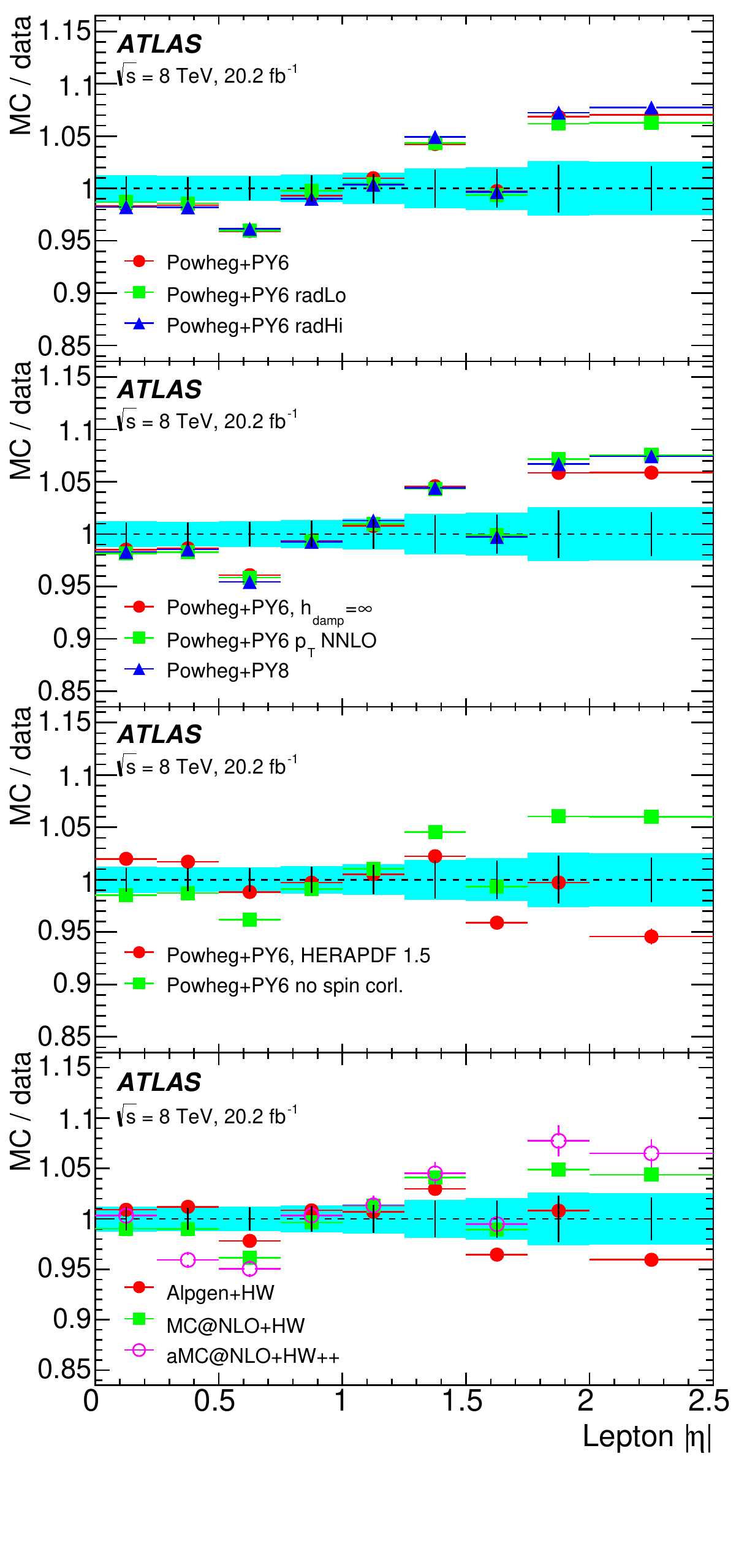}{a}{b}
\caption{\label{f:rdistresa}Ratios of predictions of normalised differential
cross-sections to data as a function of (a) \ptl\ and (b) \etal. The data
statistical uncertainties are shown by the black error bars around a ratio
of unity, and the total uncertainties are shown by the cyan bands. The 
\ttbar\ predictions are shown in four groups from top to bottom, with error bars 
indicating the uncertainties due to the limited size of the simulated samples.}
\end{figure}

\begin{figure}[tp]
\splitfigure{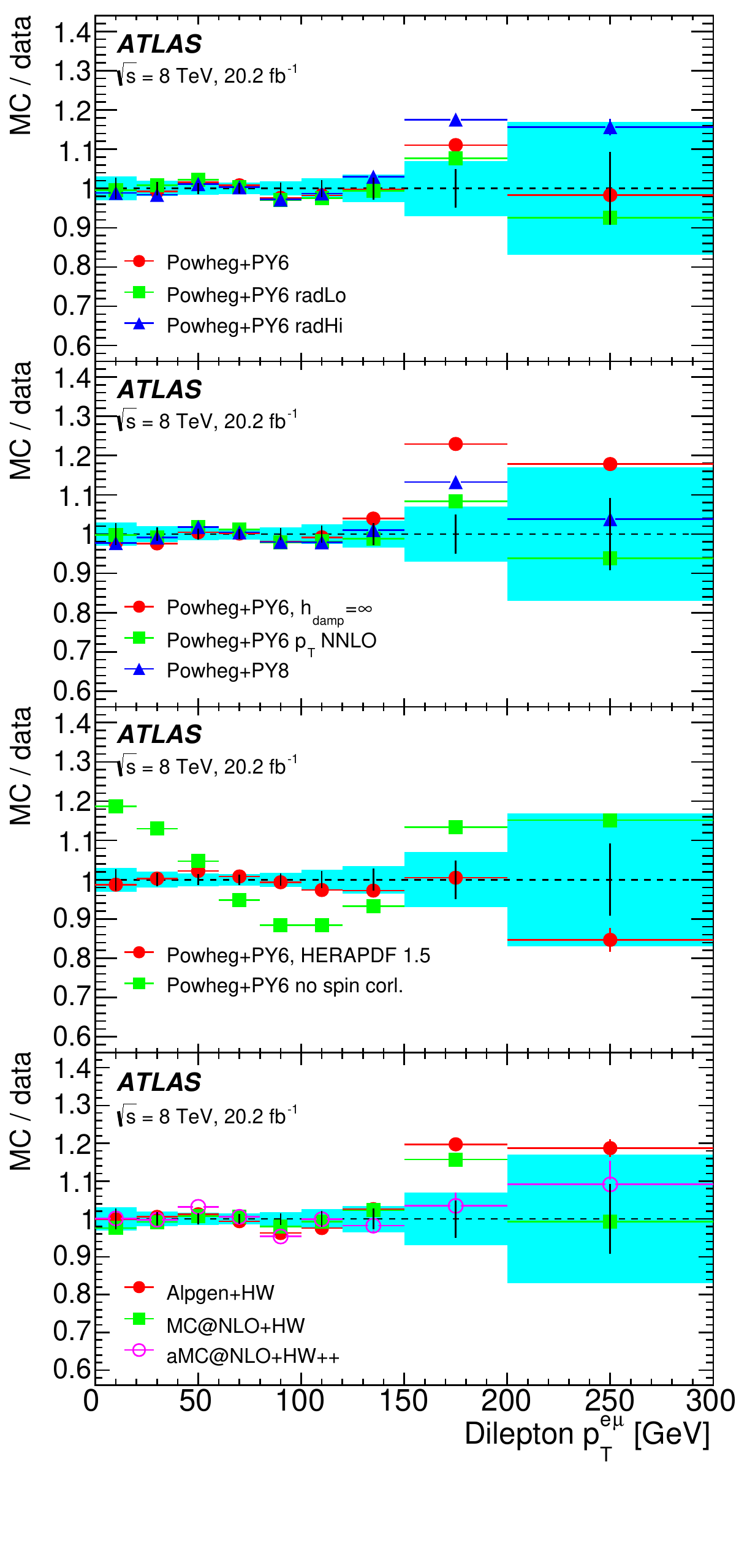}{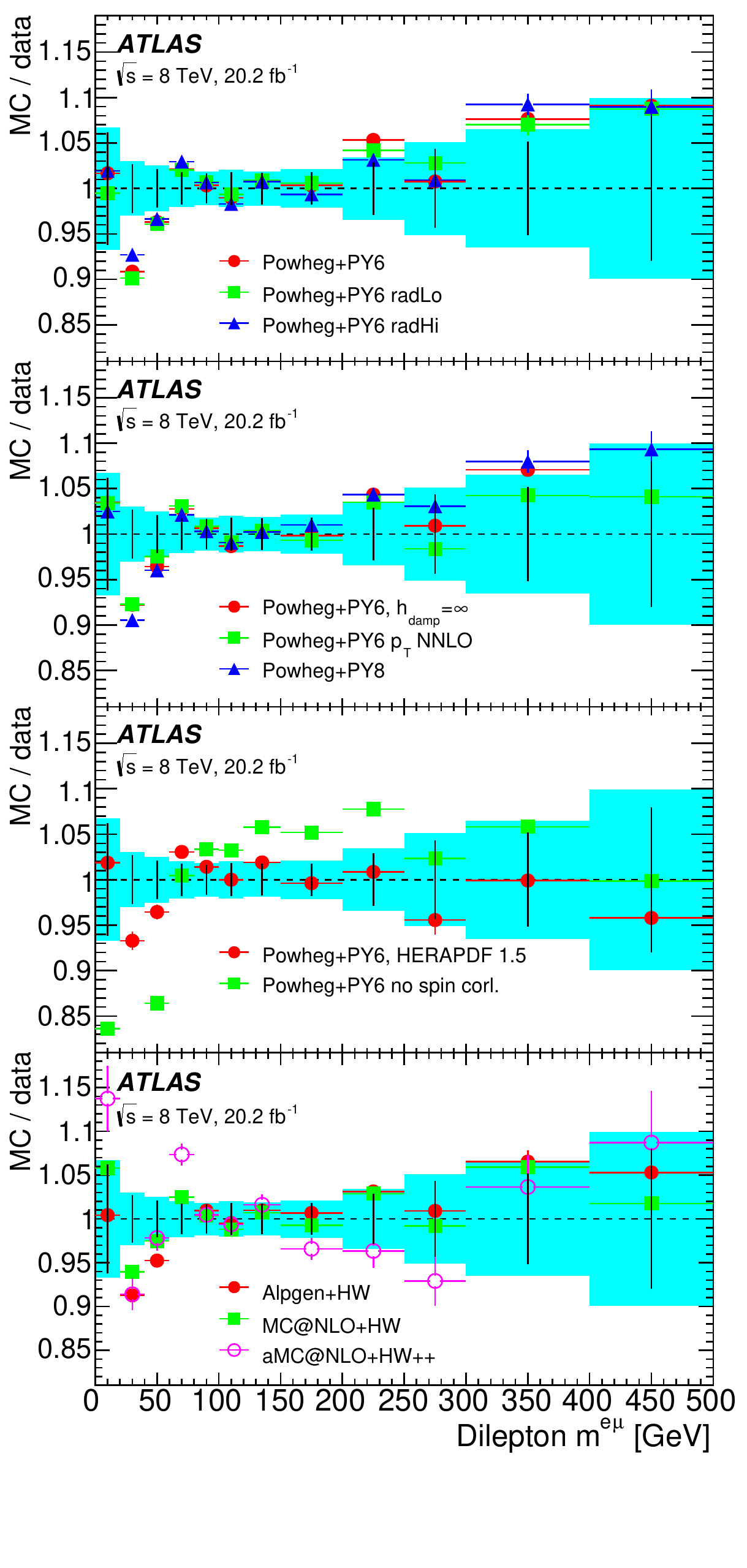}{a}{b}
\caption{\label{f:rdistresb}Ratios of predictions of normalised differential
cross-sections to data as a function of (a) \ptll\ and (b) \mll. The data
statistical uncertainties are shown by the black error bars around a ratio
of unity, and the total uncertainties are shown by the cyan bands. The 
\ttbar\ predictions are shown in four groups from top to bottom, with error bars 
indicating the uncertainties due to the limited size of the simulated samples.}
\end{figure}

\begin{figure}[tp]
\splitfigure{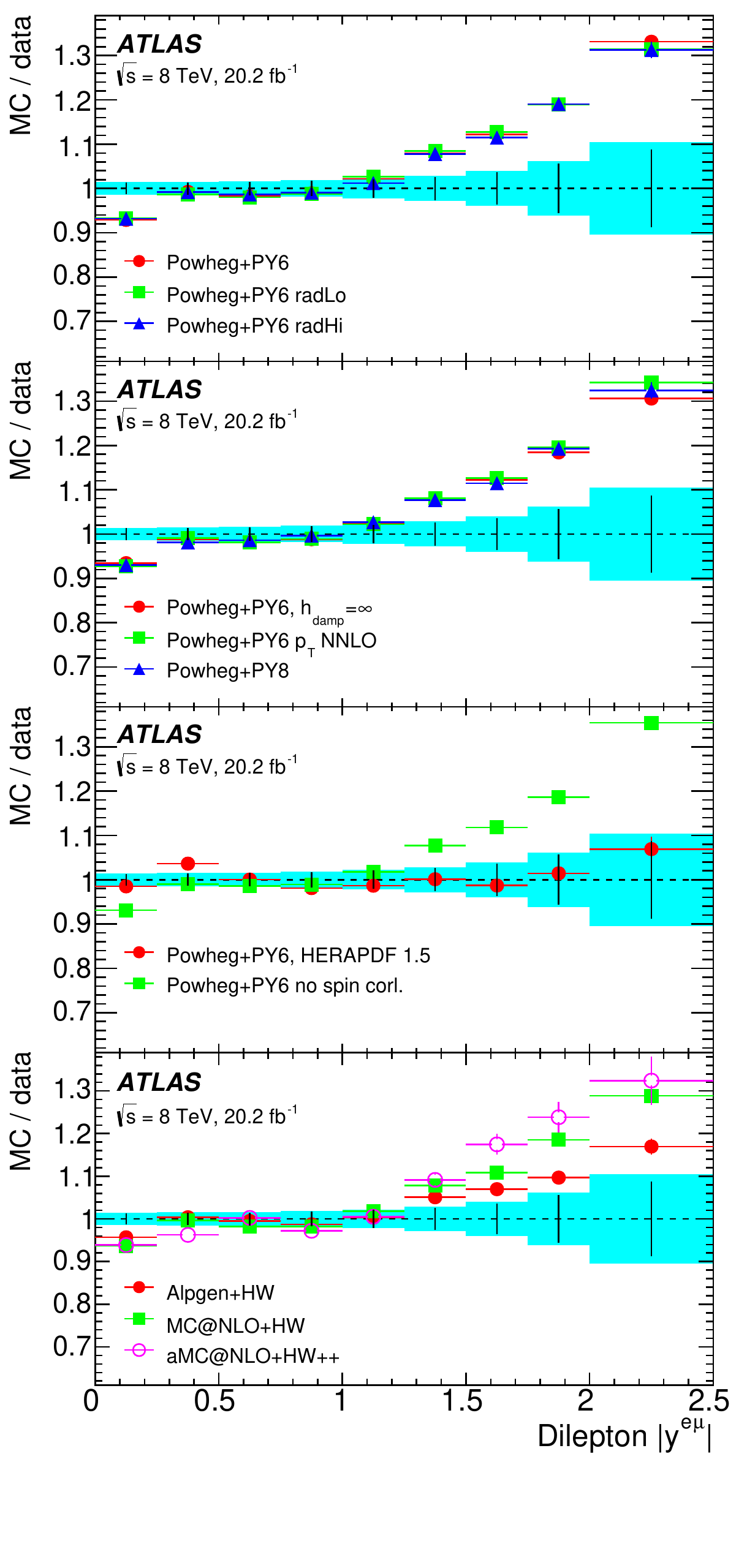}{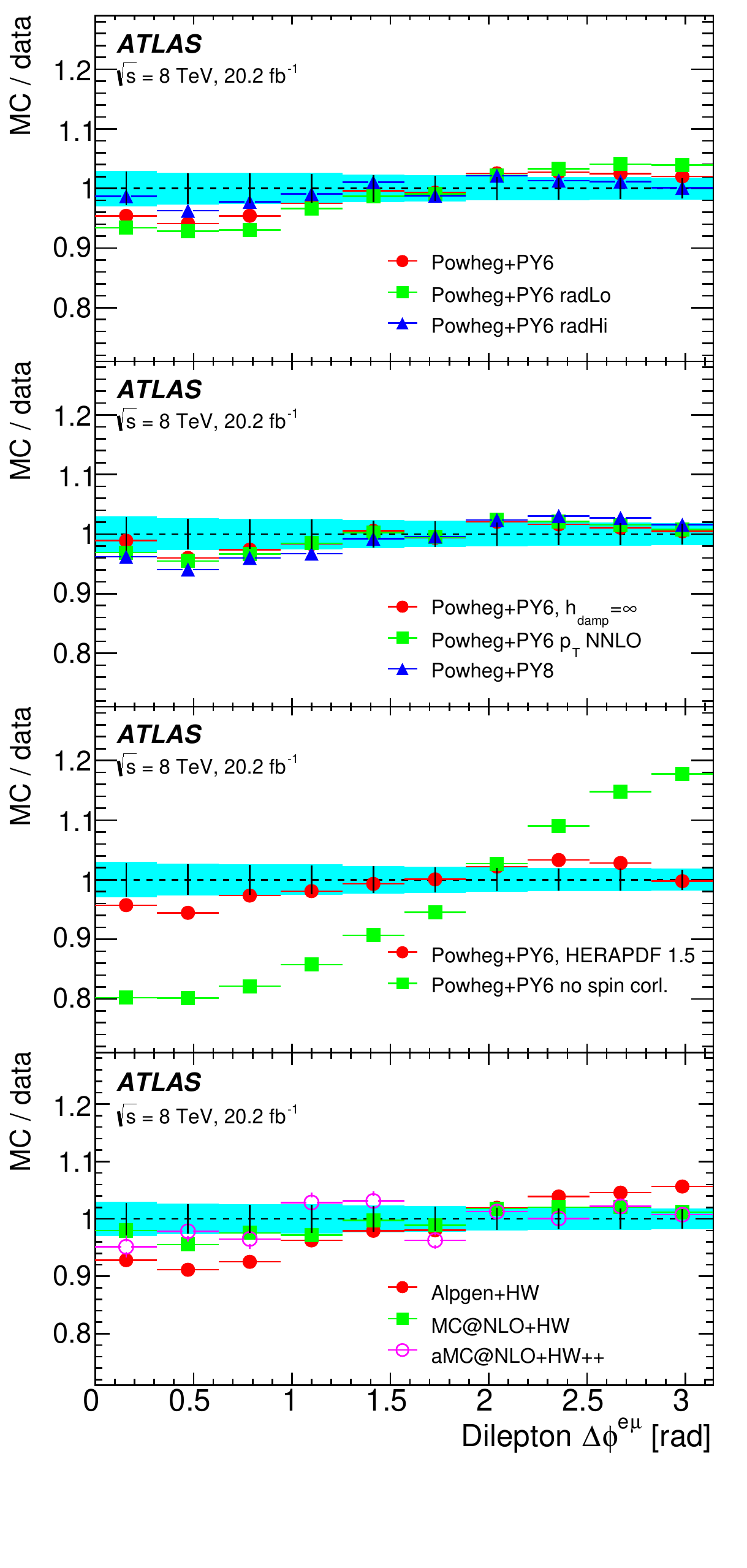}{a}{b}
\caption{\label{f:rdistresc}Ratios of predictions of normalised differential
cross-sections to data as a function of (a) \rapll\ and (b) \dphill. The data
statistical uncertainties are shown by the black error bars around a ratio
of unity, and the total uncertainties are shown by the cyan bands. The 
\ttbar\ predictions are shown in four groups from top to bottom, with error bars 
indicating the uncertainties due to the limited size of the simulated samples.}
\end{figure}

\begin{figure}[tp]
\splitfigure{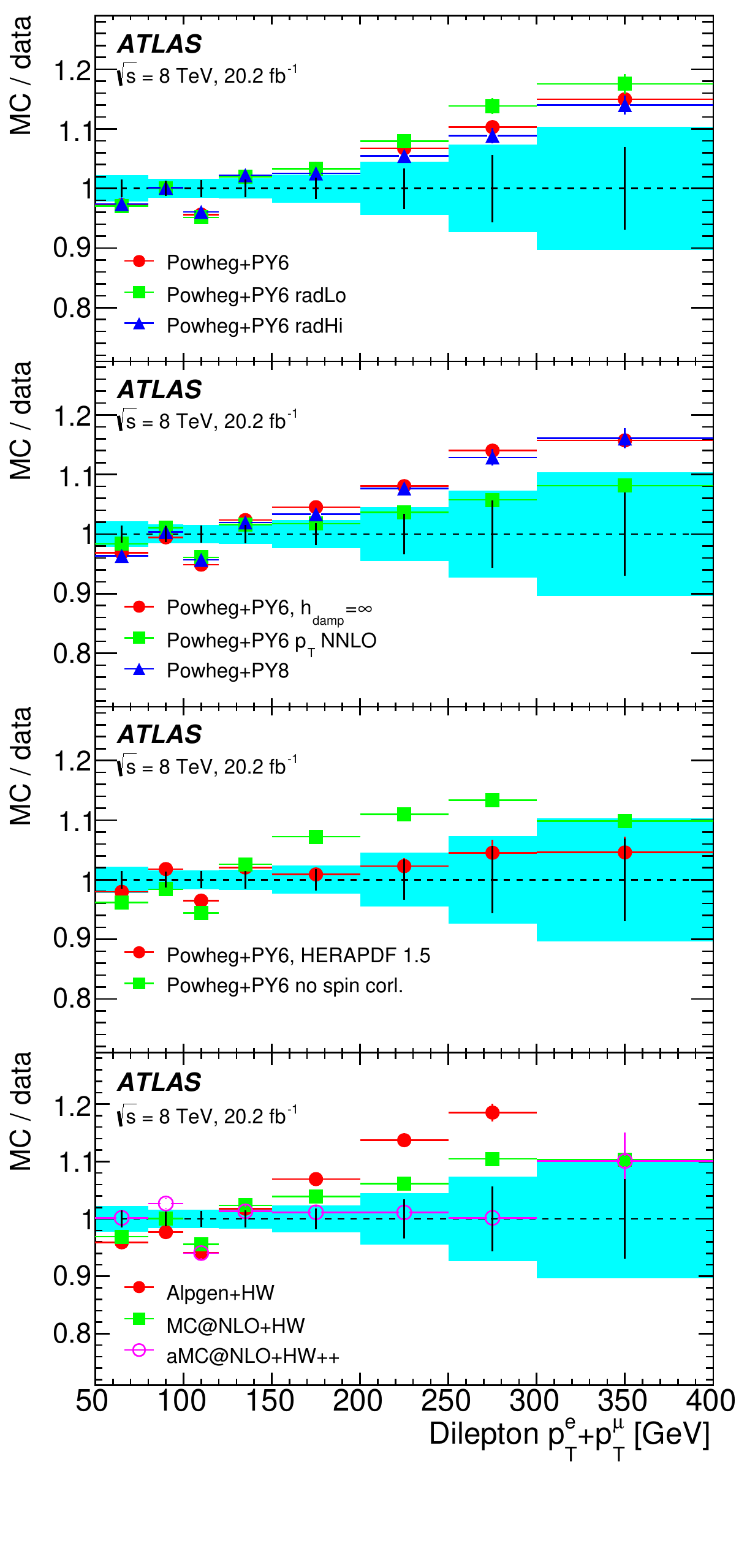}{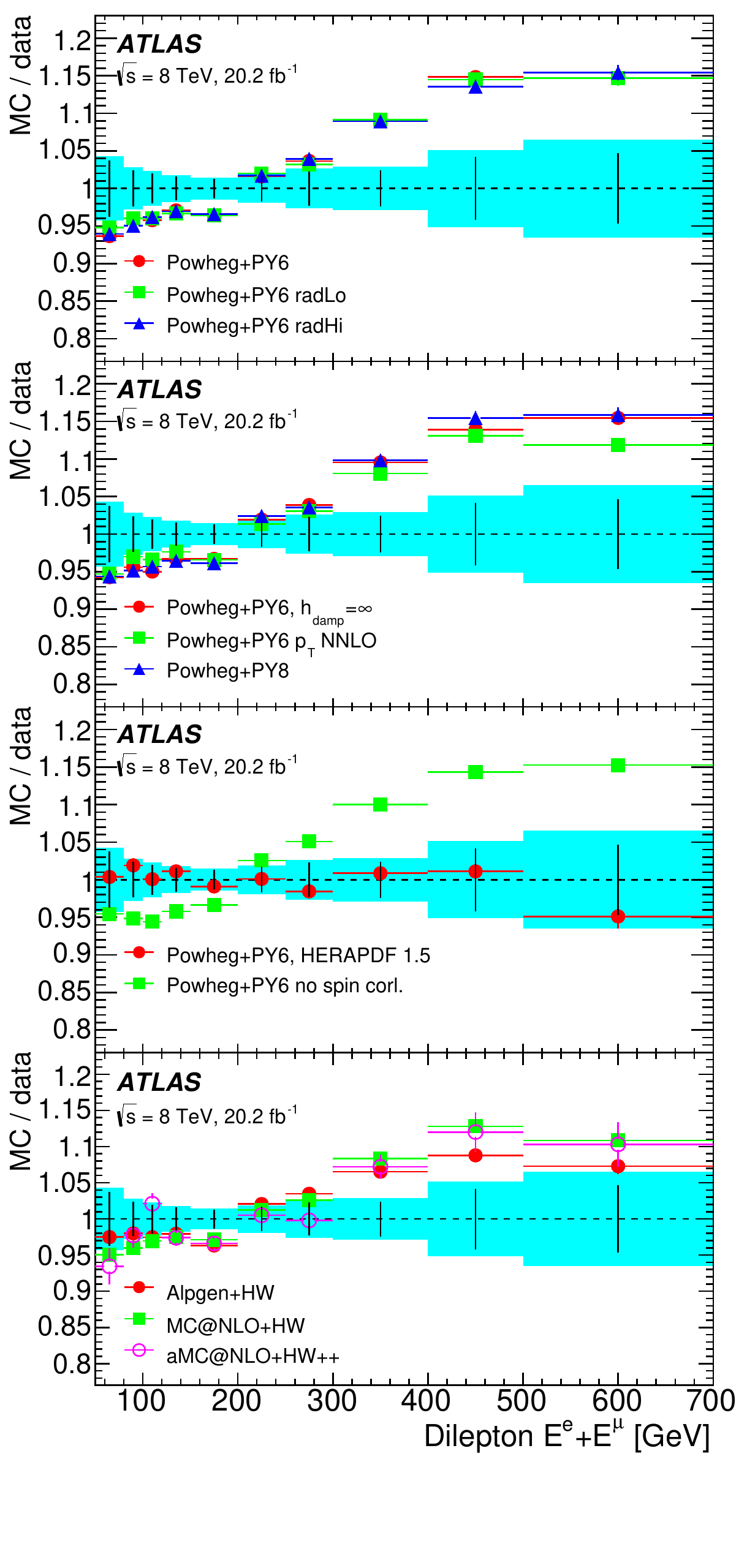}{a}{b}
\caption{\label{f:rdistresd}Ratios of predictions of normalised differential
cross-sections to data as a function of (a) \ptsum\ and (b) \esum. The data
statistical uncertainties are shown by the black error bars around a ratio
of unity, and the total uncertainties are shown by the cyan bands. The 
\ttbar\ predictions are shown in four groups from top to bottom, with error bars 
indicating the uncertainties due to the limited size of the simulated samples.}
\end{figure}

The compatibility of each prediction with each measured normalised distribution
was assessed quantitatively using a $\chi^2$ test, calculated as:
\begin{eqnarray}\label{e:redchi}
\chi^2 = {\mathbf\Delta}_{(n-1)}^T {\mathbf S}^{-1}_{(n-1)} {\mathbf\Delta}_{(n-1)}\,,
\end{eqnarray}
where ${\mathbf\Delta}_{(n-1)}$ is the vector of differences between the measured
and predicted normalised differential cross-section in each of the $n$
bins, excluding the last one, and ${\mathbf S}_{(n-1)}$ is the corresponding
covariance matrix, including both the experimental uncertainties in the
measurement and the statistical uncertainties in the predictions.
Bin-to-bin correlations in both the statistical (from the normalisation 
condition) and systematic uncertainties were taken into account
via off-diagonal 
entries. The last bin of each distribution was excluded due to
the normalisation condition, rendering the covariance matrix ${\mathbf S}_{(n-1)}$
invertible.\footnote{The $\chi^2$ value does not depend on the choice of which
bin is removed.}
The resulting $\chi^2$ values, number of degrees of freedom $(n-1)$ 
and corresponding $\chi^2$ probability $p$-values are shown for 
each distribution and prediction in Table~\ref{t:NorChiXPPData2012Data}.

\begin{table}[tp]
\centering

\begin{tabular}{l|rrrrrrrr}\hline
Generator & \ptl & \etal & \ptll & \mll & \rapll & \dphill & \ptsum & \esum \\
$N_{\mathrm dof}$ & 9 & 8 & 8 & 11 & 8 & 9 & 7 & 9 \\ \hline
{\sc Powheg\,+\,PY6}&  13.6 &  26.3 &   7.3 &  14.6 &  46.6 &  14.0 &  11.3 &  22.7 \\
{\sc Powheg\,+\,PY6} radLo&  15.9 &  22.9 &   7.6 &  14.6 &  45.6 &  25.9 &  14.0 &  22.0 \\
{\sc Powheg\,+\,PY6} radHi&  10.0 &  28.2 &  11.0 &  12.6 &  42.0 &   4.5 &   9.1 &  21.4 \\
{\sc Powheg\,+\,PY6} $\hdamp=\infty$&  17.2 &  22.5 &  14.5 &  12.9 &  42.8 &   5.0 &  15.6 &  23.4 \\
{\sc Powheg\,+\,PY6} \pt\ NNLO&   8.3 &  28.5 &   6.3 &  12.1 &  49.2 &   7.6 &   7.6 &  17.4 \\
{\sc Powheg\,+\,PY8} $\hdamp=\infty$&  15.1 &  28.9 &   8.3 &  14.4 &  44.3 &  13.0 &  12.7 &  25.8 \\
{\sc Powheg\,+\,PY6} HERAPDF 1.5&  11.4 &  11.8 &   3.6 &  11.1 &   6.7 &  10.3 &   7.0 &   1.9 \\
{\sc Powheg\,+\,PY6} no spin corl.&  21.8 &  23.2 &   152 &   100 &  45.3 &   279 &  22.4 &  27.6 \\
{\sc Alpgen\,+\,HW}&  31.2 &  11.6 &  15.5 &  13.7 &  15.3 &  36.0 &  27.4 &  12.7 \\
{\sc MC@NLO\,+\,HW}&  15.7 &  18.8 &   9.4 &   9.3 &  39.4 &   7.1 &  11.8 &  16.2 \\
{\sc aMC@NLO\,+\,HW$^{++}$}&   7.8 &  29.2 &   7.6 &  24.5 &  46.6 &   8.2 &  12.0 &  13.8 \\
\hline
{\sc Powheg\,+\,PY6}& 0.14 & 9\,$10^{-4}$ & 0.51 & 0.20 & 2\,$10^{-7}$ & 0.12 & 0.13 & 7\,$10^{-3}$ \\
{\sc Powheg\,+\,PY6} radLo& 0.070 & 3\,$10^{-3}$ & 0.48 & 0.20 & 3\,$10^{-7}$ & 2\,$10^{-3}$ & 0.052 & 9\,$10^{-3}$ \\
{\sc Powheg\,+\,PY6} radHi& 0.35 & 4\,$10^{-4}$ & 0.20 & 0.32 & 1\,$10^{-6}$ & 0.87 & 0.24 & 0.011 \\
{\sc Powheg\,+\,PY6} $\hdamp=\infty$& 0.045 & 4\,$10^{-3}$ & 0.069 & 0.30 & 1\,$10^{-6}$ & 0.83 & 0.029 & 5\,$10^{-3}$ \\
{\sc Powheg\,+\,PY6} \pt\ NNLO& 0.51 & 4\,$10^{-4}$ & 0.62 & 0.36 & 6\,$10^{-8}$ & 0.57 & 0.36 & 0.043 \\
{\sc Powheg\,+\,PY8} $\hdamp=\infty$& 0.089 & 3\,$10^{-4}$ & 0.41 & 0.21 & 5\,$10^{-7}$ & 0.16 & 0.080 & 2\,$10^{-3}$ \\
{\sc Powheg\,+\,PY6} HERAPDF 1.5& 0.25 & 0.16 & 0.89 & 0.44 & 0.57 & 0.32 & 0.43 & 0.99 \\
{\sc Powheg\,+\,PY6} no spin corl.& 0.010 & 3\,$10^{-3}$ & 0& 0& 3\,$10^{-7}$ & 0& 2\,$10^{-3}$ & 1\,$10^{-3}$ \\
{\sc Alpgen\,+\,HW}& 3\,$10^{-4}$ & 0.17 & 0.051 & 0.25 & 0.054 & 4\,$10^{-5}$ & 3\,$10^{-4}$ & 0.17 \\
{\sc MC@NLO\,+\,HW}& 0.073 & 0.016 & 0.31 & 0.60 & 4\,$10^{-6}$ & 0.62 & 0.11 & 0.063 \\
{\sc aMC@NLO\,+\,HW$^{++}$}& 0.56 & 3\,$10^{-4}$ & 0.47 & 0.011 & 2\,$10^{-7}$ & 0.52 & 0.10 & 0.13 \\
\hline
\end{tabular}
\caption{\label{t:NorChiXPPData2012Data}The $\chi^2$ values (top) and associated probabilities (bottom) for comparison of measured normalised differential fiducial cross-sections with various \ttbar\ simulation samples.
Probabilities smaller than $10^{-10}$ are shown as zero.}
\end{table}

As can be seen from Figure~\ref{f:rdistresa},
in the single-lepton \ptl\ distribution, the data are softer than the 
predictions from {\sc Powheg} with CT10 PDFs, interfaced to either 
{\sc Pythia6} or {\sc Pythia8}. The {\sc Powheg}-based predictions do not 
depend strongly on the choice of parton shower/hadronisation model or 
tune parameters controlling the amount of radiation.
However, the agreement with data is improved when using HERAPDF 1.5 or 
reweighting to the NNLO top quark \pt\ prediction from Ref. \cite{topptnnlo}. 
The predictions from the samples with alternative matrix-element event 
generators,
i.e. 
{\sc MC@NLO\,+\,Herwig} and {\sc Alpgen\,+\,Herwig},
are also harder than the data, though {\sc aMC@NLO\,+\,Herwig++} describes
the data well. 
The \ptsum\ and \esum\ distributions (Figure~\ref{f:rdistresd})
show some similar features to \ptl, being 
softer than the predictions from the {\sc Powheg\,+\,Pythia6} samples
with CT10, and better described with HERAPDF 1.5, and by 
{\sc aMC@NLO\,+\,Herwig++}.

The predictions for the single lepton \etal\  and 
dilepton \rapll\ distributions (Figures~\ref{f:rdistresa} and~\ref{f:rdistresc})
are insensitive to the choice of parton shower/hadronisation
model and tune, and are also insensitive to the top quark \pt\ reweighting.
The data distributions
are more central than the predictions of all the NLO event generators ({\sc Powheg},
{\sc MC@NLO} and {\sc aMC@NLO}) with CT10 PDFs, but are better described by 
{\sc Powheg} with HERAPDF 1.5, and to a lesser extent also by 
{\sc Alpgen\,+\,Herwig}, which uses the leading-order CTEQ6L1 PDF. These 
distributions, whose experimental measurements are limited by statistical
uncertainties over the full kinematic range, are thus particularly 
suitable for constraining PDFs, as explored further in Section~\ref{s:pdf}.

The dilepton \ptll\ and \mll\ distributions (Figure~\ref{f:rdistresb}) are
generally well described by all the NLO event generators, except for {\sc aMC@NLO}
which does not model the data well at low \mll. The \ptll\ distribution
is sensitive to the amount of parton radiation, and is better described by
the radLo than the radHi {\sc Powheg\,+\,Pythia6} sample,
and by $\hdamp=\mtop$ than $\hdamp=\infty$. Both distributions
are sensitive to the modelling of \ttbar\ spin correlations, and are 
not well-modelled by the {\sc Powheg\,+\,Pythia6} sample without 
spin correlations. 

The \dphill\ distribution (Figure~\ref{f:rdistresc})
is particularly sensitive to spin correlations,
and has been previously used to exclude \ttbar\ simulation models without spin 
correlation and the pair-production of supersymmetric top squarks with
masses close to \mtop, via template fits to reconstruction-level distributions
\cite{TOPQ-2014-07,CMS-TOP-14-023}. The particle-level \dphill\ measurements
shown here also exclude the prediction without spin correlations and
the LO implementation of spin correlations in the {\sc Alpgen\,+\,Herwig}
sample. The \dphill\ distribution is also sensitive to radiation,
this time favouring the radHi {\sc Powheg\,+\,Pythia6} sample. 

The $\chi^2$ formalism of Eq.~(\ref{e:redchi}) was extended to consider 
several normalised distributions simultaneously, by forming vectors $\Delta_i$
where the index runs over the bins of several distributions, excluding the
last bin in each one to account for the normalisation condition. The
covariance matrix {\textbf S} was extended with off-block-diagonal components 
encoding the correlations between bins of different measured distributions.
The statistical correlations between distributions were evaluated using
pseudo-experiments generated by sampling from the large simulated \ttbar\ 
sample discussed in Section~\ref{ss:valid}. The individual sources of 
systematic uncertainty were assumed to be fully correlated across
the different distributions. Five sets of combined distributions
were considered: the combination of \ptl\ and \ptll, combining all the 
information from single and dilepton \pt; the combination of \ptll, \mll\ 
and \ptsum, including all the dilepton kinematic distributions except rapidity;
the combination
of \etal\ and \rapll, combining the single and dilepton rapidity information;
the combination of \etal, \rapll\ and \esum, combining all the distributions
with longitudinal information; and the combination of all eight measured
distributions, denoted `All'. The resulting $\chi^2$ values, numbers of 
degrees of freedom and $p$-values are shown for each combination and 
prediction in Table~\ref{t:NorCombChiXPPData2012Data}. 

The results for the combinations of distributions reflect
the observations for the individual distributions. The best modelling of 
the first two combinations (involving \ptl, \ptll, \mll\ and \ptsum) is given
by {\sc Powheg\,+\,Pythia6} with either HERAPDF 1.5 or with CT10 plus 
reweighting of the top quark \pt\ distribution to the NNLO prediction; 
the radHi variation of 
{\sc Powheg\,+\,Pythia6} also does well. The combinations involving
\etal\ and \rapll\ and the combination of all eight distributions are only 
well-described by {\sc Powheg\,+Pythia6} with HERAPDF 1.5, and marginally well
described by the radHi variation. All other event generator setups (in particular
the LO multileg event generator {\sc Alpgen})
fail to describe some of the distributions, but this could potentially be 
improved by appropriate parameter tuning and switching to a different PDF set.
These results highlight the sensitivity of the differential distributions
to the choice of PDF, in particular that of the gluon, as discussed further
in Section~\ref{s:pdf}. They also indicate that NNLO corrections may be
important in describing the kinematics of the decay leptons, as well as for the
prediction of the top quark \pt\ spectrum as discussed in Ref. \cite{topptnnlo}.


\begin{table}[tp]
\centering

\begin{tabular}{l|ccccc}\hline
Generator & \ptl, \ptsum & \ptll, \mll, & \etal, \rapll & \etal, \rapll, & All \\
&  & \ptsum &  & \esum &  \\ 
$N_{\mathrm dof}$ & 16 & 26 & 16 & 25 & 69 \\ \hline
{\sc Powheg\,+\,PY6}&  20.7 &  38.2 &  57.6 &  70.0 &   120 \\
{\sc Powheg\,+\,PY6} radLo&  24.6 &  50.6 &  57.6 &  70.6 &   138 \\
{\sc Powheg\,+\,PY6} radHi&  16.4 &  29.7 &  52.3 &  62.8 &  98.7 \\
{\sc Powheg\,+\,PY6} $\hdamp=\infty$&  25.0 &  40.1 &  54.2 &  68.7 &   113 \\
{\sc Powheg\,+\,PY6} \pt\ NNLO&  15.1 &  30.0 &  60.0 &  68.2 &   109 \\
{\sc Powheg\,+\,PY8} $\hdamp=\infty$&  23.6 &  37.3 &  56.8 &  71.3 &   121 \\
{\sc Powheg\,+\,PY6} HERAPDF 1.5&  20.1 &  29.6 &  22.5 &  24.5 &  68.6 \\
{\sc Powheg\,+\,PY6} no spin corl.&  30.2 &   284 &  58.3 &  77.4 &   462 \\
{\sc Alpgen\,+\,HW}&  38.9 &  79.3 &  49.3 &  67.2 &   154 \\
{\sc MC@NLO\,+\,HW}&  23.1 &  35.2 &  54.8 &  65.7 &   110 \\
{\sc aMC@NLO\,+\,HW$^{++}$}&  19.1 &  45.2 &  63.1 &  70.2 &   128 \\
\hline
{\sc Powheg\,+\,PY6}& 0.19 & 0.058 & 1\,$10^{-6}$ & 4\,$10^{-6}$ & 1\,$10^{-4}$ \\
{\sc Powheg\,+\,PY6} radLo& 0.077 & 3\,$10^{-3}$ & 1\,$10^{-6}$ & 3\,$10^{-6}$ & 2\,$10^{-6}$ \\
{\sc Powheg\,+\,PY6} radHi& 0.43 & 0.28 & 1\,$10^{-5}$ & 4\,$10^{-5}$ & 0.011 \\
{\sc Powheg\,+\,PY6} $\hdamp=\infty$& 0.069 & 0.038 & 5\,$10^{-6}$ & 6\,$10^{-6}$ & 6\,$10^{-4}$ \\
{\sc Powheg\,+\,PY6} \pt\ NNLO& 0.51 & 0.27 & 5\,$10^{-7}$ & 7\,$10^{-6}$ & 2\,$10^{-3}$ \\
{\sc Powheg\,+\,PY8} $\hdamp=\infty$& 0.100 & 0.071 & 2\,$10^{-6}$ & 2\,$10^{-6}$ & 1\,$10^{-4}$ \\
{\sc Powheg\,+\,PY6} HERAPDF 1.5& 0.21 & 0.29 & 0.13 & 0.49 & 0.49 \\
{\sc Powheg\,+\,PY6} no spin corl.& 0.017 & 0 & 1\,$10^{-6}$ & 3\,$10^{-7}$ & 0 \\
{\sc Alpgen\,+\,HW}& 1\,$10^{-3}$ & 3\,$10^{-7}$ & 3\,$10^{-5}$ & 1\,$10^{-5}$ & 2\,$10^{-8}$ \\
{\sc MC@NLO\,+\,HW}& 0.11 & 0.11 & 4\,$10^{-6}$ & 2\,$10^{-5}$ & 1\,$10^{-3}$ \\
{\sc aMC@NLO\,+\,HW$^{++}$}& 0.26 & 0.011 & 2\,$10^{-7}$ & 4\,$10^{-6}$ & 2\,$10^{-5}$ \\
\hline
\end{tabular}
\caption{\label{t:NorCombChiXPPData2012Data}The $\chi^2$ values (top) and associated probabilities (bottom) for comparison of combinations of measured normalised differential fiducial cross-sections with various \ttbar\ simulation samples.
Probabilities smaller than $10^{-10}$ are shown as zero.}
\end{table}

\subsection{Comparison with fixed-order predictions}\label{ss:fixedpred}

The comparisons described in Section~\ref{ss:gencomp} 
show that the predictions are strongly sensitive
to the choice of PDF, and also to the QCD scale (whose variation approximates
the effects of missing higher-order corrections) and other parameters related to
the amount of radiation. In this Section, these aspects are further
explored using a set of predictions
from the MCFM program (version 6.8) \cite{mcfm}, combined with {\sc Applgrid}
(version 1.4.73) \cite{applgrid} to interface to various PDF sets available
in LHAPDF (version 6.1.5) \cite{lhapdf}. Four recent NLO PDF sets were 
considered, namely CT14 \cite{ct14}, MMHT14 \cite{mmht}, 
NNPDF 3.0 \cite{nnpdf3} and HERAPDF 2.0 \cite{herapdf20}. 
The data were also compared to HERAPDF 1.5 \cite{herapdf15} for comparison
with the results of Section~\ref{ss:gencomp}; the results from these two
PDF sets are similar.

MCFM provides an NLO fixed-order prediction
of the \ttbar\ process in the dilepton channel, including NLO QCD corrections
in both production and decay in the on-shell approximation, and full NLO spin
correlations \cite{mcfmtt}. Only the direct decays of $W\rightarrow e/\mu$
are included, so these predictions were compared to the measurements
corrected to remove the leptonic $\tau$ decay contributions. The top quark
mass \mtop\ was set to 172.5\,\GeV. Informed by the discussion in 
Ref. \cite{dyntopscale}, the central values for the
QCD renormalisation and factorisation scales were set to $\mtop/2$, the lower
than typical (\mtop) scale choice being intended to account for the 
impact of resummed soft-gluon contributions not included in the fixed-order
calculations. The MCFM predictions do not include quantum electrodynamics
(QED) final state photon 
radiation, unlike the experimental measurements where the leptons are dressed
with nearby photons as discussed in Section~\ref{s:xsecdet}. Therefore,
the MCFM predictions were corrected bin-by-bin using corrections derived
from two \ttbar\ samples generated with {\sc Pythia8} (version 8.205)
\cite{pythia82} and the ATTBAR tune \cite{ATL-PHYS-PUB-2015-007} 
with QED final-state radiation enabled and
disabled. These corrections are typically 1--2\,\% on the absolute
and always smaller than 1\,\% on the normalised differential cross-sections.
No corrections were applied to the normalised \etal\ and \rapll\ distributions,
as the determined corrections were always smaller than 0.3\,\% and consistent
with unity within the simulation statistical uncertainties.

The ratios of the MCFM normalised differential cross-section predictions with 
HERAPDF 1.5 (the PDF set found to best
fit the data when comparing with {\sc Powheg\,+\,Pythia6} samples in
Section~\ref{ss:gencomp}) to data are shown in Figure~\ref{f:fixeda}. The 
uncertainties in the predictions include effects from PDFs, QCD scales and
the value of the strong coupling constant \alphas. 
For each individual component variation, the prediction was renormalised to 
unity before
calculating the shift for each bin; the effects on the normalised 
cross-section predictions are typically significantly smaller than those on the
absolute cross-sections. The PDF uncertainties for CT14 and MMHT
were evaluated from the sum in quadrature of the symmetrised up/down 
variations from each individual eigenvector pair from the PDF error set.
For the HERAPDF sets, each pair of eigenvector or model parameter 
variations was treated as an independent variation. For
NNPDF 3.0, the 100 replica sets which represent the NNPDF uncertainty 
were used to define a full covariance matrix
taking into account correlations between the bins of each distribution.
The QCD scale uncertainties were evaluated by varying the renormalisation and
factorisation scales $\mu_R$ and $\mu_F$ separately, and adding the
variations in quadrature. Each scale was varied by factors of one-half and
two from its central value ($\mtop/2$), and the resulting variations 
symmetrised. This procedure was used instead of taking an envelope including
simultaneous variations of $\mu_F$ and $\mu_R$ in order to properly 
account for the correlations between bins of the normalised differential
cross-section predictions. Finally, the \alphas\ uncertainty was evaluated
using the HERAPDF 1.5 PDF sets with \alphas\ set to 0.116 and 0.120, rescaling
the resulting uncertainty to $\Delta\alphas=\pm 0.0015$, in line with
the corresponding PDF4LHC recommendation \cite{pdf4lhc2}.

\begin{figure}
\vspace{-7mm}
\splitfigure{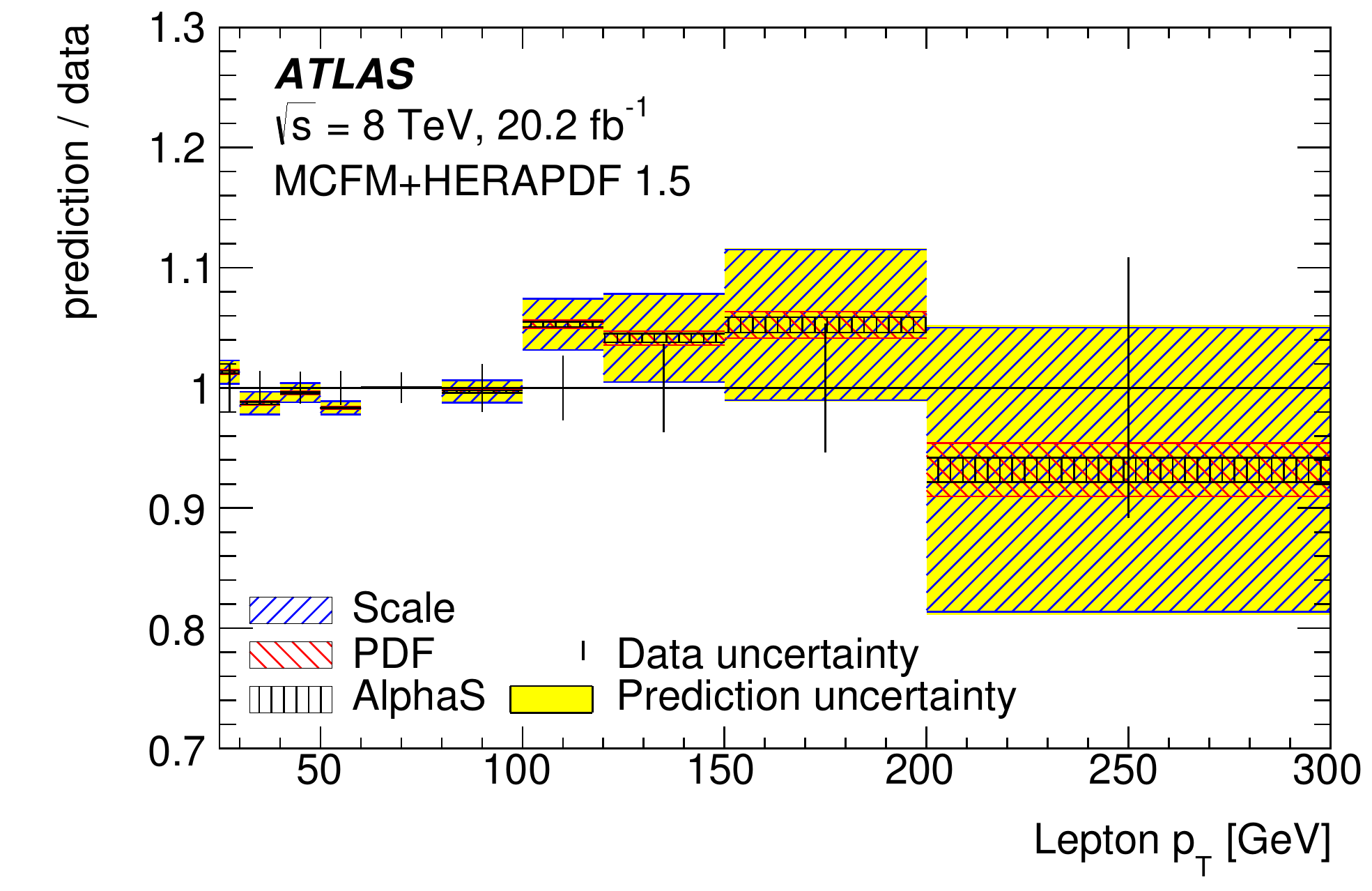}{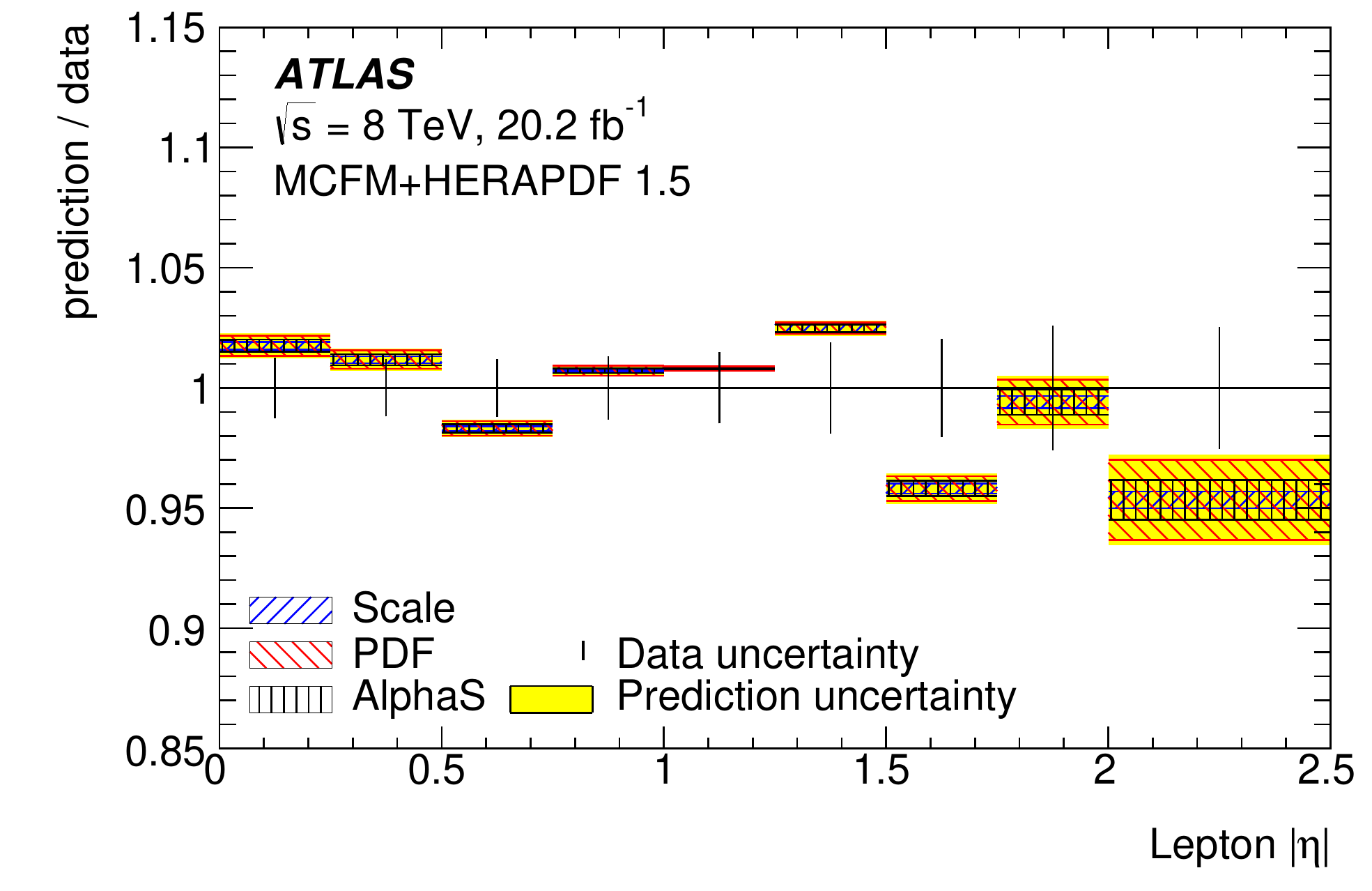}{a}{b}
\splitfigure{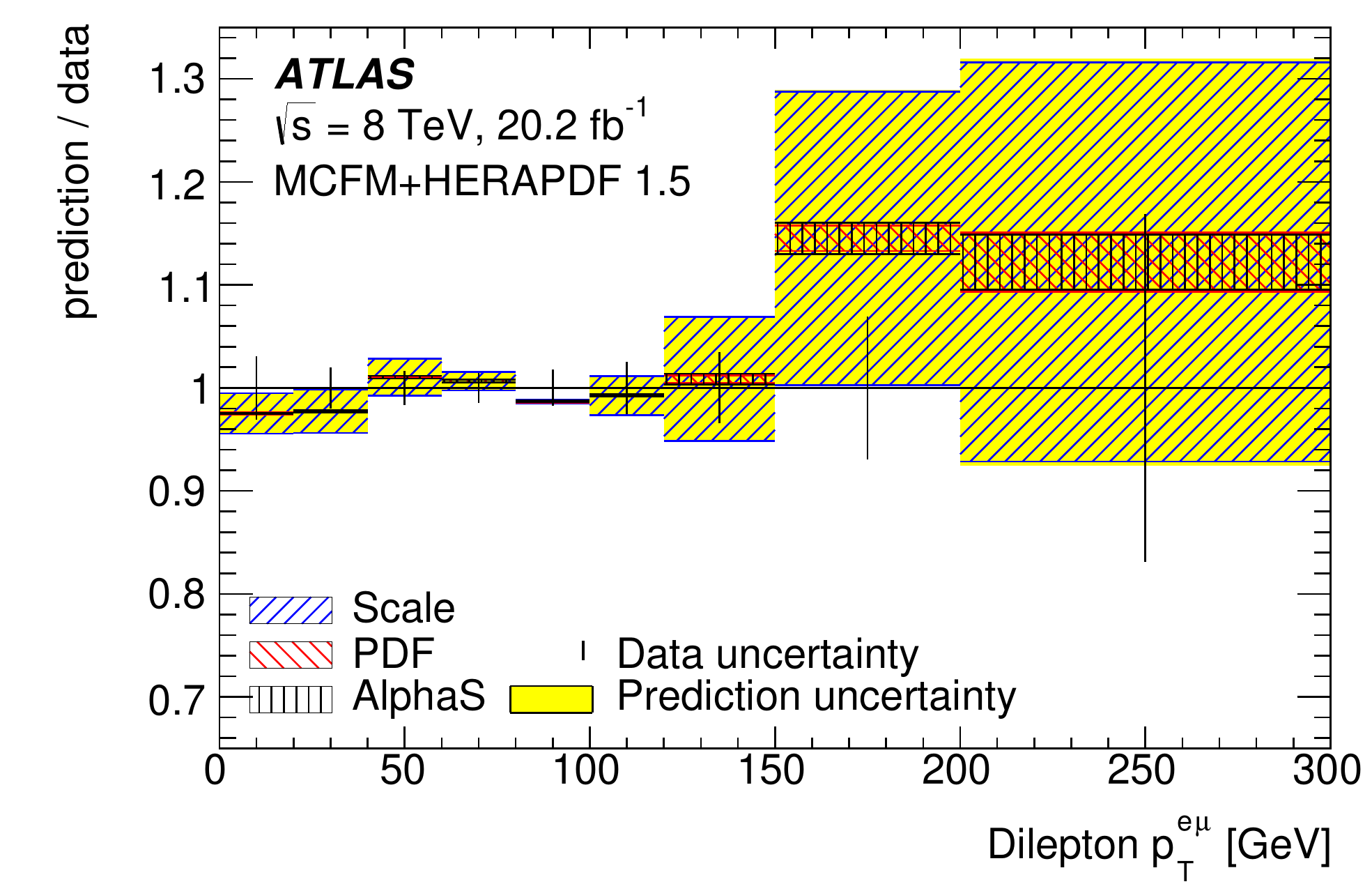}{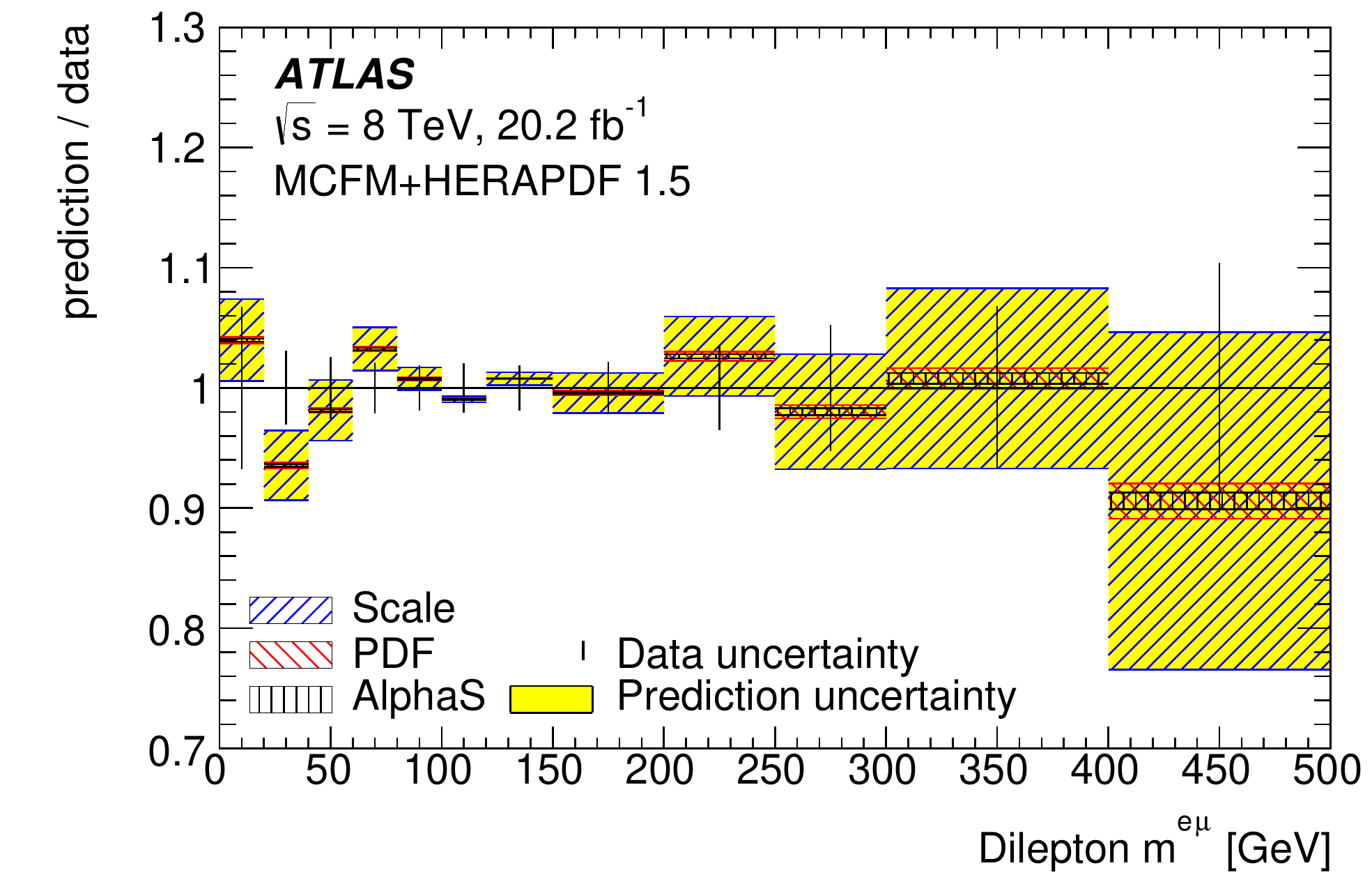}{c}{d}
\splitfigure{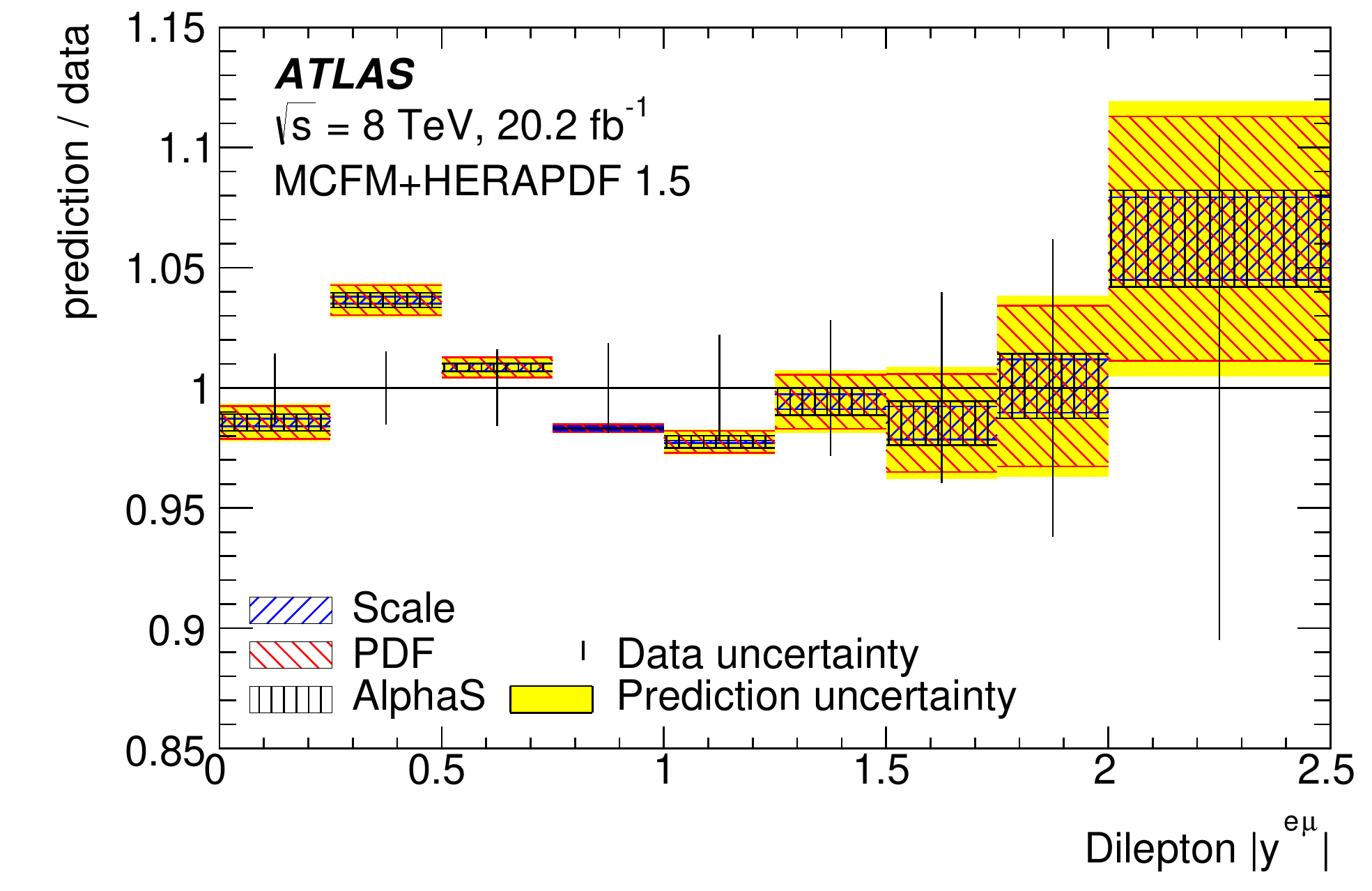}{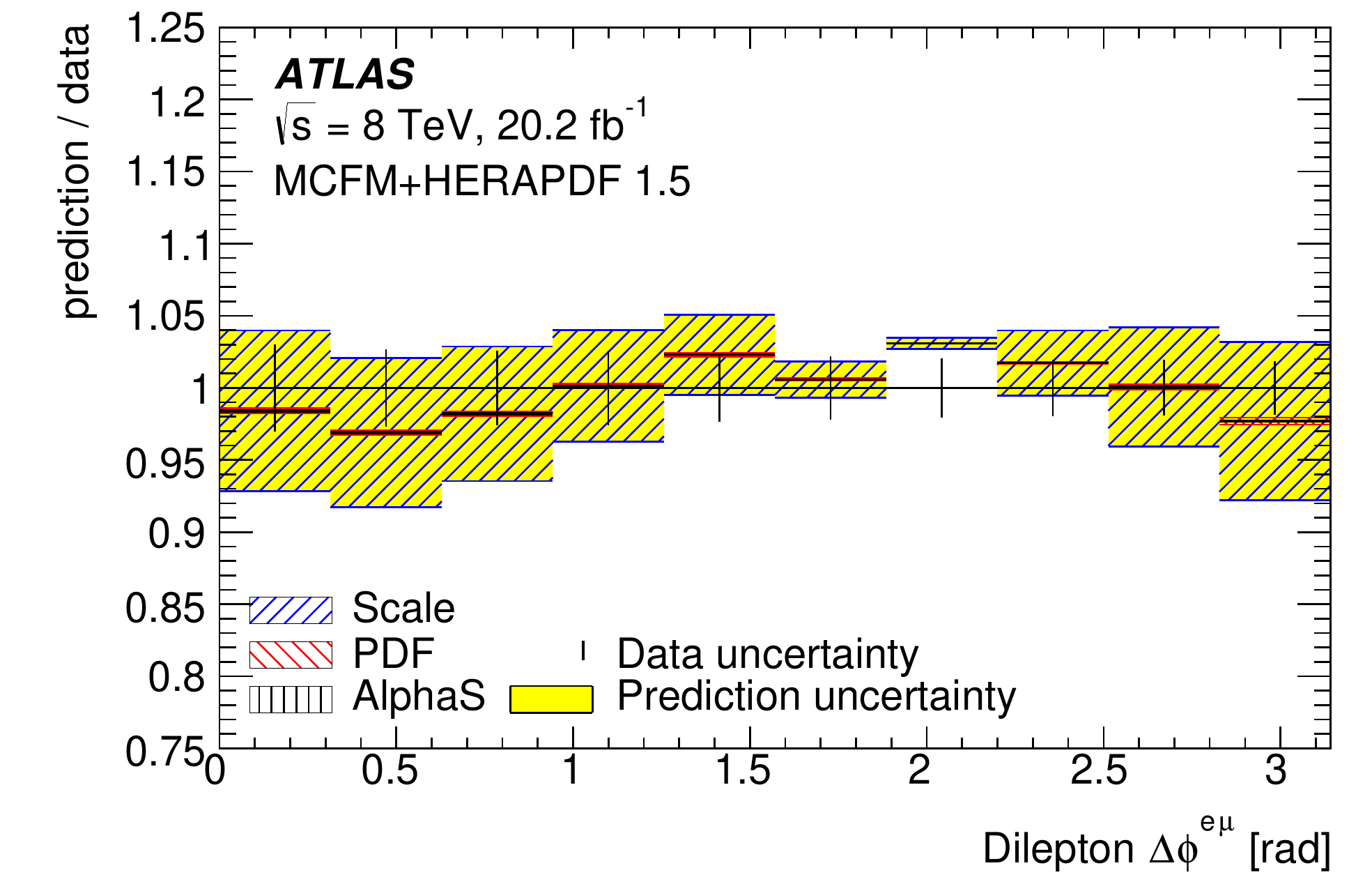}{e}{f}
\splitfigure{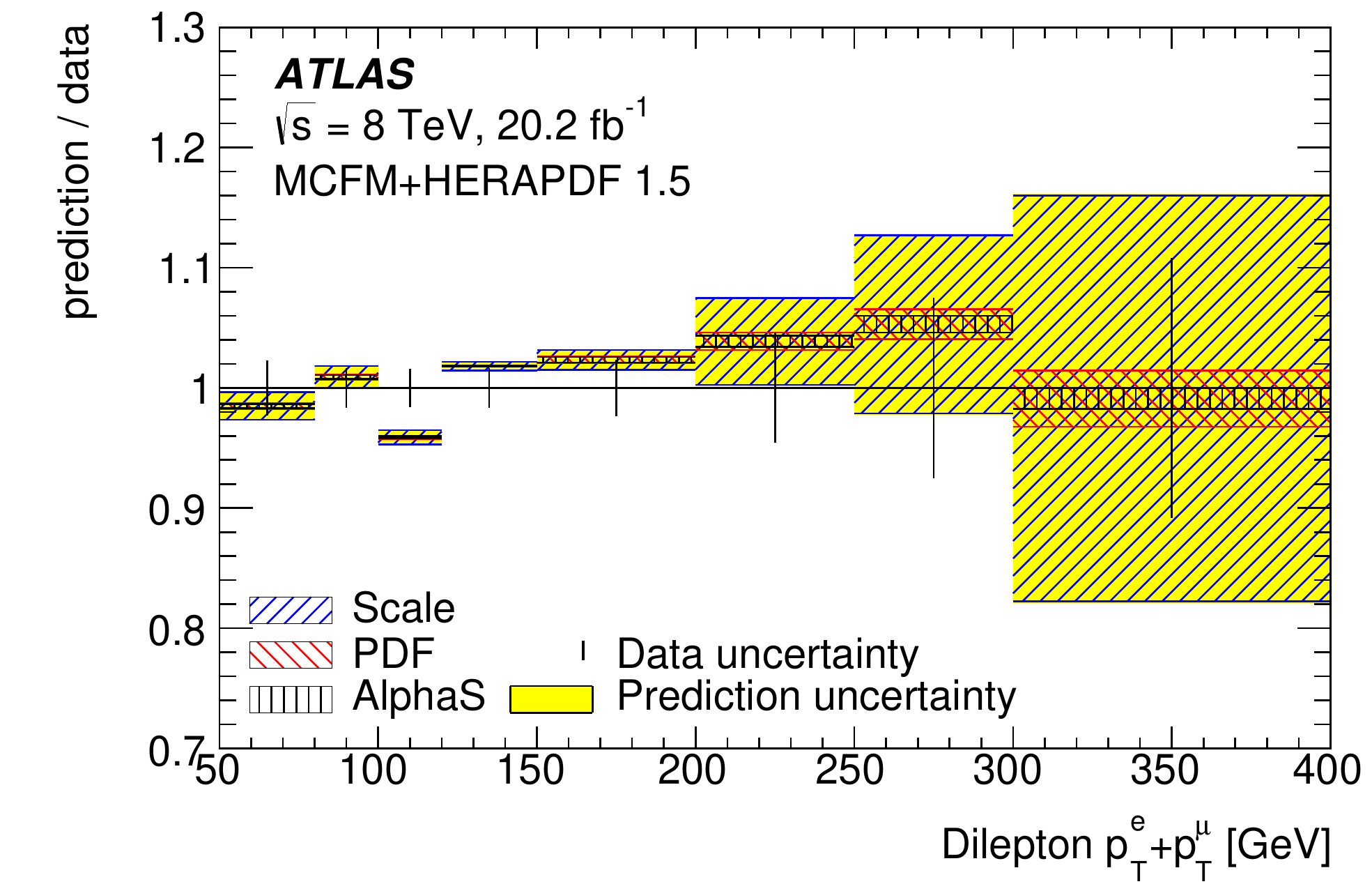}{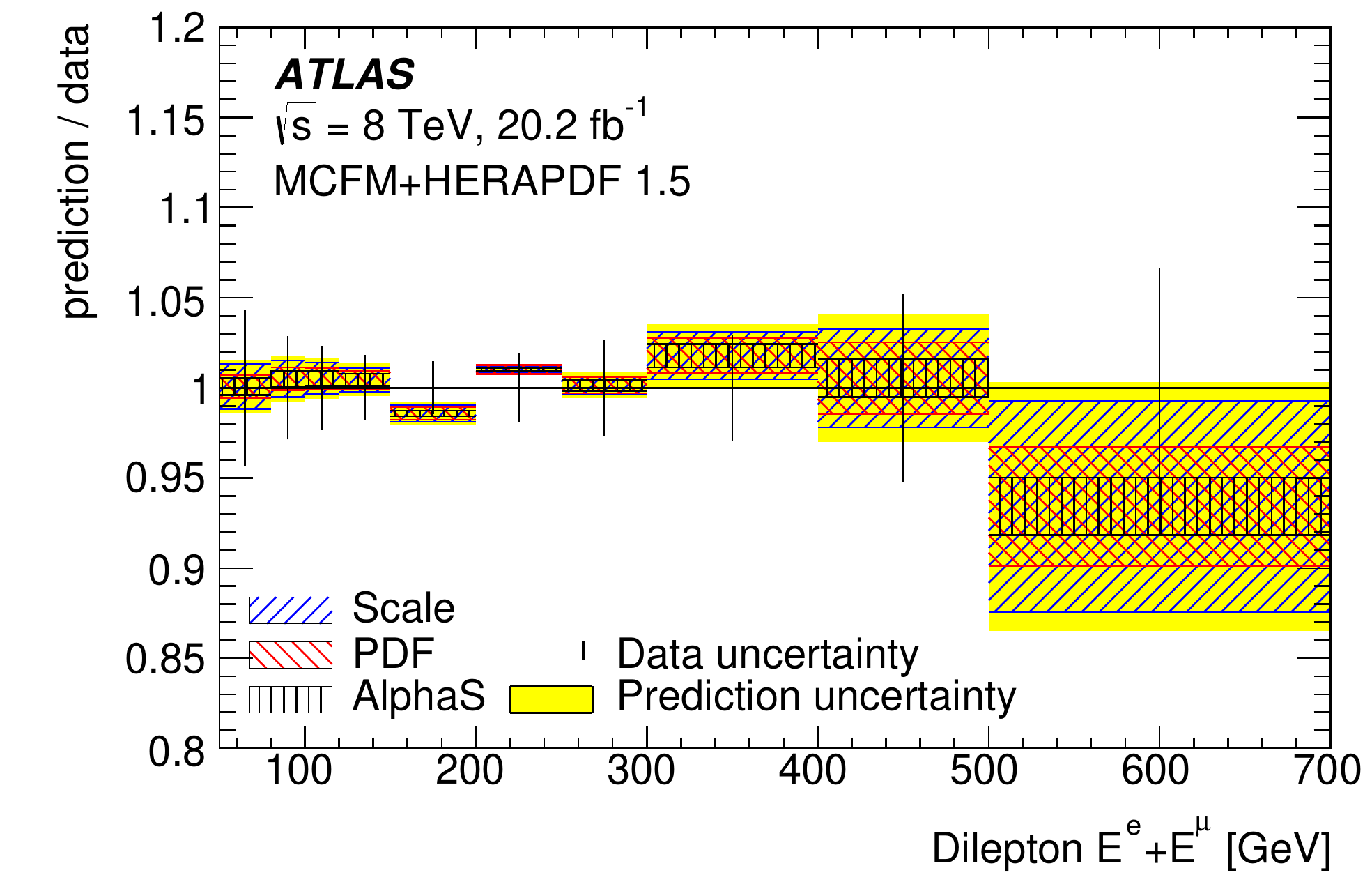}{g}{h}
\caption{\label{f:fixeda}Ratios of MCFM\,+\,HERAPDF 1.5 fixed-order 
predictions of normalised differential cross-sections to data as a function of 
lepton and dilepton variables. Contributions via 
$W\rightarrow\tau\rightarrow e/\mu$ decays are not included, and the MCFM
predictions have been corrected to include QED final-state radiation effects.
The total data uncertainties are shown by the error bars around unity. 
The separate uncertainties in the
predictions from QCD scales, PDFs and the strong coupling constant \alphas\ 
are shown by the hatched bands, and the total uncertainties in the predictions
are shown by the yellow band.}
\end{figure}

The compatibility of the predictions with the normalised cross-section 
data was tested quantitatively using the $\chi^2$ of Eq.~(\ref{e:redchi}),
updating the covariance matrix {\textbf S} to also include the theoretical 
uncertainties 
discussed above, including their bin-to-bin correlations via the off-diagonal
terms. The resulting $\chi^2$ and $p$-values are shown as the 
`MCFM\,+\,HERAPDF 1.5' entries in Table~\ref{t:NorChiXPPData2012DataMCFMksII}
for individual distributions, 
and in Table~\ref{t:NorCombChiXPPData2012DataMCFMksII}
for the combinations of distributions. As can be seen from these tables 
and from Figure~\ref{f:fixeda}, MCFM with the HERAPDF 1.5 PDF describes the
data well, once all the theoretical uncertainties are taken into account.
The predictions for \ptl, \ptll, \mll, \dphill\ and \ptsum\ have large 
scale uncertainties, which largely cover any differences between the
measurements and central predictions with scales $\mu_R=\mu_F=\mtop/2$. 
The \etal\
and \rapll\ distributions have little scale dependence and are  more sensitive 
to PDF variations, but are again well-described within the uncertainties of 
the HERAPDF 1.5 set. The
\alphas-related uncertainties are small compared to the other two classes.

\begin{table}[tp]
\centering

\begin{tabular}{l|rrrrrrrr}\hline
Generator & \ptl & \etal & \ptll & \mll & \rapll & \dphill & \ptsum & \esum \\
$N_{\mathrm dof}$ & 9 & 8 & 8 & 11 & 8 & 9 & 7 & 9 \\ \hline
MCFM\,+\,CT14&  11.5 &  14.1 &   7.2 &  11.2 &  13.0 &   7.2 &  11.4 &  11.2 \\
MCFM\,+\,MMHT&  11.3 &  12.8 &   7.2 &  11.2 &  12.6 &   7.1 &  11.2 &   9.6 \\
MCFM\,+\,NNPDF 3.0&  11.7 &  11.3 &   7.2 &  11.4 &   9.4 &   7.3 &  11.5 &   8.5 \\
MCFM\,+\,HERAPDF 1.5&   9.1 &  10.9 &   6.4 &  12.1 &   8.0 &   6.9 &   8.5 &   2.6 \\
MCFM\,+\,HERAPDF 2.0&   8.4 &  12.0 &   6.2 &  12.4 &   8.0 &   6.8 &   8.0 &   2.7 \\
\hline
MCFM\,+\,CT14& 0.24 & 0.080 & 0.51 & 0.43 & 0.11 & 0.62 & 0.12 & 0.27 \\
MCFM\,+\,MMHT& 0.26 & 0.12 & 0.51 & 0.42 & 0.13 & 0.62 & 0.13 & 0.38 \\
MCFM\,+\,NNPDF 3.0& 0.23 & 0.18 & 0.52 & 0.41 & 0.31 & 0.61 & 0.12 & 0.49 \\
MCFM\,+\,HERAPDF 1.5& 0.43 & 0.21 & 0.61 & 0.36 & 0.44 & 0.65 & 0.29 & 0.98 \\
MCFM\,+\,HERAPDF 2.0& 0.49 & 0.15 & 0.63 & 0.33 & 0.44 & 0.66 & 0.34 & 0.97 \\
\hline

\end{tabular}
\caption{\label{t:NorChiXPPData2012DataMCFMksII}The $\chi^2$ values (top) and associated probabilities (bottom) for comparison of measured normalised differential fiducial cross-sections with the
predictions of MCFM with various PDF sets. Contributions via 
$W\rightarrow\tau\rightarrow e/\mu$ decays are not included, and the MCFM
predictions have been corrected to include QED final-state radiation effects.
The results take into account
the uncertainties in both the measurements and predictions.}
\end{table}

\begin{table}[tp]
\centering

\begin{tabular}{l|ccccc}\hline
Generator & \ptl, \ptsum & \ptll, \mll, & \etal, \rapll & \etal, \rapll, & All \\
&  & \ptsum &  & \esum &  \\ 
$N_{\mathrm dof}$ & 16 & 26 & 16 & 25 & 69 \\ \hline
MCFM\,+\,CT14&  19.5 &  29.6 &  24.2 &  32.4 &  73.0 \\
MCFM\,+\,MMHT&  19.3 &  29.6 &  23.4 &  30.7 &  72.0 \\
MCFM\,+\,NNPDF 3.0&  19.9 &  29.7 &  20.1 &  27.4 &  69.3 \\
MCFM\,+\,HERAPDF 1.5&  16.1 &  28.8 &  21.5 &  26.1 &  68.8 \\
MCFM\,+\,HERAPDF 2.0&  15.3 &  30.0 &  22.7 &  27.4 &  69.0 \\
\hline
MCFM\,+\,CT14& 0.24 & 0.28 & 0.086 & 0.15 & 0.35 \\
MCFM\,+\,MMHT& 0.25 & 0.28 & 0.10 & 0.20 & 0.38 \\
MCFM\,+\,NNPDF 3.0& 0.23 & 0.28 & 0.22 & 0.34 & 0.47 \\
MCFM\,+\,HERAPDF 1.5& 0.45 & 0.32 & 0.16 & 0.40 & 0.48 \\
MCFM\,+\,HERAPDF 2.0& 0.51 & 0.27 & 0.12 & 0.34 & 0.48 \\
\hline
\end{tabular}
\caption{\label{t:NorCombChiXPPData2012DataMCFMksII}The $\chi^2$ values (top) and associated probabilities (bottom) for comparison of combinations of measured normalised differential fiducial cross-sections with the predictions of MCFM with 
various PDF sets. Contributions via 
$W\rightarrow\tau\rightarrow e/\mu$ decays are not included, and the MCFM
predictions have been corrected to include QED final-state radiation effects.
The results take into account the uncertainties in both the measurements and 
predictions.}
\end{table}

The predictions for all five PDF sets (including PDF uncertainties,
scaled to 68\,\% CL for CT14, as well as scale and \alphas\ 
uncertainties) are compared to the data
in Figure~\ref{f:fixedb}. The corresponding $\chi^2$ and $p$-values, including
the PDF, scale and \alphas\ uncertanities on the predictions, are shown in 
Tables~\ref{t:NorChiXPPData2012DataMCFMksII} 
and~\ref{t:NorCombChiXPPData2012DataMCFMksII}. The results for HERAPDF 1.5
and HERAPDF 2.0
are close to the data, whereas the CT14, MMHT
and NNPDF 3.0 PDF sets describe the data slightly less well, particularly for
\ptl, \etal, \rapll\ and \esum. These conclusions are similar to those found
for HERAPDF 1.5 and CT10 with the {\sc Powheg\,+\,Pythia6} setup discussed
in Section~\ref{ss:gencomp} above. However, the difference in $\chi^2$ between
the PDF sets is smaller for the fixed-order predictions, as the 
explicit inclusion of PDF and scale uncertainties in the predictions renders the
differences between the central predictions of each PDF less significant. 
The PDF comparisons would 
benefit from the availability of predictions including NNLO QCD effects in
both the top quark production and decay, which should substantially reduce
the scale uncertainties.

\begin{figure}
\vspace{-7mm}
\splitfigure{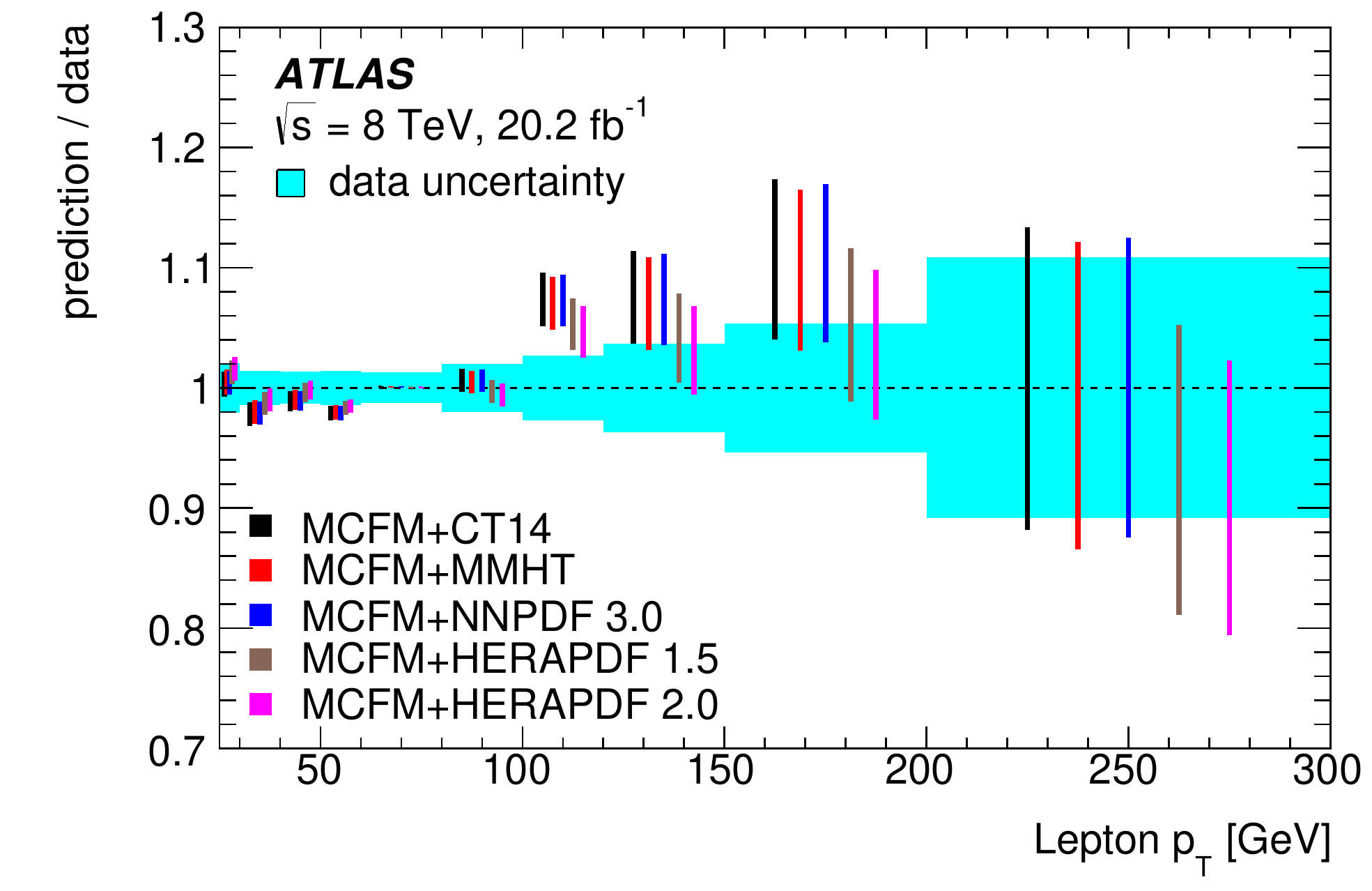}{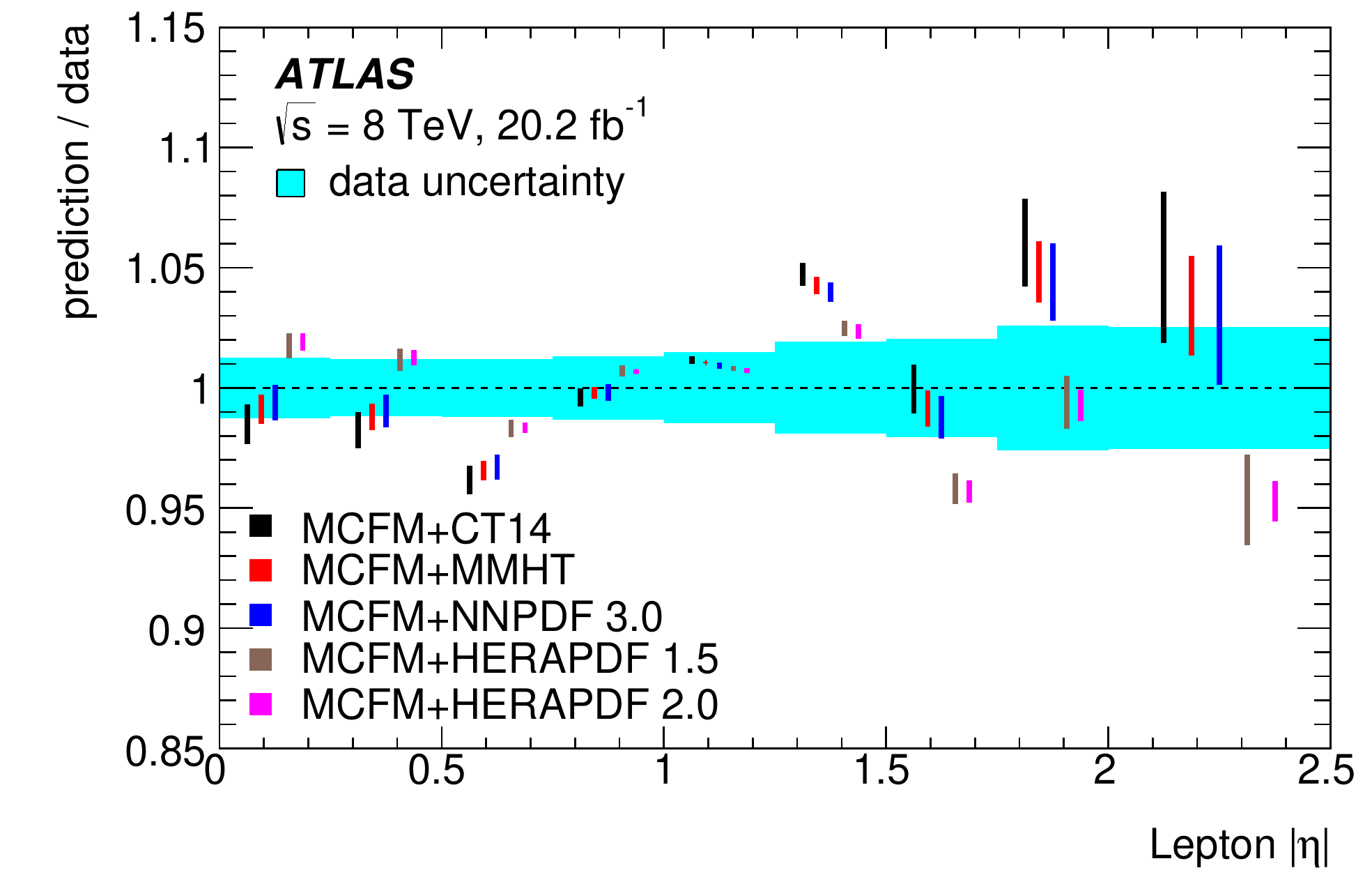}{a}{b}
\splitfigure{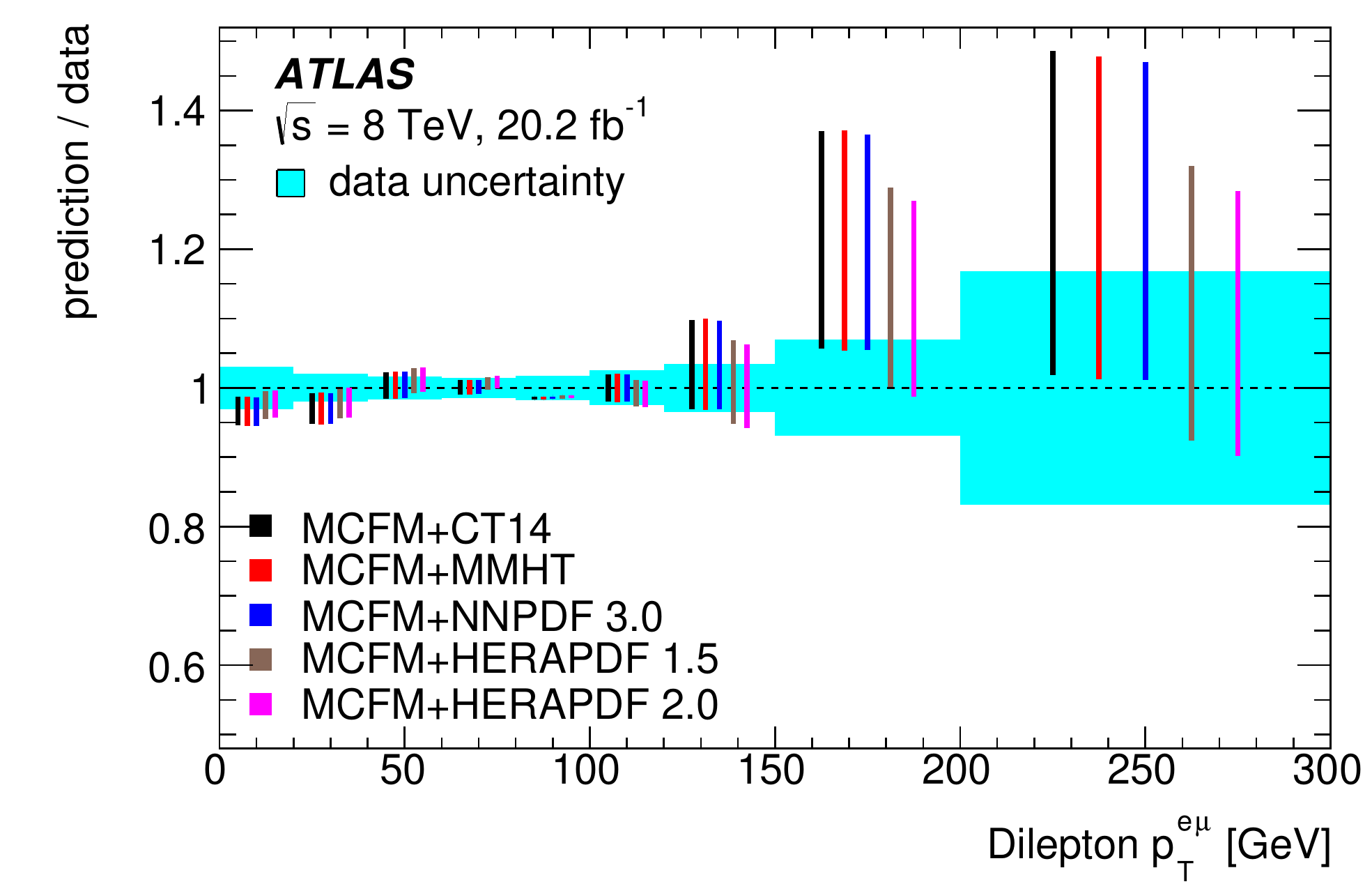}{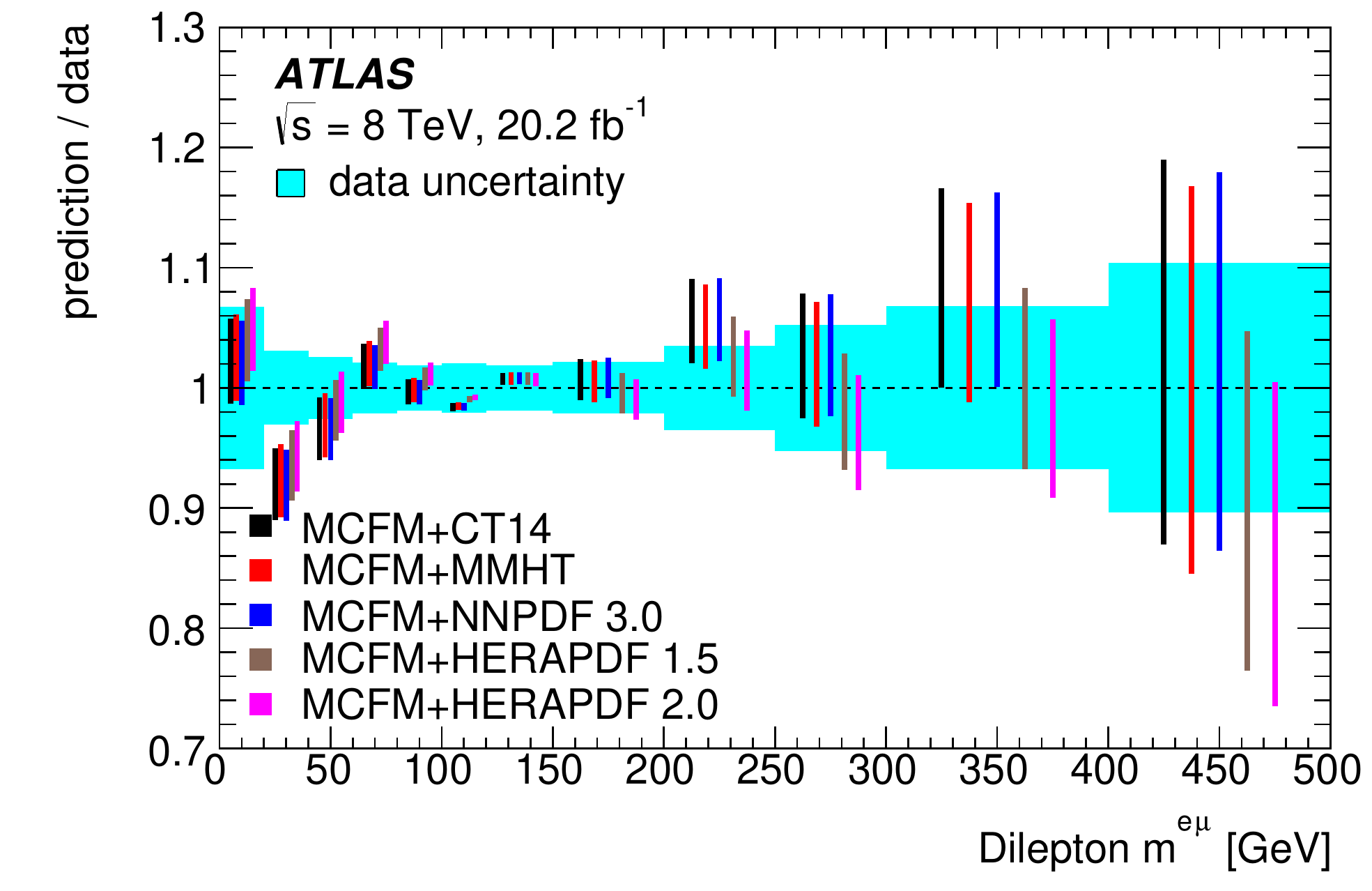}{c}{d}
\splitfigure{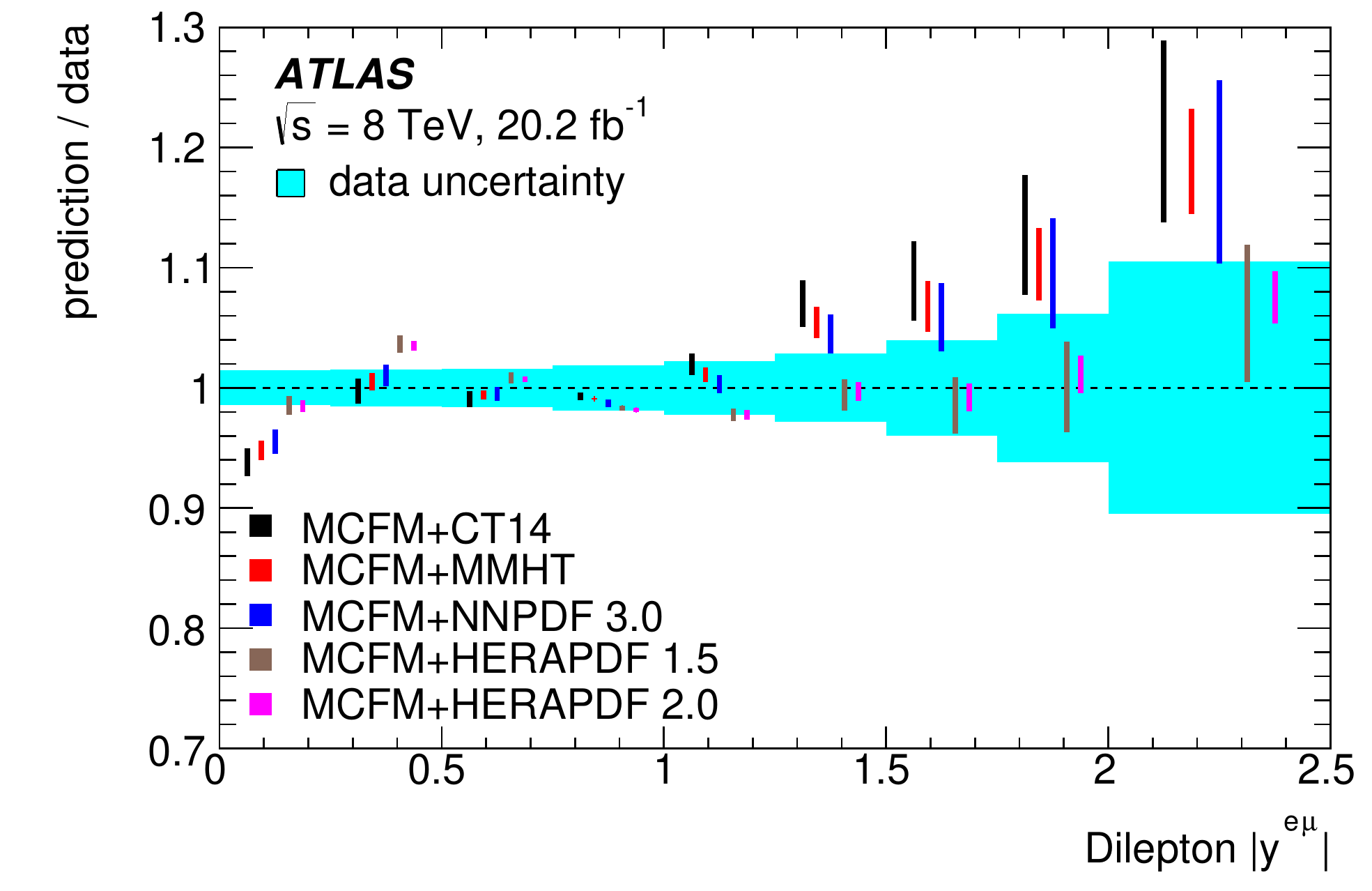}{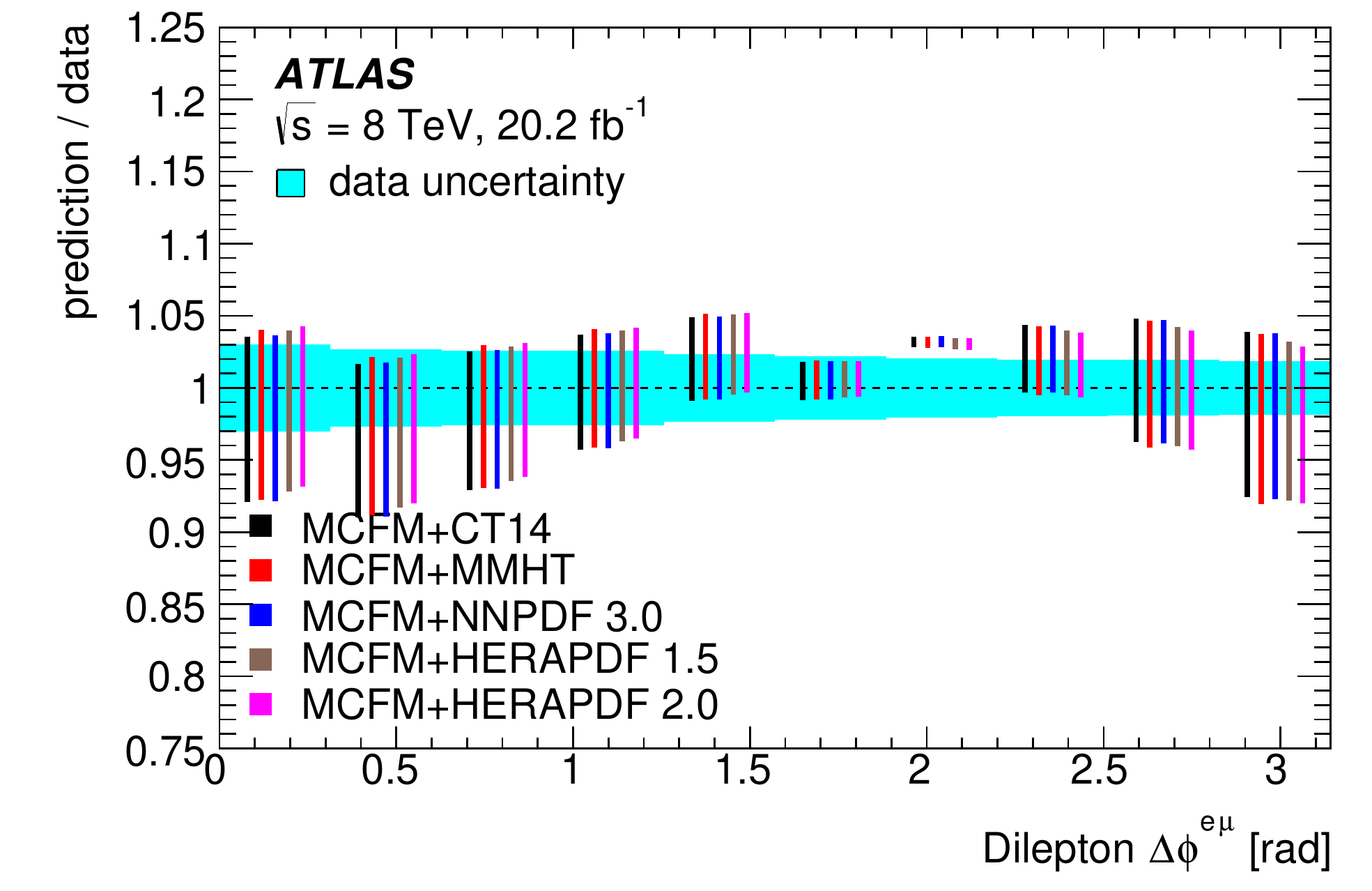}{e}{f}
\splitfigure{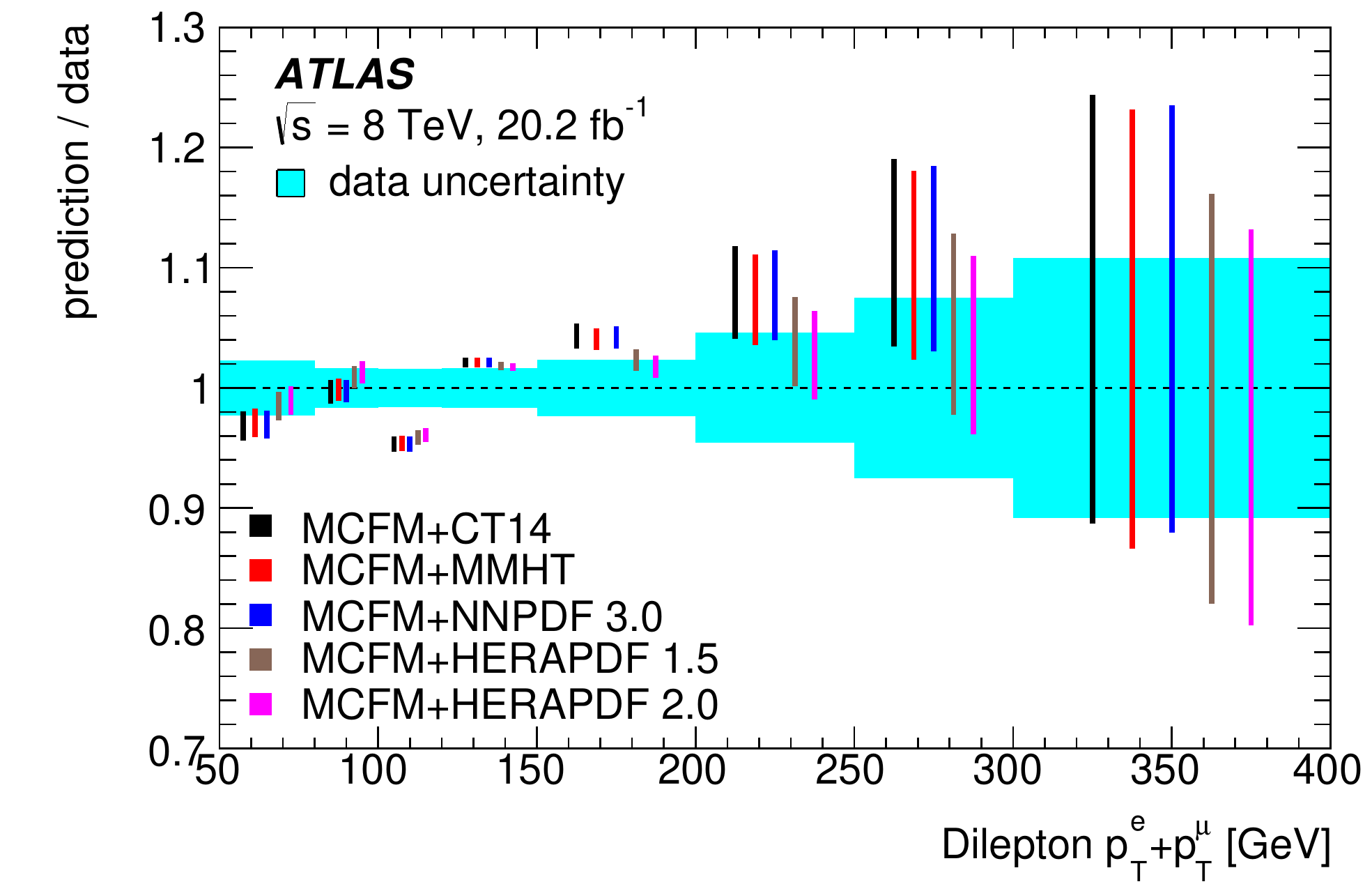}{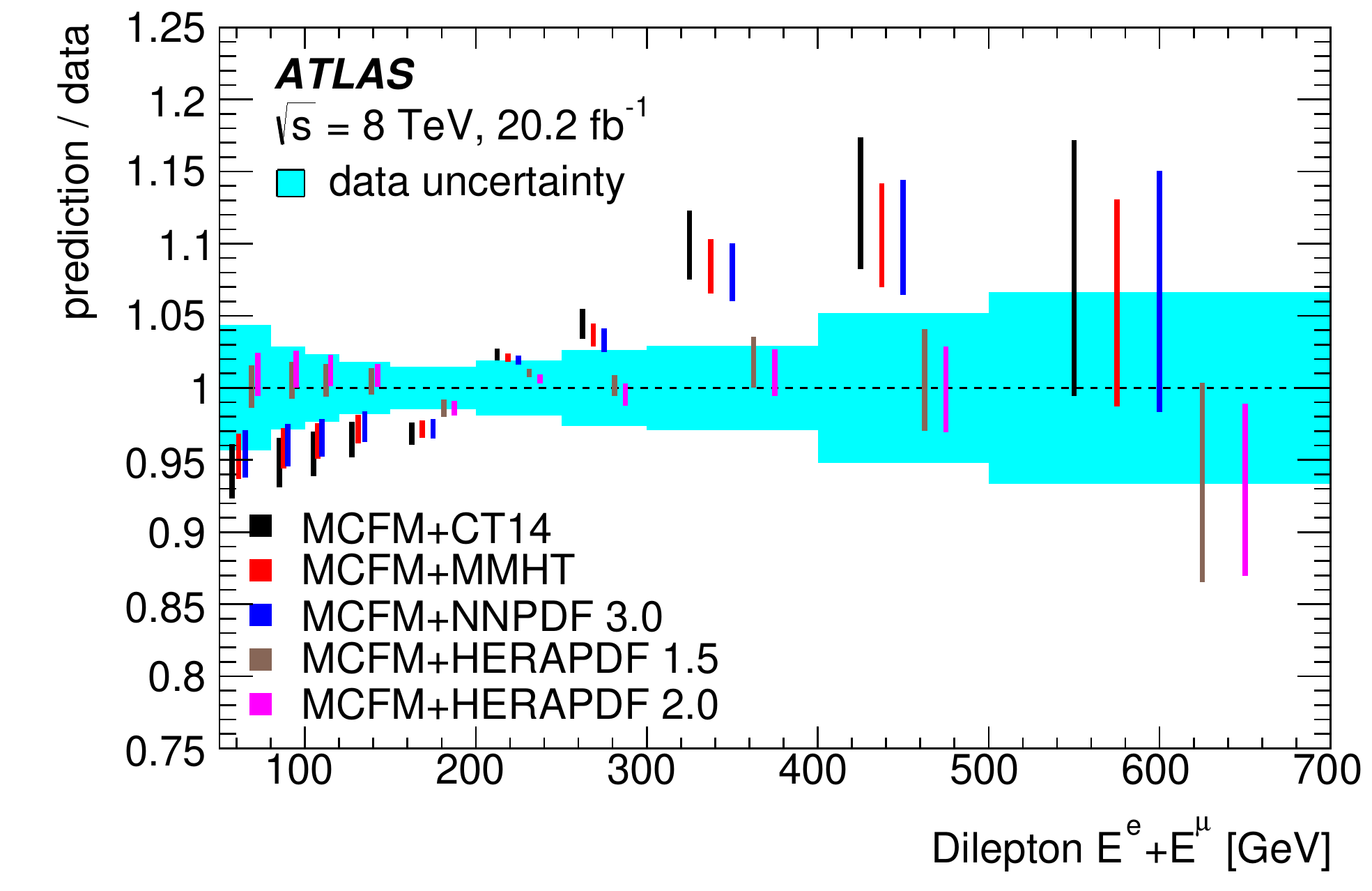}{g}{h}
\caption{\label{f:fixedb}Ratios of MCFM fixed-order 
predictions of normalised differential cross-sections to data as a function of 
lepton and dilepton variables, using the CT14, MMHT, NNPDF 3.0, HERAPDF 1.5 and HERAPDF 2.0 PDF sets for the predictions. Contributions via 
$W\rightarrow\tau\rightarrow e/\mu$ decays are not included, and the MCFM
predictions have been corrected to include QED final-state radiation effects.
The total data uncertainties
are shown by the cyan bands around unity, and the total uncertainty for 
each prediction (including QCD scales, PDFs, and the
strong coupling constant \alphas) are shown by the vertical bars.}
\end{figure}

\section{Constraints on the gluon parton distribution function}\label{s:pdf}

As a demonstration of the ability of the
normalised differential cross-section measurements to constrain the gluon PDF, 
fits were performed to deep inelastic scattering (DIS)
data from HERA I+II \cite{herapdf20}, with and without the addition of
the constraints from \ttbar\ dilepton \etal, \rapll\ and \esum\ distributions. 
As shown in Figure~\ref{f:fixeda}, these distributions are the most sensitive
to PDF variations, whilst being less sensitive to QCD scale variations and
the value of \mtop.
The fits are based on the predictions from MCFM and {\sc ApplGrid}
discussed in Section~\ref{ss:fixedpred}, allowing predictions for 
arbitrary PDF variations to be obtained much faster than if a full 
NLO plus parton shower event generator setup were to be used.
The QCD scales were set to fixed values of  $\mu_F=\mu_R=\mtop/2$.
The fits were performed using the {\sc xFitter} package 
\cite{herafitter,xfitter}, which allows
the PDF and other theoretical uncertainties to be included via asymmetric
error propagation. In this formalism, the $\chi^2$ for the compatibility
of the measurements with the prediction is expressed by:
\begin{eqnarray}
\chi^2 = \sum_{i,j} \left( \xniexp-\xnith\right)\ S^{-1}_{\mathrm{exp},ij}(\xnith,\xnjth)\ \left(\xnjexp-\xnjth\right)\ ,
\end{eqnarray}
where \xniexp\ is the measured normalised differential cross-section
in bin $i$ (equivalent to \xntti\ in Eq.~(\ref{e:normx})), \xnith\ is the
corresponding theoretical prediction, $S_{\mathrm{exp},ij}$ is the  covariance
matrix of experimental uncertainties including both statistical and systematic
contributions, and correlations between bins, and the sums for $i$ and $j$ run
over $n-1$ bins to account for the normalisation condition. Unlike in the 
formulation of Eq.~(\ref{e:redchi}), the covariance matrix is a function
of the theoretical predictions, with the statistical uncertainties being
rescaled according to the difference between the measured values and the
predictions using a Poisson distribution, and the systematic 
uncertainties being scaled in proportion to the predictions.

Following the formalism outlined in Ref. \cite{STDM-2012-20}, 
the covariance matrix was decomposed into a diagonal matrix $\mathbf D$
representing the uncorrelated parts of the uncertainties, and
a set of coefficients \gamexpij\ 
giving the one standard deviation shift in the
measurement $i$ for source $j$, where $j$ runs over the correlated part
of the  statistical uncertainties
and each source of systematic uncertainty.  
Each source of experimental uncertainty was then associated
with a `nuisance parameter' \betaexpj\ parameterising the 
associated shift in units of standard deviation. The $\chi^2$ becomes
a function of the set of PDF parameters \vp\ defining the theoretical 
prediction \xnith\ and the vector of experimental nuisance parameters
\vbexp, and is given by:
\begin{eqnarray}\label{e:fochi}
\chi^2(\vp,\vbexp)=\sum_i\frac{\left(\xniexp+\sum_j\gamexpij\betaexpj-\xnith(\vp)\right)^2}{d_{ii}^2}
+\sum_j\betaexpj^2 + L \ ,
\end{eqnarray}
where $d_{ii}$ are the non-zero elements of the diagonal matrix $\mathbf D$,
and the rescaling of the uncertainties leads to the 
logarithmic term $L$, arising from the likelihood transition to $\chi^2$
as discussed in Refs. \cite{STDM-2012-20,logpen}.
The $\chi^2$ was minimised
as a function of the PDF parameters \vp\ and the nuisance parameters \vbexp, 
and the value at the minimum provides a compatibility test of the data and 
prediction. 

For the PDF fits, the perturbative order of the 
DGLAP evolution \cite{dglap1,dglap2,dglap3}
was set to NLO, to match the order of the MCFM predictions. 
The gluon PDF $g(x)$ was parameterised as a function of Bjorken-$x$ as:
\begin{eqnarray}\label{e:gpdf}
xg(x) = Ax^B(1-x)^C(1+Ex^2)\,e^{Fx},
\end{eqnarray}
which, compared to the standard parameterisation given in 
Eq.~(27) of Ref. \cite{herapdf20}, removes the negative $A'$ term at 
low $x$ and adds more flexibility at medium and high $x$ through the
additional terms with the parameters $E$ and $F$. The standard
parameterisations were used for the quark PDFs, giving a total of 
14 free PDF parameters in the vector \vp, after imposing momentum and valance
sum rules, and the constraint that the $\bar{u}$ and $\bar{d}$ contributions
are equal at low $x$. Other parameters in the PDF fit were set as described
in Ref.~\cite{STDM-2012-20}.

The minimised $\chi^2$ values from the fits without and with the \ttbar\
data are shown in Table~\ref{t:pdffit}, which gives the partial $\chi^2$ for
each dataset included in the fit (i.e. the contribution of that dataset to the 
total $\chi^2$)
and the total $\chi^2$ for each fit. The partial $\chi^2$ values indicate that
the \ttbar\ data are well-described
by the PDF derived from the combined fit, and that the description of the
HERA I+II data is not degraded by the inclusion of the \ttbar\ data,
 i.e. there is no tension between the
two datasets. The ratios of the fitted gluon PDF central values
with and without the \ttbar\ data included are shown in 
Figure~\ref{f:pdffit}(a), 
together with the corresponding uncertainties. The ratio of relative 
uncertainties in the PDFs with and without the \ttbar\ data are shown in 
Figure~\ref{f:pdffit}(b). The inclusion of the \ttbar\ data reduces the 
uncertainty by typically 10--25\,\% over most of the relevant $x$ range.

\begin{table}[tp]
\centering

\begin{tabular}{l|cc}\hline
Datasets fitted  & HERA I+II & HERA I+II + \ttbar\ \\ \hline
Partial $\chi^2$ / $N_{\mathrm{point}}$  & &  \\
- HERA I+II & 1219 / 1056 & 1219 / 1056 \\
- \ttbar\ (\etal, \rapll, \esum) & - & 27 / 25 \\
\hline
Total $\chi^2$ / $N_{\mathrm{dof}}$ & 1219 / 1042 & 1247 / 1067 \\
\hline
\end{tabular}
\caption{\label{t:pdffit}Results of the PDF fit to HERA I+II data (left column),
 and to
HERA I+II data plus the normalised differential \ttbar\ cross-sections
as a function of \etal, \rapll\ and \esum\ (right column).
The partial $\chi^2$ and number of data points for the datasets used in 
 each fit are given, together with the overall $\chi^2$ and total number of 
degrees of freedom for each fit.}
\end{table}

\begin{figure}
\splitfigure{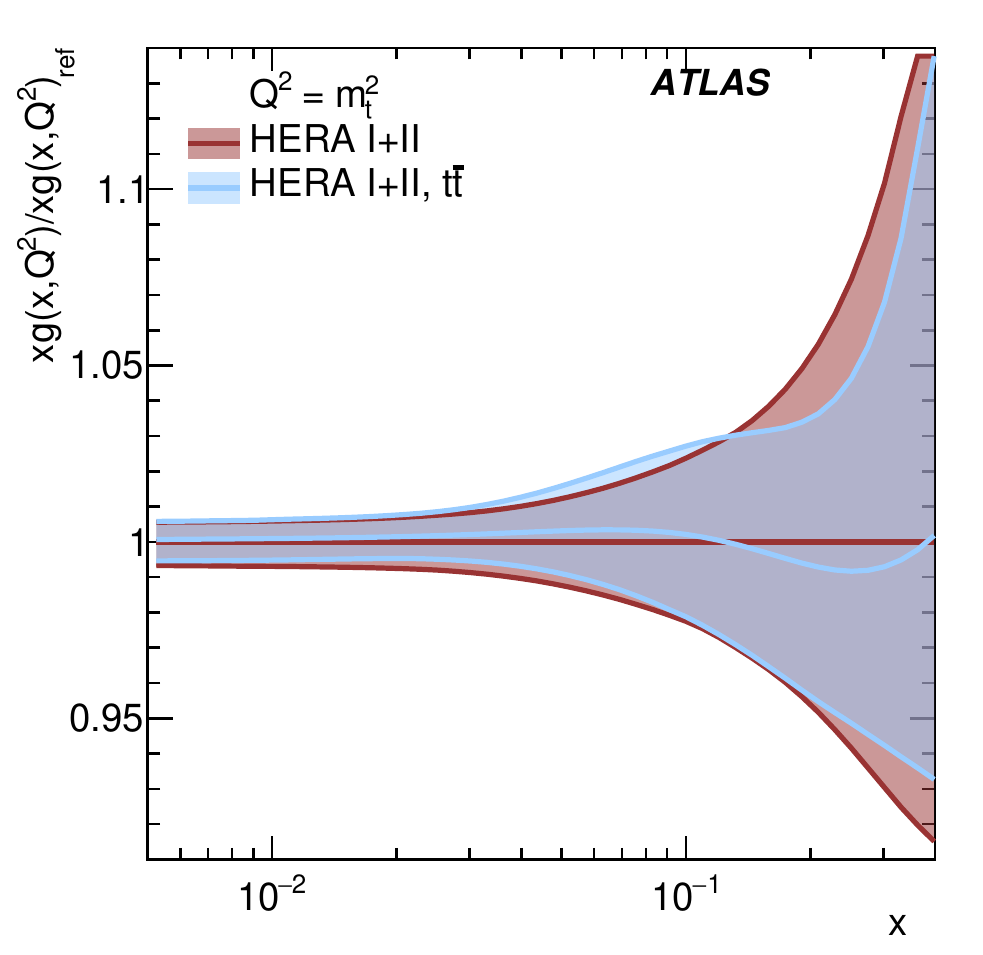}{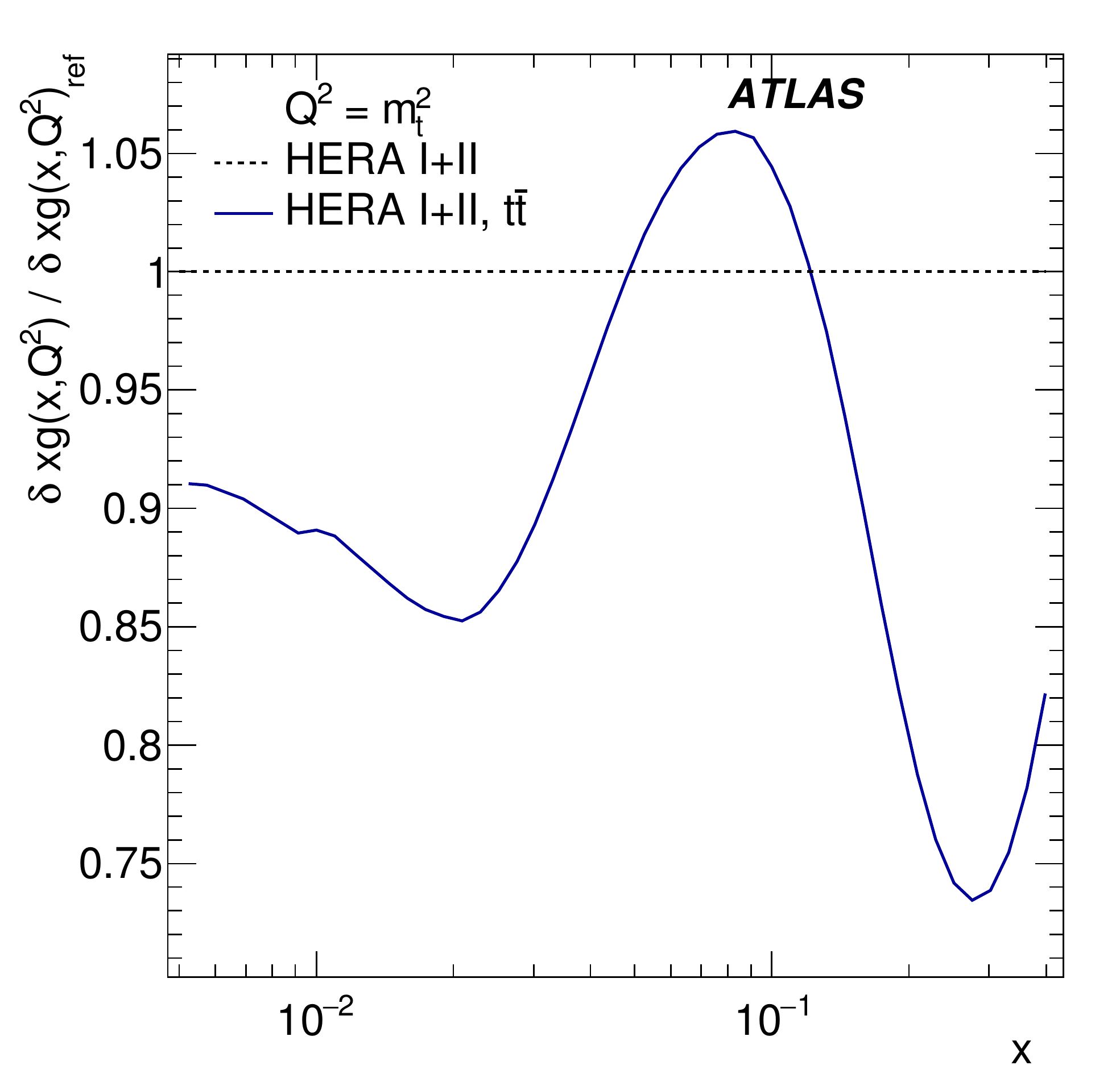}{a}{b}
\caption{\label{f:pdffit}(a) Ratio of the gluon PDF determined from the fit
using HERA I+II data plus the normalised differential cross-section 
distributions
as a function of \etal, \rapll\ and \esum\ in \ttbar\ events, to the gluon
PDF determined from the fit  using HERA I+II data alone, 
as a function of Bjorken-$x$. The uncertainty bands are shown
on the two PDFs as the blue and red shading.
(b) Ratio of the relative uncertainty in the gluon PDF determined from the fit 
to HERA I+II plus \ttbar\ data to that from 
HERA data alone. The PDFs are shown evolved to the scale $Q^2=\mtop^2$ in both
cases.}
\end{figure}

The gluon PDF obtained from this procedure is compared to the gluon PDFs
from the CT14 \cite{ct14} and NNPDF~3.0 \cite{nnpdf3} global PDF sets in
Figure~\ref{f:pdfprof}. These PDF sets, shown by the green bands,
both have a larger high-$x$ gluon than preferred by the HERA I+II data,
with or without the addition of the \ttbar\ data from this analysis. The
impact of the \ttbar\ data on the global PDF sets was investigated using
a profiling procedure \cite{STDM-2012-20,qcdtev,pdfhess}, extending
the $\chi^2$ definition of Eq.~(\ref{e:fochi}) to incorporate a vector \vbth\ of
nuisance parameters \betathk\ expressing the dependence of the theoretical
prediction \xnith\ on the uncertainties for a particular PDF set. In
this formulation, the $\chi^2$ definition becomes:
\begin{eqnarray}\label{e:fochi2}
\chi^2(\vbexp,\vbth)=\sum_i\frac{\left(\xniexp+\sum_j\gamexpij\betaexpj-\xnith-\sum_k\gamthik\betathk\right)^2}{d_{ii}^2}
+\sum_j\betaexpj^2+\sum_k\betathk^2 + L \ ,
\end{eqnarray}
where $\betathk=\pm 1$ corresponds to the $\pm 1$ standard deviation change 
of the PDF values according to the $k$th eigenvector of the PDF error set. The
values and uncertainties of the nuisance parameters $\betathk$ 
after minimisation
of the $\chi^2$ of Eq.~(\ref{e:fochi2}) give the profiled PDF with modified
central values and uncertainties according to the effect of the \ttbar\
differential cross-section distributions. These profiled PDFs are shown as the
orange bands in Figure~\ref{f:pdfprof}. Both the CT14 and NNPDF~3.0 gluon
PDFs are shifted downwards at high $x$ (corresponding to a softer gluon 
distribution). The effect is larger in the case of CT14, which has larger
uncertainties in the gluon PDF in this region.

\begin{figure}
\splitfigure{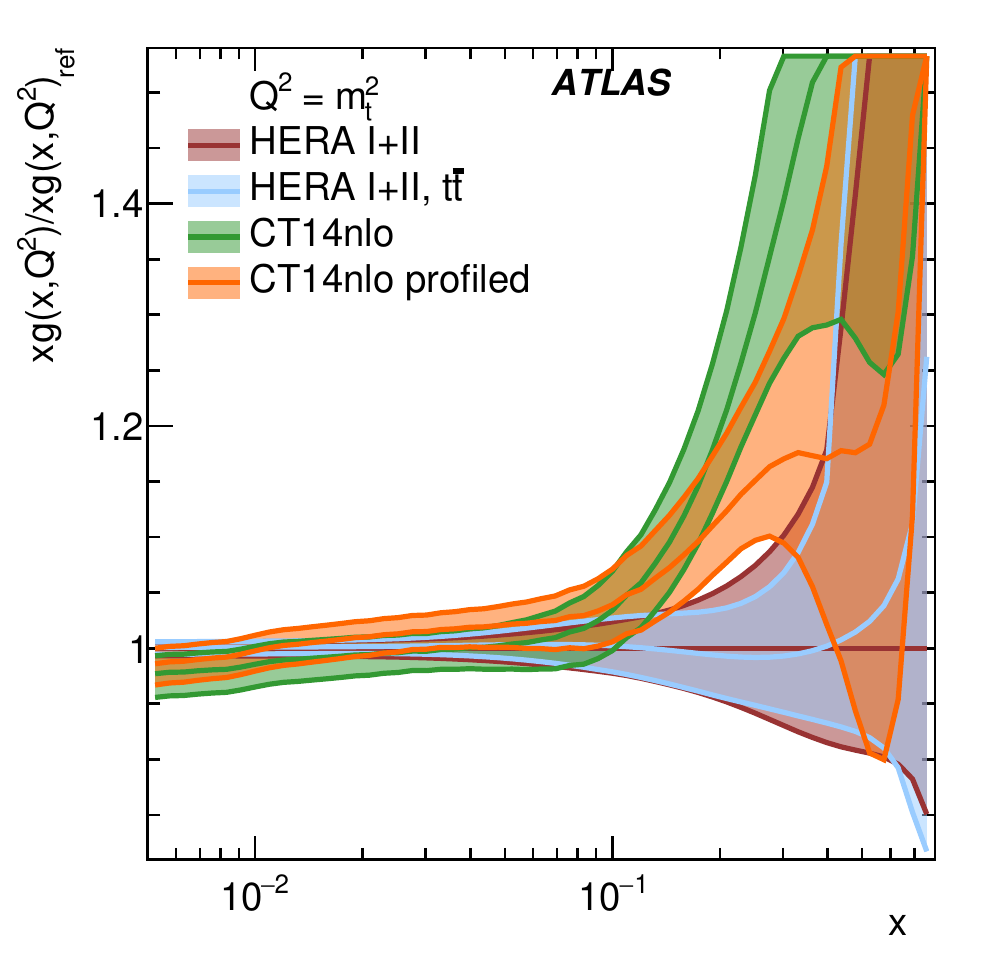}{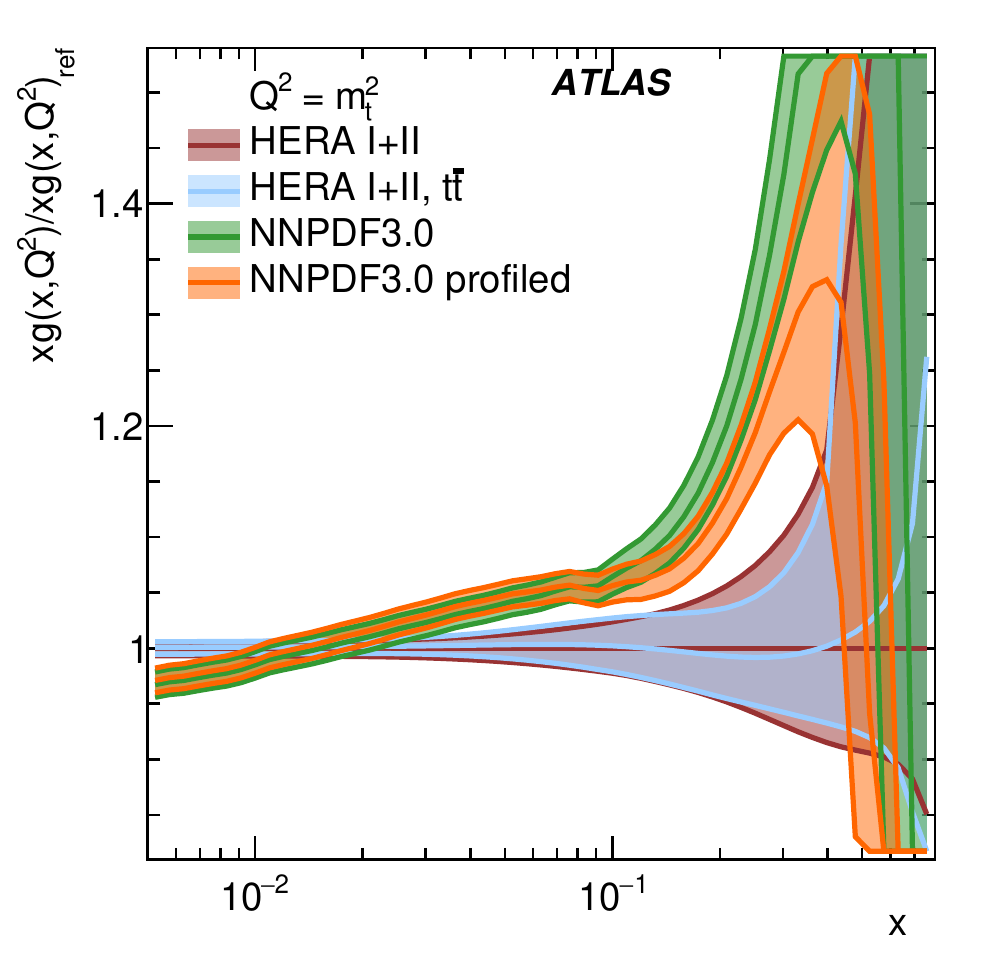}{a}{b}
\caption{\label{f:pdfprof}Ratios of various gluon PDFs and their uncertainty
bands to the gluon PDF determined from HERA I+II data alone (red shading).
The blue shaded band shows the gluon PDF from the fit to HERA I+II data plus 
the normalised differential cross-section distributions as a function
of \etal, \rapll\ and \esum\ in \ttbar\ events. The green
band shows the gluon PDF from the CT14 \cite{ct14} PDF set in (a) and
the NNPDF~3.0 \cite{nnpdf3} PDF set in (b). The orange bands show the
result of profiling these PDFs to the \ttbar\ normalised differential
cross-section data.}
\end{figure}


\section{Extraction of the top quark mass}\label{s:mtop}

The normalised lepton \ptl\ and dilepton \ptll, \mll, \ptsum\ and \esum\ 
differential distributions
are sensitive to the value of the top quark mass, as already shown
in Figure~\ref{f:biastest}(a) for \ptl\  and Figure~\ref{f:biastest}(b) 
for \ptll. Provided that other theoretical uncertainties in the predictions
(as discussed in Section~\ref{s:res})
can be kept under control, fitting these distributions offers a 
complementary way to measure \mtop\ compared to more traditional determinations
from complete reconstruction of the top quark decay products 
\cite{cdfmljets,d0mljets,TOPQ-2016-03,CMS-TOP-14-022}. Ref. \cite{topmassdiff}
explores such an approach in detail, arguing that measurements from 
normalised lepton distributions are less sensitive to the modelling of
perturbative and non-perturbative QCD, and are closer to the ideal of 
a measurement of the top quark pole mass \mtpole\ than those employing a
direct measurement of the top quark decay products. It also stresses the
importance of using several different leptonic observables to probe for
inadequacies in the theoretical descriptions of the distributions which 
may introduce biases in the extracted \mtop\ values. Experimentally,
the double-tagging technique
employed here results in measurements with little uncertainty
from the hadronic components of the \ttbar\ system, again reducing the
exposure to QCD modelling compared to
the measurements based on reconstructing the top quark decay products.

Several sets of top mass determinations are reported here, based either on
predictions from the NLO matrix element event generator {\sc Powheg} interfaced
to {\sc Pythia6} and the CT10 PDFs as described in Section~\ref{s:datasim}, 
or on fixed-order predictions with NLO descriptions of the \ttbar\
production and top quark decay from the MCFM program with various PDF sets, 
as described in Section~\ref{ss:fixedpred}. In the first case, \mtop\
is extracted either by using a template fit parameterising the predictions
as a function of \mtop\ and finding the value which minimises the $\chi^2$ 
with respect to the measured data (described in Section~\ref{ss:masstmpl}),
or by calculating moments of the distributions in data and comparing them
to the corresponding moments of the predicted distributions for
different values of \mtop\ (Section~\ref{ss:massmom}).
In the template fit method, the
comparisons between data and predictions are performed at particle level, 
in contrast to the template fits used for the ATLAS \mtop\ measurements
based on reconstruction of the top quark decay products \cite{TOPQ-2013-02},
where the comparisons are performed at detector level using the 
reconstructed distributions and fully-simulated Monte Carlo events.
The template fit method
uses the complete information from the measured distribution, taking into 
account the uncertainty in each bin, whereas the moments method, advocated in
Ref.~\cite{topmassdiff}, allows different features of the distribution
shapes to be emphasised via the comparisons of moments of different order.
The results from these two methods are discussed and compared in 
Section~\ref{ss:massresmc}.

In the mass determination from QCD fixed-order calculations, described in 
Section~\ref{ss:massfo}, $\chi^2$ values are calculated for the comparison
of data with predictions at different \mtop\ values using the formalism of
Eq.~(\ref{e:fochi2}), and the best-fit \mtop\ is 
found by polynomial interpolation.  This approach is similar to the template
fit discussed above; the use of moments was not pursued as it does not 
exploit the full information of each distribution and does not allow
the reduction of uncertainties via constrained nuisance parameters.
The \mtop\ value used in the fixed-order predictions
corresponds to a well-defined renormalisation scheme, which is the pole mass
(\mtpole) scheme within the MCFM implementation. 
Both the QCD scale uncertainties, representing the effects of missing 
higher-order
corrections beyond NLO, and the PDF uncertainties, are included in the $\chi^2$ formalism in a natural way. This formalism also 
allows \mtop\ to be determined using 
several distributions simultaneously, giving the most precise 
results from any of the techniques explored here.
The results from this method are discussed in Section~\ref{ss:massresfo}
and are used to define the final measurement of the top quark mass from
the distributions measured in this paper.

\subsection{Mass extraction using template fits}\label{ss:masstmpl}

In the template fit method, the best fit top quark mass for each
measured distribution was obtained by
minimising the $\chi^2$ for the comparison of that distribution with
predictions at different values of \mtop, defined analogously with 
Eq.~(\ref{e:redchi}):
\begin{eqnarray}\label{e:mtoptmpl}
\chi^2(\mtop) = {\mathbf\Delta}_{(n-1)}^T(\mtop)\ {\mathbf S}^{-1}_{(n-1)}\ {\mathbf\Delta}_{(n-1)}(\mtop)\ ,
\end{eqnarray}
where ${\mathbf\Delta}_{(n-1)}(\mtop)$ represents the vector of differences between
the measured normalised differential cross-section value and the prediction
for a particular value of \mtop. The latter were obtained from a set of seven
particle-level \ttbar\ samples generated using {\sc Powheg\,+\,Pythia6} with
$\hdamp=\infty$ and the CT10 PDF set, for values of \mtop\ ranging from
165--180\,\GeV\ in 2.5\,\GeV\ steps. The variation of the cross-section in each
bin was parameterised with a second-order polynomial in \mtop, allowing 
predictions for arbitrary values in the considered range to be obtained
by interpolation. An additional multiplicative correction was applied to
the predictions in each bin, based on the ratio of predictions from
{\sc Powheg\,+\,Pythia6} samples with $\hdamp=\mtop$ and $\hdamp=\infty$,
in order to correspond to the baseline event generator choice with $\hdamp=\mtop$. 
As shown in Table~\ref{t:mshifts}, the effects of this correction range
from $-1.3$ to 3.0\,\GeV\ depending on the distribution fitted, and were 
assumed to be independent of \mtop. As the predictions include the simulation 
of leptons from $W\rightarrow\tau\rightarrow e/\mu$ decays, the comparisons are
made with the experimental results including leptons from $\tau$ decays,
as in Section~\ref{ss:gencomp}. 

\begin{table}[tp]
\centering

\begin{tabular}{l|ccccc}\hline
Mass shift [\GeV] & \ptl & \ptll & \mll & \ptsum & \esum \\ \hline
{\sc Powheg} ($\hdamp=\infty)\rightarrow(\hdamp=\mtop)$ &
0.9 & 3.0 & -1.3 & 0.9 & 0.5 \\
Top \pt\ NNLO reweighting     & 1.8 & 0.3 & 2.2 & 1.3 & 1.3 \\
\hline
\end{tabular}
\caption{\label{t:mshifts}Changes in the top quark mass fitted in data
from each lepton or dilepton distribution
using the template fit method. The first row shows the shifts 
when changing the {\sc Powheg} parameter \hdamp\ from $\infty$ to \mtop,
a correction which is applied to the results quoted in Table~\ref{t:massmca}.
The second row shows additional shifts when reweighting the 
top quark \pt\ 
in {\sc Powheg\,+\,Pythia6} to the NNLO prediction of Ref. \cite{topptnnlo}.}
\end{table}

The template fit method was tested with pseudo-experiments based on
fully-simulated \ttbar\ samples with \mtop\ values in the range 165--180\,\GeV\
plus non-\ttbar\ backgrounds. The pseudo-data were processed through 
the complete analysis procedure starting from the observed event counts in
each bin, using the methodology described in 
Section~\ref{ss:valid}. The baseline {\sc Powheg\,+\,Pythia6} \ttbar\ sample
with $\mtop=172.5$\,\GeV\ was used as reference for the calculation of
\gemi, \cbi, \nibi\ and \niibi.
No statistically significant biases were found
for the fits based on the \ptl, \ptll\ and \mll\ distributions, but
biases of up to  0.6\,\GeV\ for \ptsum\ and 0.9\,\GeV\ for \esum\ were found in 
pseudo-experiments with true \mtop\ values 5\,\GeV\ away from the 172.5\,\GeV\
reference, still small compared to the expected statistical uncertainties
using these distributions. These biases were corrected in the fit results
from data discussed in Section~\ref{ss:massresmc} below. The pseudo-experiments
were also used to check the statistical uncertainties returned by the fit
via the pull distributions, which were generally found to be within 
$\pm 5$\,\% of unity.

Both the data statistical uncertainty and experimental systematic uncertainties
on the measurements of the differential distributions are included in
the matrix ${\mathbf S}_{(n-1)}$ in Eq.~(\ref{e:mtoptmpl}). Further uncertainties
in the extracted \mtop\ value arise from the choices of PDFs and event generator
setup for the predictions. The PDF uncertainties were assessed from the
variations in normalised \ttbar\ differential cross-section distributions 
predicted by {\sc MC@NLO\,+\,Herwig} reweighted using the error sets of the 
CT10, MSTW and NNPDF 2.3 PDF sets as described in Section~\ref{ss:ttsyst}.
The event generator setup uncertainties were 
assessed as the quadrature sum of a \ttbar\ generator uncertainty and
a QCD radiation uncertainty. The former was 
obtained from the comparison of results using {\sc Powheg\,+\,Pythia6} 
($\hdamp=\mtop$)
and {\sc MC@NLO\,+\,Herwig} samples (thus varying both the matrix element 
and parton shower generator). The latter was defined as
half the variation from fits using
the {\sc Powheg\,+\,Pythia6} samples with radLo and radHi tunes discussed
in Section~\ref{s:datasim}. In all cases, the uncertainties were defined
from the difference in \mtop\ values obtained when fitting the two samples
as pseudo-data, using the full experimental covariance matrix from the 
data measurement and 
the standard templates obtained from the {\sc Powheg\,+\,Pythia6} samples
as discussed above.

\subsection{Mass extraction using moments}\label{ss:massmom}

Top quark mass information can also be derived from a measured distribution
by calculating Mellin moments of the distribution, and comparing the values
observed to a calibration curve obtained from predictions with different
values of \mtop\ \cite{topmassdiff}. The $k$th order Mellin moment \mmt{k}
for a distribution $D(x)\equiv \mathrm{d}\sigma/\mathrm{d}x$ as a function of
 a kinematic variable $x$ is defined as:
\begin{eqnarray}
\mmt{k}=\frac{1}{\sigma_{\mathrm{fid}}} \int x^k D(x)\,\mathrm{d}x \,,
\end{eqnarray}
where the integral is taken over the fiducial region, and the total
fiducial cross-section 
$\sigma_{\mathrm{fid}}=\int D(x)\,\mathrm{d}x$.
These moments can in principle be evaluated without binning the data,
since for leptonic observables, the value $x$ for each individual event is 
measured with high precision. However,
for the purpose of this analysis, these moments were approximated by 
binned moments \mtheta{k} evaluated as:
\begin{eqnarray}
\mtheta{k}=\sum_i \xntti X_i\,, \ X_i=<x^k> \mathrm{in\ bin}\ i\,,
\end{eqnarray}
where \xntti\ is the fraction of the total fiducial \ttbar\
cross-section in bin $i$ (Eq.~(\ref{e:normx}))
and $X_i$ is the mean value of $x$ for all the events falling in bin $i$.
The values of $X_i$, which act as weights for each bin $i$ of each kinematic 
distribution when calculating the moment $k$, were
evaluated using the baseline {\sc Powheg\,+\,Pythia6} sample and kept 
constant when evaluating moments for the data and all simulation samples.
Calibration curves for the first, second and third moments \mtheta{1},
\mtheta{2}\ and \mtheta{3}\ were derived using the same set of
{\sc Powheg\,+\,Pythia6} samples with top quark masses in the range 
165--180\,\GeV\ as used for the template analysis. The dependencies of
$\mtheta{k}$ on \mtop\ were 
found to be well-described by second-order polynomials 
$\mtheta{k}(\mtop)=P_2(\mtop)$. A constant offset in each moment
was used to correct to the calibration appropriate for $\hdamp=\mtop$ samples,
and the polynomial inverted to obtain the \mtop\ value corresponding
to a  given measured $k$th moment \mtheta{k}. 

The extraction procedure was tested for bias with 
pseudo-experiments in the same way as for the template fit. The observed
biases were of similar size to those in the template fit, and were
corrected in the same way. Experimental systematic
uncertainties were evaluated by calculating the moments from the 
normalised cross-section distribution with each bin shifted by one standard
deviation of each systematic, and translating the resulting shift in
\mtheta{k}\ to a shift in \mtop. Uncertainties in the predictions due to the
choice of PDFs, \ttbar\ generator and radiation settings were assessed in the 
same
way, i.e. from the shifts in \mtheta{k}\ predicted by each of the alternative
samples.

\subsection{Results from the template and moment methods}\label{ss:massresmc}

The results of applying the template and first, second and third moment
methods to each of the \ptl, \ptll, \mll, \ptsum\ and \esum\ distributions
using predictions from {\sc Powheg\,+\,Pythia6} and CT10 PDFs are shown
in Table~\ref{t:massmca} and Figure~\ref{f:massmca}. The table shows
the $\chi^2$ at the best fit mass for each distribution, and the breakdown
of uncertainties into statistical, experimental systematic and theoretical
contributions, evaluated as discussed in Section~\ref{ss:masstmpl}. 
For the template fits, the
data statistical uncertainty was evaluated from a $\chi^2$ minimisation
of Eq.~(\ref{e:mtoptmpl}) with only statistical uncertainties included
in the covariance matrix $\textbf S$. The experimental systematic uncertainty
was evaluated 
as the quadrature difference between the total uncertainty (when including both
statistical and experimental systematic uncertainties in $\textbf S$), and
the data statistical uncertainty. For the moments method, the statistical and
experimental
systematic uncertainties were evaluated directly on the moments \mtheta{k}\
as discussed in Section~\ref{ss:massmom}.

\begin{table}[tp]
\centering

\begin{tabular}{l|rrrrr}\hline
Template& \ptl & \ptll & \mll & \ptsum & \esum \\\hline
$\chi^2/N_{\mathrm dof}$ &  8.1/8 &  7.5/7 & 13.9/10 &  8.0/6 & 12.5/8 \\
\mtop\,[\GeV] & $ 168.4\pm 2.3$ & $ 173.0\pm 2.1$ & $ 170.6\pm 4.2$ & $ 169.4\pm 2.0$ & $ 166.9\pm 4.0$ \\[2pt] \hline 
Data statistics & $\pm\,1.0$ & $\pm\,0.9$ & $\pm\,2.0$ & $\pm\,0.9$ & $\pm\,1.3$ \\[2pt]
Expt. systematic & $\pm\,1.6$ & $\pm\,1.0$ & $\pm\,3.1$ & $\pm\,1.6$ & $\pm\,1.5$ \\[2pt]
PDF uncertainty& $\pm\,1.0$ & $\pm\,0.2$ & $\pm\,1.6$ & $\pm\,0.6$ & $\pm\,3.4$ \\[2pt]
\ttbar\ generator& $\pm\,0.4$ & $\pm\,1.4$ & $\pm\,1.4$ & $\pm\,0.4$ & $\pm\,1.1$ \\[2pt]
QCD radiation& $\pm\,0.7$ & $\pm\,0.8$ & $\pm\,0.5$ & $\pm\,0.2$ & $\pm\,0.2$ \\[2pt]
\hline
\end{tabular}

\begin{tabular}{l|rrrrr}\hline
Moment 1& \ptl & \ptll & \mll & \ptsum & \esum \\\hline
\mtop\,[\GeV] & $ 168.2\pm 2.9$ & $ 172.4\pm 3.8$ & $ 166.6\pm 6.5$ & $ 168.4\pm 2.9$ & $ 160.8\pm 7.9$ \\[2pt] \hline 
Data statistics & $\pm\,1.0$ & $\pm\,1.0$ & $\pm\,2.4$ & $\pm\,1.1$ & $\pm\,2.2$ \\[2pt]
Expt. systematic & $\pm\,2.1$ & $\pm\,1.6$ & $\pm\,3.8$ & $\pm\,2.1$ & $\pm\,3.1$ \\[2pt]
PDF uncertainty& $\pm\,1.2$ & $\pm\,0.3$ & $\pm\,2.9$ & $\pm\,1.1$ & $\pm\,6.7$ \\[2pt]
\ttbar\ generator& $\pm\,0.2$ & $\pm\,1.3$ & $\pm\,3.4$ & $\pm\,0.2$ & $\pm\,2.0$ \\[2pt]
QCD radiation& $\pm\,1.2$ & $\pm\,3.0$ & $\pm\,1.4$ & $\pm\,1.1$ & $\pm\,0.2$ \\[2pt]
\hline
\end{tabular}

\begin{tabular}{l|rrrrr}\hline
Moment 2& \ptl & \ptll & \mll & \ptsum & \esum \\\hline
\mtop\,[\GeV] & $ 168.1\pm 3.2$ & $ 172.2\pm 4.5$ & $ 166.9\pm 6.9$ & $ 167.9\pm 3.3$ & $ 159.9\pm 9.2$ \\[2pt] \hline 
Data statistics & $\pm\,1.2$ & $\pm\,1.1$ & $\pm\,2.8$ & $\pm\,1.3$ & $\pm\,2.6$ \\[2pt]
Expt. systematic & $\pm\,2.3$ & $\pm\,2.0$ & $\pm\,4.3$ & $\pm\,2.4$ & $\pm\,3.4$ \\[2pt]
PDF uncertainty& $\pm\,1.3$ & $\pm\,0.4$ & $\pm\,3.3$ & $\pm\,1.3$ & $\pm\,7.8$ \\[2pt]
\ttbar\ generator& $\pm\,0.4$ & $\pm\,1.2$ & $\pm\,3.2$ & $\pm\,0.4$ & $\pm\,2.4$ \\[2pt]
QCD radiation& $\pm\,1.2$ & $\pm\,3.7$ & $\pm\,0.7$ & $\pm\,1.3$ & $\pm\,0.2$ \\[2pt]
\hline
\end{tabular}

\begin{tabular}{l|rrrrr}\hline
Moment 3& \ptl & \ptll & \mll & \ptsum & \esum \\\hline
\mtop\,[\GeV] & $ 168.3\pm 3.5$ & $ 172.0\pm 5.6$ & $ 166.4\pm 9.1$ & $ 167.6\pm 3.8$ & $ 160.9\pm 9.5$ \\[2pt] \hline 
Data statistics & $\pm\,1.5$ & $\pm\,1.4$ & $\pm\,4.2$ & $\pm\,1.6$ & $\pm\,3.0$ \\[2pt]
Expt. systematic & $\pm\,2.5$ & $\pm\,2.6$ & $\pm\,6.0$ & $\pm\,2.7$ & $\pm\,3.7$ \\[2pt]
PDF uncertainty& $\pm\,1.5$ & $\pm\,0.6$ & $\pm\,4.1$ & $\pm\,1.4$ & $\pm\,7.8$ \\[2pt]
\ttbar\ generator& $\pm\,0.6$ & $\pm\,1.1$ & $\pm\,3.5$ & $\pm\,0.7$ & $\pm\,2.4$ \\[2pt]
QCD radiation& $\pm\,1.1$ & $\pm\,4.6$ & $\pm\,0.2$ & $\pm\,1.4$ & $\pm\,0.2$ \\[2pt]
\hline
\end{tabular}

\caption{\label{t:massmca}
Measurements of the top quark mass from individual template fits to the
lepton \ptl\ and dilepton \ptll, \mll, \ptsum\ and \esum\ distributions,
and using the first, second and third moments of these distributions.
The data are compared to predictions from {\sc Powheg\,+\,Pythia6} with
the CT10 PDF set. The $\chi^2$
value at the best-fit mass for each distribution (for the template
fits only), the fitted mass
with its total uncertainty, and the individual uncertainty contributions
from data statistics, experimental systematics, and uncertainties in the
predictions due to the choice of \ttbar\ event generator and the modelling 
of QCD radiation are shown.}
\end{table}

\begin{figure}[tp]

\hspace{-10mm}\includegraphics[width=165mm]{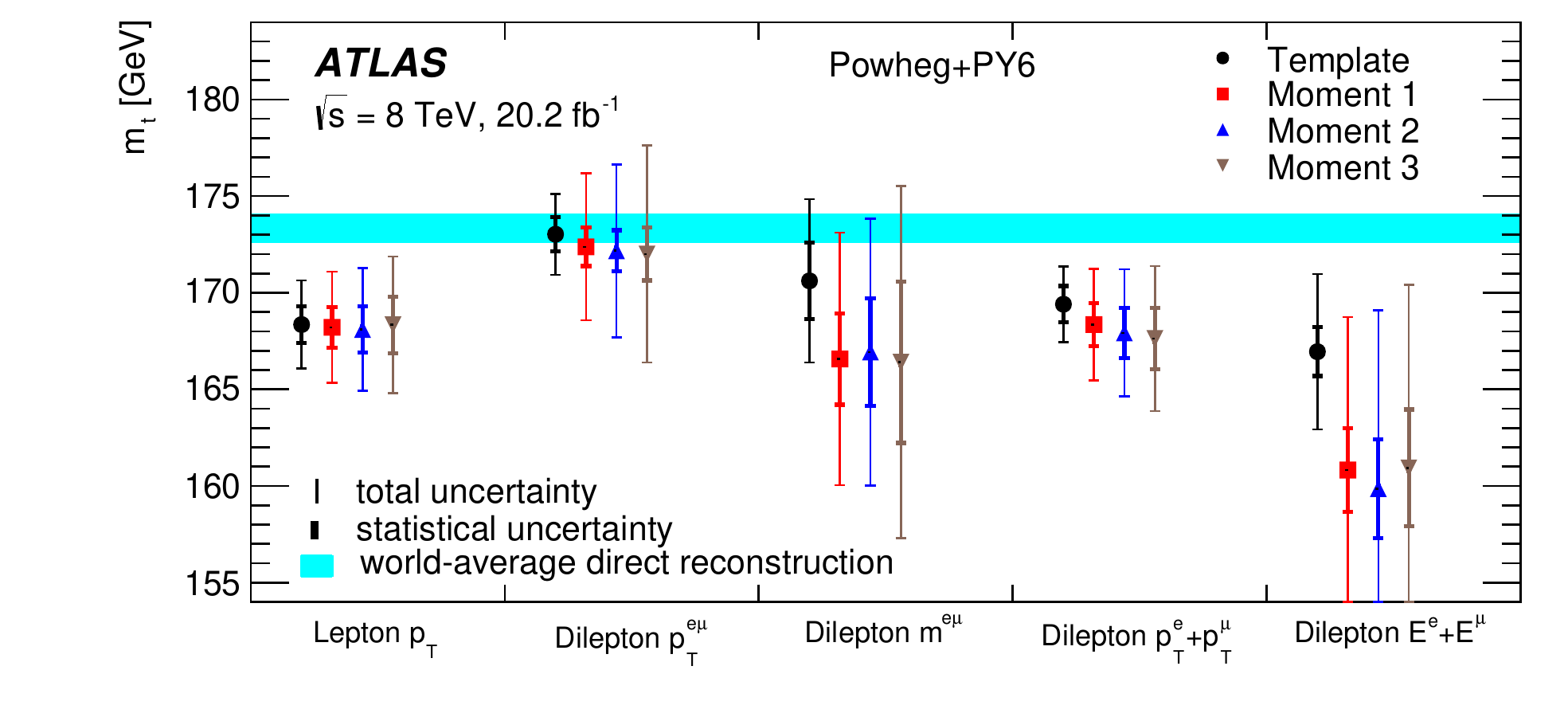}

\caption{\label{f:massmca}Measurements of the top quark mass using templates
derived from {\sc Powheg\,+\,Pythia6} with the CT10 PDF set. The results
from fitting templates of the single lepton \ptl\ and dilepton \ptll, \mll, \ptsum\ and \esum\ distributions, and from the first, second and third moments
of these distributions, are shown. For comparison, the world-average of mass 
measurements from  reconstruction of the top quark decay products and
its uncertainty \cite{masswa} is shown by the cyan band.}
\end{figure}

The ratios of predictions to data at the best-fit top quark mass found
by the application of the template fit method to  each distribution are shown in 
Figure~\ref{f:mfitmca}. The data are generally well-described by these 
predictions, as can also be seen from the $\chi^2$ values in 
Table~\ref{t:massmca}, except for the \esum\ distribution. This distribution
is quite sensitive to PDFs as well as \mtop, and is better
described by the HERAPDF PDFs than the CT10 PDFs used here to extract
\mtop, resulting in a low fitted value with a large PDF uncertainty, and
a large variation between the template and moment fit results.
Total uncertainties in \mtop\
of about 2\,\GeV\ are obtained from the template fits to the
\ptl, \ptll\ and \ptsum\ 
distributions. These results have relatively small theoretical uncertainties, 
and the experimental uncertainties are
dominated by $\ttbar-Wt$ interference and the electron energy scale. 
The \mll\ distribution is intrinsically less sensitive to \mtop, having
larger statistical, experimental and theoretical systematic uncertainties.
The results from the extraction based on moments have larger uncertainties
than those from the template fit, reflecting that the moments do not take 
into account the relative precision on the different bins of the distributions,
and that the higher moments are more sensitive to the tails of the 
distributions, which are less precisely measured and subject to larger
theoretical uncertainties.
Within each distribution, the \mtop\ values from the different moments
are close, though 3--4\,\GeV\ lower than the template fit results for \mll,
and up to 7\,\GeV\ lower in the case of \esum.

\begin{figure}[tp]
\splitfigure{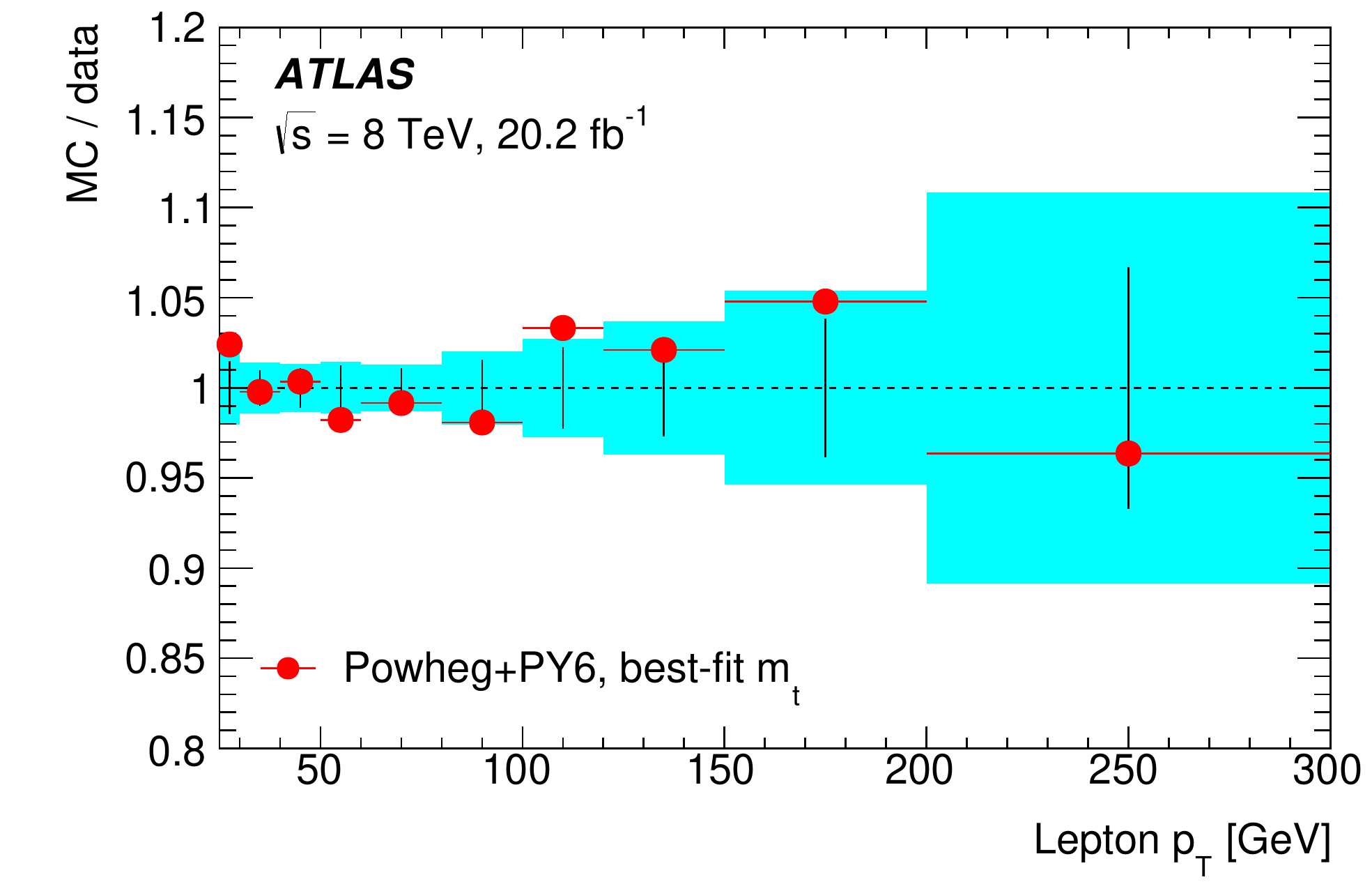}{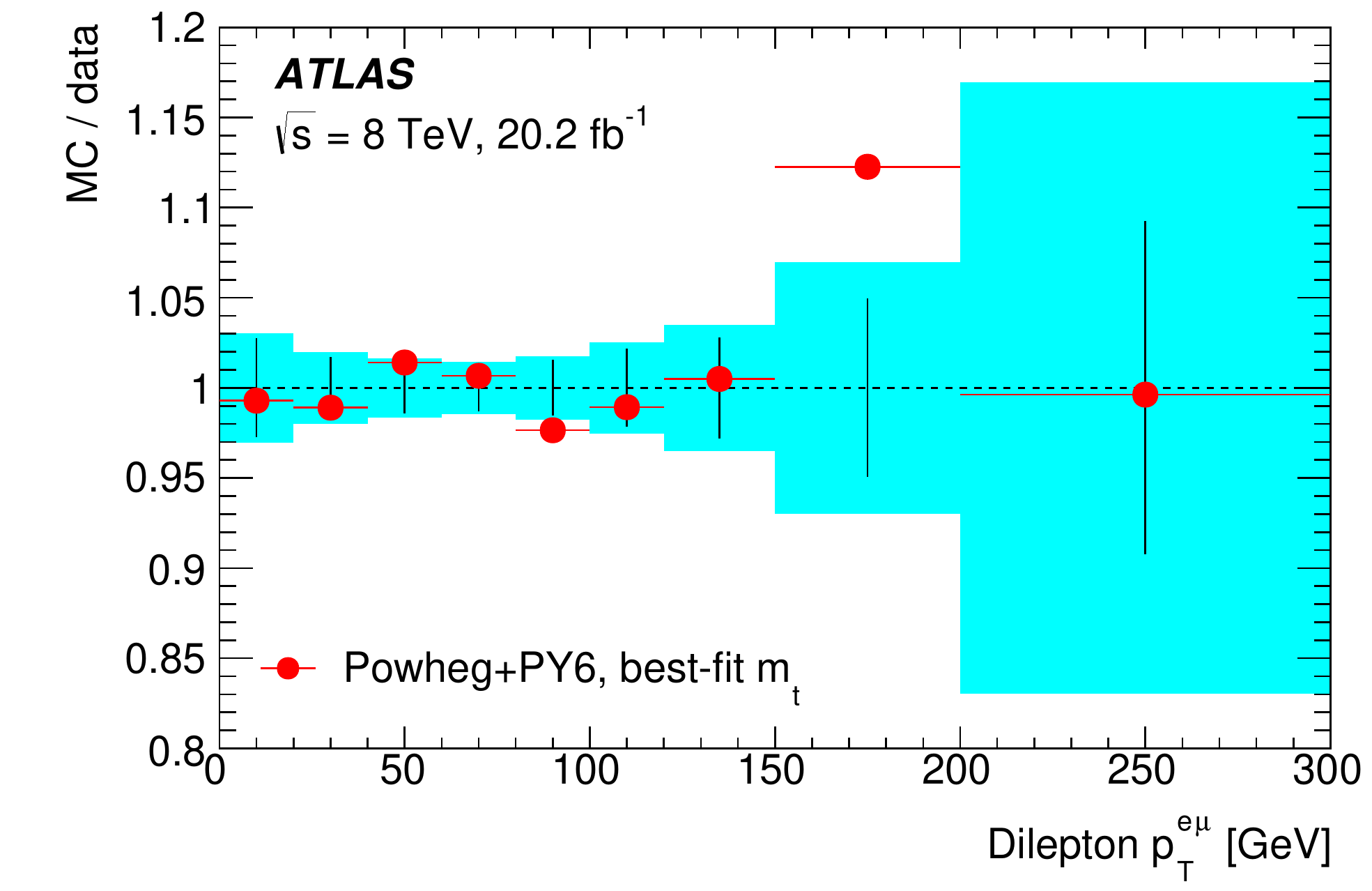}{a}{b}
\splitfigure{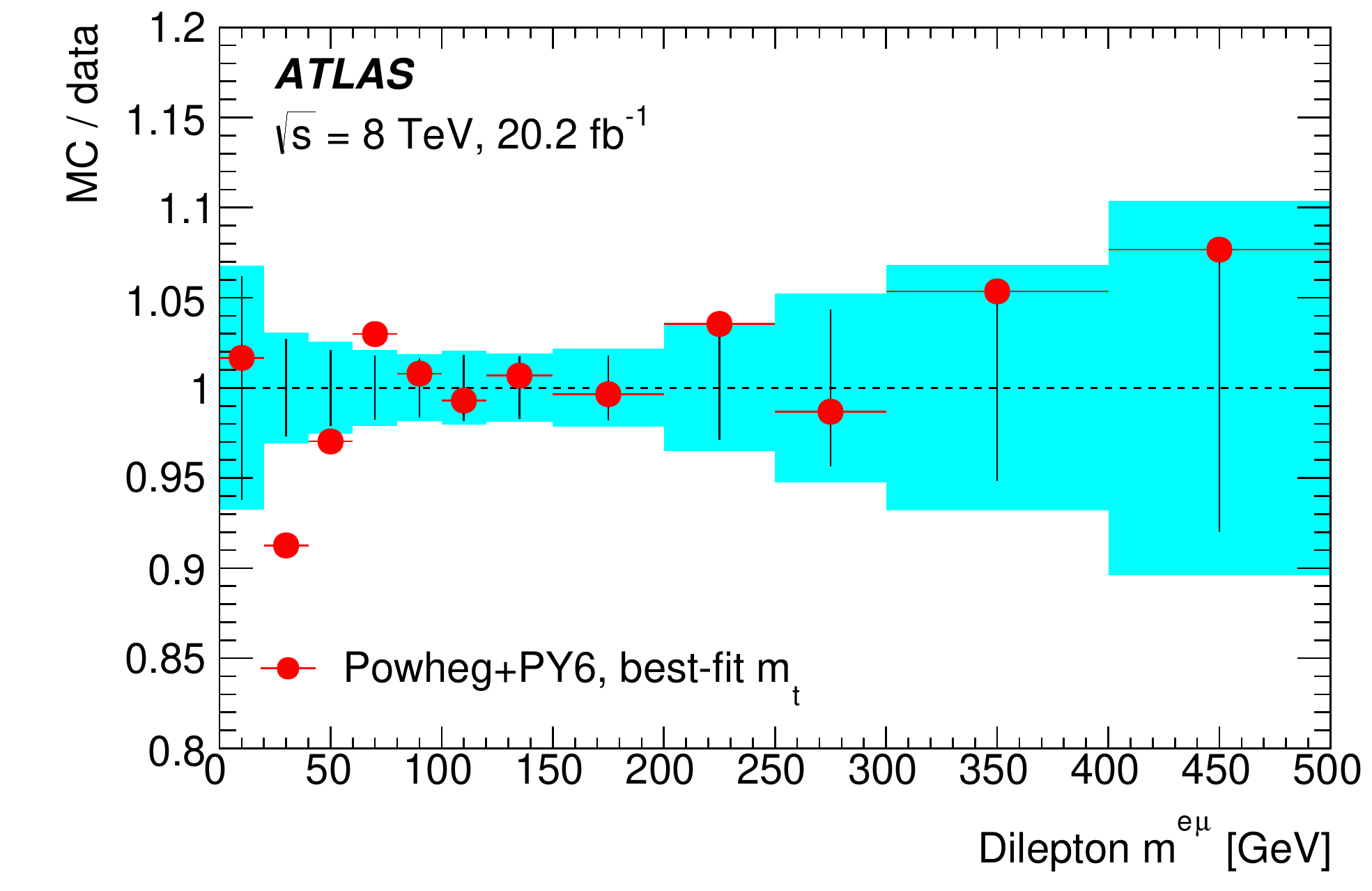}{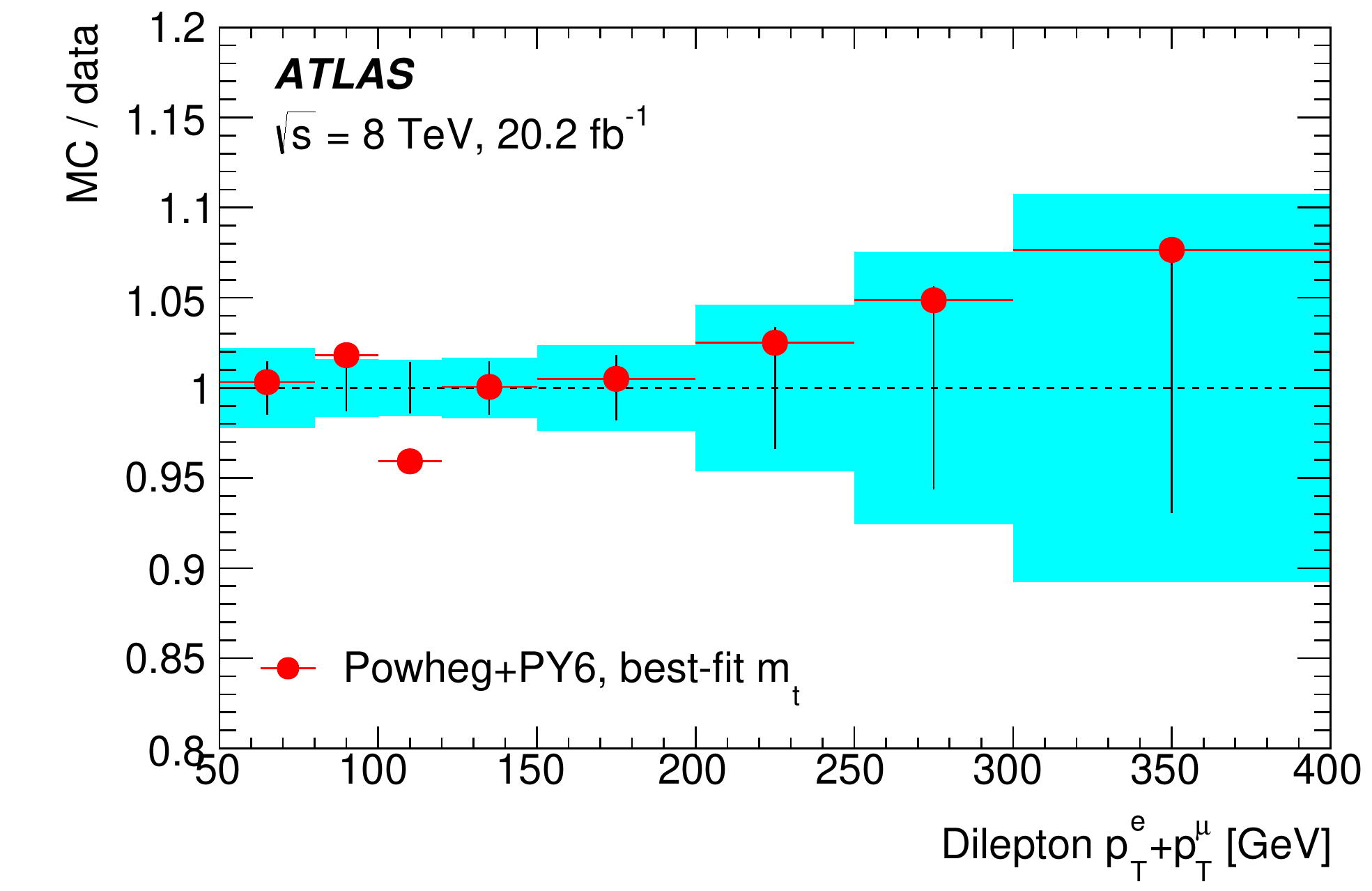}{c}{d}
\parbox{85mm}{
\includegraphics[width=78mm]{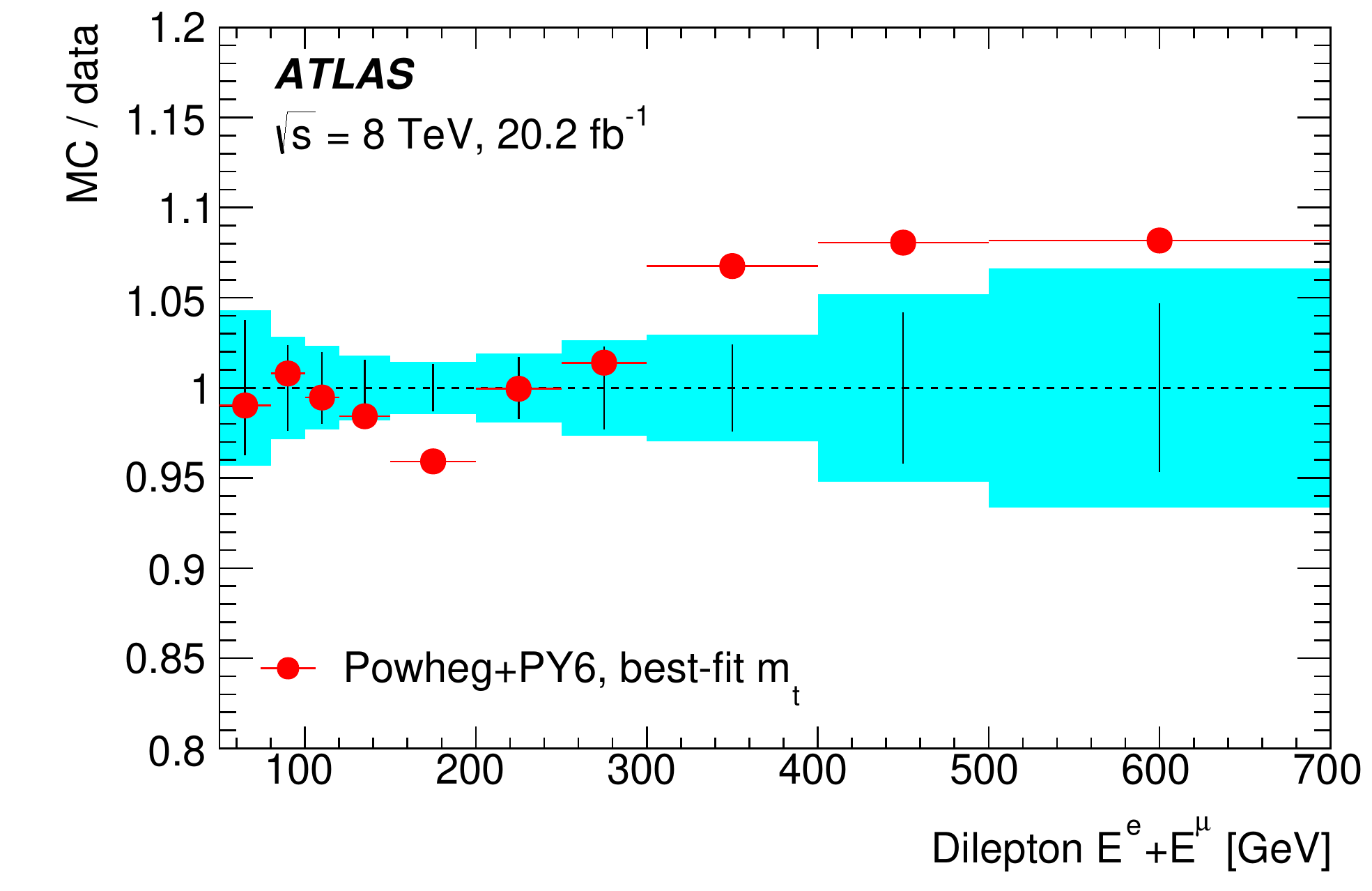}
\vspace{-7mm}
\center{(e)}
}
\caption{\label{f:mfitmca}Ratios of predictions of normalised differential 
cross-sections to data as a function of (a) \ptl, (b) \ptll, (c) \mll, (d)
\ptsum\ and (e) \esum, with the prediction taken from {\sc Powheg\,+\,Pythia6}
with the CT10 PDF
at the best-fit top quark mass \mtop\ for each distribution. The data 
statistical uncertainties are shown by the black error bars around a ratio
of unity, and the total experimental uncertainties by the cyan band.}
\end{figure}

The central values of the template fit results from the five distributions 
exhibit a spread (envelope) of about 6\,\GeV. The results from the fits of 
\ptl\ and \ptsum\ lie 4--5\,\GeV\ below that from \ptll, which is close to
the world-average mass value from reconstruction of top quark decay products of 
$173.34\pm 0.76$\,\GeV\ \cite{masswa}. 
The consistency of the fit results was assessed
by combining them using the best linear unbiased estimate (BLUE) 
technique \cite{blue}.
Correlations in the statistical uncertainties were assessed 
using pseudo-experiments as described in Section~\ref{ss:valid}. 
Correlations between systematic uncertainties were determined by
assuming the effects on \mtop\ from each individual experimental or
theoretical component to be fully correlated between distributions. 
PDF uncertainties were assessed using the eigenvector
pairs of the CT10 PDF only. The combination of all five distributions
has a $\chi^2$ probability of 4\,\%, indicating that the systematic
uncertainties may be underestimated.

The {\sc Powheg\,+\,Pythia6}
\ttbar\ samples used here do not provide a good modelling of the top quark
\pt\ spectrum 
\cite{TOPQ-2012-08,TOPQ-2013-07,TOPQ-2015-06,CMS-TOP-11-013,CMS-TOP-12-028},
potentially biasing the results. The size of this possible bias was explored
by fitting the distributions from the {\sc Powheg\,+\,Pythia6} baseline sample 
reweighted to the top quark \pt\ spectrum calculated at NNLO precision in 
Ref. \cite{topptnnlo}. The reweighted sample gives a better description of the 
\ptl\ and \ptsum\ distributions, as can be seen from the $\chi^2$ values for
`{\sc Powheg\,+\,PY6} \pt\ NNLO' in Table~\ref{t:NorChiXPPData2012Data}.
The mass shifts between the baseline and reweighted samples, representing
the amount that the top quark mass measured in data 
would be shifted upwards if the 
templates were based on reweighted samples, are shown in Table~\ref{t:mshifts}. These shifts are larger (1.3--1.8\,\GeV) for \ptsum\ and \ptl\ 
than for \ptll\ (0.3\,\GeV),
and would bring the results shown in Figure~\ref{f:massmca} into closer 
agreement with each other.
 However, given that this reweighting is relatively crude, and does
not take into account the potential NNLO effects on other distributions
important for modelling the lepton and dilepton kinematics (e.g. the invariant
mass and rapidity of the \ttbar\ system), the shifts are taken to
be purely indicative, and no attempt has been made
to correct the quoted central values for these effects. 
The predictions for the \ptl\ and \ptsum\
distributions are also sensitive to the choice of PDF. The PDF uncertainties
shown for \ptl\ and \ptsum\ in Table~\ref{t:massmca} are significantly larger
than those for \ptll, and as shown in 
Section~\ref{ss:gencomp}, the {\sc Powheg\,+\,Pythia6} sample generated using
HERAPDF 1.5 instead of CT10 gives a significantly better description of
both distributions at $\mtop=172.5$\,GeV.

The predictions from {\sc Powheg\,+\,Pythia6}, based on NLO matrix 
elements interfaced to parton showers, hence suffer from 
significant uncertainties due to missing NNLO corrections and lack of knowledge
of the  PDFs. Consequently, they do not have sufficient precision
to extract the top quark mass from individual distributions with a theoretical
uncertainty better than about 2\,\GeV, slightly larger than the uncertainties
corresponding to the precision of the experimental measurements.
These limitations are addressed by the approach discussed below, where
several distributions are fitted simultaneously to extract \mtop\ whilst
constraining the uncertainties in the theoretical predictions.

\subsection{Mass extraction using fixed-order predictions}\label{ss:massfo}

The NLO fixed-order predictions for each distribution were
generated using MCFM as discussed in Section~\ref{ss:fixedpred}, for 
top quark masses in the range 161--180\,\GeV\ in steps of 0.5\,\GeV, with 
various PDF choices. 
The $\chi^2$ for the consistency of each prediction
with the data was calculated using Eq.~(\ref{e:fochi2}), incorporating
both PDF and QCD scale uncertainties into the theoretical uncertainties
represented by the nuisance parameters \vbth. The central scales were
again chosen to be $\mu_F=\mu_R=\mtop/2$, with the values varying with
\mtop\ in the mass scan, and independent variations of
$\mu_F$ and $\mu_R$ by factors of two and one-half defining the one standard
deviation up and down scale variations. The $\chi^2$ was evaluated at
each mass point, and interpolated using a fourth-order polynomial. The
asymmetric uncertainty in the fitted value of \mtop\  was defined as the points
at which the $\chi^2$ increases by one unit either side of the minimum
point. This uncertainty naturally includes both experimental statistical
and systematic uncertainties in the measurements, and theoretical uncertainties
due to PDFs and QCD scale choices.

In this method, the top quark mass can be extracted from each measured
distribution individually, or from the combination of several distributions,
where the sum $i$ in Eq.~(\ref{e:fochi2}) runs over the bins of all considered
distributions, and the experimental covariance matrix includes both
statistical and systematic correlations between bins of the same and
different distributions, evaluated as discussed in Section~\ref{ss:gencomp}.
When fitting several distributions simultaneously, the system is 
over-constrained, profiling the various sources of theoretical uncertainty.
For example, when including all eight measured distributions, the \etal\ and 
\rapll\ distributions have little sensitivity to \mtop, but constrain
the PDF parameters. The \dphill\ distribution constrains the QCD
scale parameters $\mu_F$ and $\mu_R$, under the assumption that uncertainties
in higher-order QCD corrections are parameterised by $\mu_F$ and $\mu_R$ in a
way that can be transported from one distribution to another. Two alternative
dynamical scale choices were also tested in order to probe this assumption,
as discussed in Section~\ref{ss:massresfo} below.

Potential biases in the method were checked by using predictions with 
$\mtop=172.5$\,\GeV\ as  pseudo-data, and considering both experimental and 
theoretical uncertainties in the 
$\chi^2$ definition. The resulting fitted values of \mtop\ were within
0.1\,\GeV\ of the input value for all five fitted individual distributions
(\ptl, \ptll, \mll, \ptsum\ and \esum), and 0.01\,\GeV\ from the input
value for a combined fit of all eight distributions, also including
\etal, \rapll\ and \dphill. The widths of the pull distributions were found 
to be compatible with unity, confirming the validity of the uncertainty
estimates from the fits.

\subsection{Mass results from fixed-order predictions}\label{ss:massresfo}

The results of the fits to NLO QCD fixed-order predictions with MCFM and the 
CT14 PDF set are shown for the individual distributions in 
Table~\ref{t:massfoa}, and the results using the CT14, MMHT, NNPDF 3.0, 
HERAPDF 2.0, ABM 11 \cite{abmpdf} and NNPDF 3.0\_nojet \cite{nnpdf3} PDF sets
are shown in Figure~\ref{f:massfoa}. As shown in Section~\ref{s:pdf}, the
constraint on the gluon PDF from the leptonic \ttbar\ measurements is
consistent with the PDF determination from DIS data. The use of the 
NNPDF 3.0\_nojet PDF set, which does not include Tevatron and LHC jet production
data, allows the effects on \mtop\ of any possible tension between
DIS and jet data in the determination of the gluon PDF to be tested.
The results from combined fits to
all eight distributions, using predictions from all six PDF sets, are
shown in Table~\ref{t:massfob} and Figure~\ref{f:massfoa}.
In Tables~\ref{t:massfoa} and~\ref{t:massfob}, the decomposition
of the total uncertainty from each mass fit
into statistical, experimental and theoretical
(PDF and QCD scales) uncertainties was obtained in analogy to the 
numerical procedure outlined in Ref. \cite{valassiblue}. For each individual
source of statistical or systematic uncertainty (corresponding to a 
nuisance parameter \betaexpj\ or \betathk\ in Eq.~(\ref{e:fochi2})), the data
were shifted by plus or minus one standard deviation, and a new \mtop\ value
obtained by re-minimising the $\chi^2$ function. The resulting shifts in
\mtop\ were added in quadrature to obtain the decomposition into the various
categories. The quadrature
sum of the decomposed uncertainties agrees with the total to  within 10\,\% 
in all cases, the residual differences being
due to non-linearity between the uncertainty sources and the extracted
values of \mtop.

\begin{table}[tp]
\centering

\begin{tabular}{l|rrrrr}\hline
& \ptl & \ptll & \mll & \ptsum & \esum \\\hline
$\chi^2/N_{\mathrm dof}$ &    9/8 &    5/7 &   11/10 &   11/6 &    8/8 \\
\mtpole\,[\GeV] & $ 169.7\,^{+2.9}_{-2.7}$ & $ 175.1\pm 1.9$ & $ 174.5\,^{+5.1}_{-5.3}$ & $ 170.3\pm 2.1$ & $ 168.5\,^{+3.2}_{-3.3}$ \\[2pt] \hline 
Data statistics & $\pm\,2.0$ & $\pm\,1.4$ & $\,^{+3.8}_{-4.0}$ & $\pm\,1.4$ & $\pm\,2.3$ \\[2pt]
Expt. systematic & $\,^{+2.5}_{-2.3}$ & $\pm\,0.9$ & $\,^{+2.9}_{-3.3}$ & $\,^{+1.5}_{-1.6}$ & $\pm\,2.0$ \\[2pt]
PDF uncertainty& $\pm\,0.5$ & $\pm\,0.1$ & $\pm\,1.1$ & $\pm\,0.5$ & $\pm\,1.4$ \\[2pt]
QCD scales& $\pm\,1.1$ & $\,^{+0.7}_{-0.8}$ & $\pm\,2.6$ & $\,^{+0.4}_{-0.5}$ & $\pm\,0.7$ \\[2pt]
\hline
\end{tabular}

\caption{\label{t:massfoa}
Measurements of the top quark mass from individual fits to the
lepton \ptl\ and dilepton \ptll, \mll, \ptsum\ and \esum\ distributions,
using fixed-order predictions from MCFM with the CT14 PDF set. The $\chi^2$
value at the best-fit mass for each distribution, the fitted mass
with its total uncertainty, and the individual uncertainty contributions
from data statistics, experimental systematics, and uncertainties in the
predictions from PDF and QCD scale effects are shown.}
\end{table}

\begin{table}[tp]
\centering

\begin{tabular}{l|rrrrrr}\hline
& CT14 & MMHT & NNPDF 3.0 & HERAPDF 2.0 & ABM 11 & NNPDF nojet \\\hline
$\mu_F=\mu_R=\mtop/2$ & \\
$\chi^2/N_{\mathrm dof}$ & 71/68 & 70/68 & 67/68 & 67/68 & 71/68 & 64/68 \\
\mtpole\,[\GeV] & $ 173.5\pm 1.2$ & $ 173.4\pm 1.2$ & $ 173.2\pm 1.2$ & $ 172.9\pm 1.2$ & $ 172.8\,^{+1.3}_{-1.2}$ & $ 173.1\pm 1.2$ \\[2pt] \hline 
Data statistics & $\pm\,0.9$ & $\pm\,0.9$ & $\pm\,0.9$ & $\pm\,0.9$ & $\pm\,0.9$ & $\pm\,0.9$ \\[2pt]
Expt. systematic & $\,^{+0.7}_{-0.8}$ & $\pm\,0.8$ & $\pm\,0.8$ & $\pm\,0.9$ & $\,^{+0.9}_{-0.8}$ & $\pm\,0.8$ \\[2pt]
PDF uncertainty& $\pm\,0.1$ & $\pm\,0.1$ & $\,^{+0.1}_{-0.2}$ & $\pm\,0.1$ & $\pm\,0.1$ & $\pm\,0.4$ \\[2pt]
QCD scales& $\pm\,0.1$ & $\pm\,0.1$ & $\,^{+0.1}_{-0.0}$ & $\pm\,0.1$ & $\pm\,0.1$ & $\pm\,0.0$ \\[2pt] \hline
\hline
$\mu_F=\mu_R=H_T/4$ & \\
$\chi^2/N_{\mathrm dof}$ & 69/68 & 67/68 & 64/68 & 61/68 & 66/68 & 60/68 \\
\mtpole\,[\GeV] & $ 173.6\pm 1.3$ & $ 173.4\pm 1.3$ & $ 173.2\pm 1.3$ & $ 173.6\pm 1.3$ & $ 173.7\,^{+1.3}_{-1.2}$ & $ 173.2\,^{+1.3}_{-1.4}$ \\[2pt] \hline 
\hline
$\mu_F=\mu_R=E_T/2$ & \\
$\chi^2/N_{\mathrm dof}$ & 71/68 & 70/68 & 66/68 & 64/68 & 68/68 & 64/68 \\
\mtpole\,[\GeV] & $ 174.7\pm 1.4$ & $ 174.5\,^{+1.5}_{-1.4}$ & $ 174.3\,^{+1.5}_{-1.4}$ & $ 173.6\,^{+1.3}_{-1.2}$ & $ 173.4\,^{+1.2}_{-1.1}$ & $ 174.0\,^{+1.5}_{-1.4}$ \\[2pt] \hline 
\end{tabular}

\caption{\label{t:massfob}
Measurements of the top quark mass from combined fits to all eight 
lepton and dilepton distributions,
using fixed-order predictions from MCFM with the CT14, MMHT, NNPDF 3.0, 
HERAPDF 2.0, ABM 11 and NNPDF 3.0\_nojet PDF sets, and various choices for the
central QCD factorisation and  renormalisation scales $\mu_F$ and $\mu_R$. 
The upper section of the table gives the results for $\mu_F=\mu_R=\mtop/2$, 
showing the $\chi^2$
values at the best-fit mass for each PDF set, the fitted mass
with its total uncertainty, and the breakdown of individual uncertainty 
contributions
from data statistics, experimental systematics, and uncertainties in the
predictions from PDF and QCD scale effects. Uncertainties given as `0.0' are
smaller than 0.05\,\GeV.
The lower parts of the table
give the $\chi^2$ values, fitted mass and total uncertainty for alternative
scale choices of $\mu_F=\mu_R=H_T/4$ and $E_T/2$.}
\end{table}

\begin{figure}[tp]

\hspace{-10mm}\includegraphics[width=165mm]{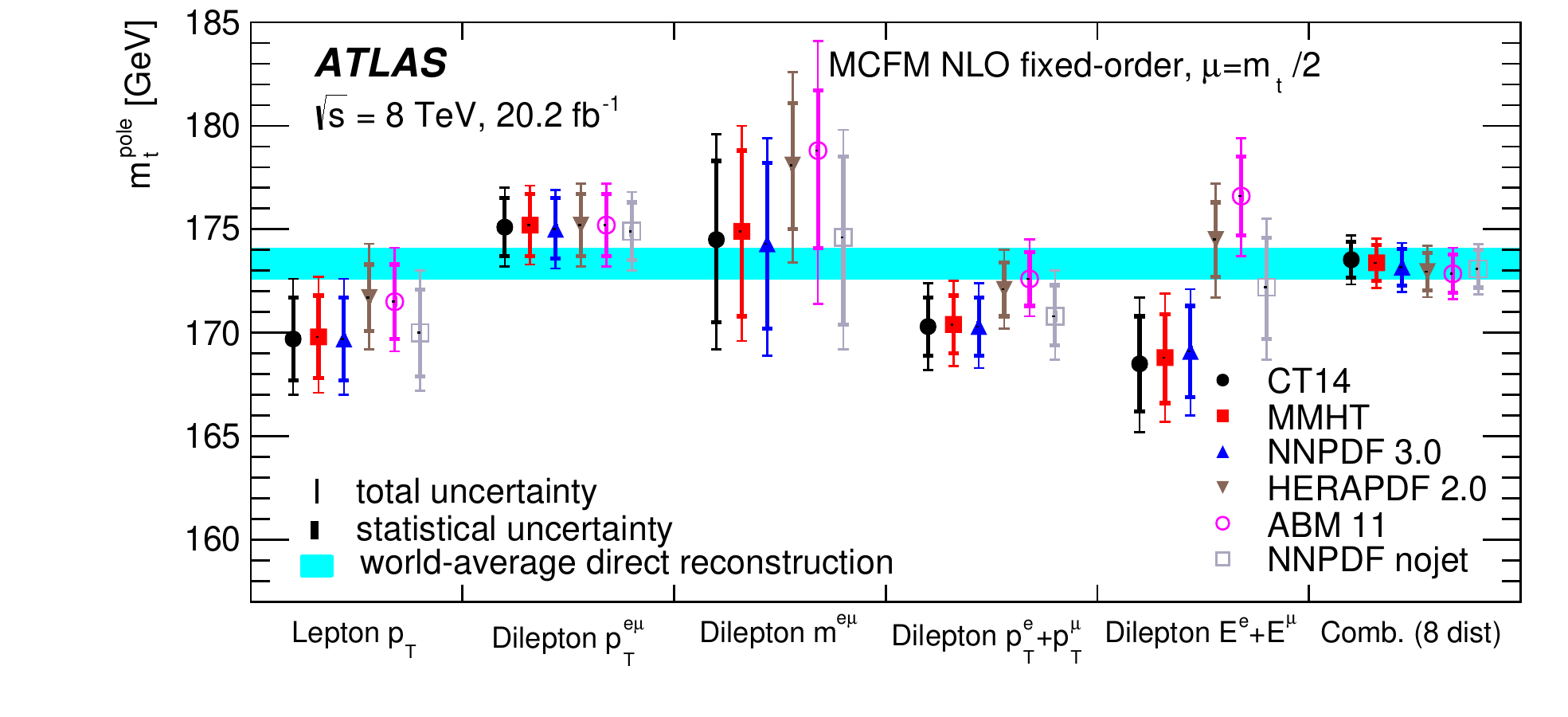}

\caption{\label{f:massfoa}Measurements of the top quark mass using predictions
derived from MCFM with the CT14, MMHT, NNPDF 3.0, HERAPDF 2.0, ABM 11 and
NNPDF 3.0\_nojet PDF sets. The central factorisation and renormalisation
scales are set to $\mu_F=\mu_R=\mtop/2$. The results
from fitting templates of the single lepton \ptl\ and dilepton \ptll, \mll, \ptsum\ and \esum\ distributions one at a time, and of a combined fit to
these five distributions plus the \etal, \rapll\ and \dphill\ distributions
together, are shown. For comparison, the world-average of mass 
measurements from reconstruction of the top quark decay products 
and its uncertainty \cite{masswa} is shown by the cyan band.}
\end{figure}

The MCFM fixed-order results for individual distributions shown in 
Table~\ref{t:massfoa} and
Figure~\ref{f:massfoa} show some similar patterns to those from the 
{\sc Powheg\,+\,Pythia6}-based template fits shown in Table~\ref{t:massmca}
and Figure~\ref{f:massmca}. The results from \ptl\ and \ptsum\ are close, 
the largest \mtop\ values come from \ptll, the smallest from \esum\ and the
least precise determination is obtained from \mll.
The envelope of the central values 
is similar (6\,\GeV), but all values are shifted up by a few \GeV\ compared
to the corresponding {\sc Powheg\,+\,Pythia6}-based 
template fit results for the same distribution. 
The $\chi^2$ values are reasonable, indicating a satisfactory description
of the data by the predictions at the best-fit \mtop\ values. The various
distributions show different relative sensitivities to the PDF and QCD
scale uncertainties. 

As shown in Table~\ref{t:massfob}, 
the combination of all eight measured distributions (including \etal, \rapll\
and \dphill\ which are not sensitive to \mtop)  significantly reduces the
theoretical uncertainties due to both PDF and QCD scale effects. 
The $\chi^2$
values for the combined description of all eight distributions are reasonable
for all PDFs, implying that there is no significant tension between the mass fit
results from the individual distributions, once the correlations between
the distributions are taken into account. Several additional tests
using the predictions based on NNPDF 3.0
were performed to probe the compatibility of the top quark mass values
extracted from the different distributions, and the accuracy of the
physics modelling used to perform the extraction.
The combined fit was repeated removing
one distribution at a time. The largest shift of $-1.4\pm 1.1$\,\GeV\ 
was observed when removing the \ptll\ distribution,
where the uncertainty corresponds to the quadrature difference of the fit
uncertainties with and without the \ptll\ distribution included. The removal
of any other single distribution changed the result by less than 0.3\,GeV,
and a fit to only the five distributions directly sensitive to \mtop\
(excluding \etal, \rapll\ and \dphill) gave a result of
$173.1\pm 1.2$\,\GeV, corresponding to a shift of $-0.1$\,\GeV\ with 
respect to the eight-distribution result.
Finally, the individual measurements from the five directly-sensitive 
distributions were combined using the {\sc HAverager} program 
\cite{heraverager,heradisav}. Correlated statistical and 
systematic uncertainties were taken into account using nuisance parameters, 
but post-fit correlations between these nuisance parameters were neglected, 
unlike in the simultaneous fit approach with
{\sc xFitter}. The average of the five measurements is
$173.4\pm 1.6$\,\GeV\ with a $\chi^2$ of $6.4/4$, 
in reasonable agreement with the result from the simultaneous fit of the 
five distributions. No additional uncertainty was included as a result of 
these tests.

The combined-fit $\chi^2$ values in Table~\ref{t:massfob} are smallest
for the HERAPDF 2.0 and NNPDF~3.0\_nojet PDF sets, which do not include the
constraints on the gluon PDF from LHC and Tevatron jet data in the region 
relevant for \ttbar\ production. However, the \mtop\ values resulting from
the NNPDF~3.0 and NNPDF~3.0\_nojet PDFs are close, 
indicating that the results are not sensitive to whether 
the jet data are included or not. Amongst the `global fit' PDF sets incorporating
a larger set of experimental data, the smallest 
$\chi^2$ values result from the fit with NNPDF 3.0, though the values from
the other PDFs are also reasonable.
The results using NNPDF 3.0 were therefore used to define
the central \mtop\ value from the combined fit to all eight distributions,
and an additional uncertainty of 0.3\,\GeV, corresponding to half
the difference of the envelope encompassing all the other PDFs, 
was added in quadrature to the PDF uncertainty from NNPDF 3.0 alone. 
The effect of the uncertainty in the value of \alphas\ was found to be 
0.01\,\GeV. The residual dependence of the measured differential
cross-sections on the top quark mass assumed in the simulation 
(see Section~\ref{ss:ttsyst}) is very small. A $\pm 5$\,\GeV\ variation
around the baseline value of $\mtop=172.5$\,\GeV\
was assumed, giving a 0.1\,\GeV\ change on the result of the
combined fit.

The choice of a fixed central scale, $\mu_F=\mu_R=\mtop/2$ is expected to 
provide a good description of the inclusive \ttbar\ cross-section and 
differential distributions in the kinematic regions dominated by top quarks
with relatively low \pt. However, dynamical scales, which vary as a function
of the top quark kinematics, are expected to be more appropriate for modelling
the regions with high \pt \cite{dyntopscale}.
Two alternative dynamical central scale choices 
for the \ttbar\ production process were 
explored to test the sensitivity of the results to this choice:
\begin{itemize}
\item $\mu_F=\mu_R=H_T/4$ where $H_T$ is defined as 
$\sqrt{\mtop^2+\pt(t)^2}+\sqrt{\mtop^2+\pt(\bar{t})^2}$ and $\pt(t)$ and 
$\pt(\bar{t})$ are the transverse momentum of the top quark and antiquark,
corresponding to one of the dynamical scales suggested in 
Ref. \cite{dyntopscale}.
\item $\mu_F=\mu_R=E_T/2$ where $E_T$ is defined as $\sqrt{\mtop^2+\pt(\ttbar)^2}$ and $\pt(\ttbar)$ is the \pt\ of the \ttbar\ system, analogously to a
scale $\sqrt{m_W^2+\pt(W)^2}$ used in the description of jet production in 
association with $W$ bosons \cite{vjscale,cdfwjets}.
\end{itemize}
In both cases, the central scale for the top quark decay process
$t\rightarrow b\ell\nu+X$ was fixed at $\mtop/2$. The corresponding
predictions for the top quark \pt\ spectrum from MCFM with NNPDF 3.0 and 
these scale choices are shown in Figure~\ref{f:mcfmtoppt}, 
and compared to the ATLAS \sxvt\ measurement
using \ttbar\ events with a lepton and at least four jets \cite{TOPQ-2015-06}.
Unlike the predictions of {\sc Powheg\,+\,Pythia6} used in 
Section~\ref{ss:massresmc}, the MCFM predictions with central scale choices
of $\mu_F=\mu_R=\mtop/2$, $H_T/4$ and $E_T/2$ provide good descriptions of 
the measured top quark \pt\ spectrum, whereas $\mu_F=\mu_R=\mtop$ is too hard.

\begin{figure}[tp]
\centering

\includegraphics[width=90mm]{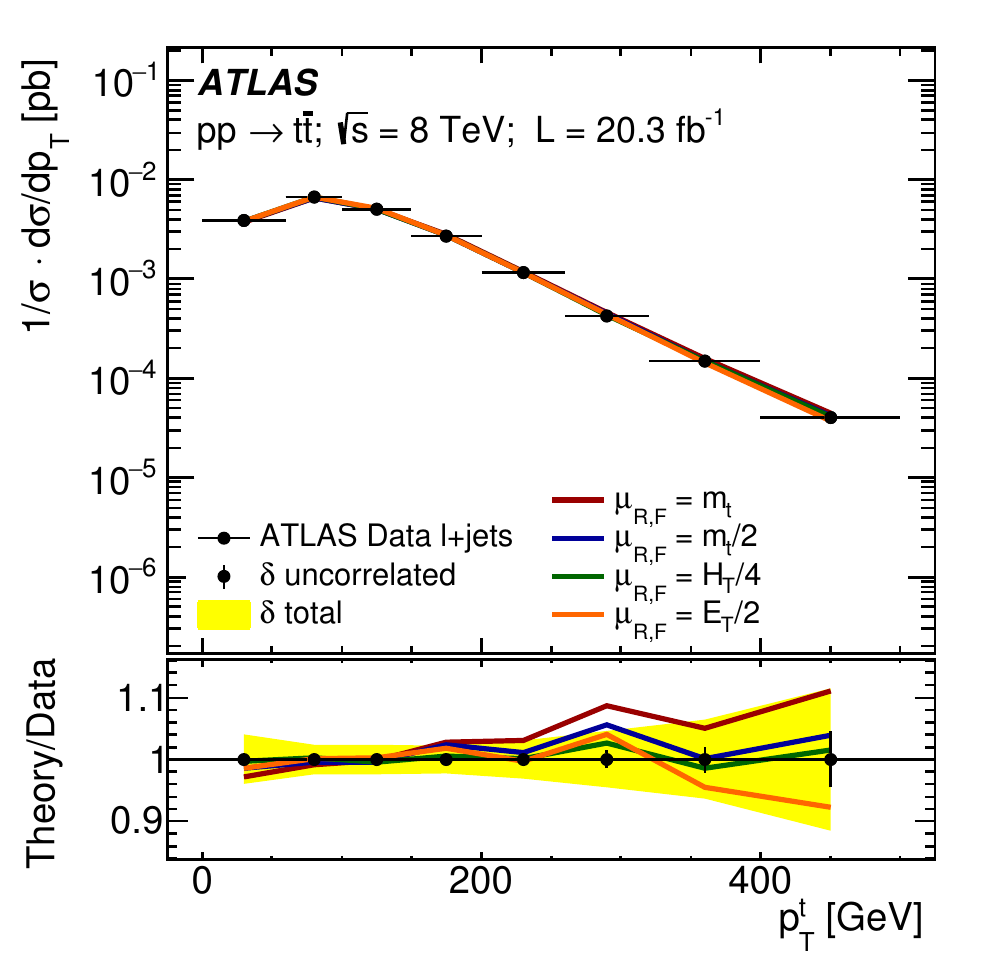}

\caption{\label{f:mcfmtoppt}Measurement of the top quark \pt\ spectrum
in $pp$ collisions at \sxvt\ from ATLAS events with a lepton and at least
four jets \cite{TOPQ-2015-06}, compared to the predictions from MCFM as used
in this analysis with NNPDF~3.0, $\mtop=173.3$\,\GeV, 
and QCD scale choices of $\mu_F=\mu_R=\mtop/2$,
 $H_T/4$ and $E_T/2$, as well as with $\mu_F=\mu_R=\mtop$. 
The measurement uncertainties
are represented by the yellow band, with the uncorrelated component 
shown by the black error bar. The lower plots show the ratios of the
different predictions to the data.}
\end{figure}

The results from the combined fit to all eight distributions with these
scale choices and all six PDF sets are shown in the lower part of 
Table~\ref{t:massfob}, and displayed graphically in Figure~\ref{f:massfob}.
In the same way as for the fixed central scale, the actual factorisation and 
normalisation scales used in the predictions
were allowed to vary independently
around the dynamical central scales, with one standard
deviation variations corresponding to factors of two and one-half.
The $\chi^2$ values for the fits with a central scale of $H_T/4$ are all
improved compared to those for $\mtop/2$, reflecting a generally better
description of the high-\pt\ tails of the distributions. The $\chi^2$ values
from the $E_T/2$ fits lie between the other two choices. The largest difference
in the \mtop\ values from a dynamical scale and the fixed scale with any
PDF (1.1\,GeV for $E_T/2$ vs. $\mtop/2$ with the CT14 PDF) was used to define
an additional theoretical uncertainty due to the choice of the 
functional form of the QCD scales.

\begin{figure}[tp]

\hspace{-10mm}\includegraphics[width=165mm]{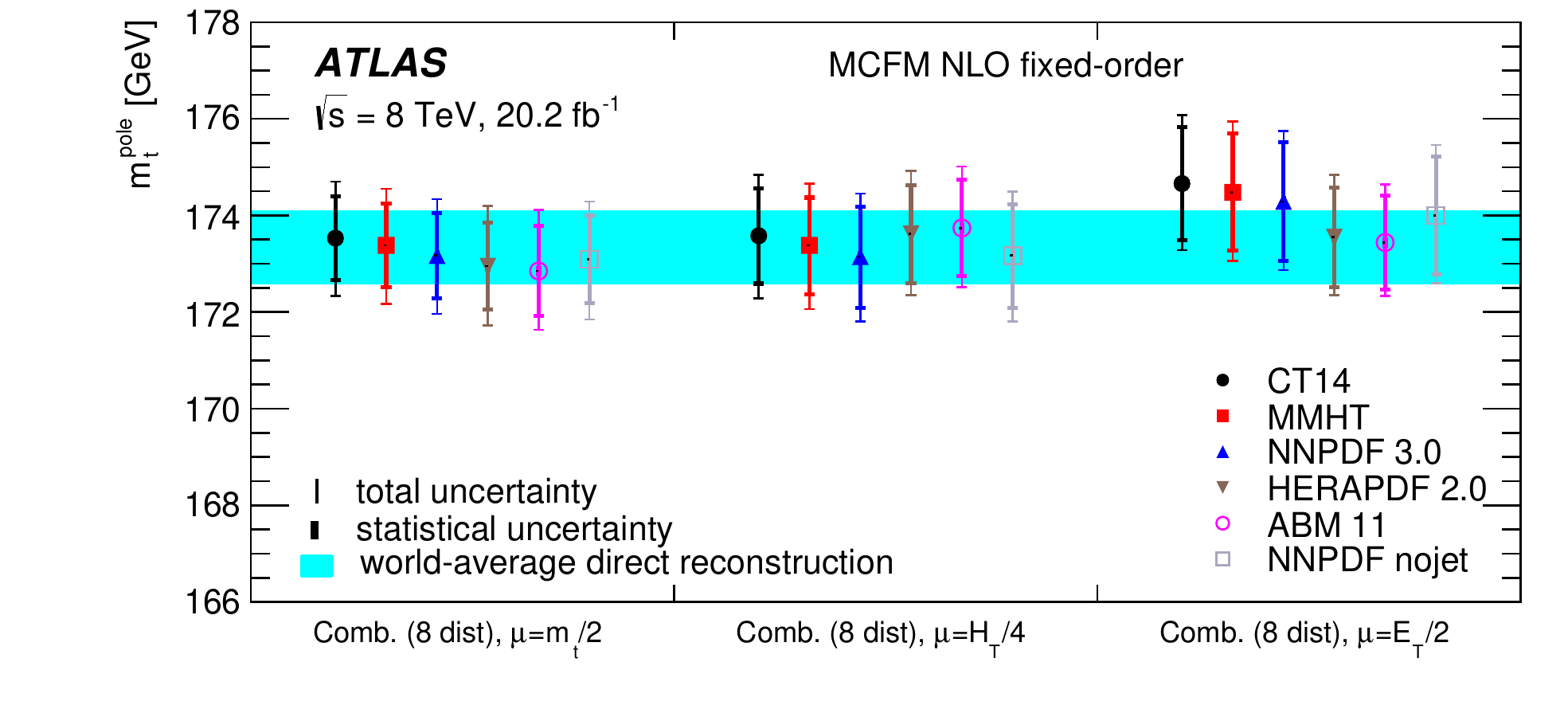}

\caption{\label{f:massfob}Measurements of the top quark mass using predictions
derived from MCFM with the CT14, MMHT, NNPDF 3.0, HERAPDF 2.0, ABM 11 and
NNPDF 3.0\_nojet PDF sets, and the central QCD factorisation and renormalisation
scales $\mu_F$ and $\mu_R$ set to $\mtop/2$, $H_T/4$ and $E_T/2$. 
The results are derived from a combined fit to all eight lepton and
dilepton distributions. For comparison, the world-average of mass 
measurements from reconstruction of the top quark decay products \cite{masswa}
is shown by the cyan band.}
\end{figure}

The final top quark mass value
from the combination of all distributions is:
\[
\mtpole=\mtfoval\pm\mtfostat\pm\mtfoexp\pm\mtfoth\,\GeV,
\]
where the three uncertainties arise from data statistics, experimental 
systematic effects, and uncertainties in the theoretical predictions,
giving a total uncertainty of $\mtfotot$\,\GeV. The theoretical uncertainty is
dominated by the comparison of results with different QCD central scale choices.
Figure~\ref{f:mtoppole} shows a comparison with previous determinations of the 
top quark pole mass from the inclusive \ttbar\ production cross-section
\cite{d0mtpoleincl,TOPQ-2013-04,CMS-TOP-13-004}
and from the invariant mass distribution of the \ttbar\ plus
one jet system \cite{TOPQ-2014-06}. The present
result is in agreement with these other results, 
all of which have larger uncertainties. It is also in
agreement with the Tevatron and LHC average measurement of 
$173.34\pm 0.76$\,\GeV\ from reconstruction
of the top quark decay products \cite{masswa}, as well as with more precise
recent results using similar techniques 
\cite{tevnewmtop,TOPQ-2016-03,CMS-TOP-14-022}.
However, the precision of the present pole mass result
is not sufficient to probe potential differences between it and the other
techniques at the 1\,\GeV\ level.

\begin{figure}[tp]
\centering

\includegraphics[width=110mm]{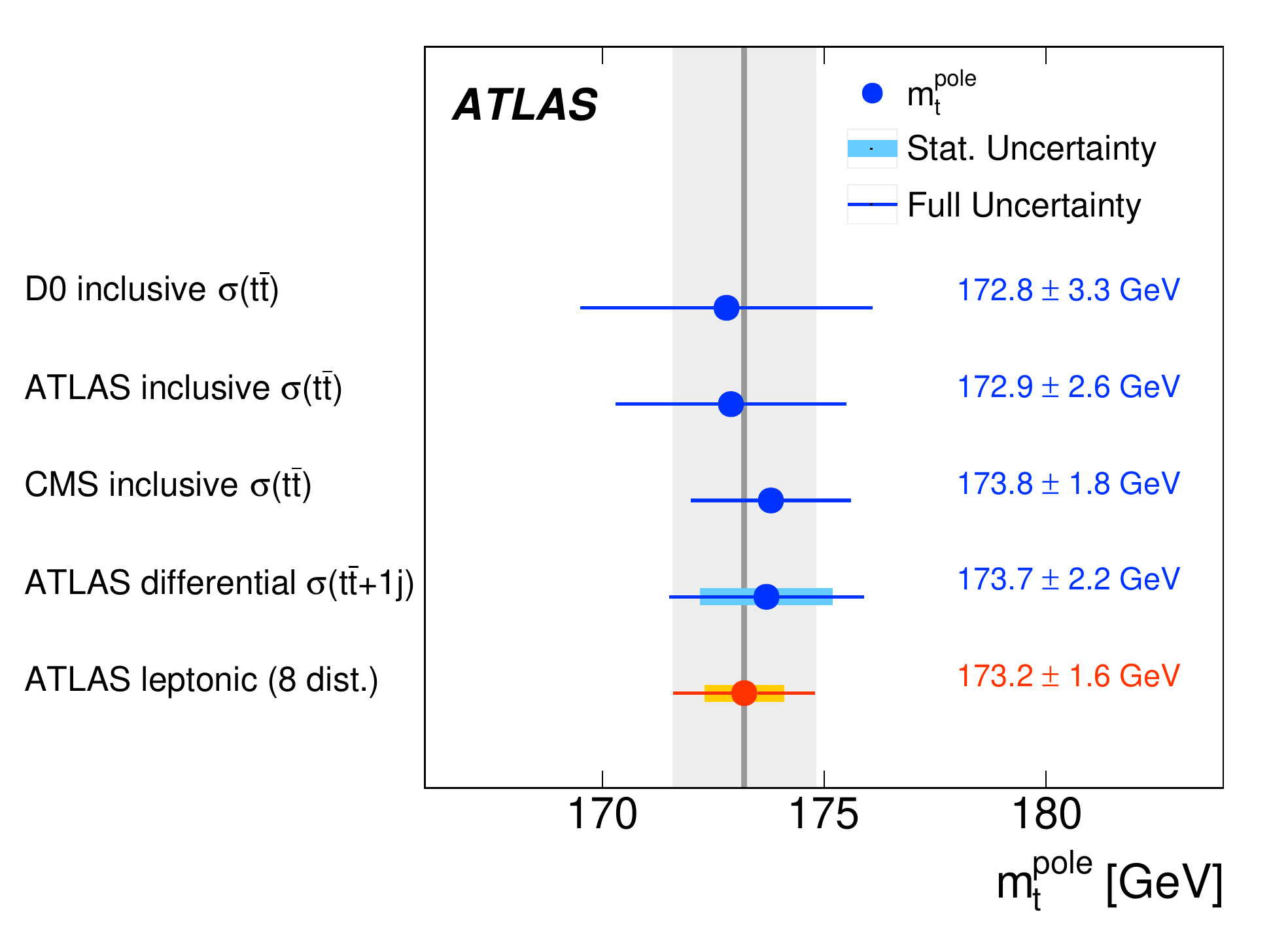}
\caption{\label{f:mtoppole}Result of the top quark pole mass determination
from the combined fit to eight leptonic distributions (shown by the red point and grey band), 
compared to  other determinations from inclusive and differential cross-section
measurements in \ttbar\ events
\cite{d0mtpoleincl,TOPQ-2013-04,CMS-TOP-13-004,TOPQ-2014-06}. The
statistical uncertainties are shown separately by the thick error bars where
available.}
\end{figure}

The theoretical uncertainty of \mtfoth\,\GeV\ on the final result using
fixed-order predictions
is significantly smaller than the uncertainties due to \ttbar\ modelling 
and potential NNLO effects in the top quark \pt\ spectrum for the fits
based on {\sc Powheg\,+\,Pythia6} templates. In the fixed-order approach,
the potential missing NNLO corrections are absorbed into the variations of
the QCD scales $\mu_F$ and $\mu_R$, which are significantly constrained by
the fit to the complete set of distributions, including those with little
sensitivity to \mtop. However, there remains a significant uncertainty
of about 1\,\GeV\ due to the choice of the functional form of the QCD scales,
limiting the gain from the combined fit.
This approach would therefore benefit significantly 
from the availability of fixed-order calculations including NNLO effects
in the top quark production and decay \cite{topdecaynnlo},
which should reduce the uncertainties due to scale choices. 
Off-shell and interference effects in the 
$pp\rightarrow WW\bbbar\rightarrow e\mu\nu\bar{\nu}\bbbar+X$
process (including both \ttbar\ and single top $Wt$ contributions)
\cite{oshell1,oshell2,oshell3,oshell4,oshell5,oshell6,oshell7}, as
well as NLO electroweak corrections \cite{weak1,weak2},
were not considered in this analysis. They are expected to be small 
compared to the theoretical uncertainties of the current result, but
likely cannot be neglected in a determination of \mtop\ based on 
NNLO QCD predictions.
These theoretical advances would allow the power of the
full set of distributions to be utilised more effectively, especially
in view of the likely reduction in the experimental statistical and 
systematic uncertainties from the larger \ttbar\ samples now becoming
available from LHC running at \sxyt.


\FloatBarrier
\section{Conclusions} \label{s:conc}

Lepton and dilepton differential cross-section distributions have been measured
in $\ttbar\rightarrow e\mu\nu\bar{\nu}\bbbar$ events selected from
\intlumi\,\ifb of $pp$ collisions at \sxvt\ recorded by the ATLAS detector
at the LHC. The absolute and normalised 
cross-sections were measured using opposite-charge $e\mu$ 
events with one or two $b$-tagged jets, and corrected to a fiducial volume
corresponding to the experimental acceptance of the leptons and no requirements
on jets. Eight single lepton and dilepton differential distributions
were measured, with relative uncertainties varying in the range 1--10\,\%,
and presented with and without the contribution from leptonic decays of
$\tau$-leptons produced in the $W$ decays.

The results were compared to the predictions of various \ttbar\ NLO and 
LO multileg
matrix element event generators interfaced to several parton shower
and hadronisation models. These generally give a good description of the
distributions, though some distributions are modelled poorly by certain event
generators. Those involving rapidity information are better described
by the HERAPDF PDF sets than the CT10 set used as default. The distributions
also show some
sensitivity to NNLO corrections in the description of the top quark \pt\ 
spectrum. The data are sensitive to the gluon PDF around $x\approx 0.1$
and have the potential to reduce PDF uncertainties in this region.

Several of the measured distributions are sensitive to the top quark mass, in
a way which is complementary to traditional measurements of \mtop\ using the
invariant mass of the reconstructed top quark decay products. Various
techniques for extracting the top quark mass from the measured distributions
were explored, including fits using templates from {\sc Powheg\,+\,Pythia6} 
simulated samples, mass determinations based on moments of the distributions,
and fits to fixed-order NLO QCD predictions, giving access to the
top quark pole mass in a well-defined renormalisation scheme as implemented
in MCFM.
The most precise result was obtained from a fit of fixed-order predictions 
to all eight measured distributions simultaneously,
extracting \mtpole\ whilst simultaneously constraining uncertainties due to 
PDFs and QCD scales. The final result is:
\begin{eqnarray}
\mtpole=\mtfoval\pm\mtfostat\pm\mtfoexp\pm\mtfoth\,\GeV,\nonumber
\end{eqnarray}
where the three uncertainties arise from data statistics, experimental 
systematic effects, and uncertainties in the theoretical predictions. This 
result is in excellent agreement with other determinations of \mtpole\
from inclusive and differential cross-sections, and
traditional measurements based on reconstruction of the top quark 
decay products.

\section*{Acknowledgements}

We thank CERN for the very successful operation of the LHC, as well as the
support staff from our institutions without whom ATLAS could not be
operated efficiently.

We acknowledge the support of ANPCyT, Argentina; YerPhI, Armenia; ARC, Australia; BMWFW and FWF, Austria; ANAS, Azerbaijan; SSTC, Belarus; CNPq and FAPESP, Brazil; NSERC, NRC and CFI, Canada; CERN; CONICYT, Chile; CAS, MOST and NSFC, China; COLCIENCIAS, Colombia; MSMT CR, MPO CR and VSC CR, Czech Republic; DNRF and DNSRC, Denmark; IN2P3-CNRS, CEA-DSM/IRFU, France; SRNSF, Georgia; BMBF, HGF, and MPG, Germany; GSRT, Greece; RGC, Hong Kong SAR, China; ISF, I-CORE and Benoziyo Center, Israel; INFN, Italy; MEXT and JSPS, Japan; CNRST, Morocco; NWO, Netherlands; RCN, Norway; MNiSW and NCN, Poland; FCT, Portugal; MNE/IFA, Romania; MES of Russia and NRC KI, Russian Federation; JINR; MESTD, Serbia; MSSR, Slovakia; ARRS and MIZ\v{S}, Slovenia; DST/NRF, South Africa; MINECO, Spain; SRC and Wallenberg Foundation, Sweden; SERI, SNSF and Cantons of Bern and Geneva, Switzerland; MOST, Taiwan; TAEK, Turkey; STFC, United Kingdom; DOE and NSF, United States of America. In addition, individual groups and members have received support from BCKDF, the Canada Council, CANARIE, CRC, Compute Canada, FQRNT, and the Ontario Innovation Trust, Canada; EPLANET, ERC, ERDF, FP7, Horizon 2020 and Marie Sk{\l}odowska-Curie Actions, European Union; Investissements d'Avenir Labex and Idex, ANR, R{\'e}gion Auvergne and Fondation Partager le Savoir, France; DFG and AvH Foundation, Germany; Herakleitos, Thales and Aristeia programmes co-financed by EU-ESF and the Greek NSRF; BSF, GIF and Minerva, Israel; BRF, Norway; CERCA Programme Generalitat de Catalunya, Generalitat Valenciana, Spain; the Royal Society and Leverhulme Trust, United Kingdom.

The crucial computing support from all WLCG partners is acknowledged gratefully, in particular from CERN, the ATLAS Tier-1 facilities at TRIUMF (Canada), NDGF (Denmark, Norway, Sweden), CC-IN2P3 (France), KIT/GridKA (Germany), INFN-CNAF (Italy), NL-T1 (Netherlands), PIC (Spain), ASGC (Taiwan), RAL (UK) and BNL (USA), the Tier-2 facilities worldwide and large non-WLCG resource providers. Major contributors of computing resources are listed in Ref.~\cite{ATL-GEN-PUB-2016-002}.


\printbibliography


\newpage 
\begin{flushleft}
{\Large The ATLAS Collaboration}

\bigskip

M.~Aaboud$^\textrm{\scriptsize 137d}$,
G.~Aad$^\textrm{\scriptsize 88}$,
B.~Abbott$^\textrm{\scriptsize 115}$,
O.~Abdinov$^\textrm{\scriptsize 12}$$^{,*}$,
B.~Abeloos$^\textrm{\scriptsize 119}$,
S.H.~Abidi$^\textrm{\scriptsize 161}$,
O.S.~AbouZeid$^\textrm{\scriptsize 139}$,
N.L.~Abraham$^\textrm{\scriptsize 151}$,
H.~Abramowicz$^\textrm{\scriptsize 155}$,
H.~Abreu$^\textrm{\scriptsize 154}$
% ...
\bigskip
\\
% ...
$^{12}$ Institute of Physics, Azerbaijan Academy of Sciences, Baku, Azerbaijan\\
% ...
$^{88}$ CPPM, Aix-Marseille Universit{\'e} and CNRS/IN2P3, Marseille, France\\
% ...
$^{115}$ Homer L. Dodge Department of Physics and Astronomy, University of Oklahoma, Norman OK, United States of America\\
% ...
$^{119}$ LAL, Univ. Paris-Sud, CNRS/IN2P3, Universit{\'e} Paris-Saclay, Orsay, France\\
% ...
$^{137}$ $^{(a)}$ Facult{\'e} des Sciences Ain Chock, R{\'e}seau Universitaire de Physique des Hautes Energies - Universit{\'e} Hassan II, Casablanca; $^{(b)}$ Centre National de l'Energie des Sciences Techniques Nucleaires, Rabat; $^{(c)}$ Facult{\'e} des Sciences Semlalia, Universit{\'e} Cadi Ayyad, LPHEA-Marrakech; $^{(d)}$ Facult{\'e} des Sciences, Universit{\'e} Mohamed Premier and LPTPM, Oujda; $^{(e)}$ Facult{\'e} des sciences, Universit{\'e} Mohammed V, Rabat, Morocco\\
% ...
$^{139}$ Santa Cruz Institute for Particle Physics, University of California Santa Cruz, Santa Cruz CA, United States of America\\
% ...
$^{151}$ Department of Physics and Astronomy, University of Sussex, Brighton, United Kingdom\\
% ...
$^{154}$ Department of Physics, Technion: Israel Institute of Technology, Haifa, Israel\\
$^{155}$ Raymond and Beverly Sackler School of Physics and Astronomy, Tel Aviv University, Tel Aviv, Israel\\
% ...
$^{161}$ Department of Physics, University of Toronto, Toronto ON, Canada\\
% ...
$^{a}$ Also at Department of Physics, King's College London, London, United Kingdom\\
$^{b}$ Also at Institute of Physics, Azerbaijan Academy of Sciences, Baku, Azerbaijan\\
$^{c}$ Also at Novosibirsk State University, Novosibirsk, Russia\\
$^{d}$ Also at TRIUMF, Vancouver BC, Canada\\
$^{e}$ Also at Department of Physics {\&} Astronomy, University of Louisville, Louisville, KY, United States of America\\
% ...
$^{*}$ Deceased
\end{flushleft}


\end{document}